\documentclass[12pt]{cit_thesis}
\preprint
\input epsf.tex
\input amssymb.sty 

\def\vslash{v\!\!\!\slash}
\def\Slash#1#2{{\rlap{\hspace{#2em}/}#1}}

\newcommand{\nn}{\nonumber}
\newcommand{\tensor}{\stackrel{\!\!\!\leftarrow\!\!\!\rightarrow}{\mbox{\normalsize $\nabla^2$}}}

\author{Iain W. Stewart}
\university{\centerline{\epsfbox{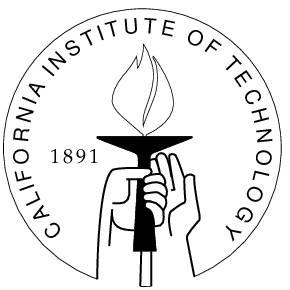}}
California Institute of Technology}
\address{Pasadena, California}
\copyyear{\the\year}
\title{Applications of Chiral Perturbation Theory \\
	in Reactions with Heavy Particles}

\begin{document}

\maketitle

\begin{frontmatter}
\setcounter{page}{2}

\begin{acknowledgements}

I would like to thank the people on the fourth floor of Lauritsen, S. Frautschi, D. 
Politzer, J. Preskill, J. Schwarz, M. Wise, M. Aganagic, M. Becker, K. Becker, O. 
Bergman,  S.  Cherkis, H. Davoudiasl, E. Gimon, W. Goldberger, M.  Gremm, P. 
Horava, E.  Keski-Vakkuri, P. Kraus, A.  Kapustin,  A. Leibovich, Z. Ligeti, D. Lowe,
T. Mehen, S.  Ouellette, C. Popescu, K.  Rajagopal, K.  Scaldeferri, H. Tuck, and E. 
Westphal for providing a friendly atmosphere in which to work.  I am particularly
grateful to my advisor Mark Wise and to people with whom I collaborated during
my stay at Caltech: Martin Gremm, Mark Wise, Adam Leibovich, Zoltan Ligeti, Tom
Mehen, and Sean Fleming. Special thanks go to my wife Susan and parents for
their constant love and support.  This work was supported in part by the
Department of Energy under grant number DE-FG03-92-ER 40701.

\end{acknowledgements}

\begin{abstract}

Effective field theory techniques are used to describe the interaction of heavy
hadrons in a model independent way.  Predictability is obtained by exploiting the
symmetries of QCD.  Heavy hadron chiral perturbation theory is reviewed and
used to describe $D^*$ decays.  The phenomenologically important $D^*D\pi$
coupling is extracted from data working to first order in the chiral and heavy
quark symmetry breaking parameters.  A method is described for determining
$|V_{ub}|$ from exclusive semileptonic $B$ and $D$ decays with $10\%$
uncertainty.  An effective field theory for two-nucleon systems is then
discussed.  The large S-wave scattering lengths necessitate expanding around a
non-trivial fixed point.   A detailed discussion of the interplay between
renormalization and the power counting is given.  In power counting pion
interactions with nucleons it is useful to consider three classes of pion:
potential, radiation, and soft.  A power counting for massive radiation is
developed.  Finally, it is shown that the leading terms in the effective theory for
nucleon-nucleon interactions are invariant under Wigner's SU(4) spin-isospin
symmetry in the infinite scattering length limit.

\end{abstract}
  
\tableofcontents

\listoffigures
\listoftables

\end{frontmatter}


\chapter{Introduction}

The minimal standard model is an appealing theory which describes the strong,
weak, and electromagnetic interactions in terms of 19 input parameters.  This
renormalizable quantum field theory gives quite an accurate description of nature,
as shown through precision tests of QED \cite{Yennie,Kinoshita}, the Electro-Weak
sector \cite{Martinez,Kim}, and to a lesser extent QCD
\cite{deBoer,Graf,FusterVerdu}.  These tests examine observables for which a
perturbative treatment of the couplings is applicable.  For QCD, this is a valid
approach for high energy processes due to asymptotic freedom
\cite{Politzer,Gross}.  However, at low energy or large distance the coupling
becomes strong and the quarks and gluons are confined into the observed mesons
and baryons.  At these energies a non-perturbative approach, such as lattice
gauge theory, is necessary.  Since lattice calculations are still fairly crude, it is
reasonable to ask if the non-perturbative nature of QCD can be handled in another
model independent fashion.  In certain situations the answer is yes, because the
symmetries and dynamics of QCD provide other expansion parameters besides the
strong coupling.  Expanding about a symmetry limit provides us with a means for
describing non-perturbative effects by a series of low energy parameters (matrix
elements or effective couplings) which can be determined from experimental data. 
This approach is predicative since there are typically several observables that
depend on a given parameter.  In many ways this is complimentary to lattice QCD
calculations, which can then concentrate on calculating these parameters.

Effective field theory is a useful tool for implementing these ideas.  We begin by
writing down fields for the relevant degrees of freedom, and constructing an
effective Lagrangian.  The Lagrangian includes all possible terms that transform
correctly under the symmetries, and is typically non-renormalizable with an
infinite number of terms.  These terms are organized in importance by power
counting in a small parameter.  Identifying a small expansion parameter usually
depends on having scales which are widely separated.  The low energy
Lagrangian is a sum of terms of the form ${\cal L}\sim C(\mu) {\cal O}(\mu)$. The
coefficients $C(\mu)$ and operators ${\cal O}(\mu)$ encode short and long
distance physics respectively, and the renormalization point $\mu$ separates the
two regimes. This is essentially a Wilsonian operator product expansion
\cite{Wilson}.   The effective field theory approach is important for several
reasons. In an effective field theory different scales in the problem are
separated, so that one can concentrate on the most interesting physics at a
particular scale.  Furthermore, the power counting gives us a way to estimate the
uncertainty in working at a given order.  Finally, calculations are often much
simpler in the effective theory.   Depending on the situation, effective field
theories are used in two somewhat distinct ways, either from the \emph{top
down} or from the \emph{bottom up}.  

In a top down approach the high energy theory is understood, but we find it
useful or necessary to use a simpler theory at lower energies.   Since the high
energy theory is known, the $C(\mu)$ couplings can be calculated by performing
a perturbative matching at the high scale.  The theory is then run down to the
desired low energy scale using the renormalization group.   Solving the
renormalization group equations for the running of the coefficients, we sum
potentially large logarithms between the two scales. At the low scale, matrix
elements of the operators are natural in size.  A standard example is the
calculation of QCD corrections to weak processes at momenta $p \ll 90\,{\rm
GeV}$ (see Ref.~\cite{buchalla} for a review).  Here integrating out the $W$ and
$Z$ leaves four-fermion interactions and an expansion in $p^2/m_W^2$.  A
second example is non-relativistic QED (NRQED) \cite{Caswell}, which is used in
describing the electromagnetic interactions of non-relativistic leptons.  NRQED
is especially useful in describing Coulombic bound states such as positronium,
where a pure coupling constant expansion is inappropriate.  Instead, a dual
expansion is performed in the electromagnetic coupling and the velocity of the
non-relativistic leptons.  This effective theory is especially tractable since both
the coefficients and matrix elements can be calculated.   In QCD the quark
masses are such that $\Lambda_{\rm QCD} \ll m_{c,b}=m_Q $.  In the limit
$m_Q\to \infty$, QCD exhibits additional flavor and spin symmetries, called heavy
quark symmetry (HQS) \cite{HQS1,HQS2}.  Heavy Quark Effective Theory (HQET)
uses these symmetries and an expansion in $\Lambda_{\rm QCD}/m_Q$ to make
predictions for processes involving hadrons containing one heavy quark.  At high
energies this effective field theory is matched onto QCD.  In this case the matrix
elements are typically not calculable, but are still related by HQS.  For systems 
with two heavy quarks, the appropriate effective theory is called non-relativistic
QCD (NRQCD) \cite{Bodwin}.

A second approach to effective field theory is from the bottom up.  In this case the
high energy theory is either unknown or not calculable.  A well known example is
SU(3) chiral perturbation theory, which exploits the pattern of dynamical chiral
symmetry breaking observed in QCD.  In the limit $m_q\to 0$, QCD has additional
chiral symmetries, giving enhanced predictive power.  Approximate chiral symmetry
is a result of the small light quark masses, $m_q=m_{u,d,s} \ll \Lambda_{\rm
QCD}$.  Phenomenologically, this approach is valid for energy and momenta $\ll
\Lambda$ where $\Lambda\sim 1.2\,{\rm GeV}$ is the chiral symmetry breaking
scale.  The relevant degrees of freedom here are the pions, kaons, and eta which
are the pseudo-Goldstone bosons of the $SU(3)_L\times SU(3)_R\to SU(3)_V$
breaking.  At low energy, matching onto QCD is not possible, so the couplings
$C(\mu)$ in this low energy theory must be determined from experimental data. 
However, because our fields correspond to the asymptotically observed particles,
the matrix elements are calculable.  Processes involving a single heavy hadron can
also be incorporated in this approach by combining the power counting in HQET
and chiral perturbation theory into heavy hadron chiral perturbation theory
\cite{ChptW,Burdman,yan,jm}.  The effective field theory approach has also been
extended to processes with two or more heavy particles, such as nucleon-nucleon
interactions \cite{W1,W2,vanKolck,Lepage,ksw1,ksw2,ksw3}.  The latter theory will
be discussed in some detail.

In the modern view, the standard model itself is a low energy effective theory.  As
an effective field theory it includes the usual Lagrangian as well as operators of
dimension five and higher built out of standard model fields.  Such operators are
suppressed by powers of a scale $\Lambda$, where $\Lambda$ is a measure of
the energy at which the new physics becomes relevant.  At energies $\sim
\Lambda$ the standard model effective field theory must be replaced by
something more fundamental.  The fact that the standard model is renormalizable
is significant since it implies that the scale $\Lambda$ is not generated by
standard model interactions, and is therefore, in principal, unconstrained.   Large
values of $\Lambda$ then explain several of the beautiful features of the
standard model, such as baryon and lepton number conservation and the
absence of flavor changing neutral currents.

In this thesis several applications of effective field theory are discussed.  The focus
will be on using chiral perturbation theory for processes with heavy particles. 
Much of this material has now been published \cite{i1,lsw,ms0,ms1, ms2,msw}, but
unpublished material appears in sections~\ref{NNbeta} and chapter 8.  Chapter 2
reviews the necessary theoretical tools and establishes notation.  In chapter 3,
$D^*$ decays are investigated.  The $D^{*0}$, $D^{*+}$, and $D_s^*$ branching
fractions are used to extract the $D^*D\pi$ and $D^*D\gamma$ couplings working
to first order in the symmetry breaking parameters, $m_q$ and $1/m_c$.  Important
effects due to the heavy meson mass splittings and unknown order $m_q$
couplings are included.  Predictions for the $D^*$ and $B^*$ widths are given. 
Chapter 4 discusses a method for determining $|V_{ub}|$ from exclusive $B$
semileptonic decay.  The calculable deviation from unity of the double ratio of form
factors $(f^{(B\to\rho)}/f^{(B\to K^*)})/$ $(f^{(D\to\rho)}/f^{(D\to K^*)})$ is
determined using chiral perturbation theory and is found to be small.  It is shown
that combining experimental data from $B\to\rho\,\ell\,\bar\nu_\ell$, $B\to
K^*\ell\,\bar\ell$ and $D\to\rho\,\bar\ell\,\nu_\ell$ can lead to a model independent
determination of $|V_{ub}|$ with an uncertainty from theory of about 10\%.  

In chapter 5 an effective field theory for nucleon-nucleon interactions is discussed.
The power counting in this theory is controlled by the presence of a non-trivial
ultraviolet fixed point or, equivalently, a bound state near threshold.  Two
renormalization schemes which have manifest power counting are discussed in
detail, the Power Divergence Subtraction scheme (PDS)\cite{ksw1} and an off-shell
momentum subtraction scheme which we call the OS scheme.  Comparing results
in these schemes gives us a method for determining if a statement about the
behavior of the theory is scheme dependent. The effect of low energy poles on the
organization of the perturbation series is explained.  Comments are also made
regarding the constraints that ultraviolet divergences make on the power counting.
Theoretical and empirical arguments are then given about the range of this theory.  

In chapter 6, radiative pion effects are discussed.  It is shown that for the purpose
of power counting the pion interactions should be divided into three classes:
potential, radiation, and soft.  A power counting is introduced for systematically
including radiation pion effects.  The leading order radiation pion graphs for
nucleon-nucleon scattering are evaluated.  The power counting for soft pions is
also discussed.  

Chapter 7 discusses the symmetries of the lowest order nucleon effective field
theory.  It is shown that in the limit where the $NN$ $^1S_0$ and $^3S_1$
scattering lengths, $a^{(^1S_0)}$ and $a^{(^3S_1)}$, go to infinity, the leading
terms in the effective field theory for strong $NN$ interactions are invariant under
Wigner's SU(4) spin-isospin symmetry.  This explains why the leading effects of
radiation pions on the S-wave $NN$ scattering amplitudes vanish as $a^{(^1S_0)}$
and $a^{(^3S_1)}$ go to infinity.  Implications of this symmetry are also discussed
for $NN \to NN\, \mbox{axion}$ and $\gamma\, d\to n\, p$.

A brief discussion of predictions for the $^3S_1-^3D_1$ mixing parameter
$\epsilon_1$ is given in chapter 8.  Working in the theory with pions at NNLO gives
a one parameter prediction for $\epsilon_1(p)$.  The accuracy of this prediction is
compared to results in the theory without pions.  

Chapter 9 contains concluding remarks.
  
\chapter{Theoretical Background}   \label{Bgrnd}

This chapter introduces the chiral perturbation theory formalism for theories
with zero, one or two heavy particles.  We begin with the QCD Lagrangian,
\begin{eqnarray}\label{LQCD}
  {\cal L}_{\rm QCD} = -\frac14 G_{\mu\nu}^A G^{A\mu\nu} + 
    \bar q (i \Slash{D}{.2}-m_q) q + \bar Q  (i \Slash{D}{.2}-m_Q) Q + \mbox{g.f.} 
   + \mbox{c.t.}\,,
\end{eqnarray}
where $G_{\mu\nu}^A$ is the field strength for the gluon field $A_\mu^A$,
$D^\mu=\partial^\mu + i g A_\mu^A T^A$ is the color covariant derivative, g.f. 
stands for gauge-fixing and ghost terms, and c.t. stands for counterterms. The
field $q$ includes the three light quark fields $u,d,s$ with masses
$m_q=m_{u,d,s}$, while $Q$ includes the three heavy quarks $c,b,t$ with masses
$m_Q=m_{c,b,t}$.  The quark-gluon interaction is flavor blind so in QCD only the
masses distinguish the quarks.  The Lagrangian in Eq.~(\ref{LQCD}) is
renormalizable, Lorentz invariant, and is also invariant under parity, charge
conjugation, and time reversal\footnote{Motivated by instanton configurations,
a term of the form $\theta/(64\pi^2)\, G_{\mu\nu}^A\tilde G^{A\mu\nu}$ can be
added to ${\cal L}_{\rm QCD}$.  This term violates parity and time-reversal
invariance.  In nature $\theta$ is tiny, limits on the neutron electric dipole moment
\cite{nedm} give $\theta \lesssim10^{-9}$.  The occurrence of this unnaturally small
value is known as the strong CP problem.}.  The quantum theory of QCD depends
on another dynamically generated scale, $\Lambda_{\rm QCD}\sim 250\,{\rm MeV}$,
where 
\begin{eqnarray}
   \alpha_s(\mu) = \frac{g(\mu)^2}{4\pi} = 
	{4\pi \over \beta_0 \ln(\mu^2/\Lambda_{\rm QCD}^2) } +\ldots \,.
\end{eqnarray}
Here $\mu$ is the renormalization point, and $\beta_0=11 N_c/3-2 n_f/3$ is the
lowest order coefficient of the QCD beta function for $n_f$ flavors and $N_c(=3)$
colors.  QCD is asymptotically free \cite{Politzer,Gross}, $\alpha_s(\mu\to \infty)\to
0$, making perturbation theory valid at large energies.  At low energy
$\alpha_s(\mu)$ becomes large, and the quarks and gluons become confined. 
Confinement is a non-perturbative phenomenum and a direct proof from QCD has
not been given.

In the limit $m_q\to 0$ the light quark term in Eq.~(\ref{LQCD}) is 
invariant under the chiral symmetry transformation
\begin{eqnarray}
   q_L \to L\, q_L\,, \qquad q_R \to R\, q_R\,, \qquad \mbox{where  } L \in SU(3)_L 
   \mbox{  and  } R \in SU(3)_R \,.
\end{eqnarray}
Since $m_q \ll \Lambda_{\rm QCD}$ this is an approximate symmetry of QCD. 
This symmetry is spontaneously broken, $SU(3)_L \times SU(3)_R \to SU(3)_V$,
by the vacuum expectation value
\begin{eqnarray} \label{vev}
   \langle 0| \,\bar q_R^{\,a} \,q_L^{\,b} \,|0 \rangle = \upsilon\, \delta^{ab} \,, \qquad
  \mbox{ where  } \upsilon \sim \Lambda_{\rm QCD}^3\,.
\end{eqnarray}
The breaking of chiral symmetry is another non-perturbative effect, and occurs
at a scale $\Lambda_\chi\sim 1\,{\rm GeV}$.  The up and down quarks are much
lighter than the strange quark, so $SU(2)_L \times SU(2)_R$ is an even better
symmetry.  In this case the unbroken $SU(2)_V$ subgroup is isospin.  Chiral
symmetry has important implications for the interaction of pions, kaons, and the
eta with each other as well as with the heavier hadrons.

In the limit $m_Q\to \infty$ the heavy quark sector in Eq.~(\ref{LQCD}) also
exhibits additional symmetries.  Consider a heavy quark with momentum $p=m_Q
v$ which interacts with a gluon with momentum $k$, so that the final momentum
of the heavy quark is $p'= m_Q v+k$.  For $k\sim\Lambda_{\rm QCD}$ the
velocity $v$ of the heavy quark is conserved up to small terms of order
$\Lambda_{\rm QCD}/m_Q$ and becomes a useful label for the heavy quark field.
To construct a Lagrangian with a good $m_Q\to \infty$ limit we set
\begin{eqnarray}\label{HQF}
  Q(x) = e^{-i m_Q v\cdot x}  h_v(x) + \ldots \,,
\end{eqnarray}
where $\frac12({1+\vslash})\, h_v =h_v$ and $v^2=1$.  Momenta of order
$\Lambda_{\rm QCD}$ cannot produce a heavy anti-quark, so in the heavy
quark sector the anti-particles can be integrated out as indicated by the ellipsis
in Eq.~(\ref{HQF}).  The number of heavy quarks is then conserved in the 
effective theory.  After some straightforward algebra, the Lagrangian for heavy 
quarks with velocity $v$ becomes\cite{Georgiv}
\begin{eqnarray}
  {\cal L}_v = \bar h_v \, i v\cdot D\, h_v \,+ {\cal O}(1/m_Q) \,.
\end{eqnarray}
At leading order this Lagrangian is independent of the spin and flavor of the
heavy quark.  For $N$ heavy quarks we have an $SU(2N)_v$ symmetry, known as
heavy quark symmetry\cite{HQS1,HQS2}.  This symmetry has important 
implications for exclusive and inclusive decays involving hadrons with a heavy
quark, such as the $D, D^*, B, B^*, \Lambda_b, \Lambda_c,$ etc.  For a more
detailed discussion of the implications of heavy quark symmetry see
Ref.~\cite{MW}.  

The next few sections explain the implication of these symmetries in the
formulation of the low energy effective field theories for interaction of the light
pseudoscalars $\pi, K, \eta$, with heavy hadrons including the charm and bottom
mesons $D, D^*, B, B^*$, vector mesons $\rho, K^*, \phi, \omega$, and nucleons
$N=p,n$.  For reviews of the formalism for zero and one heavy particle see
\cite{MW,Manoharrev,ulf}.  For effective field theory with two heavy particles see
\cite{Bodwin,ksw2,labelle,lm,ls,vanKolck}.

\section{Chiral symmetry and chiral perturbation theory}  \label{ChS}

In this section $SU(3)$ chiral perturbation theory is reviewed. The formalism for
the more accurate but less predictive $SU(2)$ chiral perturbation theory follows
in a similar manner.  The eight pseudo-Goldstone bosons $\pi^i$ that arise from
the breaking $SU(3)_L \times SU(3)_R \rightarrow SU(3)_V$ are identified with
the observed light pseudoscalar mesons, $\pi^\pm, \pi^0, K^\pm, K^0, \bar K^0$,
and $\eta$.  These will be encoded in the exponential representation 
\begin{eqnarray} \label{Sig}
 \Sigma = \xi^2 =\exp\left({2i \pi^i \lambda^i \over f} \right) \,,
\end{eqnarray}
where $\Sigma^\dagger \Sigma=\xi^\dagger \xi = 1$ and the $\lambda^i$ are 
$3\times 3$ matrices such that
\begin{eqnarray}{
  \Pi = \pi^i \lambda^i = \left( \begin{array}{ccc}
   \pi^0/\sqrt{2}+\eta/\sqrt{6}  & \pi^+   & K^+ \\
   \pi^-   & -\pi^0/\sqrt{2} +\eta/\sqrt{6}  & K^0 \\
   K^-	   & \bar K^0 & -2 \eta/\sqrt{6} \end{array} \right) \,. } \label{pgb}
\end{eqnarray} 
In Eq.~(\ref{Sig}) $f\sim f_\pi = 131 \,{\rm MeV}$, where $f_\pi$ is the pion decay 
constant,
\begin{eqnarray}
  \langle 0 |\, \bar u\, \gamma^\mu\, \gamma_5\, d\, | \pi^-(p)\rangle = 
     -i f_\pi p^\mu \,.
\end{eqnarray}
Note that $\Pi$ transforms as an octet under the unbroken $SU(3)_V$.  The $\xi$
field in Eq.~(\ref{Sig}) will become useful when we add heavy matter fields in
section~\ref{HQS}.  Under an $SU(3)_L\times SU(3)_R$ transformation 
\begin{eqnarray}
\Sigma &\to& L \Sigma R^\dagger \,, \nn \\
\xi &\to& L \xi U^\dagger = U \xi R^\dagger \,,\qquad \mbox{where    }
  U = \sqrt{\Sigma} R^\dagger \sqrt{R \Sigma^\dagger L^\dagger} \,.
\end{eqnarray} 

The non-zero quark masses $m_q={\rm diag}(m_u,m_d,m_s)$ break the chiral
symmetry.  To include $m_q$ in our low energy Lagrangian we need to write a term
that transforms in the same way as the light quark mass term in ${\cal L}_{\rm
QCD}$ in Eq.~(\ref{LQCD}).  To do this we pretend that $m_q \to L m_q R^\dagger$
under a $SU(3)_L\times SU(3)_R$ transformation,  and then form invariants with
$m_q$.  The Lagrangian with the fewest derivatives and powers of $m_q$ that
satisfies the symmetry constraints is 
\begin{eqnarray} \label{Lchi}
 {\cal L}_\chi^{(2)} &=& \frac{f^2}8 {\rm Tr}\, \partial^\mu\Sigma\, \partial_\mu
   \Sigma^\dagger +{f^2 B_0 \over 4}{\rm Tr} (m_q \Sigma + m_q \Sigma^\dagger) 
 \,, \label{Lag00}
\end{eqnarray}
where ${f^2 B_0 / 4} =v $ in Eq.~(\ref{vev}).  Note that expanding the $\Sigma$
fields in terms of $\Pi$ gives a canonically normalized kinetic term for $\Pi$ plus 
an infinite number of interaction terms with determined coefficients.  In the
interaction terms the pseudo-scalar fields are derivatively coupled, which is a
general feature of chirally invariant couplings involving $\Pi$.  This follows from 
the fact that constant goldstone boson fields $\pi_i$ are a rotation of $\Sigma$,
and correspond to an equivalent vacuum for the spontaneous symmetry breaking.   
When neglecting isospin violation it is conventional to define $\hat m = (m_u + 
m_d)/2$.  In this case the meson masses are
\begin{eqnarray}
   m_\pi^2 = 2 B_0 \hat m\,, \qquad m_K^2 = B_0 (\hat m+ m_s)\,, 
	\qquad m_\eta^2 = \frac{2 B_0}{3} (2 m_s +  \hat m) \,.
\end{eqnarray}
Therefore, one power of a quark mass corresponds to two powers of a meson 
mass.  This is the most general leading order behavior, given that these squared
masses have a Taylor series in $m_q$, and that a constant term is forbidden by the
fact that in the limit $m_q\to 0$ the mesons are massless Goldstone bosons.


The Lagrangian in Eq.~(\ref{Lchi}) is not the most general one that is invariant
under the desired symmetries.  In particular, we can add an operator with
dimension $m$ that involves more derivatives or powers of $m_q$ and a
coupling of dimension $4-m$. After using the equations of motion\footnote{Note
that after using the equations of motion the remaining Lagrangian can also be 
used in loop calculations \cite{Politzer2,Georgieom,arzt}.} there are 
$10$ linearly independent terms with dimension 0 coefficients\cite{GLrev,GL2,GL}. 
For example,
\begin{eqnarray} \label{Lchi4}
  {\cal L}_\chi^{(4)} = \alpha_1 \left[ {\rm Tr}\, \partial^\mu\Sigma\, \partial_\mu
   \Sigma^\dagger \right]^2 + \ldots \,.
\end{eqnarray}
Couplings like $\alpha_1$ encode information about the short distance physics
which was integrated out, so their scale is set by short distance scales like the
chiral symmetry breaking scale $\Lambda_\chi$.  If $p$ is a typical momentum,
then higher dimension operators are suppressed by powers of
$p^2/\Lambda_\chi^2$, $m_\pi^2/\Lambda_\chi^2$, and
$m_K^2/\Lambda_\chi^2$.  This is the chiral power counting. It is convenient to
consider $p^2\sim m_\pi^2\sim m_K^2 \sim m_q$ and then call Eq.~(\ref{Lchi})
the ${\cal O}(p^2)$ Lagrangian.   Since the particles in ${\cal L}_\chi$ are
relativistic, $E^2=p^2+m^2$, and counting powers of the energy and powers of
momenta are equivalent here.

Along with higher dimension operators we must also consider loop corrections.
These corrections are necessary, for instance, to restore unitarity to the S-matrix.
The chiral power counting can also be applied to loop diagrams.  Consider a 
graph with $L$ loops, and $n_m$ vertices that are ${\cal O}(p^m)$.  Weinberg 
\cite{Wein1} proved that this diagram is ${\cal O}(p^D)$ where
\begin{eqnarray} \label{cpc}
  D = 2 (L+1) + \sum_m (m-2) n_m \  \ge\: 2 \,.
\end{eqnarray}
Each additional loop adds two powers of $p$.  Instead of remembering the
formula in Eq.~(\ref{cpc}) we can power count an arbitrary loop graph by
assigning appropriate powers of $p$ to the vertices, a $p^4$ for each loop
integration, and factors of $1/p^2$ for each propagator.  If loop integrals are
regulated in a mass independent way, then this counting is not affected by the
divergences.  It is convenient to use dimensional regularization where we
continue the dimension of space time to $d=4-2\epsilon$.  At ${\cal O}(p^4)$ we
must include the ${\cal O}(p^4)$ vertices at tree level and the ${\cal O}(p^2)$
vertices at one loop.  Because of renormalization, these two contributions can
not be separately specified in a unique manner.  A logarithmic divergence in a
loop integral induces a $1/\epsilon + \ln(\mu^2/p^2)$ dependence in the result. 
The $1/\epsilon$ pole is subtracted or absorbed into a ${\cal O}(p^4)$ coefficient,
so in this sense these couplings act as counterterms.  The fact that divergences
are polynomial in the mass or momentum squared \cite{Collins} along with the
chiral power counting implies that divergences can always be absorbed in this 
way.  This theory is said to be renormalizable order-by-order in the power
counting.  The finite ($\epsilon$ independent) part of the ${\cal O}(p^4)$ coupling
depends on $\mu$ in such a way that it cancels the $\mu$ dependence from the
loop.  Changing $\mu$ changes the value of the loop with a compensating change
in the value of the ${\cal O}(p^4)$ coupling.  It might seem strange that the $\mu$
dependence exactly cancels (which is different than the situation in perturbative
QCD where the $\mu$ dependence only cancels to a given order in
$\alpha_s(\mu)$).  However, this is nothing more than the statement that if we
could calculate the full amplitude then it would be independent of the
renormalization point.  Therefore, expanding this amplitude in a power series in
$p^2$ and $m_q$ gives coefficients which are termwise independent of $\mu$.

The $\mu$ dependence of the loops and counterterms gives us a method for
determining the size of $\Lambda_\chi$, called naive dimensional analysis 
\cite{GM}.  Consider the graphs for $\pi \pi$ scattering at ${\cal O}(p^4)$.  For 
simplicity we use SU(2) chiral perturbation theory.  Setting constants of order 
unity equal to $1$, the amplitude takes the form
\begin{eqnarray}  \label{pipi}
  {p^2 \over f^2}+{p^4 \over f^2 (4\pi f)^2} \bigg[ \ln{\Big(\frac{\mu^2}{p^2}\Big)} 
	+ K \bigg] + {p^4\over f^4} \, \alpha_1(\mu) + \ldots \,,
\end{eqnarray}
where $K$ is a number and $p$ is the center of mass momentum.  At tree level the
first term in Eq.~(\ref{Lchi}) gives an order $p^2$ contribution which is the first
term in Eq.~(\ref{pipi}).  The second term is the contribution from the loop
graphs, and the factor of $1/(4\pi)^2$ arises from the loop integration.  The third
term in Eq.~(\ref{pipi}) is the contribution of the operator in Eq.~(\ref{Lchi4}).  For
simplicity, terms with $m_\pi$ dependence have been left out as indicated by the
ellipses.   If the value of $\mu$ is changed then the second and third terms
change in size while the sum stays the same.  If there is no fine-tuning of
parameters, then $\alpha_1(\mu)$ must be at least as large as the change to the
loop graph induced by rescaling $\mu$ by an amount of order 1 \cite{GM}.  Thus, 
naive dimensional analysis implies that the second and third terms will be roughly 
the same size, and $\alpha_1(\mu)\sim 1/(4\pi)^2\sim 0.006$.  It also implies that a
natural size for the chiral symmetry breaking scale is\footnote{Some authors use
$F_\pi=93\,{\rm MeV}$ rather than $f\simeq 131\,{\rm MeV}$, so $\Lambda_\chi=4\pi F_\pi$.  Dimensional analysis can not tell the difference.} $\Lambda_\chi \sim 4\pi f$. 
When the ${\cal O}(p^4)$ couplings are fit to data, they are found to be $\sim
10^{-3}$ with $\mu=m_\rho$, which agrees with the dimensional analysis 
argument.  The choice of $\mu$ reflects the fact that $\alpha_1(\mu)$ knows
only about short distance scales (and in particular is independent of $m_\pi$) so 
\begin{eqnarray}
  \alpha_1(\mu) \sim -{1 \over (4\pi)^2} \bigg[ \ln{\Big(\frac{\mu^2}{\Lambda^2}
	\Big)}+K' \bigg] \,,
\end{eqnarray}
where $K'$ is a constant and $\Lambda \sim m_\rho \mbox{ or } \Lambda_\chi$.  
To avoid large logarithms in the coefficients we pick $\mu \sim \Lambda$.  This
leaves potentially large logarithms in the matrix elements, $\ln{(\mu^2/m_\pi^2)}$.
However since the theory is finite in the chiral limit $m_\pi \to 0$ these come 
multiplied by a power of $m_\pi^2$, so although they are enhanced relative to 
other ${\cal O}(m_\pi^2)$ terms, they are not particularly large.

\section{Dynamics with one heavy particle} 

\subsection{The $D^{(*)}$ and $B^{(*)}$ and heavy quark symmetry} \label{HQS}

The use of heavy quark symmetry\cite{HQS1,HQS2} results in a dramatic
improvement in our understanding of the spectroscopy of hadrons containing a
single heavy quark.  In the limit where the heavy quark mass goes to infinity,
$m_Q\to\infty$, such hadrons are classified not only by their total spin $J$, but
also by the spin of their light degrees of freedom (i.e., light quarks and gluons),
$s_l$ \cite{spect}.  In this limit, hadrons containing a single heavy quark come in
degenerate doublets with total spin, $J_\pm = s_l \pm \frac12$, coming from
combining the spin of the light degrees of freedom with the spin of the heavy
quark, $s_Q = \frac12$.  (An exception occurs for baryons with $s_l = 0$, where
there is only a single state with $J = \frac12$.)  The ground state mesons with
$Q\,\bar q$ flavor quantum numbers contain light degrees of freedom with
spin-parity $s_l^{\pi_l}=\frac12^-$, yielding a doublet containing a spin zero and
spin one meson.  For $Q=c$ these mesons are the $D$ and $D^*$, while $Q=b$
gives the $B$ and $B^*$ mesons.  The observed doublets are indeed very close in
mass, $m_D= 1.867\,{\rm GeV}$, $m_{D^*}=2.008\,{\rm GeV}$, $m_B= 5.279\,{\rm
GeV}$, and $m_{B^*}=5.325\,{\rm GeV}$\cite{PDG96}.   The heavy quark flavor
symmetry gives further relations between the $D^{(*)}$ and $B^{(*)}$.  

The heavy mesons come in triplets under the $SU(3)_V$ symmetry, ($D^0$,
$D^+$, $D_s$), ($D^{*0}$, $D^{*+}$, $D_s^*$), ($B^-$, $B^0$, $B_s$), and
($B^{*-}$, $B^{*0}$, $B_s^*$).  We will use the dimension $3/2$ HQET velocity 
dependent fields $P_a^{(Q)}(v)$ and $P_a^{*(Q)\mu}(v)$\ (a=1,2,3), where 
$P^{(c)}(v)$ destroys a $D$ with velocity $v$, etc.  It is convenient to include 
$P_a$ and $P^*_a$ in a $4\times4$ matrix
\begin{equation}\label{Hdef}
H_a^{(Q)} = \frac{1+\vslash}2\, \Big[ P_a^{*(Q)\mu} \gamma_\mu 
  - P_a^{(Q)}\, \gamma_5 \Big] \,. \label{HQf}
\end{equation} 
$H_a^{(Q)}$ transforms linearly under both a heavy quark spin transformation 
$D(R)$, and under a heavy quark flavor transformation $U\in SU(2)$ \cite{MW},
\begin{eqnarray}
  H_a^{(Q)} \to D(R) H_a^{(Q)} \,, \qquad H_a^{(Q_i)} \to U_{ij} H_a^{(Q_j)} \,,
\end{eqnarray}
and satisfies $\vslash H_a=H_a=-H_a\vslash$.  The conjugate field is defined
as $\bar H_a = \gamma^0 H_a^\dagger \gamma^0$.  It is also convenient to 
define vector and axial vector currents, 
\begin{eqnarray} \label{avcur}
 V^\mu=\frac12 (\xi^\dagger\partial^\mu\xi + \xi\partial^\mu \xi^\dagger)\,,  
     \qquad \mbox{and}\qquad 
 A^\mu=\frac{i}2 (\xi^\dagger\partial^\mu\xi -\xi\partial^\mu \xi^\dagger)\,,
\end{eqnarray}
which contain an even and odd number of $\Pi$ fields respectively. Under a 
$SU(3)_L\times SU(3)_R$ transformation 
\begin{eqnarray}
  H_a \to H_b U^\dagger_{ba}\,, \qquad V^\mu \to U V^\mu U^\dagger + 
     i\, U \partial^\mu U^\dagger \,,\quad \mbox{  and  }\quad 
  A^\mu \to U A^\mu U^\dagger \,.
\end{eqnarray}
The lowest order Lagrangian invariant under these symmetries is
\begin{eqnarray} \label{LH}
 {\cal L}_H^{(1)} &=& 
 -{\rm Tr}\, \bar H_a i v\cdot D_{ba} H_b + 
   g\, {\rm Tr}\, \bar H_a H_b \gamma_\mu \gamma_5 A^\mu_{ba} \,, \label{Lag0H}
\end{eqnarray}
where the chiral covariant derivative is $D^\mu_{ab}=\delta_{ab}\, \partial^\mu
-V_{ab}^\mu$.   The $H_a$ propagator derived from the kinetic term in
Eq.~(\ref{LH}) is often referred to as static, since in the rest frame $v=(1,\vec 0)$ the
equations of motion give zero energy for an onshell particle. (Recall that analagous
to Eq.~(\ref{HQF}), the $v$ dependent fields already have a factor of $m_H$
subtracted from their energy.) Like ${\cal L}_\chi$ in Eq.~(\ref{Lchi}), ${\cal L}_H$ is
organized by an expansion in derivatives and powers of $m_q$.  ${\cal L}_H$ also
involves an expansion in powers of $1/m_Q$, where terms at order $1/m_Q$ break
heavy quark symmetry.  

Since $D^\mu \sim A^\mu \sim p$ the Lagrangian in Eq.~(\ref{LH}) is ${\cal
O}(p)$.  It contains one coupling $g$ for $P^{(Q)*}P^{(Q)}\Pi$ and
$P^{(Q)*}P^{(Q)*}\Pi$.  This coupling will be discussed in greater detail in
Chapter 3.  The propagators for the heavy pseudo-scalar and vector mesons,
\begin{eqnarray}
   { i\, \delta_{ab} \over 2 (v\cdot k + i\epsilon) } \,, \qquad \mbox{  and  }\qquad
   { -i\, \delta_{ab}\, (g_{\mu\nu}-v_\mu v_\nu) \over 2 (v\cdot k + i\epsilon) } \,,
\end{eqnarray}
are ${\cal O}(1/p)$.  When power counting graphs, the meson propagators give 
$1/p^2$ and the loop measure gives a $p^4$ as before.  Because of the form of 
the heavy propagators and couplings, the power counting involves powers 
of $p$ and $m_\pi$.  Higher order corrections to the Lagrangian in Eq.~(\ref{LH}) 
are discussed in chapter 3.

\subsection{The heavy-vector meson chiral Lagrangian} \label{HV}

In this section we extend the heavy matter formalism to the vector mesons,
$\rho^\pm$, $\rho^0$, $\phi$, $\omega$, $K^{*\pm}$, $K^{*0}$, and $\bar K^{*0}$,
following the presentation in Ref.~\cite{jmw}.  One might ask if these mesons
should be treated relativistically; the lightest has mass $m_\rho=770\,{\rm MeV}$
which is starting to approach the low momentum regime of interest.  However, if
these particles are not treated as heavy then predictive power is lost.  Unlike the
pion, the vector mesons are not pseudo-Goldstone bosons, so they do not have to
be derivatively coupled and their self-interactions are not constrained by chiral
symmetry.  When they are treated as heavy, the interaction terms in the Lagrangian
can be expanded in derivatives giving an expansion in powers of $p/m_\rho$.  

The vector meson fields are introduced as a $3\times 3$ octet matrix and a 
singlet
\begin{eqnarray} \label{vm}
   {\cal O}_\mu &=& \phi_\mu^i \lambda^i = \left( \begin{array}{ccc}
   \rho^0_\mu/\sqrt{2}+\phi^{(8)}_\mu/\sqrt{6}  & \rho^+_\mu   & K^{*+}_\mu \\
   \rho^-_\mu   & -\rho^0_\mu/\sqrt{2} +\phi^{(8)}_\mu/\sqrt{6}  & K^{*0}_\mu \\
   K^{*-}_\mu   & \bar K^{*0}_\mu & -2 \phi^{(8)}_\mu/\sqrt{6} \end{array} \right) 
    \,, \nn \\
   S_\mu &=& \phi^{(0)}_\mu \,,  
\end{eqnarray}
where $v\cdot {\cal O} = v\cdot S =0$.
The dependence of these fields on the fixed four-velocity $v$ has been 
suppressed.   Under $SU(3)_L \times SU(3)_R$ the fields in Eq.~(\ref{vm}) 
transform as
\begin{eqnarray}
    {\cal O}_\mu \to U  {\cal O}_\mu U^\dagger  \,,\qquad S_\mu \to S_\mu \,.
\end{eqnarray}
The ${\cal O}(p)$ Lagrangian is \cite{jmw}
\begin{eqnarray}  \label{LV}
 {\cal L}_V &=& -\, S_\mu^\dagger\, i v\cdot \partial\, S^\mu - 
     {\rm Tr} {\cal O}_\mu^\dagger\, i v\cdot D\, {\cal O}^\mu \\
 && + i g_1\, S_\mu^\dagger\,  {\rm Tr} ({\cal O}_\nu A_\lambda)\, v_\sigma 
 \epsilon^{\mu\nu\lambda\sigma} + h.c. +  i g_2\, {\rm Tr} ( \{ 
 {\cal O}_\mu^\dagger, {\cal O}_\nu \} A_\lambda )\, v_\sigma 
 \epsilon^{\mu\nu\lambda\sigma} \,, \nn
\end{eqnarray}
where the chiral covariant derivative is $D^\nu {\cal O}^\mu = \partial^\nu {\cal
O}^\mu + [V^\nu, {\cal O}^\mu]$ and $A^\mu$ and $V^\mu$ are given in
Eq.~(\ref{avcur}).  The octet and singlet originally have masses $\mu_0$ and
$\mu_8$.  When the velocity dependent fields are constructed we rescale both
${\cal O}^\mu$ and $S^\mu$ by a common factor, $\sqrt{2 \mu_8}\, e^{i\mu_8
v\cdot x}$.  This leaves a term involving the mass difference, $\Delta \mu =
\mu_0-\mu_8 < 200\,{\rm MeV}$ which may be treated as order $m_q$.  
Corrections to Eq.~(\ref{LV}) involving the quark mass matrix $m_q$ induce mass 
differences between the vector mesons.  The mass eigenstates of the 
$\phi^{(0)}-\phi^{(8)}$ mass matrix are 
\begin{eqnarray}
  | \phi \rangle = \sin\theta\, |\phi^{(0)}\rangle - \cos\theta\, |\phi^{(8)}\rangle\,,
\qquad
  | \omega \rangle = \cos\theta\, |\phi^{(0)}\rangle + \sin\theta\, |\phi^{(8)}\rangle\,,
\end{eqnarray}
where the $SU(3)_V$ prediction for the mixing angle is 
$\tan \theta \simeq \pm 0.76$.  

Further predictive power can be obtained by considering the limit of large $N_c$
\cite{jmw}.  In this limit $\Delta \mu=0$, $\tan\theta=1/\sqrt{2}$ and the octet and 
singlet mesons can be combined into a single \emph{nonet} matrix
\begin{eqnarray}  \label{HVN}
  N_\mu &=& {\cal O}_\mu + {\mathbb{I}\over \sqrt{3}} S_\mu 
 =  \left( \begin{array}{ccc}
   \rho^0_\mu/\sqrt{2}+\omega_\mu/\sqrt{2}  & \rho^+_\mu   & K^{*+}_\mu \\
   \rho^-_\mu   & -\rho^0_\mu/\sqrt{2} + \omega_\mu/\sqrt{2} & K^{*0}_\mu \\
   K^{*-}_\mu   & \bar K^{*0}_\mu & \phi_\mu \end{array} \right) . \ \ \
\end{eqnarray}
At leading order in $N_c$ the  Lagrangian in Eq.~(\ref{LV}) becomes
\begin{eqnarray}
  {\cal L}_V &=& - {\rm Tr} N_\mu^\dagger\, i v\cdot D\, N^\mu 
 +  i g_2\, {\rm Tr} ( \{ N_\mu^\dagger, N_\nu \} A_\lambda )\, v_\sigma 
 \epsilon^{\mu\nu\lambda\sigma} \,.  
\end{eqnarray}  

Chiral and heavy quark symmetries can be used to relate the  form factors
describing the semileptonic decays: $D\to K^*\,\bar\ell\,\nu_\ell$,
$D\to\rho\,\bar\ell\,\nu_\ell$, $B\to K^*\ell\,\bar\ell$, and $B \to
\rho\,\ell\,\bar\nu_\ell$.  In chapter 4 the heavy vector meson formalism described in
this section will be used to estimate symmetry breaking corrections to the heavy
quark and chiral symmetry relations between these form factors.
 
\section{Dynamics with two heavy particles}  \label{NN}

\subsection{Two nucleon effective field theory}  \label{NNintro}

This section considers an effective field theory for two heavy particles. The
application in chapter 5 involves nucleon-nucleon scattering, so the particles are
taken to be nucleons.  For processes with one nucleon a formalism similar to that
in Section~\ref{HQS} may be used.  For simplicity, pion-nucleon interactions will
not be considered in this section, but will be considered in chapter 5. This
simplification will allow us to emphasize the qualitatively new features of the
two-nucleon theory.  Without pions the effective field theory is a valid description
of nucleon interactions for momenta $p \ll m_\pi$.  
 
Below the scale $m_\pi$, the pion can be integrated out, leaving a theory of
non-relativistic nucleons interacting via contact interactions. The nucleon field
$N$ is a doublet under isospin. The full Lagrangian in the two nucleon sector is
given by:
\begin{equation} \label{LN0}
  {\cal L}_{NN} =  N^\dagger \Big[ i\partial_t + \overrightarrow\nabla^2/(2M) 
   + \ldots \Big] N - \sum_s\,\sum_{m=0}^{\infty} C^{(s)}_{2m}\,
   {\cal O}^{(s)}_{2m}  + \ldots \,, 
\end{equation} 
where $M$ is the nucleon mass, and the ellipsis refers to relativistic corrections. 
The transformation to non-relativistic fields is analogous to Eq.~(\ref{HQF}) 
where here it is convenient to choose $v=(1,\vec 0)$.  In Eq.~(\ref{LN0}), 
${\cal O}^{(s)}_{2m}$ is an operator with $2 m$ spatial derivatives and
four-nucleon fields. We will work in a basis in which these operators
mediate transitions between ingoing and outgoing two-nucleon states of definite
total angular momentum. The superscript $s$ will give the angular momentum
quantum numbers of these states in the standard spectroscopic notation,
$^{2S+1}L_J$.  States with $(-1)^{S+L}$ even are isospin triplets, while those
with $(-1)^{S+L}$ odd are isosinglets.  If we denote the incoming and outgoing
orbital angular momentum by $L$ and $L^{\prime}$, then any operator mediating
a transition between these states must contain at least $L+L^{\prime}$
derivatives.  For states with $S=0$, $|L-L^{\prime}| = 0$, while for states with
$S=1$, $|L-L^{\prime}|= 0~{\rm or}~2$.  

In this section only S-wave transitions ($L=L'=0$) will be discussed.  For 
$s={}^1\!S_0$ or $^3S_1$ the first two terms in the series are
\begin{eqnarray}  \label{L2s}
\lefteqn {\sum_{s,m} C^{(s)}_{2m} \, {\cal O}^{(s)}_{2m} } \\  
  &=& {C_0^{(s)}} ( N^T P^{(s)}_i N)^\dagger ( N^T P^{(s)}_i N) 
   - {C_2^{(s)}\over 8} \left[ ( N^T P^{(s)}_i N)^\dagger ( N^T P^{(s)}_i 
    \:\tensor N) + h.c. \right] + \ldots  \,, \nn 
\end{eqnarray}
where the matrices $P_i^{(s)}$ project onto the correct
spin and isospin states
\begin{eqnarray} \label{Sproj}
  P_i^{({}^1\!S_0)} = \frac1{\sqrt{8}} \, (i\sigma_2) \, (i\tau_2 \tau_i ) \ ,\qquad 
  P_i^{({}^3\!S_1)} = \frac1{\sqrt{8}} \, (i\sigma_2 \sigma_i  ) \, (i\tau_2)  \ .
\end{eqnarray}
The Galilean invariant derivative in Eq.~(\ref{L2s}) is ${\tensor} =
\overleftarrow{\nabla}^2 - 2\overleftarrow\nabla\cdot \overrightarrow\nabla +
\overrightarrow{\nabla}^2$, and the ellipsis denote contributions with more
derivatives.  The normalization in Eq.~(\ref{L2s}) implies that between S-wave 
states the Feynman rules are
\begin{eqnarray}
  \begin{picture}(15,10)(1,1)
      \put(1,3){\line(1,1){10}} \put(1,3){\line(1,-1){10}} 
      \put(1,3){\line(-1,1){10}} \put(1,3){\line(-1,-1){10}}
      \put(-4,15){\mbox{\footnotesize $C_0$}}
  \end{picture} &=& -i\, C_0 \,,\\[15pt] 
  \begin{picture}(15,10)(1,1)
      \put(1,3){\line(1,1){10}} \put(1,3){\line(1,-1){10}} 
      \put(1,3){\line(-1,1){10}} \put(1,3){\line(-1,-1){10}}
      \put(-4,15){\mbox{\footnotesize $C_2$}}
      \put(-16,10){\mbox{\tiny 1}}  \put(15,10){\mbox{\tiny 3}}
      \put(-16,-7){\mbox{\tiny 2}}  \put(15,-7){\mbox{\tiny 4}}
  \end{picture} 
   &=& -i\, \frac{C_2}{8}\ [(p_1-p_2)^2+(p_3-p_4)^2]  =-i\,C_2\, p^2 \,, \nn 
\end{eqnarray}
where the last equality is true when the nucleons are onshell in the center of 
mass frame, and $p$ is the magnitude of the center of mass momentum. 

In a theory with two heavy particles using a static propagator is problematic.
The loop graph
\begin{eqnarray}  \label{pinch}
  \begin{picture}(15,10)(1,1)
      \put(1,3){\line(-1,1){10}} \put(1,3){\line(-1,-1){10}} 
      \put(11,3){\circle{20}}
      \put(21,3){\line(1,1){10}} \put(21,3){\line(1,-1){10}}
      \put(-6,18){\mbox{\footnotesize $C_0$}} 
      \put(14,18){\mbox{\footnotesize $C_0$}}
  \end{picture} \qquad
 = (-i C_0)^2  \int {d^d q \over (2\pi)^d}  \:{i \over q_0 + i\epsilon} \:
	{i \over -q_0 + i\epsilon}
\end{eqnarray}
has a pinch-singularity in the $q_0$ integration at small $q_0$.  This infrared
singularity indicates that the static propagator is missing some essential
physics.  In theories with two heavy particles the kinetic energy term in
Eq.~(\ref{LN0}) becomes a relevant operator of the same order as the $\partial_t$
term.  Including this term removes the singularity. The equations of motion for
external nucleons are then $p_0=p^2/(2M)$, so for power counting, the nucleon
energies and momentum are not equal in size.  In dimensional regularization the
loop graph in Eq.~(\ref{pinch}) is finite
\begin{eqnarray}  \label{bubble}
  \begin{picture}(15,10)(1,1)
      \put(1,3){\line(-1,1){10}} \put(1,3){\line(-1,-1){10}} 
      \put(11,3){\circle{20}}
      \put(21,3){\line(1,1){10}} \put(21,3){\line(1,-1){10}}
      \put(-6,18){\mbox{\footnotesize $C_0$}} 
      \put(14,18){\mbox{\footnotesize $C_0$}}
  \end{picture} \qquad
 &=& (-i C_0)^2  \int {d^d q \over (2\pi)^d}  \:
	{i \over \frac{E}2+q_0 -\frac{\vec q\,^2}{2M}+ i\epsilon} \:
	{i \over \frac{E}2-q_0 -\frac{\vec q\,^2}{2M}+ i\epsilon} \nn\\[5pt]
 &=& i\, (C_0)^2 \int {d^{d-1}q \over (2\pi)^{d-1}} \
	{M \over {\vec q\,^2 -ME -i\epsilon} }\\[5pt]
 &=& -i\, (C_0)^2 \  { M \sqrt{-M E -i\epsilon}  \over 4\pi} 
   =  -i\, (C_0)^2 \  \bigg( { -i M p \over 4\pi} \bigg) \,. \nn
\end{eqnarray}
Here $E=p^2/M$ is the center of mass energy.  In the second line of 
Eq.~(\ref{bubble}) the $q_0$ integration was done by contour integration.  To
power count this graph note that $q_0\sim p^2/M$, ${\vec q}\sim p$, so $d^4q \sim
p^5/M$, and each of the nucleon propagators gives an $M/p^2$.  The graph is 
therefore order $p$ in agreement with Eq.~(\ref{bubble}).

The result in Eq.~(\ref{bubble}) has a factor of the nucleon mass $M$ in the
numerator.  Since each loop with two nucleons gives an additional factor of $M$
one might worry that these large factors will spoil the power counting.   The
reason for using a non-relativistic expansion in the first place was that each graph
scales as a definite power of $M$, so we can keep track of these large
factors\footnote{What we are keeping track of is the \emph{explicit} $M$
dependence in the Lagrangian.  The Lagrangian does have further implicit $M$
dependence since $M$ is a function of $\Lambda_{\rm QCD}$ and the scales that
appear in the short distance couplings depend on $\Lambda_{\rm QCD}$ as well.
For this reason saying we know the $M$ dependence of an amplitude is not as
strong a statement as saying we know the $m_\pi$ dependence.}.  From
Eq.~(\ref{LN0}) the coupling $C_0$ has dimension $-2$. To count factors of $M$
we must determine how the $C_{2m}$ couplings scale\footnote{If the contact
interactions were replaced with Coulombic photon exchange then the interaction
would not involve any powers of $M$.  In this case the graph in Eq.~(\ref{bubble})
would scale as $\alpha^2 M/p^3\sim \alpha^2/(p^2 v)$ where $\alpha=e^2/4\pi$ is 
the fine structure constant.  This is a factor of $\alpha/v$ times a single photon 
exchange.  For a Coulombic bound state $\alpha/v\sim 1$.  Summing the most 
singular $\alpha/v$ terms is equivalent to solving the Schroedinger equation in a 
Coulomb potential.} with $M$.  To determine this, rescale all energies, $q^0 \to 
\tilde q^0 /M$, and time coordinates, $t \to M \tilde t $, so that dimensionful 
quantities have the same size (ie., are measured in units of $p$).  If we demand that 
the action is independent of $M$, then since the measure $d^4 x\sim M$, the 
Lagrange density ${\cal L}\sim 1/M$.  The kinetic term determines that our nucleon 
fields scale as $N(x)\sim M^0$, so from Eq.~(\ref{LN0}) the coupling
\begin{eqnarray}
      C_{2m} \sim 1/M \,.
\end{eqnarray}
With the $M$ scaling for the couplings determined, the scaling of any Feynman
graph can be found.  A nucleon propagator gives one power of $M$, and each
momentum space loop integration gives a $1/M$.   For bubble graphs that have
insertions of the four-nucleon operators, $N_P=N_L+N_V-1$, where $N_P$, $N_L$,
$N_V$ are the number of propagators, loops and vertices.  Thus, when relativistic
corrections are neglected any graph built out of the interactions in Eq.~(\ref{LN0})
scales as $M^{-1}$ since $N_P-N_L-N_V = -1$.  Therefore, the $2\to 2$ scattering
amplitude ${\cal A}\sim 1/M$.  With the definition of ${\cal A}$ used here, this
scaling gives a finite cross-section in the $M\to \infty$ limit which is physically
sensible. (Note that there is a limit of QCD where $M\to \infty$ but $\Lambda_{\rm
QCD}$ is finite, namely large $N_c$ with $N_c \alpha_s$ held fixed \cite{thooft}.)
Since all graphs scale the same way with $M$, $M$ is irrelevant to the power
counting.  Relativistic corrections are included perturbatively\cite{Mnopi}, and are 
generally suppressed by $p^2/M^2$ relative to the leading contribution to an 
observable.  

Applying dimensional analysis to the short distance coupling 
constants now gives
\begin{eqnarray}  \label{Cnat}
  C_0 \sim \frac1{M\Lambda}\,, \qquad C_2 \sim \frac1{M\Lambda^3}\,, 
  \qquad C_{2m} \sim \frac1{M\Lambda^{2m+1}} \,,
\end{eqnarray}
where $\Lambda$ denotes the scale of short distance physics that was not
included explicitly (and we are assuming $\Lambda<M$).  Treating the $C_{2m}$ 
couplings perturbatively gives an expansion in $p/\Lambda$.  In the current case
we expect that $\Lambda\sim m_\pi$.  

In nature, however, the dimensional analysis in Eq.~(\ref{Cnat}) fails.  This is
because the nucleon-nucleon system is fine tuned to have bound states near
threshold\cite{W1,Bira,ksw0}.  In the $^3S_1$ channel this bound state is the
deuteron with binding energy $B=2.2245\pm 0.0002\,{\rm MeV}$.  This energy
corresponds to the momentum $\gamma = \sqrt{M B} =45.7\,{\rm MeV} \ll
\Lambda_{\rm QCD}$.  This bound state gives a pole in the two-to-two scattering
amplitude which can not be reproduced perturbatively and therefore limits the
range of the momentum expansion to $p< \gamma$.  In the $^1S_0$
channel the situation is even worse.  In this case there is a pseudo-bound state
sitting $8\,{\rm MeV}$ above threshold which limits the momentum expansion to
$p< 8\,{\rm MeV}$.  These bound states are related to the occurrence of unnaturally
large scattering lengths.  Recall that low energy scattering can be described by an
effective range expansion 
\begin{eqnarray} \label{ere0}
    p\cot{\delta^{(s)}} =- \frac{1}{a^{(s)}} + \frac12\, r_0^{(s)}\, p^2 + \ldots \,,
\end{eqnarray}
where $\delta^{(s)}$ is the phase shift, $a^{(s)}$ is the scattering length and
$r_0^{(s)}$ is the effective range.  In quantum mechanics, with a well-behaved 
potential, it can be shown that $r_0^{(s)}$ is of order the range of the interaction,
$r_0\sim 1/\Lambda$, but $a^{(s)}$ can differ from $\Lambda$ by orders of 
magnitude\cite{Blatt}.  From the formula for the amplitudes\footnote{Strictly 
speaking  Eq.~(\ref{Apcotd0}) only holds in the $^1S_0$ channel.  The $^3S_1$ 
channel is more complicated because of $^3S_1-^3D_1$ mixing, but this mixing 
is a small effect which we will ignore for the time being, but discuss it in Chapter 8.}
\begin{eqnarray}  \label{Apcotd0}
    {\cal A}^{(s)} = \frac{4\pi}{M}\, \frac1{p\cot{\delta^{(s)}} -i p} \,,
\end{eqnarray}
it is straightforward to show that large negative or positive scattering lengths
correspond to bound states just above or below threshold.  
Experimentally\cite{burcham},
\begin{eqnarray}
a^{(^1S_0)}&=&-23.714 \pm 0.013 \,{\rm fm}\,, \qquad\mbox{and}\qquad 
   a^{(^3S_1)}=5.425 \pm 0.001\,{\rm fm}, \qquad\
\end{eqnarray}
or $1/a^{(^1S_0)}=-8.3\,{\rm MeV}$ and $1/a^{(^3S_1)}= 36\,{\rm MeV}$.  These
values of $1/a$ are small relative to $m_\pi$ and $\Lambda_{\rm QCD}$.

If the scale for the bound states or scattering lengths were set by $\Lambda$
then the scaling in Eq.~(\ref{Cnat}) would be correct.  The next section will
discuss how the $C_{2m}$ coefficients scale for a theory with large scattering
lengths.  This power counting was worked out by Kaplan, Savage, and Wise
\cite{ksw1,ksw2} (KSW).  

\subsection{Power counting and ultraviolet fixed points} \label{NNbeta}

In this section we will examine how unnaturally large scattering lengths affect
the importance of four-nucleon operators.  We also explain how linear ultraviolet
divergences play a role in determining the KSW power counting, and why it is 
useful to recast this in the framework of the renormalization group.  

\begin{figure}[!t]
  \centerline{\epsfxsize=14.5truecm \epsfbox{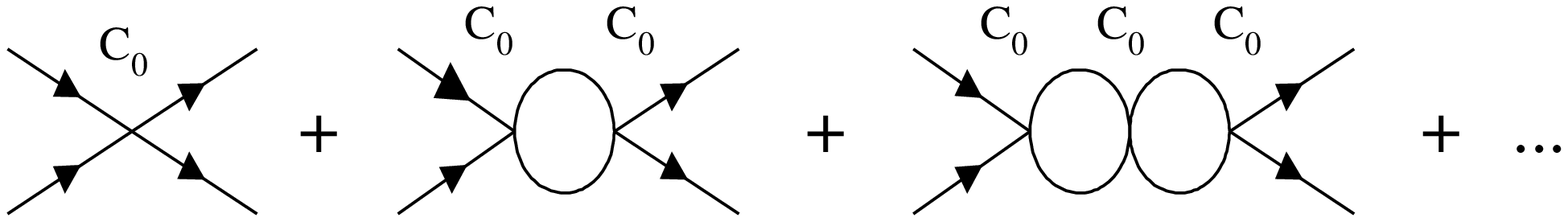}  }
 {\caption{The leading order contribution to the $NN$ scattering amplitude.} 
\label{C0bub} }
\end{figure}

Treating the effective range term and higher powers of $p^2$ in Eq.~(\ref{ere0}) 
as perturbations, the amplitude is
\begin{eqnarray}  \label{LOA}
   {\cal A}^{(s)} = -\frac{4\pi}{M}\, \frac1{ 1/a^{(s)} +i p} \,,
\end{eqnarray}
and has a good $a\to \infty$ limit.  The pole at $p=i/a^{(^3S_1)}$ corresponds to
the deuteron bound state. The amplitude in Eq.~(\ref{LOA}) is reproduced in the 
effective field theory by summing the loop graphs in Fig.~\ref{C0bub} 
using the result in Eq.~(\ref{bubble})
\begin{eqnarray}  \label{MSamp}
  i {\cal A} = -i\, \bar{C_0} \sum_{m=0}^\infty \bigg( { -i  p M \bar{C_0} \over 4\pi} 
  \bigg)^m   = { -i\, \bar{C_0} \over 1 + \frac{i p M}{4\pi}\, \bar{C_0} }\,.
\end{eqnarray}
Matching onto the effective range expansion then gives the values
\begin{eqnarray}  \label{MS}
  \bar{C_0}^{(s)} = \frac{4\pi a^{(s)}}{M} \,.
\end{eqnarray}
Thus, the $C_0$ coupling must be fine tuned to a large value to reproduce the 
observed scattering length\cite{W2}.  This result depends on our definition for the
renormalized coupling constant $C_0$, or in other words our choice of
renormalization scheme (which is distinct from choosing a regularization method). 
The bars in Eqs.~(\ref{MSamp}) and (\ref{MS}) indicate that only divergent terms
have been subtracted, which is the Minimal Subtraction (MS) scheme.  The sum in
Eq.~(\ref{MSamp}) only converges for $p<1/a$.  Each loop graph in the sum grows
with $p$, and the radius of convergence is $1/a$.  For $p>1/a$ we analytically
continue and use the result on the right hand side of Eq.~(\ref{MSamp}).

Looking back at Eq.~(\ref{bubble}), this one-loop graph is linearly divergent with
$d=4$,  however in dimensional regularization this power divergence is not
present.  If we had calculated this loop graph with a finite momentum cutoff $L$ 
then we would have
\begin{eqnarray}  \label{bubble2}
  \begin{picture}(15,10)(1,1)
      \put(1,3){\line(-1,1){10}} \put(1,3){\line(-1,-1){10}} 
      \put(11,3){\circle{20}}
      \put(21,3){\line(1,1){10}} \put(21,3){\line(1,-1){10}}
      \put(-6,18){\mbox{\footnotesize $C_0$}} 
      \put(14,18){\mbox{\footnotesize $C_0$}}
  \end{picture} \qquad
  = i\, (C_0)^2 \int_0^L {d^3q \over (2\pi)^3} 
	{M \over {\vec q\,^2 -p^2 -i\epsilon} } 
   =   {i\,M (C_0)^2 \over 4\pi} \  \bigg( ip+\frac{2L}{\pi} + \ldots \bigg) \,, \quad
\end{eqnarray}
where the ellipses denote terms that vanish as $L\to\infty$.  Defining a counterterm
to cancel the term proportional to $L$, and taking $L\to\infty$ gives back the result
in Eq.~(\ref{bubble}).
Note that the limit $p\to\infty$ and the integral over $q$ do not commute because
the integral is divergent.  The growth of Eq.~(\ref{bubble2}) with $p$ is tied to this
linear ultraviolet divergence.  It is useful to use our freedom in defining the
renormalization scheme to keep track of this.  In the power divergence subtraction
scheme (PDS)\cite{ksw1} additional finite subtractions are made in one-to-one
correspondence with the linear divergences.  In this scheme the coupling
constants depend on the renormalization point $\mu_R$ and the value of the loop
in Eq.~(\ref{bubble2}) becomes\footnote{The notation $\mu_R$ is used for the
renormalization point in this section to agree with the notation used in chapter 5.}
\begin{eqnarray}
  \begin{picture}(15,10)(1,1)
      \put(1,3){\line(-1,1){10}} \put(1,3){\line(-1,-1){10}} 
      \put(11,3){\circle{20}}
      \put(21,3){\line(1,1){10}} \put(21,3){\line(1,-1){10}}
      \put(-6,18){\mbox{\footnotesize $C_0$}} 
      \put(14,18){\mbox{\footnotesize $C_0$}}
  \end{picture} \qquad
 =  {i\,M  \over 4\pi} \ [C_0(\mu_R)]^2\  \bigg( ip+\mu_R \bigg) \,.
\end{eqnarray} 
In fact, the same result is obtained in a more physical scheme where the
renormalized coupling is defined to be the value of the four point function
evaluated at $p=i\mu_R$ \cite{W1,Gegelia1,ms0,ms1}.  These renormalization
schemes will be discussed in greater detail in Chapter 5. 

In PDS, the renormalized coupling is \cite{ksw1,ksw2}
\begin{eqnarray} \label{C0_pds}
   C_0^{(s)}(\mu_R) = -\frac{4\pi}{M} \frac1{\mu_R -1/a^{(s)}} \,.
\end{eqnarray}
In this scheme the fine tuning as $a\to\infty$, is that $C_0(\mu_R)$ gets closer to 
its $\mu_R\to \infty$ value. $C_0(\mu_R)$ has the beta function
\begin{eqnarray}  \label{bC0}
  \beta_0 = \mu_R \frac{\partial}{\partial \mu_R} C_0(\mu_R) = 
     {M\mu_R \over 4\pi} \: C_0(\mu_R)^2 \,.
\end{eqnarray}
The renormalization group scaling gives us information about the behavior of
the theory at the scale $\mu_R$.  For $\mu_R \sim p \ll 1/a$, $C_0(\mu_R)$
behaves like a constant and $p\,C_0(\mu_R)$ can be treated perturbatively.  For
$\mu_R \sim p > 1/a$, $\ p\, C_0(\mu_R)\sim 1$ and the sum in
Eq.~(\ref{MSamp}) must be done.   The factors of $\mu_R$ make each term in 
the sum roughly the same size which is good from the point of view of power 
counting.  For $\mu_R \sim p \gg 1/a$ the summation of $C_0$ bubble graphs is 
always necessary and should be considered to be infrared physics that has been
built into the theory.  In Ref.~\cite{birse} it was shown that the scaling in
Eq.~(\ref{C0_pds}) is also reproduced with a Wilsonian renormalization group
approach.

\begin{figure}[ht!]
  \centerline{\epsfxsize=15.0truecm \epsfbox{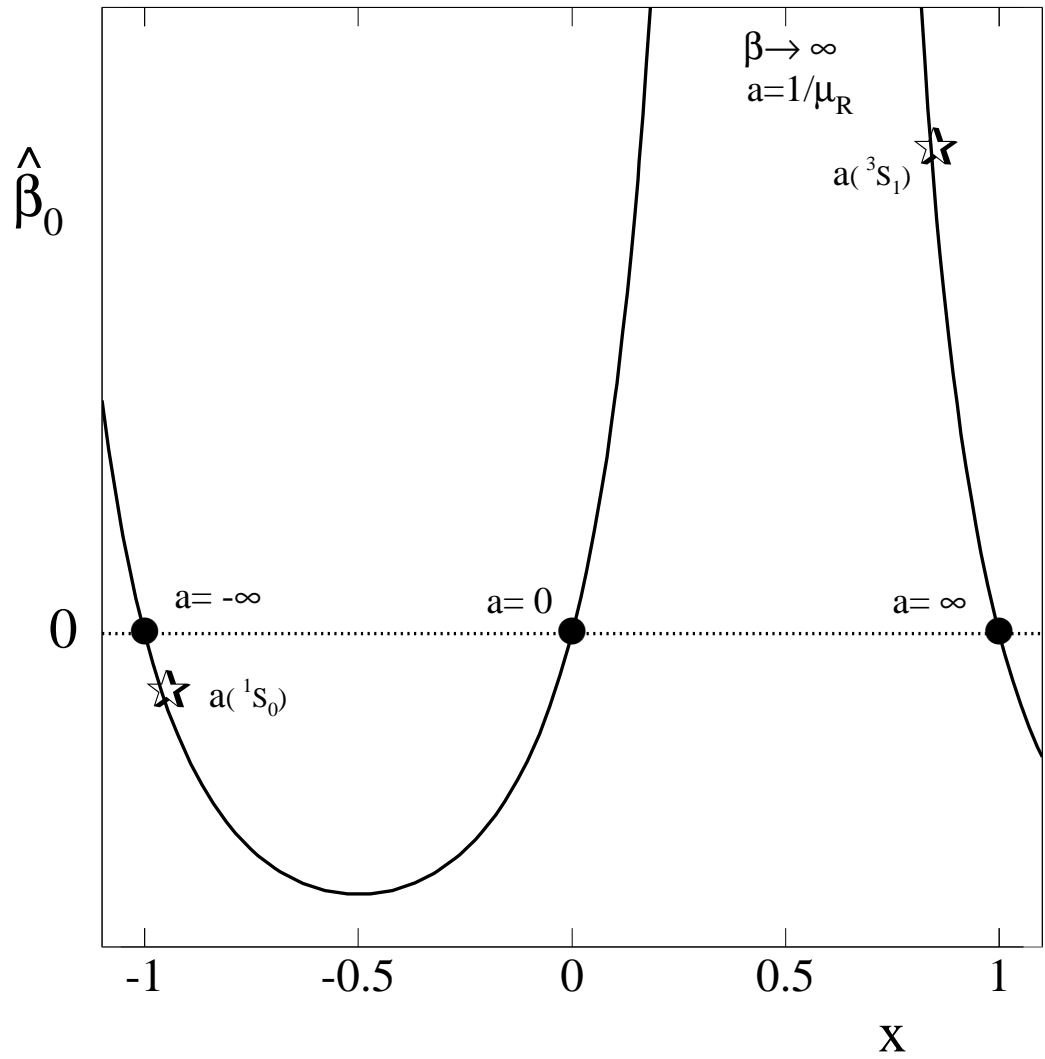}}
 {\caption{Fixed point structure of the beta function for $C_0(\mu_R)$ in the
  PDS renormalization scheme.} 
 \label{fig_bC0} }
\end{figure}

The significance of the beta function in Eq.~(\ref{bC0}) is shown more clearly in 
Fig.~\ref{fig_bC0}\footnote{This figure was inspired by a similar plot shown by
David B. Kaplan in a physics colloquium.}.  To make the beta function 
dimensionless it has been rescaled by $M\Lambda/(4\pi)$,
\begin{eqnarray}
  \hat \beta_0 &=& {M\Lambda \over 4\pi}\: \beta_0 = 
	a \Lambda \ {a\,\mu_R \over (1-a\, \mu_R)^2 } \,.
\end{eqnarray}
The prefactor $a \Lambda$ can be safely ignored. The important dependence
is in $a\,\mu_R$ since this factor measures the scale of interest relative to the 
scattering length.  To study values of $a\,\mu_R$ from $-\infty$
to $\infty$,  a variable $x$ is defined to map this range onto a compact interval,
\begin{eqnarray}
  a\,\mu_R &=& \tan{\left(\frac{\pi\, x}2\right)} \,.
\end{eqnarray}
The dependence of $\hat \beta(C_0)$ on $x$ is plotted in Fig.~\ref{fig_bC0}. 
Consider fixing the value of $\mu_R>0$ and varying the value of $a$.  The points
$a=-\infty, 0$, and $\infty$ are fixed points of the beta function.  Classically this 
makes sense since the scattering length is a measure of the interaction size.  For
$a\simeq 0$ or $\pm \infty$ the size is so small or big that the interaction looks 
the same at all scales.  In fact $a=0$ is a trivial non-interacting fixed point,
whereas $a=\pm \infty$ are non-trivial interacting fixed points where the theory is
scale invariant at lowest order.  Another feature in Fig.~\ref{fig_bC0} is that
$\beta\to \infty$ for $\mu_R=1/a>0$.  This corresponds to the deuteron bound
state.  Performing perturbation theory about $a=0$ we can never describe this
bound state, so we are limited to describing the region $\mu_R<1/a$.  If
perturbation theory is performed near $a=\infty$, then the deuteron is a physical
state in the spectrum of the theory.  If we are interested in physics at
$\mu_R\simeq m_\pi$ then the observed $^1S_0$ and $^3S_1$ scattering lengths
place us at the location of the stars in Fig.~\ref{fig_bC0}.  Looking at the distance
along the $x$ axis we are much closer to the $a=\pm \infty$ fixed points than to
$a=0$.  

For $p\gg 1/a$ powers $Q\sim p \sim \mu_R$ are counted instead of just powers 
of momenta.  The graphs in Fig.~(\ref{C0bub}) are leading order and
all scale as $1/Q$. In the PDS renormalization scheme this power counting is
manifest.   By solving the beta functions for the coefficients of operators with
more derivatives it can be shown that \cite{ksw2}
\begin{eqnarray}
  C_{2m}(\mu_R)\, p^{2m} \sim {4\pi\over M} {p^{2m} \over 
       \Lambda^m \mu_R^{m+1} } \sim {Q^{(m-1)} \over M\Lambda^m} \,,
\end{eqnarray}
for $p\sim \mu_R \gg 1/a$.  Since insertions of these operators scale with
non-negative powers of $Q$, they may be treated perturbatively.  This 
renormalization group scaling is also important in determining the power 
counting of operators that involve four nucleon fields and fields like the 
photon\cite{ksw3,kevin} and pion.
Note that since the deuteron corresponds to the pole in our scattering amplitude, 
deuteron properties can be systematically calculated in this field theory\cite{ksw3}.

When pion interactions are included they enter at order $Q^0$, which is one higher
order than insertions of $C_0$.  Therefore, the discussion in this section is also
relevant to the leading order theory with pions.  In the theory with pions the
physics encoded in the scale $\Lambda$ in Eq.~(\ref{fig_bC0}) changes.  However,
the power counting for the $C_{2m}$ coefficients remains the same because pion
effects are subleading corrections to the running of these operators.  Pion
interactions in the two-nucleon theory will be discussed in more detail in chapter
5.  



\chapter{Extraction of the $D^*D\pi$ coupling from $D^*$ decays}
\label{DsDpi}

Combining chiral perturbation theory with heavy quark effective theory (HQET)
gives a good description of the low energy strong interactions between the
pseudo-Goldstone bosons and mesons containing a single heavy quark. In this
chapter we extend the formalism in Chapter 2 to describe $D^*$ and $B^*$ decays
at subleading order.  Due to heavy quark symmetry there is one coupling, $g$, for
$D^*D\pi$, $D^*D^*\pi$, $B^*B\pi$, and $B^*B^*\pi$, and one coupling, $\beta$, for
$D^*D\gamma$, $D^*D^*\gamma$, $B^*B\gamma$, and $B^*B^*\gamma$ at leading
order{\footnote{Where it is meaningful we use $\pi$ to denote any member of the
pseudo-Goldstone boson SU(3) octet, and $D^*$ and $D$ for any member of the
triplets $(D^{*0},D^{*+},D_s^*)$ and $(D^0,D^+,D_s)$ with a similar notation for
$B^*$ and $B$.}}.  The value of the coupling $g$ is important, since it appears in
the expressions for many measurable quantities at low energy.  These include the
rate $B\to D^{(*)}\pi\ell\bar\nu_\ell$ \cite{LLW,Cheng,kramer,lee,Goity1}, form
factors for weak transitions between heavy and light pseudo-scalars
\cite{ChptW,Burdman,yan,Bpi1,BLNY,Fleischer,Casalbuoni,Xu,FG}, decay constants
for the heavy mesons \cite{Goity2,BGetal,Neub,Gtdr,BG}, weak transitions to
vector mesons \cite{Hooman}, form factors for $B\to D^{(*)}\ell \bar\nu_\ell$
\cite{Randall,Chow,Boyd}, and heavy meson mass splittings
\cite{Rosner,RS,Jenkins} (for a review see \cite{HMreview}).  However, the value of
$g$ has remained somewhat elusive, with values in the literature ranging from
$\sim 0.2$ to $1.0$.  Recently, a CLEO measurement \cite{CLEO1} of $D^{*+}\to
D^+\gamma$ brought the experimental uncertainties down to a level where a 
model independent extraction of $g$ is possible from $D^*$ decays.  

As a consequence of HQS, the mass splitting between $D^*$ and $D$ mesons is
small (of order $\Lambda_{\rm QCD}^2/m_c$), leaving only a small amount of
phase space for $D^*$ decays. In the dominant modes, $D^*\to D\pi$, and
$D^*\to D\gamma$, the outgoing pion and photon are soft, making the chiral
expansion a valid framework.  The branching ratios for $D^{*+}$ decay are
$D^0\,\pi^+$~(67.6\%), $D^+\,\pi^0$~(30.7\%) and $D^+\,\gamma$~(1.7\%)
\cite{CLEO1}.  A $D^{*0}$ can only decay into $D^0\,\pi^0$~(61.9\%) and
$D^0\gamma$~(38.1\%) \cite{PDG96} since there is not enough phase space for
$D^+\,\pi^-$.  The $D^*_s$ decays predominantly to $D_s\,\gamma$~(94.2\%)
with a small amount going into the isospin violating mode $D_s\,\pi^0$~(5.8\%)
\cite{PDG96}.  Since a measurement of the widths of the $D^*$ mesons has not yet
been made, it is only possible to compare the ratios of branching fractions with
theoretical predictions.   The ratio 
\begin{eqnarray}
    R_\pi^+={\cal B}(D^{*+}\to D^0\pi^+)/{\cal B}(D^{*+}\to D^+\pi^0)
\end{eqnarray}
is fixed by isospin to be $R_\pi^+ = 2 |\vec k_{\pi^+}|^3 / |\vec k_{\pi^0}|^3 = 2.199
\pm 0.064$ \cite{CLEO1} (where $\vec k_{\pi^{+,0}}$ are three momenta for the
outgoing pions in the $D^*$ rest frame).  This constraint is often used in 
experimental extractions of the branching ratios to reduce systematic errors.

It is interesting to note that the quark model predictions\cite{qmod,Angelos} for
$D^{*0}$ and $D^{*+}$ decays agree qualitatively with the data.  One can
understand, for instance, why the branching ratio ${\cal B}(D^{*+}\to
D^+\gamma)$ is small compared to ${\cal B}(D^{*0}\to D^0\gamma)$.  In the
quark model the photon couples to the meson with a strength proportional to
the sum of the magnetic moments of the two quarks, $\mu_2 =
2/(3\,m_c)-1/(3\,m_d)$ for $D^{*+}\to D^+\gamma$ and $\mu_1
=2/(3\,m_c)+2/(3\,m_u)$ for $D^{*0}\to D^0\gamma$.  The rate for the former is
then suppressed by a factor 
\begin{eqnarray} 
  \left| \frac{\mu_2}{\mu_1} \right|^2 = 
       (m_u/m_d)^2 { (m_d/m_c-1/2)^2 \over (m_u/m_c+1)^2 } \simeq 0.04 \:,
\end{eqnarray} 
where mass ratios appropriate for constituent quarks have been used, $m_u/m_d
\simeq 1$, $m_d/m_c \simeq m_u/m_c \simeq 1/4$.  This suppression results from
the opposite signs in $\mu_1$ and $\mu_2$, which in turn follow from the (quark)
charge assignments and spin wavefunctions for the heavy mesons.  

In the quark model $g=1$ and $\beta\simeq 3\,{\rm GeV^{-1}}$, while for the chiral
quark model $g=0.75$ \cite{GM}.  Relativistic quark models tend to give smaller
values, $g\sim 0.4$ \cite{Colangelo1,Colangelo2}, as do QCD sum rules, $g \sim 0.2
- 0.4$ \cite{Belyaev,Colangelo3,Aliev}.

Our purpose here is to use heavy meson chiral perturbation theory at one-loop to
extract the couplings $g$ and $\beta$ from $D^*$ decays. In other words, we wish
to examine the sensitivity of a model independent extraction of $g$ and $\beta$ to
higher order corrections.  For $D^*\to D\gamma$, analyses beyond leading order
have included the heavy quark's magnetic moment which arises at $1/m_c$
\cite{aetal,cg}, and the leading non-analytic effects from chiral loops proportional
to $\sqrt{m_q}$ \cite{aetal}.  $\sqrt{m_q}$ terms proportional to both $m_K$ and
$m_\pi$ were found to be important.  These effects do not introduce any new
unknown quantities into the calculation of the decay rates.  For $D^*\to D\gamma$
and the isospin conserving $D^*\to D\pi$ decays the effect of chiral logarithms,
$m_q \ln{(\mu/m_q)}$, have also been considered \cite{ccsu3}.  These are formally
enhanced over other $m_q$ corrections in the chiral limit, $m_q\to 0$, however, the
choice of the scale $\mu$ leads to some ambiguity in their contribution.  (This
scale dependence is cancelled by unknown couplings which arise at order $m_q$
in the chiral Lagrangian.)  The isospin violating decay $D_s^*\to D_s\pi^0$ has
only been considered at leading order, where it occurs through $\eta-\pi^0$
mixing\cite{chowise}.  

Here the investigation of all $D^*$ decays is extended to one-loop,
including symmetry breaking corrections to order $m_q$ and $1/m_c$.  Further
$1/m_c$ and $m_q$ contributions considered here include the effect of nonzero
$D^*$--$D$ and $D_s$--$D^0$ mass splittings, and the exact kinematics
corresponding to nonzero outgoing pion or photon energy in the loop diagrams. 
(Their inclusion is motivated numerically since $m_{\pi^0} \sim m_{D^*}-m_D
\sim m_{D_s}-m_{D}$, and the decay $D^*\to D\pi^0$ only occurs if
$m_{D^*}-m_D > m_{\pi^0}$.) To simplify the organization of the calculation
these splittings will be included as residual mass terms in our heavy meson
propagators. This gives new non-analytic contributions to the $D^*\to D\pi^0$
and $D^*\to D\gamma$ decay rates.  (To treat the mass splittings as
perturbations one can simply expand these non-analytic functions.) At order
$m_q$ there are also analytic contributions due to new unknown couplings
which are discussed.  These new couplings can, in principle, be fixed using other
observables.  We estimate the effect these unknown couplings have on the
extraction of $g$ and $\beta$.  

The calculation of the decay rates to order $m_q$ and $1/m_c$ is taken up in
sections~\ref{DDp} and \ref{DDg}.  In section~\ref{extract} we compare the
theoretical partial rates with the data to extract the $D^*D\pi$ and
$D^*D\gamma$ couplings and discuss the uncertainty involved.  Predictions for
the widths of the $D^{*}$ and $B^{*}$ mesons are also given.  Conclusions can
be found in section~\ref{discuss}.  


\section{$D^*_a \to D_a$ $\pi$ decays}  \label{DDp}

In this section we construct the effective chiral Lagrangian that describes the
decays $D^* \to D\pi$ to first order in the symmetry breaking parameters $m_q$
and $1/m_c$. Going beyond leading order also involves including loops with the
pseudo-Goldstone bosons.  From Chapter 2 section 2 recall that the lowest 
order Lagrangian is 
\begin{eqnarray}
 {\cal L}_0 &=& \frac{f^2}8 {\rm Tr}\, \partial^\mu\Sigma\, \partial_\mu
   \Sigma^\dagger +{f^2 B_0 \over 4}{\rm Tr} (m_q \Sigma + m_q \Sigma^\dagger)
  \nn \\ 
 && -{\rm Tr}\, \bar H_a i v\cdot D_{ba} H_b + 
   g\, {\rm Tr}\, \bar H_a H_b \gamma_\mu \gamma_5 A^\mu_{ba} \,. \label{Lag0}
\end{eqnarray}
The last term in Eq.~(\ref{Lag0}) couples $P^*P\pi$ and $P^*P^*\pi$ with strength
$g$ (where $P=D,B$) and determines the decay rate $D^* \rightarrow D \pi$ at
lowest order.    At order $m_q \sim 1/m_c$ the following mass correction terms 
appear
\begin{eqnarray}
 {\cal L}_{m} = \frac{\lambda_2}{4\,m_Q} {\rm Tr}\,\bar H_a 
  \sigma^{\mu\nu} H_a \sigma_{\mu\nu} + 2\lambda_1 {\rm Tr}\,\bar H_a H_b 
  m^\xi_{ba} + 2\lambda_1' {\rm Tr}\,\bar H_a H_a m^\xi_{bb}, \label{Lagm}
\end{eqnarray}
where $m^\xi = \frac12 (\xi m_q \xi^\dagger + \xi^\dagger m_q \xi)$.  The
$\lambda_1'$ term can be absorbed into the definition of $m_H$ by a phase
redefinition of $H$.  The $\lambda_2$ term is responsible for the $D^*$-$D$ mass
splitting at this order, 
\begin{eqnarray}
  \Delta=m_{D^{*}}-m_{D}= -2\lambda_2/m_c \,.
\end{eqnarray}
The term involving $\lambda_1$ splits the mass of the triplets of $D$ and $D^*$ 
states.  Ignoring isospin violation this splitting is characterized by
\begin{eqnarray}
  \delta=m_{D_s^*}-m_{D^*} = m_{D_s}-m_{D}=2\lambda_1(m_s-\hat m)
\end{eqnarray}
where $\hat m=m_u=m_d$.  For the purpose of our power counting $\delta 
\sim m_q \sim 1/m_c
\sim \Delta$.  The effect of these mass splitting terms can be taken into
account by including a residual mass term  in each heavy meson propagator. 
Since we are interested in decay rates we choose the phase redefinition for our
heavy fields to scale out the decaying particle's mass.  For $D^{*0}$ and
$D^{*+}$ decays the denominator of our propagators are: $2v\cdot k$ for
$D^{*0}$ and $D^{*+}$, $2(v\cdot k -\delta)$ for $D_s^*$, $2(v\cdot k+\Delta)$
for $D^0$ and $D^+$, and $2(v\cdot k+\Delta-\delta)$ for $D_s$.  For the
$D_s^*$ decays the denominators are the above factors plus $2\delta$.  (If we
had scaled out a different mass then the calculation in the rest frame of the initial
particle would involve a residual `momentum' for the initial particle, but would
yield the same results.)   This results in additional non-analytic contributions
from one-loop diagrams which are functions of the quantities $\Delta/m_{\pi_i}$
and $\delta/m_{\pi_i}$, where $m_{\pi_i}\in \{m_\pi,m_K,m_\eta\}$.  Formally,
$m_{\pi_i}^2 \sim m_q \sim \Delta \sim \delta$ and one can expand these
contributions to get back the result of treating the terms in Eq.~(\ref{Lagm}) as
perturbative mass insertions.  

Another type of $1/m_c$ corrections are those whose coefficients are fixed by
velocity reparameterization invariance \cite{vpi,BG}
\begin{eqnarray}
   \delta{\cal L}_v &=& - \frac1{2m_Q}{\rm Tr}\,\bar H_a (i \vec D)^2_{ba} H_b + 
  \frac{g}{m_Q}{\rm Tr} \bar H_c (i \overleftarrow D^\mu_{ac} v\cdot A_{ba} 
  - i v\cdot A_{ac}\overrightarrow D^\mu_{ba}) H_b \gamma_\mu\gamma_5 
  \nn \,. \\  \label{Lagv}
\end{eqnarray}
The first term here is analogous to the HQET kinetic operator, $O_{\rm
kin}=\frac1{2m_Q}\bar h_v\,(i \vec D)^2\, h_v$, but written in terms of the
interpolating fields $P_a$ and $P^{*\mu}_a$.  In Eq.~(\ref{Lagv}) the derivatives
give powers of the heavy meson's momentum.  There are also contributions from
$O_{\rm kin}$ that break the flavor symmetry where the derivatives are order
$\Lambda_{\rm QCD}$.   In conjunction with the HQET chromomagnetic operator,
\mbox{$O_{\rm mag}= \frac1{2m_Q}\bar h_v\,\frac{g_s}2\,\sigma_{\alpha\beta}
G^{\alpha\beta}\,h_v$}, these contributions to the Lagrangian modify the dynamics
of the heavy meson states.  They give $\Lambda_{\rm QCD}/m_Q$ corrections in
the form of time ordered products with the leading order current \cite{cchqs},
which induce spin and flavor symmetry violating corrections to the form of the
$D^*D\pi$ coupling.  We account for these corrections by introducing the
couplings $g_1$ and $g_2$ in Eq.~(\ref{Lagg}) below.  The last term in
Eq.~(\ref{Lagv}) contributes at higher order in our power counting since it is
suppressed by both a derivative and a power of $1/m_c$.  

Further terms that correct the Lagrangian in Eq.~(\ref{Lag0}) at order $m_q 
\sim 1/m_c$ include \cite{BG}\footnote{The $\kappa_1'$ term was not 
present in \cite{BG}. The factor $B_0/\Lambda_\chi^2$ is introduced here for 
later convenience.}
\begin{eqnarray}
 \delta{\cal L}_g &=& \frac{g\kappa_1\,B_0}{\Lambda_\chi^2}{\rm Tr}\, 
 \bar H_a H_b \gamma_\mu\gamma_5 A^\mu_{bc} m^\xi_{ca} + 
  \frac{g\kappa_1'\,B_0}
 {\Lambda_\chi^2} {\rm Tr}\, \bar H_a H_b \gamma_\mu\gamma_5 m^\xi_{bc} 
 A^\mu_{ca} \nn\\
 &+& \frac{g\kappa_3\,B_0}{\Lambda_\chi^2}{\rm Tr}\, \bar H_a H_b 
  \gamma_\mu\gamma_5 A^\mu_{ba} m^\xi_{cc} + \frac{g\kappa_5\,B_0}
  {\Lambda_\chi^2}{\rm Tr}\, \bar H_a H_a \gamma_\mu\gamma_5 A^\mu_{bc} 
  m^\xi_{cb} \nn\\ 
 &+& \frac{\delta_2}{\Lambda_\chi}{\rm Tr}\,\bar H_a H_b 
  \gamma_\mu\gamma_5 iv\cdot D_{bc} A_{ca}^\mu + 
  \frac{\delta_3}{\Lambda_\chi}
  {\rm Tr}\,\bar H_a H_b \gamma_\mu\gamma_5 i D^\mu_{bc} v\cdot A_{ca} \nn\\
 &+& \frac{g_1}{m_Q}{\rm Tr}\,\bar H_a H_b \gamma_\mu\gamma_5 A^\mu_{ba} +
  \frac{g_2}{m_Q}{\rm Tr}\,\bar H_a \gamma_\mu \gamma_5 H_b A^\mu_{ba}
  + \ldots \,, \label{Lagg}
\end{eqnarray}
where $D^\alpha_{bc}A^\beta_{ca}=\partial^\alpha A^\beta_{ba}+
[V^\alpha,A^\beta]_{ba}$ and $\Lambda_\chi=4\pi f$.  The ellipses here denote
terms linear in $m^{\xi}_-=\frac12 (\xi m_q \xi^\dagger -\xi^\dagger m_q\xi)$
which contribute to processes with more than one pion, as well as terms with
$(iv\cdot D)$ acting on an $H$.  For processes with at most one pion and $H$
on-shell the latter terms can be eliminated at this order, regardless of
their chiral indices, by using the equations of motion for $H$.  The $\kappa_i$
coefficients contain infinite and scale dependent pieces which cancel the
corresponding contributions from the one-loop $D^* \to D\pi$ diagrams.  For the
$\kappa_1$ and $\kappa_1'$ terms only the combination $\tilde\kappa_1 =
\kappa_1+\kappa_1'$ will enter in an isospin conserving manner here. (The
combination $\kappa_1-\kappa_1'$ will contribute an isospin violating
correction to $R^+_\pi$.) At a given scale $\mu$, the finite part of $\kappa_3$
can be absorbed into the definition of $g$.  The decays $D^*\to D\pi$ have
analytic contributions from $\tilde\kappa_1$ and $\kappa_5$ at order $m_q$.

For $m_Q=m_c$ the term in Eq.~(\ref{Lagg}) involving $g_1$ can be absorbed
into $g$ (this term only enters into a comparison with $B^*$ decays).  The term
$g_2$ breaks the equality of the $D^*D\pi$ and $D^*D^*\pi$ couplings.  Since we
only need the coupling $D^*D^*\pi$ in loops we can also absorb $g_2$ into the
definition of $g$.  Thus, our $g$ is defined as the $D^*D\pi$ coupling with
$1/m_Q$ corrections arising in relating it to the couplings for $D^*D^*\pi$ and
$B^{(*)}B^*\pi$.  

The terms in Eq.~(\ref{Lagg}) involving $\delta_2$ and $\delta_3$ contribute
to $D^*\to D\pi^0$, entering in a fixed linear combination with the tree level
coupling $g$ of the form $g - (\delta_2+\delta_3) v\cdot k/ \Lambda_\chi$. 
These are $\sim 10\%$ corrections for the decays $D^* \rightarrow D\pi$.  The
energy of the outgoing pion is roughly the same for all three decays, $v\cdot
k\sim .144 \,{\rm GeV}$.  Therefore, it is impossible to disentangle the
contribution of $\delta_{2,3}$ from that of $g$ for these decays, and the
extraction of $g$ presented here will implicitly include their contribution. 
For other processes involving pions with different $v\cdot k$ these
counterterms can give a different contribution.  This should be kept in mind
when this value of $g$ is used in a different context.

Techniques for one-loop calculations in heavy hadron chiral perturbation theory
are well known and will not be discussed here.  Dimensional regularization is 
used and the renormalized counterterms are defined by subtracting the pole 
terms $1/\epsilon -\gamma+\log{(4\pi)}$.  In doing this type of one-loop 
calculation an important integral is
\begin{eqnarray}
  \int\!\! {d^{4-2\epsilon}q \over (2\pi)^{4-2\epsilon} } { \mu^{2\epsilon}
 \over (q^2-m^2+i\varepsilon)\,2( q\cdot v - b + i\varepsilon) } = 
 {-i\, b \over (4\pi)^2} \left[ \frac1{\hat\epsilon} +\ln{(\frac{\mu^2}{m^2})} 
 + 2 - 2 F(\frac{m}{b})  \right] \,, \nn \\ \label{nastint}
\end{eqnarray}
where $1/\hat\epsilon=1/\epsilon-\gamma+\log{(4\pi)}$.
$F$ is needed for both positive and negative $b$, so 
\begin{eqnarray}
 F\left(\frac1{x}\right) &=& \left\{ \begin{array}{l}
  -\mbox{\large ${\sqrt{1-x^2}\over x}$} \left[\mbox{\large $\frac{\pi}2$} 
  -\tan^{-1}\left( \mbox{\large $\frac{x}{\sqrt{1-x^2}}$} \right) \right] 
  \quad\qquad |x|  \le 1 \\
  \phantom{-} \mbox{\large ${\sqrt{x^2-1}\over x}$}\: \ln{\left(x+\sqrt{x^2-1} 
  \,\right)} \quad\qquad\quad\ \ \ |x| \ge 1 \end{array} \right.\,. \label{F}
\end{eqnarray}
For $b>0$ the function $F$ was derived in \cite{manjen,FG} and agrees with the
above formula{\footnote{Eq.~(\ref{F}) for $F$ disagrees with
\cite{BG} for $x<0$.  Their $F(1/x)$ is even under $x\to -x$ making
Eq.~(\ref{nastint}) discontinuous at $\Delta=0$.  Furthermore, their $F$ has no
imaginary part corresponding to the physical intermediate state.}}.  For
$x=b/m < -1$ the logarithm in Eq.~(\ref{F}) has an imaginary part. This
corresponds to the physical intermediate state where a heavy meson of mass
$m_H$ produces particles of mass $m_H+b$ and $m$.  For the calculation
here the imaginary part only contributes from $F(m_\pi/(d_0-\Delta))$, and was
found to always be numerically insignificant.  Note that the real part of
$x\,F(1/x)$ is continuous everywhere, and differentiable everywhere except
$x=-1$.  Also $F(1)=F(-1)=0$.

The decays $D^{*0}\to D^0 \pi^0$ 
and $D^{*+}\to D^+ \pi^0$, and $D_s^* \to D_s \pi^0$ have decay rates
$\Gamma^1_{\pi}$, $\Gamma^2_{\pi}$, and $\Gamma^3_{\pi}$ given by
\begin{eqnarray}
  \Gamma^a_{\pi} &=& {g^2 \over 12\,\pi\,f^2} \left| {Z_{\rm wf}^a \over 
	Z_\pi^{a}} \right|^2 \: |{\vec k}^{\,a}_\pi|^3 \,. \label{ratepi}
\end{eqnarray} 
Here ${\vec k}^{\,a}_\pi$ is the three momentum of the outgoing pion,
$Z_\pi^{a}$ contains the vertex corrections, and $Z_{\rm wf}^a= \sqrt{
Z_{D^*}^a Z_{D}^a }$ contains the wavefunction renormalization for the $D^*$ 
and $D$.  When the ratio of $\Gamma^a_\pi$ to the $D^*\to D\gamma$ rate is
taken $Z_{\rm wf}^a$ will cancel out. However, $Z_{\rm wf}^a$ does contribute
to our predictions for the $D^*$ widths, where the ratio $Z_{\rm
wf}^a/Z_\pi^{a}$ will be kept to order $g^2$. 
\begin{figure}[t!]  
  \centerline{\epsfxsize=15.0truecm \epsfbox{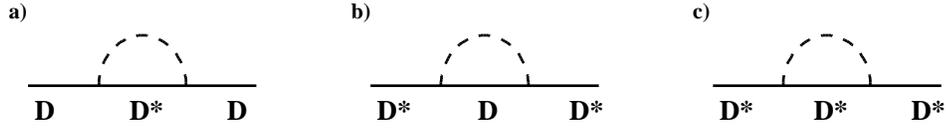}}
{  \caption{ $D$ and $D^*$ wavefunction renormalization graphs.  The dashed
	line represents a pseudo-Goldstone boson.} \label{fig_wf} }
\end{figure}  
The graphs in Fig.~\ref{fig_wf} give
\begin{eqnarray}
 Z^a_{D} &=& 1 + {g^2 \over (4\pi f)^2} (\lambda^i_{ab} \lambda^{i\dagger}_{ba})
  \bigg\{ [3 m_i^2-6(\Delta+d_0)^2] \log{(\frac{\mu^2}{m_i^2})} +\,
  3\, G_1(m_i,\Delta+d_0) \bigg\} \,, \nn\\ 
 Z^a_{D^*} &=& 1 + {g^2 \over (4\pi f)^2} (\lambda^i_{ab} 
  \lambda^{i\dagger}_{ba}) \bigg\{ [3m_i^2-4d_0^2-2(d_0-\Delta)^2] 
  \log{(\frac{\mu^2}{m_i^2})} + 2\, G_1(m_i,d_0)  \nn\\
 & &\qquad\qquad\qquad\qquad\quad +\, G_1(m_i,d_0-\Delta) 
  \bigg\} \,, \label{wfnren0} 
\end{eqnarray}
where $m_i$ is the mass of $\pi^i$, $d_0=\delta^{b3}\delta$ for $D^{*0}$ and
$D^{*+}$ decays and $d_0=(\delta^{b3}-1)\delta$ for $D^*_s$ decays.  The
notation in Eq.~(\ref{wfnren0}) assumes that we sum over $b=1,2,3$ and
$i=1,\ldots,8$. The logarithms agree with \cite{ccsu3}, except that we have
kept terms of order $\Delta^2\sim d_0^2$ in the prefactor since these terms are
enhanced for $m_q\to 0$.  Analytic terms of order $\Delta^2 \sim d_0^2$ are
neglected since they are higher order in our power counting.  The function
$G_1(a,b)$ in Eq.~(\ref{wfnren0}) has mass dimension $2$,
\begin{eqnarray}
 G_1(a,b) &=& \frac53\,a^2 + (4b^2-\frac43 a^2) F(a/b)
   + \frac43 (a^2-b^2) \frac{a}{b} F'(a/b) \,. \label{G1G2} 
\end{eqnarray}
It contains an analytic part proportional to $a^2$, and a non-analytic part 
which is a function of the ratio $b/a$.  In the limit $\Delta\to 0$
Eq.~(\ref{wfnren0}) gives $Z_{D}=Z_{D^*}$ in agreement with HQS.  To obtain
HQS in the finite part of the dimensionally regularized calculation of the
graphs in Fig.~\ref{fig_wf} it was necessary to continue the $D^*$ fields to
$d=4-2\epsilon$ dimensions (so the $D^*$ polarization vector
$\epsilon_\alpha=(1-\frac{\epsilon}3) \tilde\epsilon_\alpha$ where $\sum
\tilde\epsilon_\alpha^{\,*} \tilde\epsilon^\alpha = -3$).  

\begin{figure}[t]  
  \centerline{\epsfxsize=18.0truecm \epsfbox{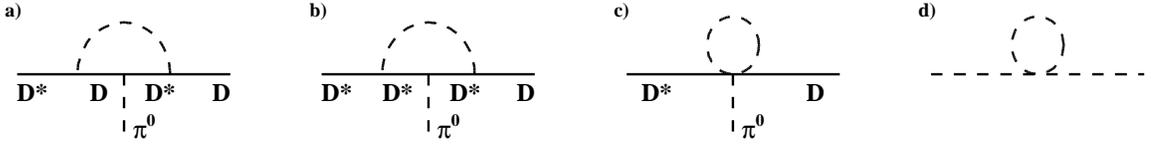}}
{  \caption{Nonzero one-loop vertex corrections for the decays $D^{*0}\to 
     D^0\pi^0$ and $D^{*+}\to D^+\pi^0$ (a,b,c) and the pseudo-Goldstone boson 
     wave function renormalization graph (d). } \label{fig_pi12} }
\end{figure}

For $a=1,2$ the decay proceeds directly so that at tree level $Z_{\rm wf}^{1,2}
/Z_\pi^{1,2}=1$.  At one loop we have non-zero vertex corrections from the
graphs in Fig.~\ref{fig_pi12}a,b,c.  As noted in \cite{ccsu3}, the two one-loop
graphs that contain a $D^{(*)}D^*\pi\pi$ vertex (not shown) vanish, and the
graph in Fig.~\ref{fig_pi12}c cancels with the $\pi^0$ wavefunction
renormalization in Fig.~\ref{fig_pi12}d (this is also true for $D^{*+}\to
D^0\pi^+$ and $D_s^*\to D_s\pi^0$).  For $a=1,2$ the vertex corrections are
\begin{eqnarray}
 {1\over Z_\pi^{\,a}} &=& 1 + {g^2 \over (4\pi f)^2} {\lambda^i_{ab}
  \lambda^1_{bb} \lambda^{i\dagger}_{ba} \over \lambda^1_{aa}} \Bigg\{ 
  \log{(\frac{\mu^2}{m_i^2})} \Big[ m_i^2 +\frac23(-d_1^2+d_1\,d_2+d_2^2
  -2 d_1\,d_0-2\,d_0^2) \Big] \nn \\
 & &+ 2\,F_1(m_i,d_1,d_2)-4\,F_1(m_i,d_1,d_0) \Bigg\} 
  + \varrho_{\pi}^a(\tilde\kappa_1,\kappa_5) \,, \label{pi12vertex}
\end{eqnarray}
where here $d_0=\delta^{b3}\delta$, $d_1=k\cdot v + d_0$, $d_2=-\Delta+d_0$,
and $k$ is the outgoing momentum of the $\pi^0$.  The coefficient of the
$m_i^2\log{(\mu^2/m_i^2)}$ term agrees with \cite{ccsu3}.  The function $F_1$
in Eq.~(\ref{pi12vertex}) has mass dimension $2$ and contains both analytic 
and non-analytic parts.  $\varrho_{\pi\,ct}^a$ contains the dependence of the 
rate on the (renormalized) counterterms $\tilde\kappa_1(\mu)$ and 
$\kappa_5(\mu)$.
\begin{eqnarray}\label{varrhopi}
  F_1(a,b,c) &=& -\frac76 \,a^2 + \frac{2}{3(b-c)}\left[
   b(a^2-b^2)F(a/b) - c(a^2-c^2)F(a/c) \right] \,,  \nn \\
 \varrho_{\pi}^{a=1,2} &=&  {m_\pi^2\over (4\pi f)^2} {\tilde\kappa_1(\mu) 
   \over 2}  \,.
\end{eqnarray}
We have ignored isospin violating counterterm corrections in $\varrho^{1,2}_\pi$.
In this case $\varrho_{\pi}^{1,2}$ do not depend on $\kappa_5$, and furthermore are
proportional to $m_\pi^2/(4\pi f)^2$, so these counterterms are small. 

\begin{figure}[tb]  
  \centerline{\epsfxsize=18.0truecm \epsfbox{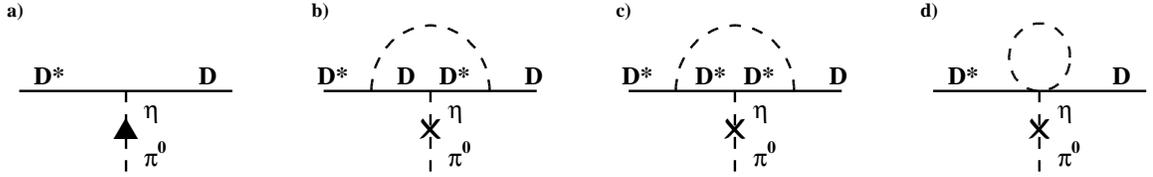}}
{  \caption{Nonzero vertex corrections  for the decay $D_s^*\to D_s\pi^0$ 
     which involve  $\pi^0-\eta$ mixing. The cross denotes leading order 
      mixing while the triangle denotes mixing at next to leading order.} 
 \label{fig_pi3} }
\end{figure}

The decay $D_s^*\to D_s\pi^0$ is isospin violating, and the leading
contribution occurs through $\eta-\pi^0$ mixing\cite{chowise}.  To first order
in the isospin violation the decay is suppressed at tree level by the mixing 
angle $\theta = (1.00\pm 0.05)\times 10^{-2} $ \cite{GLrev}
\begin{eqnarray}
 {1\over Z_\pi^{3}}= {(m_u-m_d)\over 2\,(m_s-\hat m)} = -\frac{2}{\sqrt{3}}
  \theta \simeq -\frac1{87.0} \,. \label{pi3vertex0}
\end{eqnarray}
Beyond tree level we have corrections to the $\eta-\pi^0$ mixing angle
parameterized by $\delta_{mix}=0.11$ \cite{GL} (Fig.~\ref{fig_pi3}a), loop
corrections to the $\eta-\pi^0$ mixing graph (Figs.~\ref{fig_pi3}b,c,d), as
well as loop graphs with decay directly to $\pi^0$ that occur in an isospin
violating combination (Figs.~\ref{fig_pi12}a,b).  The contribution of
Fig.~\ref{fig_pi3}d is again cancelled by the pseudo-Goldstone boson wave
function renormalization graph (Fig.~\ref{fig_pi12}d). 
Note that the decay $D_s^*\to D_s\pi^0$ cannot occur via a single virtual 
photon in the effective theory.  In the quark model, decay to the spin and
color singlet $\pi^0$ can occur if the single photon is accompanied by at least
two gluons (with suppression $\alpha/\pi\simeq 1/430$ \cite{chowise}).  We will
neglect the possibility of such a single photon mediated transition here. Thus,
\begin{eqnarray}  \label{pi3vertex}
 {1\over Z_\pi^{\,3}} &=& {(m_u-m_d) \over 2\,(m_s-\hat m)} \Bigg[ 1 + 
  \delta_{mix} + {g^2 \over (4\pi f)^2} { \lambda^i_{3b}
  \lambda^8_{bb} \lambda^{i\dagger}_{b3} \over \lambda^8_{33} } \Bigg\{  
  \log{(\frac{\mu^2}{m_i^2})} \Big[ m_i^2 +\frac23(-d_1^2+d_1\,d_2 \nn\\
 & & \qquad\qquad +d_2^2-2 d_1\,d_0-2\,d_0^2) \Big] + 
  2\,F_1(m_i,d_1,d_2)-4\,F_1(m_i,d_1,d_0) \Bigg\} \Bigg] \nn \\ 
 & &+{g^2 \over (4\pi f)^2} { \lambda^i_{3b} \lambda^1_{bb} 
  \lambda^{i\dagger}_{b3} \over (1/\sqrt{2})} \Bigg[ \tilde m_i^2
  \log{(\frac{\mu^2}{\tilde m_i^2})} + 2\,F_1({\tilde m}_i,d_1,d_2)-4\,
  F_1({\tilde m}_i,d_1,d_0) \Bigg] \nn\\
 & & + \varrho_{\pi}^3(\tilde\kappa_1,\kappa_5)  \,,  
\end{eqnarray}
where for $D_s^*$ decay $d_0=(\delta^{b3}-1)\delta$, $d_1=k\cdot v + d_0$, and
$d_2=-\Delta+d_0$.  The tilde on the mass, $\tilde m_i$, indicates that isospin
violation is taken into account.  Also note that
$\sqrt{2}\:\sum_{i,b}\lambda^i_{3b} \lambda^1_{bb} \lambda^{i\dagger}_{b3} \tilde
m_i^2 =m_{K^\pm}^2 - m_{K^0}^2$.  In Eq.~(\ref{pi3vertex}), the function
$\varrho_\pi^3$ depends on $\tilde\kappa_1$ and $\kappa_5$, and at leading 
order in the isospin violation is
\begin{eqnarray}
 \varrho_{\pi}^3 &=& {1\over (4\pi f)^2}\: {(m_u-m_d)\over 2(m_s-\hat m)}\: 
    \bigg[ (m_K^2-\frac{m_\pi^2}2)\,\tilde\kappa_1+(m_K^2-m_\pi^2)\,\kappa_5
    \bigg] + {(m_{K^\pm}^2-m_{K^0}^2) \over (4\pi f)^2}\:  \kappa_5 \,. \nn\\
\end{eqnarray}
In deriving this equation use has been made of $m_\pi^2 = 2B_0 m_u = 2B_0
m_d =2B_0 \hat m$, $m_K^2-m_\pi^2/2=B_0 m_s$, and
$m_{K^\pm}^2-m_{K^0}^2=(m_u-m_d) B_0$.


\section{$D^*_a \to D_a$ $\gamma$ decays}  \label{DDg}

 To describe $D^* \to D\gamma$, electromagnetic effects must be
included, so the Lagrangian in Eq.~(\ref{Lag0}) is gauged with a $U(1)$
photon field $B^\mu$.  With octet and singlet charges, $Q={\rm
diag}(\frac23,-\frac13,-\frac13)$ and $Q'= \frac23$ (for the $c$), the
covariant derivative ${\cal D_\mu}$ is \cite{ccem} 
\begin{eqnarray}
 {\cal D_\mu}\, \xi &=& \partial_\mu\xi + ieB_\mu[Q,\xi] \,,  \\
 {\cal D_\mu}\, H &=& \partial_\mu H +ieB_\mu(Q'H - H Q) - {\cal V_\mu}H \,,\nn
\end{eqnarray}
where the vector and axial vector currents are now 
\begin{eqnarray}
  {\cal V_\mu} &=& \frac12 (\xi^\dagger {\cal D_\mu}\xi + \xi{\cal D_\mu}
      \xi^\dagger) \,,\\ 
  {\cal A_\mu} &=& \frac{i}2 (\xi^\dagger{\cal D_\mu}\xi - \xi{\cal D_\mu} 
      \xi^\dagger) \,. \nn
\end{eqnarray}  However, this procedure does not induce a
coupling between $D^*$, $D$ and $B_\mu$ without additional pions.  Gauge
invariant contact terms should also be included, and it is one of these that
gives rise to the $D^*D\gamma$ coupling (and a $D^*D^*\gamma$ coupling)
\begin{eqnarray}
 {\cal L}_\beta = \frac{\beta\,e}{4}{\rm Tr}\,\bar H_a H_b 
  \sigma^{\mu\nu} F_{\mu\nu} Q^\xi_{ba} . \label{Lagb}
\end{eqnarray}
Here $\beta$ has mass dimension $-1$, $Q^\xi = \frac12 (\xi^\dagger Q \xi
+ \xi Q \xi^\dagger)$, and $F_{\mu\nu}=\partial_\mu B_\nu-\partial_\nu B_\mu$. 
The terms which correct this Lagrangian at order $m_q \sim 1/m_c$ have a 
similar form to those in Eq.~(\ref{Lagg})
\begin{eqnarray}
 \delta{\cal L}_\beta &=&\frac{\alpha_1\,B_0}{\Lambda_\chi^2}\frac{\beta\,e}{4}
  {\rm Tr}\, \bar H_a H_b \sigma^{\mu\nu}F_{\mu\nu} Q^\xi_{bc} m^\xi_{ca} 
  + \frac{\alpha_1'\,B_0}{\Lambda_\chi^2} \frac{\beta\,e}{4} {\rm Tr}\, 
  \bar H_a H_b \sigma^{\mu\nu}F_{\mu\nu}  m^\xi_{bc} Q^\xi_{ca} \nn\\
 &+& \frac{\alpha_3\,B_0}{\Lambda_\chi^2}\frac{\beta\,e}{4}{\rm Tr}\, 
  \bar H_a H_b \sigma^{\mu\nu} F_{\mu\nu} Q^\xi_{ba} m^\xi_{cc} 
  + \frac{\alpha_5\,B_0}{\Lambda_\chi^2}\frac{\beta\,e}{4}{\rm Tr}\,
  \bar H_a H_a \sigma^{\mu\nu} F_{\mu\nu} Q^\xi_{bc} m^\xi_{cb} \nn\\
 &+& \frac{\tau_2\,e}
  {4\Lambda_\chi^2}{\rm Tr}\,\bar H_a H_b \sigma^{\mu\nu} Q^\xi_{bc}
  iv\cdot D_{ca} F_{\mu\nu}
  + \frac{\tau_3\,e}{4\Lambda_\chi^2}
  {\rm Tr}\,\bar H_a H_b \sigma^{\mu\nu}  Q^\xi_{bc} i D^\mu_{ca} v^\lambda 
  F_{\nu\lambda} \nn\\
 &+& \frac{\beta_1}{m_Q}\frac{e}{4}{\rm Tr}\,\bar H_a H_b 
  \sigma^{\mu\nu} F_{\mu\nu} Q^\xi_{ba}
  + \frac{\beta_2}{m_Q}\frac{e}{4}{\rm Tr}\,
  \bar H_a \sigma^{\mu\nu} H_b  F_{\mu\nu} Q^\xi_{ba} \nn\\
 &-& \frac{e}{4\,m_Q} Q'\, 
  {\rm Tr}\, \bar H_a \sigma^{\mu\nu} H_a F_{\mu\nu} + \ldots\,. \label{Lagdb} 
\end{eqnarray} 
The ellipses denote terms that do not contribute for processes without
additional pions and/or can be eliminated using the equations of motion for $H$. 
For our purposes $Q^\xi$ and $m^\xi$ in Eq.~(\ref{Lagdb}) are diagonal so only
$\tilde \alpha=\alpha_1+\alpha_1'$ contributes.  The finite part of $\alpha_3$ will
be absorbed into the definition of $\beta$.  For $m_Q=m_c$, the $\beta_1$ term
can be absorbed, and we absorb the part of the $\beta_2$ term that contributes
to $D^*D\gamma$ since $D^*D^*\gamma$ only contributes in loops for us.  Thus,
$\beta$ is defined to be the $D^*D\gamma$ coupling at order $1/m_c$.  The last
term in Eq.~(\ref{Lagdb}) is the contribution from the photon coupling to the $c$
quark and has a coefficient which is fixed by heavy quark symmetry \cite{cg,aetal}. 
This term is numerically important. However, here the leading order contribution
to $D^*_a \to D_a \gamma$ is taken to be the $m_Q\to \infty$, $m_q\to 0$ effect
from Eq.~(\ref{Lagb}), so this $1/m_c$ term is part of the first order corrections.  
The $\tau_{1,2}$ terms are similar to the $\delta_{2,3}$ terms in Eq.~(\ref{Lagg}),
and appear with $\beta$ in the combination $\beta - (\tau_1+\tau_2) v\cdot k/
\Lambda_\chi^2$.  Here $\tau_1+\tau_2$ will have an infinite part necessary for
the one-loop renormalization.  Again it is not possible to isolate the finite part of
the $(\tau_1+\tau_2)$ contribution from that of $\beta$, so the extraction at this
order includes the renormalized $\tau_{1,2}$ with $v\cdot k \sim 0.137\,{\rm
GeV}$.  

\begin{figure}[t!]  
  \centerline{\epsfxsize=18.0truecm \epsfbox{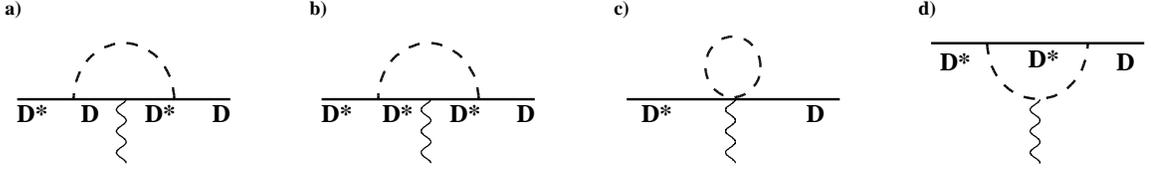}}
 {\caption{Nonzero vertex corrections for the decays $D^*\to D\gamma$. }  
  \label{diag_gam} }
\end{figure}

The decays $D^{*0}\to D^0\gamma$, $D^{*+}\to D^+\gamma$, and $D_s^*\to 
D_s\gamma$ have decay rates $\Gamma_\gamma^1$, $\Gamma_\gamma^2$, and 
$\Gamma_\gamma^3$ given by
\begin{eqnarray}
 \Gamma^a_\gamma &=& \frac{\alpha}3\: |\mu_a|^2\: |{\vec k}^{\,a}_\gamma|^3, 
  \qquad \qquad
  \mu_a = Z_{\rm wf}^a\left( \beta \,{ Q_{aa} \over Z_\gamma^a } +{Q' \over m_c}
  \right) \,, \label{rateg}
\end{eqnarray}
where $\alpha \simeq 1/137$, ${\vec k}^{\,a}_\gamma$ is the three momentum of
the outgoing photon, and the wavefunction renormalization, $Z_{\rm wf}^a$, is
given by Eq.~(\ref{wfnren0}).  To predict the $D^*$ widths, $Z_{\rm
wf}^a/Z_\gamma^a$ is kept to order $g^2$ and we take $Z_{\rm wf}^a\times
1/m_c = 1/m_c$. The vertex correction factor $Z_\gamma^a$ has nonzero 
contributions
from the graphs in Fig.~\ref{diag_gam}.  Note that the two one-loop graphs that
contain a $D^{(*)}D^*\pi\gamma$ vertex (not shown) do not contribute
\cite{ccsu3}.  
Furthermore, the graph in Fig.~\ref{diag_gam}b has no contribution from the
$D^*D^*\gamma$ coupling which arises from gauging the lowest order 
Lagrangian in Eq.~(\ref{Lag0}).  Thus
\begin{eqnarray}
 {1\over Z_\gamma^{\,a}} &=& 1 + {g^2 \over (4\pi f)^2} {\lambda^i_{ab} Q_{bb} 
  \lambda^{i\dagger}_{ba} \over Q_{aa}} \Bigg\{ 
  \log{(\frac{\mu^2}{m_i^2})} \Big[ m_i^2 +\frac23(-d_1^2+d_1\,d_2+d_2^2
  -2 d_1\,d_0-2\,d_0^2) \Big] \nn \\*
 & &\qquad\qquad\qquad\qquad\qquad
    + 2\,F_1(m_i,d_1,d_2)-4\,F_1(m_i,d_1,d_0) \Bigg\}\nn \\
 & &- {1\over 
  (4\pi f)^2} {[\lambda^{i\dagger},[Q,\lambda^i]]_{aa} \over 2\,Q_{aa}} 
  \bigg[m_i^2\,\log{(\frac{\mu^2}{m_i^2})+m_i^2 \bigg] } \nn\\*
 & &+ {4\,g^2 \over (4\pi f)^2} {(\lambda^i_{ab}\lambda^{i\dagger}_{ba}) 
  \,q^i \over \beta Q_{aa}} \Big[ -
  \log{(\frac{\mu^2}{m_i^2})}(d_0+\frac{k\cdot v}2) + 
  F_2(m_i,d_0,k\cdot v) \Big] \nn\\*
 & & +  \varrho_{\gamma}^a(\tilde\alpha_1,\alpha_5) \,, \label{vertexg}
\end{eqnarray}
where $q^i$ is the charge of meson $\pi^i$, $k$ is now the outgoing photon
momentum, and the $d_i$ are as above (again they differ depending on whether it
is $D^*_s$ or one of $D^{*0}$, $D^{*+}$ that is decaying).  The coefficients of
the $m_i^2\log{(\mu^2/m_i^2)}$ terms agree with \cite{ccsu3}. The new function
$F_2$ has mass dimension $1$,
\begin{eqnarray}
 F_2(a,b,c) &=& -2\,b-c-\frac{2a^2}{c}\int^{a/(b+c)}_{a/b} dt {F(t)\over t^3} 
  \nn\\
  &=&-2\,b-c- \frac{2a^2}{c} \left[ { 1\over 4x^2 }- {F(x)\over 2\,x^2} 
  - {F(x)^2 \over 4\,(x^2-1) } \right] \Bigg|_{x=a/b}^{x=a/(b+c)} \,.
\end{eqnarray}
It contains an analytic part proportional to
$2d_0+v\cdot k$, and a non-analytic part which is a function of $\delta/m_i$
and $v\cdot k/m_i$.  In Eq.~(\ref{vertexg}) $\varrho^a_\gamma$ contains the 
dependence of the rate
on the (renormalized) counterterms $\tilde\alpha_1(\mu)$ and $\alpha_5(\mu)$. 
Assuming isospin to be conserved we have
\begin{eqnarray}
 \varrho_{\gamma}^{\,a=1,2} &=& {m_\pi^2\: \tilde\alpha_1 \over 
  2\,(4\pi f)^2}\: - {(m_K^2-m_\pi^2)\:\alpha_5 \over 3\,Q_{aa}\,(4\pi f)^2} 
  \,,\nn\\
 \varrho_{\gamma}^3 &=& {(2\,m_K^2-m_\pi^2)\:\tilde \alpha_1 \over 
  2\,(4\pi f)^2} +{(m_K^2-m_\pi^2)\:\alpha_5\over (4\pi f)^2} \,. 
\end{eqnarray}

By examining Eqs.~(\ref{wfnren0}), (\ref{pi12vertex}), (\ref{pi3vertex}), and
(\ref{vertexg}) we can get an idea of the size of the various one-loop
corrections to $\Gamma_\pi^a$ and $\Gamma_\gamma^a$.  With our power
counting $\Delta \sim \delta \sim v\cdot k \sim m_q \sim m_i^2$ so we can 
consider expanding in $\Delta/m_i$, $\delta/m_i$, and $v\cdot k/m_i$
\begin{eqnarray}
  G_1(m_i,b) &=& \frac{m_i^2}3 \left[ 1- \frac{6\pi b}{m_i} + 
    \frac{16 b^2}{m_i^2} + \ldots \right] \:\,, \nn\\
  F_1(m_i,b,c) &=& -\frac{m_i^2}2 \left[ 1 - \frac{\pi(b+c)}{m_i} +
    \frac{16 (b^2+b c+c^2)}{9 m_i^2} +\ldots \right]  \:\,,   \nn\\
  F_2(m_i,d_0,k\cdot v) &=& -\pi m_i \left[ 1 -\frac{3d_0^2+3 d_0 k\cdot v 
    +(k\cdot v)\,^2}{6 m_i^2}+\ldots 
    \right] \:\,.  \label{expns}
\end{eqnarray}
The leading terms in $G_1$ and $F_1$ are $m_q$ corrections to the rates.  The
second terms are order $m_q^{3/2}$ and $\sqrt{m_q}/m_c$, and can be kept since
they are unambiguously determined at the order we are working.  The third and
remaining terms in $G_1$ and $F_1$ are subleading in our power counting.  The
term $-\pi m_i$ in $F_2$ is the formally enhanced contribution discovered in
\cite{aetal}.  Note that there are no contributions to $F_2$ proportional to
$\delta$ or $k\cdot v$.  The second term in $F_2$ in Eq.~(\ref{expns}) has
contributions of order $m_q^{3/2}$, $\sqrt{m_q} k\cdot v$, and $(k\cdot
v)^2/\sqrt{m_q}$ which again can be kept since they are unambiguously 
determined.

The above power counting is sensible when $m_i$ is $m_K$ or $m_\eta$.  We
know that numerically $m_{\pi} \sim \Delta \sim \delta \sim k\cdot v$, so for
$m_i=m_\pi$ the series in Eq.~(\ref{expns}) are not sensible.  In \cite{aetal} the
term $-\pi m_\pi$ in $F_2$ was found to be important, so we want to keep
corrections with $m_\pi$ dependence.  Therefore, instead of expanding the
non-analytic functions we choose to keep them in the non-analytic forms given
in the Appendix. Numerically the one-loop corrections to $\Gamma_\pi^1$ and
$\Gamma_\pi^2$ are very small; with $g=1$ they are of order $\sim 2\%$.  For
$\Gamma_\pi^3$, $\delta_{mix}$ is a $11\%$ correction to the tree level result in
Eq.~(\ref{pi3vertex0}).  Individually the terms proportional to $g^2 F_1$ and $g^2
\log{(\mu/m_q)}$ in Eq.~(\ref{pi3vertex}) are $\sim 10\%$ corrections for $g=1$. 
However, the loops graphs with $\eta-\pi^0$ mixing tend to cancel those without
$\eta-\pi^0$ mixing leaving a $\sim 2\%$ correction.  The one-loop corrections to
$\Gamma_\gamma^a$ are larger, for instance the graph in Fig.~\ref{diag_gam}c
gives sizeable corrections that are not suppressed by $g^2$.  Corrections to the
coefficient of the leading $g^2/\beta$ term range from $\sim 3\%$ for $D^*_s$ and
$\sim 20\%$ for $D^{*0}$ decay, to $\sim 50\%$ for the $D^{*+}$.  (The latter
percentage is large because the only contribution for this decay comes from a
charged pion in the loop of Fig.~\ref{diag_gam}d.) Corrections proportional to
$g^2$ are only sizeable for $D^*_s\to D_s\gamma$ where they are $\sim 10\%$
for $g=1$.


\section{Extraction of the couplings $\lowercase{g}$ and $\beta$}  \label{extract}

Using the calculation of the decay rates from the previous section, the
couplings $g$ and $\beta$ can be extracted from a fit to the experimental data.
Input parameters include $m_c=1.4\,{\rm GeV}$ \cite{glkw,gs}, the meson masses
from \cite{PDG96}, $\Delta=m_{D^*}-m_D=0.142\,{\rm GeV}$,
$\delta=m_{D_s^{(*)}}-m_{D^{(*)}}=0.100\,{\rm GeV}$, and $v\cdot k$ which is
determined from the masses.  When isospin is assumed we use
$m_K=0.4957\,{\rm GeV}$ and $m_\pi=0.1373\,{\rm GeV}$.  $f$ is extracted from
$\pi^-$ decays. At tree level we use $f=f_\pi=0.131\,{\rm GeV}$ \cite{PDG96},
while when loop contributions are included we use the one-loop relation between
$f$ and $f_\pi$ \cite{GL} to get $f=0.120\,{\rm GeV}$.  The ratio of the decay
rates $\Gamma_\gamma^a$ and $\Gamma_\pi^a$ are fit to the experimental 
numbers
\begin{eqnarray}
 {{\cal B}(D^{*0}\to D^0\gamma) /{\cal B}(D^{*0}\to D^0\pi^0)} &=&
   0.616 \pm 0.076 \ \cite{PDG96} \,, \nn\\
 {{\cal B}(D^{*+}\to D^+\gamma) / {\cal B}(D^{*+}\to D^+\pi^0)} &=&
   0.055 \pm 0.017 \ \cite{CLEO1} \,, \nn\\
 {{\cal B}(D^{*\ }_s\to D_s\pi^0) / {\cal B}(D^{*\ }_s\to D_s\gamma)} &=&
   0.062 \pm 0.029 \ \cite{PDG96} \,, \label{exptratios}
\end{eqnarray}
where the errors combine both statistical and systematic. Using the 
masses $m_{D^{*0}}$, $m_{D^{*+}}$, $m_{D^{*}_s}$, and mass splittings
$m_{D^{*0}}-m_{D^0}$, $m_{D^{*+}}-m_{D^+}$, $m_{D^{*}_s}-m_{D_s}$ from
\cite{PDG96} gives the momentum ratios that appear in 
$\Gamma_\gamma^a/\Gamma_\pi^a$:
\begin{eqnarray}
  {|\vec k_\gamma^1|^3\over |\vec k_\pi^1|^3} = 32.65 \pm 0.44, \qquad  
  {|\vec k_\gamma^2|^3\over |\vec k_\pi^2|^3} = 45.2 \pm 1.0, \qquad  
  {|\vec k_\gamma^3|^3\over |\vec k_\pi^3|^3} = 24.4 \pm 1.5 \,. \label{exptmom}
\end{eqnarray}
The errors here are clearly dominated by those in Eq.~(\ref{exptratios}).
Equating the numbers in Eq.~(\ref{exptratios}) to the ratio of rates from
Eqs.~(\ref{ratepi}) and (\ref{rateg}) gives a set of three nonlinear equations
for $g$ and $\beta$ (where we ignore for the moment the unknown counterterms). 
In general any pair of these equations will have several possible solutions. 
To find the best solution we take the error from Eq.~(\ref{exptratios}) and
minimize the $\chi^2$ for the fit to the three measurements. We will restrict
ourselves to the interesting range of values, $0 < g < 1$ and $0 < \beta < 6$,
discarding any solutions that lie outside this range. (The sign of $g$ will not
be determined here since it only appears quadratically in $\Gamma_\pi^a$ and
$\Gamma_\gamma^a$.)

{
\begin{table}[!t]
\begin{center}
\begin{tabular}{lcccccccccc}  
 \multicolumn{2}{c}{Order} & $g$ & $\beta$ & $\chi^2$ & &
      $g$ &  $\beta$ & $\chi^2$ &  \\ \hline 
  tree level &&  \multicolumn{2}{c}{$\beta/g=3.6$} & $30.$  &   &  &  \\
  +$Q'/m_c$ + one-loop with $\sqrt{m_q}$ && $0.23$ & $0.89$ & $4.3$
   && $0.45$ & $2.8$ & $3.7$ & \\
  + chiral logs && $0.25$ & $0.78$ & $4.1$ && $0.56$ & $3.2$ & $1.4$ & \\
  one-loop with nonzero $\Delta$,$\delta$,$v\cdot k$, 
   && $0.25$ & $0.86$ & $3.9$ && $0.83$ & $6.0$ & $2.5$ & \\[-2pt]
  \quad without analytic $m_q$ terms \\
  order $m_q\sim 1/m_c$ with  && $0.265$ & $0.85$ & $3.0$ & 
  & $0.756$ & $4.9$ & $3.9$ & \\
  \quad $\tilde\kappa_1=\kappa_5=\tilde\alpha_1=\alpha_5=0$
\end{tabular}
\end{center}
{  \caption{Solutions for $g$ and $\beta({\rm GeV^{-1}})$ which minimize the 
  $\chi^2$ associated with a fit to the three ratios in Eq.~(\ref{exptratios}). There 
  are two solutions in the region of interest.}
\label{table_solns} }
\end{table} }
To test the consistency of the chiral expansion we will first check how the
extraction of $g$ and $\beta$ differs at various orders.  The results are given
in Table~\ref{table_solns}.  At tree level only the ratio $\beta/g$ is
determined, and the $\chi^2$ is rather large.  We might next consider adding
the contribution from the chiral loop corrections to $D^*\to D\gamma$ which go
as $\sqrt{m_q}$.  However, this does not lead to a consistent solution between
the three data points unless $\beta$ is negative.  This signals the importance
of the $Q'/m_c$ contribution in Eq.~(\ref{rateg}) corresponding to a nonzero
heavy quark magnetic moment.  Adding this contribution gives the results in the
second row of Table~\ref{table_solns}, where there are now two solutions with
similar $\chi^2$ in the region of interest.  Adding the chiral logarithms,
$m_q\log{(\mu/m_q)}$, at scale $\mu=1\,{\rm GeV}$ gives the solutions in the
third row.  Taking nonzero $\delta$, $\Delta$, and $v\cdot k$ in the
non-analytic functions $F_1$ and $F_2$ gives the solutions in the fourth row of
Table~\ref{table_solns}, where the value of $g$ in the second solution has
increased by $\sim 50\%$.  For these two solutions only the analytic $m_i^2$
dependence has been neglected.  Finally, the solutions in row five include the
analytic $m_i^2$ dependence with the counterterms set to zero (at $\mu=1\,{\rm
GeV}$).  The uncertainty associated with these counterterms will be
investigated below.   It is interesting to note that the extracted value of $g$
in the second column of Table~\ref{table_solns} changes very little with the
addition of the various corrections.

\begin{figure}[!t]  
  \centerline{\epsfxsize=9.0truecm \epsfbox{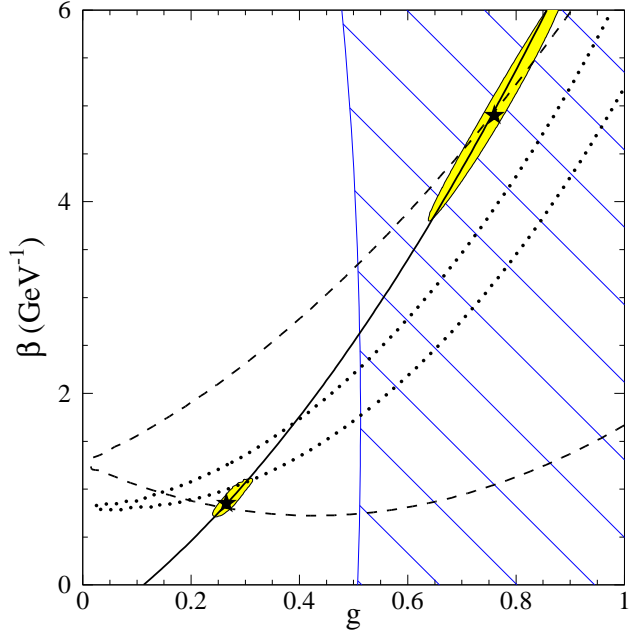} }
 { \caption{Solution contours in the $g$-$\beta$ plane for the situation in
row 5 of Table~\ref{table_solns}. The solid, dashed, and dotted lines
correspond to solution lines for the $D^{*0}$, $D^{*+}$, and $D^*_s$ decay rate
ratios respectively.  The stars correspond to the minimal $\chi^2$ solutions 
and the shaded regions correspond to the 68\% confidence level 
of experimental error in the fit.  The hatched region is excluded by the 
experimental limit $\Gamma(D^{*+}) < 0.13\,{\rm MeV}$ \cite{ACCMOR}.} 
\label{fig_solns} }
\end{figure}
\begin{figure}[!t]  
  \centerline{\epsfxsize=8.0truecm \epsfbox{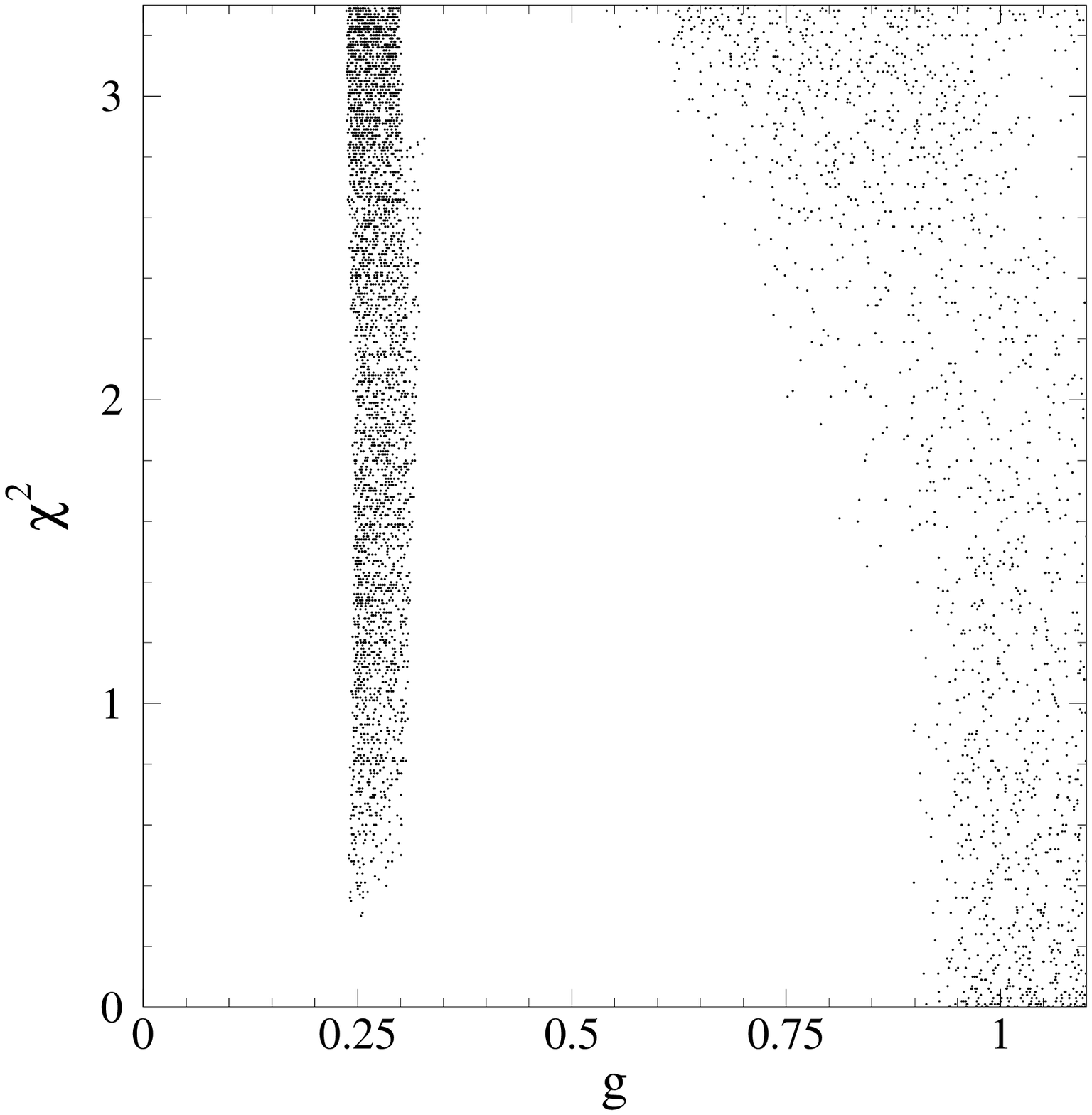}
	\epsfxsize=8truecm \epsfbox{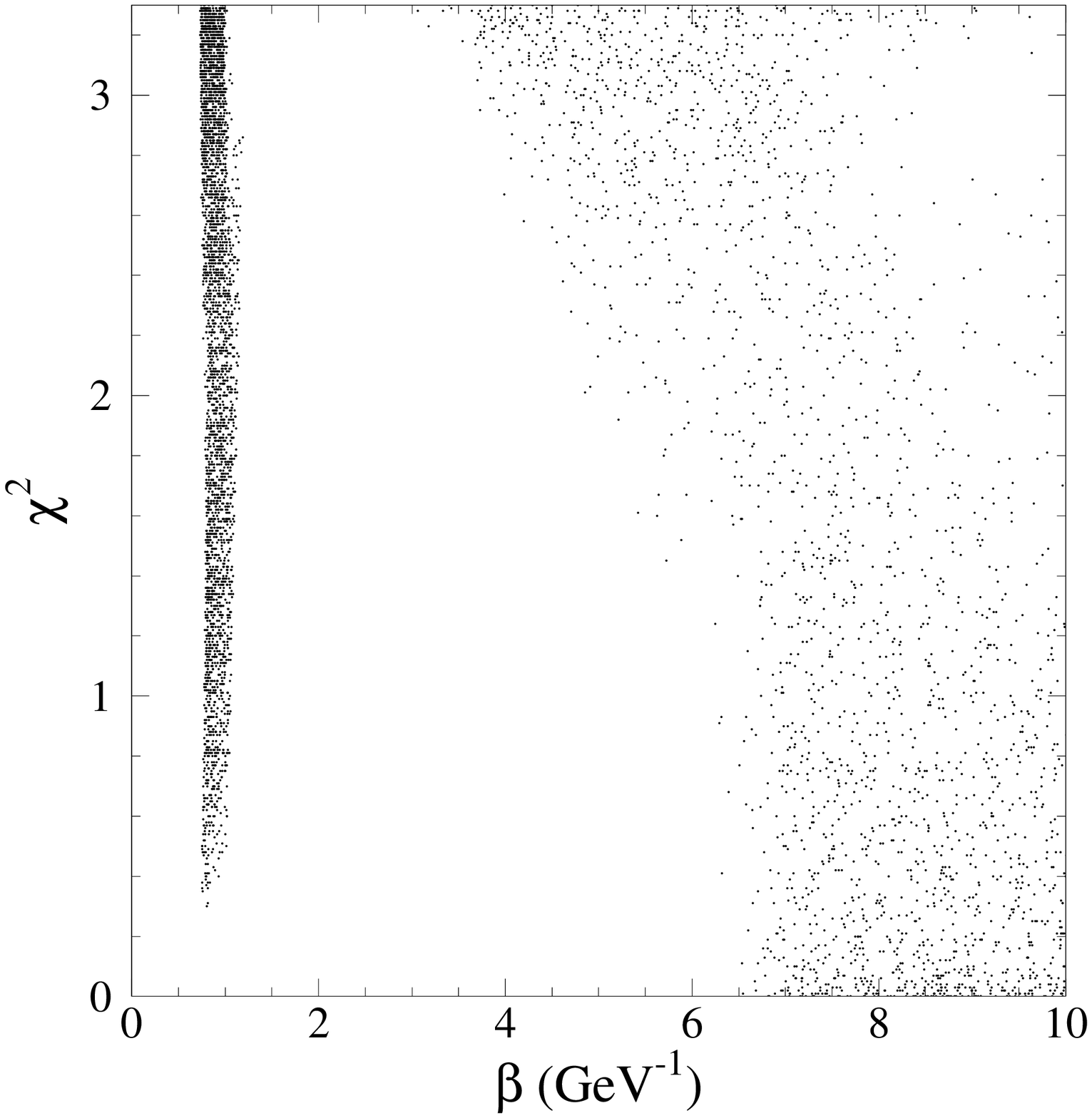} }
 {  \caption{Effect of the order $m_q$ counterterms ($\tilde\kappa_1$,
$\kappa_5$, $\tilde\alpha_1$, and $\alpha_5$) on the solutions in
Eq.~(\ref{soln1}).  The counterterms are taken to be randomly distributed with
$-1 < \tilde\kappa_1,\kappa_5 < 1$, $-2 < \alpha_1,\alpha_5 < 2$.  For each set
of counterterms $g$ and $\beta$ were determined at the new minimal $\chi^2$.
$5000$ sets were generated near each of the two solutions.} \label{fig_rand} }
\end{figure}

One can see more clearly how these solutions are determined by looking at
Fig.~\ref{fig_solns}.  The central value for each ratio of decay rates in
Eq.~(\ref{exptratios}) gives a possible contour in the $g$-$\beta$ plane, as
shown by the solid ($D^{*0}$), dashed ($D^{*+}$), and dotted ($D_s^*$) lines. 
An exact solution for two of the ratios occurs at the intersection of two of
these contour lines.  However, a good solution for all three ratios requires a
point that is close to all three lines.  The solutions in the fifth row of
Table~\ref{table_solns} are indicated by stars in Fig.~\ref{fig_solns}.    The
size of the experimental uncertainties can be seen in the 68\% confidence level
ellipses which are shown as shaded regions in the figure (for two degrees of
freedom they correspond to $\chi^2 \le \chi^2_{min}+2.3$).  These regions are
centered on the solid line since the $D^{*0}$ ratio has the smallest
experimental error.  The errors in Eq.~(\ref{exptratios}) give the following
one sigma errors on the two solutions
\begin{eqnarray}
 g&=&0.265^{+.036}_{-.018} \ \ \ \beta=0.85^{+.21}_{-.10}\,{\rm GeV^{-1}} \:, \nn\\
 g&=&0.756^{+0.028}_{-0.027} \ \ \ \beta=4.90^{+.27}_{-.26}\,{\rm GeV^{-1}} \:.
  \label{soln1}
\end{eqnarray}
Both solutions fit the first two ratios in Eq.~(\ref{exptratios}), but do not
do as well for the third.  Minimizing the $\chi^2$ has biased against the third
ratio as a result of its large experimental error.  For this ratio the
$g=0.265$ and $g=0.76$ solutions give values which are $4$ and $13$ times too
small respectively.  For the first solution it is possible to improve the fit
to the third ratio with reasonably sized counterterms.  For instance, simply
taking $\tilde\alpha_1=2$ gives ${{\cal B}(D^{*\ }_s\to D_s\pi^0) / {\cal
B}(D^{*\ }_s\to D_s\gamma)}= 0.036$.  As we will see below, a large $g$
solution with $\chi^2 \lesssim 1$ is only possible if $g$ increases to $\sim
0.9$ and $\beta$ increases to $\sim 6.0\,{\rm GeV^{-1}}$ (c.f. 
Fig.~\ref{fig_rand}).

The experimental limit $\Gamma(D^{*+}) < 0.13\,{\rm MeV}$ \cite{ACCMOR}
translates into an upper bound on the value of $g$.  Since ${\cal B}(D^{*+}\to
D^+\gamma)$ is small, this bound is almost $\beta$ independent and to a good
approximation is
\begin{eqnarray}
  g < 0.52\,\sqrt{\,\sqrt{1+3.01\,x}-1} \qquad\qquad 
     x=\Gamma(D^{*+})^{\:\rm limit}/(0.13\,{\rm MeV}) \,.  \label{ulmt}
\end{eqnarray}
For the situation in row five of Table~\ref{table_solns} this excludes the
hatched region in Fig.~\ref{fig_solns}.  The limit on $\Gamma(D^{*+})$
therefore eliminates the $g\simeq 0.76$ solution at the two sigma level.  Since
this limit has not been confirmed by other groups it would be useful to have
further experimental evidence that could exclude this solution.  

The central values in Eq.~(\ref{soln1}) have uncertainty associated with
the parameter $m_c$.  Taking $m_c=1.4\pm0.1\,{\rm GeV}$ gives $0.25 < g < 0.28$
and $0.79\,{\rm GeV^{-1}} < \beta < 0.93\,{\rm GeV^{-1}}$ for the first
solution, and $0.72 < g < 0.80$ and $4.6\,{\rm GeV^{-1}} < \beta < 5.3\,{\rm
GeV^{-1}}$ for the second solution (in both cases the $\chi^2$ changes very
little).  There is also ambiguity in the solution in Eq.~(\ref{soln1}) due to
the choice of scale $\mu$ (ie., the value of the counterterms $\alpha_1$,
$\alpha_5$, $\tilde\kappa_1$ and $\kappa_5$).  Increasing $\mu$ to $1.3\,{\rm
GeV}$  gives solutions $(g=0.28, \beta=0.91\,{\rm GeV^{-1}}, \chi^2=1.4)$ and
$(g=0.78,\beta=5.0\,{\rm GeV^{-1}},\chi^2=4.1)$, while decreasing $\mu$ to
$0.7\,{\rm GeV}$ gives solutions $(g=0.25, \beta=0.83\,{\rm GeV^{-1}},
\chi^2=3.7)$ and $(g=0.72, \beta=4.7\,{\rm GeV^{-1}}, \chi^2=3.1)$.  Note that
the $\chi^2$ of the second solution remains large, while the $\chi^2$ of the
first solution is reduced significantly by an increased scale.

Another method of testing the effect of the unknown counterterms
$\tilde\alpha_1$, $\alpha_5$, $\tilde\kappa_1$ and $\kappa_5$ is to take their
values at $\mu=1\,{\rm GeV}$ to be randomly distributed within some reasonable
range of values.  We take $-1 < \tilde\kappa_1, \kappa_5 < 1$ and $-2 <
\tilde\alpha_1, \alpha_5 < 2$, with the motivation that the counterterms change
the tree level value of $Z_{\pi}^a$ and $Z_\gamma^a$ by less than $30\%$, and
give corrections that are not much bigger than those from the one-loop graphs. 
Near each of the two solutions $5000$ values of $g$ and $\beta$ were then
generated by minimizing the $\chi^2$.  This gives the distributions in
Fig.~\ref{fig_rand}.  The solution with $g=0.265$ and $\beta=0.85\,{\rm
GeV^{-1}}$ has fairly small uncertainty from the counterterms.  The $g=0.76$,
$\beta=4.9\,{\rm GeV^{-1}}$ solution has much larger uncertainty because the
corresponding contour lines in Fig.~\ref{fig_solns} are almost parallel.  For
this solution the upper bounds are determined by the limits of a few 
${\rm MeV}$ \cite{PDG96} on the $D^*$ widths.  
From this analysis we estimate the theoretical uncertainty of the solutions in 
Eq.~(\ref{soln1}) to be roughly
\begin{eqnarray}
 g&=&0.265^{+0.05}_{-0.02} \ \ \ \beta=0.85^{+0.3}_{-0.1}\:{\rm GeV^{-1}} \:, \nn \\
 g&=&0.76^{+0.2}_{-0.1} \ \ \ \beta=4.9^{+5.0}_{-0.7}\:{\rm GeV^{-1}} \,,
  \label{soln2}
\end{eqnarray}
at this order in chiral perturbation theory.  The errors on $g$ and $\beta$ are
positively correlated since the values of $g$ and $\beta$ are constrained in
one direction by the small error on the $D^{*0}$ rate ratio in 
Eq.~(\ref{exptratios}).  

From Eq.~(\ref{ulmt}) and Fig.~\ref{fig_rand}, we see that if the error in
\begin{eqnarray}
{{\cal B}(D^{*\ }_s\to D_s\pi^0)\over {\cal B}(D^{*\ }_s\to D_s\gamma)}
\end{eqnarray}
can be decreased by a factor of two, in conjunction with a limit of $\Gamma(D^{*+})
\lesssim 0.6\,{\rm MeV}$ then this could provide strong evidence that the $g=0.76$
solution is excluded.  On the other hand if the central values of the second and
third ratios in Eq.~(\ref{exptratios}) decrease, then a width measurement or
stronger limit on $\Gamma(D^{*+})$ will be needed to distinguish the two solutions.

Using the extracted values of $g$ and $\beta$ gives the widths shown in
Table~\ref{table_widths}.  The couplings were extracted at one-loop and order
$m_q \sim 1/m_c$, so the predictions for the $D^*$ widths are made at this
order.  The experimental uncertainty in the $D^*$ widths is estimated by
setting $g$ and $\beta$ to the extremal values in Eq.~(\ref{soln1}), which
gives the range shown in the second and fourth rows of the Table.  The
uncertainty from the unknown counterterms in the third and fifth rows is
estimated in the same way using the uncertainties from Eq.~(\ref{soln2}).  Note
that for the $g=0.265$ solution the $D_s^*$ width is small due to a delicate
cancellation in $\mu_3$ resulting from setting $Z_{\rm wf}^a \times 1/m_c =
1/m_c$.  Keeping $Z_{\rm wf}^a/m_c$ to order $m_q$ gives a $D^*_s$ width of
$0.28\,{\rm keV}$ with a range of $0.1-0.4\,{\rm keV}$ for both the 
experimental and the counterterm uncertainties.

Making use of HQS allows us to predict the width of the $B^*$ mesons from their
dominant mode $B^* \to B\gamma$.  Eq.~(\ref{rateg}) gives the rate for $B^*\to
B\gamma$ with $Q'=-1/3$ and $m_c\to m_b$.  Since the couplings $\beta_1$ and
$\beta_2$ are unknown these rates can not be determined at order $1/m_{c,b}$,
but we can include the order $m_q$ corrections. The $B$ meson masses are
taken from \cite{PDG96} and we use $m_b=4.8\,{\rm GeV}$ \cite{glkw}.  We set
$\delta=0.047\,{\rm GeV}$\, and $\Delta=k\cdot v=0$, but since the contribution
$Q'/m_b$ in Eq.~(\ref{rateg}) is numerically important it is kept in our estimate. 
For comparison the widths obtained with the $g=0.76$ and $\beta=4.9\,{\rm
GeV^{-1}}$ solution are also shown.  
\begin{table}[!t]
\begin{center}
{
\begin{tabular}{lccc|ccc}  
 Widths (${\rm keV}$)  &
  $D^{*0}$ & $D^{*+}$ & $D_s^{*}$ & $B^{*+}$ & $B^{*0}$ & $B_s^{*}$ \\ \hline 
  $g=0.265$,  & 18 & 26 & 0.06 & $\sim$ 0.06 & $\sim$ 0.03 & $\sim$ 0.04 \\
  \quad $\beta=0.85\,{\rm GeV^{-1}}$ &&&&&\\
  uncertainty from  & 16 - 24 & 23 - 35 & 0.01 - 0.13\ \  & 
    $-$ & $-$ & $-$ \\
 \quad experiment &&&&&\\
  uncertainty from   & 16 - 27 & 22 - 39 & 0.04 - 0.13\ \ &
    $-$ & $-$ & $-$ \\ 
 \quad counterterms &&&&&\\ \hline 
  $g=0.76$,  &  323 &  448 & 103 & 
    $\sim$ 2.1 & $\sim$ 2.0 & $\sim$ 1.6 \\
  \quad $\beta=4.9\,{\rm GeV^{-1}}$ &&&&&\\
  uncertainty from & 285 - 367 & 396 - 508 & 83 - 128\phantom{7} & 
    $-$ & $-$ & $-$ \\
 \quad experiment &&&&&\\
  uncertainty from  & 215 - 1318 & 281 - 1157 & 53 - 1078 \ \ &
    $-$ & $-$ & $-$ \\ 
 \quad counterterms &&&&&\\ \hline 
\end{tabular} }
\end{center}
{ \caption{Predicted widths in ${\rm keV}$ for the $D^*$ and $B^*$ mesons.  The
experimental and counterterm ranges are determined by the extremal values of
$g$ and $\beta$ in Eqs.~(\ref{soln1}) and (\ref{soln2}).  For $g=0.265$ the
$D_s^*$ width is small due to a delicate cancellation in $\mu_3$ as
explained in the text. The uncertainty in the $B^*$ widths is large due to
unknown $1/m_{c,b}$ corrections.}  \label{table_widths} }
\end{table} 

As a final comment, we note that heavy meson chiral perturbation theory can also
be used to examine excited $D^{(*)}$ mesons, such as the p-wave states, $D_0^*$,
$D_1^*$, $D_1$, and $D_2^*$ \cite{excited1,excited2,Kilian,HMreview}.  To do so,
explicit fields for these particles may be added to the Lagrangian giving a new
effective theory. For interactions without external excited mesons (such as the
ones considered here) these new particles can then contribute as virtual particles. 
However, since we have not included these heavier particles they are assumed to
be `integrated out', whereby such contributions are absorbed into the definitions of
our couplings.


\section{Summary}  \label{discuss}

For the $D^{*0}$, $D^{*+}$, and $D^*_s$, the decays $D^* \to D\pi$ and $D^* \to
D\gamma$ are well described by heavy meson chiral perturbation theory.  Using
the recent measurement of ${\cal B}(D^{*+} \to D^+ \gamma)$ \cite{CLEO1}, the
ratios of the $D \gamma$ and $D\pi^0$ branching fractions were used to extract
the couplings $g$ and $\beta$.  Here $g$ and $\beta$ are the $D^*D\pi$ and
$D^*D\gamma$ couplings since order $m_q$ and $1/m_Q$ corrections have been
absorbed into their definitions.  Two solutions were found
\begin{eqnarray}
 g&=&0.265\,^{+.04}_{-.02} \,^{+.05}_{-.02}\ \ 
	\beta=0.85^{+.2}_{-.1}\,^{+.3}_{-.1}\,{\rm GeV^{-1}}\, \nn\\
 g&=&0.76\,^{+.03}_{-.03} \,^{+.2}_{-.1}\ \ 
 	\beta=4.9^{+.3}_{-.3}\,^{+5.0}_{-.7}\,{\rm GeV^{-1}}\,.
\label{soln_concl}
\end{eqnarray}
The first error here is the one sigma error associated with a minimized
$\chi^2$ fit to the three experimental branching fraction ratios (see
Fig.~\ref{fig_solns}).  The second error is our estimate of the uncertainty in
the extraction due to four unknown counterterms $\tilde\alpha_1$, $\alpha_5$,
$\tilde\kappa_1$ and $\kappa_5$ that arise at order $m_q$ (see
Fig.~\ref{fig_rand}).  

It is possible that the uncertainty from these counterterms can be reduced by
determining them from other processes. For these corrections to contribute at
low enough order in the chiral expansion we need processes with outgoing
photons or pseudo-Goldstone bosons, such as semileptonic $D$ decays to $K$,
$\eta$, or $\pi$.  Here there are also SU(3) corrections to the left handed current
which involve an unknown parameter $\eta_0$ \cite{BG}.  Information on
$\kappa_1$ and $\kappa_1'$ can be determined from the pole part of the $D_s\to
K\ell\nu_\ell$ form factor \cite{BG}.  In a similar manner $D_s \to \eta\ell\nu_\ell$
can constrain $\tilde\kappa_1$ and $\kappa_5$, and a comparison of the form
factors for $D^+\to \bar K^0\ell\nu_\ell$ and $D_s\to \eta\ell\nu_\ell$ gives
information on $\kappa_1'$ and $\kappa_5$.  These investigations were beyond
the scope of this study. In principle, information about the constants
$\tilde\alpha_1$, and $\alpha_5$ could be obtained from a measurement of $B\to
\gamma\ell\nu_\ell$. The CLEO experimental bound on $B\to \ell\nu_\ell$
($\ell=e,\mu$)\cite{CLEO2} is roughly two orders of magnitude above the
theoretical prediction, but due to the helicity suppression for $B\to \ell\nu_\ell$
the branching ratio for $B\to \gamma\ell\nu_\ell$ may be up to an order of
magnitude bigger\cite{Colangelo4,Eilam}.  

Another possible approach would be to use large $N_c$ scaling for the
counterterms in $\delta{\cal L}_g$ and $\delta{\cal L}_\beta$.  Terms that have
two chiral traces are suppressed by a power of $N_c$ compared to those with
only one trace.  In the large $N_c$ limit the counterterms $\tilde\kappa_1$ and
$\tilde\alpha_1$ would dominate, and $\kappa_5$ and $\alpha_5$ could be
neglected, thus reducing the theoretical uncertainty.

The smaller solution for $g$ in Eq.~(\ref{soln_concl}) is fairly insensitive to the
addition of the one-loop corrections (see Table~\ref{table_solns}).  However,
corrections at order $m_q\sim 1/m_c$, including the heavy meson mass
splittings, were important in determining the solution with larger $g$.  The limit
$\Gamma(D^{*+}) < 0.13\,{\rm MeV}$ \cite{ACCMOR} gives an upper bound on the
coupling $g$ (see Eq.~(\ref{ulmt}) and Fig.~\ref{fig_solns}), and eliminates the
$g=0.76$, $\beta=4.9\,{\rm GeV^{-1}}$ solution.  Experimental confirmation of this
limit is therefore desirable.  Note that the largest experimental uncertainty in our
extraction comes from the measurement of ${\cal B}(D_s^*\to D_s \pi^0)$, and
dominates the theoretical uncertainty due to decay via single photon exchange. 
A better measurement of ${{\cal B}(D^{*\ }_s\to D_s\pi^0) / {\cal B}(D^{*\ }_s\to
D_s\gamma)}$ along with a limit $\Gamma(D^{*+})\lesssim 0.6\,{\rm MeV}$ could
provide further evidence that the $g=0.76$ solution is excluded.  However, if the
central values of the second and third ratios in Eq.~(\ref{exptratios}) decrease
then a width measurement or stronger limit on $\Gamma(D^{*+})$ will be needed
to distinguish the two solutions.  An improved measurement of ${\cal B}(D^{*\
}_s\to D_s\pi^0)$ may also give valuable information on the unknown couplings
$\tilde\kappa_1$, $\kappa_5$, $\tilde\alpha_1$, and $\alpha_5$.  

The extraction of $g$ has important consequences for other physical quantities
[2-11].  For example\footnote{Glenn Boyd and Ben Grinstein, private
communication.}, for the $B\to \pi\ell\bar\nu_\ell$ form factors with $E_\pi <
2\,m_\pi$, analyticity bounds combined with chiral perturbation theory give $g\,f_B
\lesssim 50\,{\rm MeV}$ \cite{disp}.  The solution $g=0.265$ gives $f_B \lesssim
190\,{\rm MeV}$ for the $B$ decay constant.  However, for $g=0.76$ we have $f_B
\lesssim 66\,{\rm MeV}$, which is roughly a factor of three smaller than lattice QCD
values, $f_B \simeq 160-205$
\cite{lattice1,lattice2,lattice3,lattice4,lattice5,lattice6,lattice7}.



\chapter{$V_{ub}$ from Exclusive Semileptonic $B$ and $D$ Decays}
\label{Vub}

The next generation of $B$ decay experiments will test the flavor sector of the
standard model at high precision.  The basic approach is to determine the
elements of the CKM matrix using different methods and then check for the
consistency of these results.  At the present time $CP$ non-conservation has
only been observed in kaon decay arising from $K^0-\bar K^0$ mixing.  Many
extensions of the minimal standard model (e.g.,~models with several Higgs
doublets or low energy supersymmetry) have new particles with weak scale
masses that contribute to flavor changing neutral current processes like
$K^0-\bar K^0$ mixing, $B^0-\bar B^0$ mixing, $B\to K^* \gamma$, etc., at a
level comparable to the standard model.

At the present time, the magnitude of the $b \to u$ CKM matrix element is
determined by comparing experimental results on the inclusive electron spectrum
in the endpoint region with phenomenological models \cite{incl}, or by
comparing experimental results on $B\to\rho\,\ell\,\bar\nu_\ell$ and
$B\to\pi\,\ell\,\bar\nu_\ell$ with phenomenological models and lattice QCD
results \cite{excl}.  These two approaches yield remarkably consistent
determinations of $|V_{ub}|$, but have large uncertainties.

In this chapter we discuss the proposal to determine $|V_{ub}|$ \cite{IsWi,lw}
using a combination of heavy quark symmetry \cite{HQS1,HQS2} and $SU(3)$ flavor
symmetry.  The basic idea is to compare $D\to K^*\,\bar\ell\,\nu_\ell$ with the
Cabibbo suppressed decay $D\to\rho\,\bar\ell\,\nu_\ell$.  Using heavy quark
symmetry the $SU(3)$ violations in the form factors that occur in these decays
are related to those that occur in a comparison of $B\to K^*\ell\,\bar\ell$ (or
$B\to K^*\,\nu_\ell\,\bar\nu_\ell$) with $B \to \rho\,\ell\,\bar\nu_\ell$. 
Therefore, experimental data on $B\to K^*\ell\,\bar\ell$ in conjunction with
data on $D\to\rho\,\bar\ell\,\nu_\ell$ and $D\to K^*\,\bar\ell\,\nu_\ell$ can
be used to determine $|V_{ub}|$.  This proposal is complementary to other
approaches for determining $|V_{ub}|$, since it relies on the standard model
correctly describing the rare flavor changing neutral current process $B\to
K^*\ell\,\bar\ell$.  

In this chapter we compute corrections to these form factor relations which
violate both chiral and heavy quark symmetry, and are non-analytic in the
symmetry breaking parameters.  We also reconsider the influence of long
distance effects on the extraction of the $B\to K^*$ form factors from $B\to
K^*\ell\,\bar\ell$.

We denote the form factors relevant for semileptonic transitions
between a pseudoscalar meson $P^{(Q)}$, containing a heavy quark $Q$, and a
member of the lowest lying multiplet of vector mesons, $V$, by $g^{(H\to V)}$,
$f^{(H\to V)}$ and $a_\pm^{(H\to V)}$, where
\begin{eqnarray}\label{ffdef}
\langle V(p',\epsilon) |\,\bar q\,\gamma_\mu\, Q\,| H(p)\rangle
&=& i\,g^{(H\to V)}\, \varepsilon_{\mu\nu\lambda\sigma}\, \epsilon^{*\nu}\,
  (p+p')^\lambda\, (p-p')^\sigma \,, \nn \\*
\langle V(p',\epsilon) |\,\bar q\,\gamma_\mu\gamma_5\, Q\,| H(p)\rangle
&=& f^{(H\to V)}\,\epsilon^*_\mu 
  + a_+^{(H\to V)}\,(\epsilon^*\cdot p)\,(p+p')_\mu \nn \\
  && + a_-^{(H\to V)}\,(\epsilon^*\cdot p)\,(p-p')_\mu \,,
\end{eqnarray}
and $\varepsilon^{0123}=-\varepsilon_{0123}=1$.  We view the form factors $g$,
$f$ and $a_\pm$ as functions of the dimensionless variable $y=v\cdot v'$, where
$p=m_H\,v$, $p'=m_V\,v'$, and $q^2=(p-p')^2=m_H^2+m_V^2-2m_H\,m_V\,y$. 
(Although we are using the variable $v\cdot v'$, we are not treating the quarks
in $V$ as heavy.)  The experimental values for the $D\to
K^*\,\bar\ell\,\nu_\ell$ form factors assuming nearest pole dominance for the
$q^2$ dependences are \cite{E791b}
\begin{eqnarray}\label{ffexp}
f^{(D\to K^*)}(y) &=& {(1.9\pm0.1)\,{\rm GeV}\over 1+0.63\,(y-1)}\,, 
  \nonumber\\*
a_+^{(D\to K^*)}(y) &=& -{(0.18\pm0.03)\,{\rm GeV}^{-1}\over 1+0.63\,(y-1)}\,, 
  \nonumber\\*
g^{(D\to K^*)}(y) &=& -{(0.49\pm0.04)\,{\rm GeV}^{-1}\over 1+0.96\,(y-1)}\,.
\end{eqnarray}
The shapes of these form factors are beginning to be probed experimentally
\cite{E791b}.  The form factor $a_-$ is not measured because its contribution to
the $D\to K^*\,\bar\ell\,\nu_\ell$ decay amplitude is suppressed by the lepton
mass.  The minimal value of $y$ is unity (corresponding to the zero recoil point)
and the maximum value of $y$ is $(m_D^2+m_{K^*}^2)/(2m_D\,m_{K^*}) 
\simeq1.3$ (corresponding to $q^2=0$).  Note that $f(y)$ changes by less than
20\% over the whole kinematic range $1<y<1.3$.  In the following analysis we will
extrapolate the measured form factors to the larger region $1<y<1.5$.  The full
kinematic region for $B\to\rho\,\ell\,\bar\nu_\ell$ is $1<y<3.5$.

The differential decay rate for semileptonic $B$ decay (neglecting the lepton
mass, and not summing over the lepton type $\ell$) is
\begin{equation}\label{SLrate}
{{\rm d}\Gamma(B\to\rho\,\ell\,\bar\nu_\ell)\over{\rm d}y} 
  = {G_F^2\,|V_{ub}|^2\over48\,\pi^3}\, m_B\, m_\rho^2\, S^{(B\to\rho)}(y) \,.
\end{equation}
Here $S^{(H\to V)}(y)$ is the function
\begin{eqnarray}\label{shape}
S^{(H\to V)}(y) &=& \sqrt{y^2-1}\, \bigg[ \Big|f^{(H\to V)}(y)\Big|^2\,
  (2+y^2-6yr+3r^2) \\*
&&\phantom{} + 4{\rm Re} \Big[a_+^{(H\to V)}(y)\, f^{(H\to V)}(y)\Big]
  m_H^2\, r\, (y-r) (y^2-1) \nonumber\\*
&&\phantom{} + 4\Big|a_+^{(H\to V)}(y)\Big|^2 m_H^4\, r^2 (y^2-1)^2 \nn\\*
&&\phantom{} + 8\Big|g^{(H\to V)}(y)\Big|^2 m_H^4\, r^2 (1+r^2-2yr)(y^2-1)\, 
  \bigg] \nonumber\\*
&=& \sqrt{y^2-1}\, \Big|f^{(H\to V)}(y)\Big|^2\, (2+y^2-6yr+3r^2)\, 
  [1+\delta^{(H\to V)}(y)] \,, \nn
\end{eqnarray}
with $r=m_V/m_H$.  The function $\delta^{(H\to V)}$ depends on the ratios of
form factors $a_+^{(H\to V)}/f^{(H\to V)}$ and $g^{(H\to V)}/f^{(H\to V)}$. 
$S^{(B\to\rho)}(y)$ can be estimated using combinations of $SU(3)$ flavor
symmetry and heavy quark symmetry.  $SU(3)$ symmetry implies that the $\bar
B^0\to\rho^+$ form factors are equal to the $B\to K^*$ form factors and the
$B^-\to\rho^0$ form factors are equal to $1/\sqrt2$ times the $B\to K^*$ form
factors.  Heavy quark symmetry implies the relations \cite{IsWi}
\begin{eqnarray}\label{BDrel}
f^{(B\to K^*)}(y) &=& \left({m_B\over m_D}\right)^{1/2} 
  \bigg[{\alpha_s(m_b)\over \alpha_s(m_c)}\bigg]^{-6/25}\, 
  f^{(D\to K^*)}(y)\,, \nonumber\\*
a_+^{(B\to K^*)}(y) &=& \left({m_D\over m_B}\right)^{1/2} 
  \bigg[{\alpha_s(m_b)\over \alpha_s(m_c)}\bigg]^{-6/25}\,
  a_+^{(D\to K^*)}(y) \,, \nonumber\\*
g^{(B\to K^*)}(y) &=& \left({m_D\over m_B}\right)^{1/2} 
  \bigg[{\alpha_s(m_b)\over \alpha_s(m_c)}\bigg]^{-6/25}\, 
  g^{(D\to K^*)}(y)\,.
\end{eqnarray}
The second relation is obtained using $a_-^{(D\to K^*)}=-a_+^{(D\to K^*)}$, 
valid in the large $m_c$ limit.

Using Eq.~(\ref{BDrel}) and $SU(3)$ to get $\bar
B^0\to\rho^+\,\ell\,\bar\nu_\ell$ form factors (in the region $1<y<1.5$) from
those for $D\to K^*\bar\ell\,\nu_\ell$ given in Eq.~(\ref{ffexp}) yields
$S^{(B\to\rho)}(y)$ plotted in Fig.~\ref{fig_R} of Ref.~\cite{lw}.  The numerical values
in Eq.~(\ref{ffexp}) differ slightly from those used in Ref.~\cite{lw}.  This
makes only a small difference in $S^{(B\to\rho)}$, but changes
$\delta^{(B\to\rho)}$ more significantly.  In the region $1<y<1.5$,
$|\delta^{(B\to\rho)}(y)|$ defined in Eq.~(\ref{shape}) is less than 0.06,
indicating that $a_+^{(B\to\rho)}$ and $g^{(B\to\rho)}$ make a small
contribution to the differential rate in this region.

This prediction for $S^{(B\to\rho)}$ can be used to determine $|V_{ub}|$ from a
measurement of the $B\to\rho\,\ell\,\bar\nu_\ell$ semileptonic decay rate in
the region $1<y<1.5$.  This method is model independent, but cannot be expected
to yield a very accurate value of $|V_{ub}|$.  Typical $SU(3)$ violations are
at the $10-20$\% level and one expects similar violations of heavy quark
symmetry.  

Ref.~\cite{lw} proposed a method for getting a value of $S^{(B\to\rho)}(y)$
with small theoretical uncertainty.  They noted that the ``Grinstein-type"
\cite{Gtdr} double ratio
\begin{equation}\label{Gtdr}
R(y) = \Big[ f^{(B\to\rho)}(y) / f^{(B\to K^*)}(y) \Big] \Big/
  \Big[ f^{(D\to\rho)}(y) / f^{(D\to K^*)}(y) \Big]
\end{equation}
is unity in the limit of $SU(3)$ symmetry or in the limit of heavy quark
symmetry.  Corrections to the prediction $R(y)=1$ are suppressed by
$m_s/m_{c,b}$ ($m_{u,d} \ll m_s$) instead of $m_s/\Lambda_{\rm QCD}$ or
$\Lambda_{\rm QCD}/m_{c,b}$.  Since $R(y)$ is very close to unity, the 
relation 
\begin{equation}\label{magic}
S^{(B\to\rho)}(y) = S^{(B\to K^*)}(y)\,
  \left|{f^{(D\to\rho)}(y)\over f^{(D\to K^*)}(y)}\right|^2\,
  \bigg({m_B-m_\rho\over m_B-m_{K^*}}\bigg)^2\,,
\end{equation}
together with measurements of $|f^{(D\to K^*)}|$, $|f^{(D\to\rho)}|$, and
$S^{(B\to K^*)}$ will determine $S^{(B\to\rho)}$ with small theoretical
uncertainty.  The last term on the right-hand-side makes Eq.~(\ref{magic})
equivalent to Eq.~(\ref{Gtdr}) in the $y\to1$ limit.  The ratio of the
$(2+y^2-6yr+3r^2)\,[1+\delta^{(B\to V)}(y)]$ terms makes only a small and
almost $y$-independent contribution to $S^{(B\to\rho)}/S^{(B\to K^*)}$ in the
range $1<y<1.5$.  Therefore, corrections to Eq.~(\ref{magic}) are at most a few
percent larger than those to $R(y)=1$.

$|f^{(D\to K^*)}|$ has already been determined.  $|f^{(D\to\rho)}|$ may be
obtainable in the future, for example from experiments at $B$ factories, where
improvements in particle identification help reduce the background from the
Cabibbo allowed decay.  The measurement ${\cal
B}(D\to\rho^0\,\bar\ell\,\nu_\ell)/{\cal B}(D\to\bar
K^{*0}\,\bar\ell\,\nu_\ell) = 0.047\pm0.013$ \cite{E791a} already suggests that
$|f^{(D\to\rho)}/f^{(D\to K^*)}|$ is close to unity.  Assuming $SU(3)$ symmetry
for the form factors, but keeping the explicit $m_V$-dependence in $S^{(D\to
V)}(y)$ and in the limits of the $y$ integration, the measured form factors in
Eq.~(\ref{ffexp}) imply ${\cal B}(D\to\rho^0\,\bar\ell\,\nu_\ell)/{\cal
B}(D\to\bar K^{*0}\,\bar\ell\,\nu_\ell) = 0.044$.\footnote{This prediction
would be $|V_{cd}/V_{cs}|^2/2\simeq0.026$ with $m_\rho=m_{K^*}$.  Phase space
enhances $D\to\rho$ compared to $D\to K^*$ to yield the quoted prediction.}
$S^{(B\to K^*)}$ is obtainable from experimental data on $B\to
K^*\,\nu_\ell\,\bar\nu_\ell$ or $B\to K^*\ell\,\bar\ell$.  While the former
process is very clean theoretically, it is very difficult experimentally.  A
more realistic goal is to use $B\to K^*\ell\,\bar\ell$, since CDF expects to
observe $400-1100$ events in the Tevatron run II (if the branching ratio is in
the standard model range) \cite{CDF2}.  There are some uncertainties associated
with long distance nonperturbative strong interaction physics in this
extraction of $S^{(B\to K^*)}(y)$.  To use the kinematic region $1<y<1.5$, the
form factor ratio $f^{(D\to\rho)}/f^{(D\to K^*)}$ in Eq.~(\ref{magic}) must be
extrapolated to a greater region than what can be probed experimentally.  For
this ratio, the uncertainty related to this extrapolation is likely to be
small.

The main purpose of this study is to examine the deviation of $R$ from unity
using chiral perturbation theory.  We find that it is at the few percent level.
The uncertainty from long distance physics in the extraction of $S^{(B\to
K^*)}$ is also reviewed.  On average, in the region $1<y<1.5$, this is probably
less than a 10\% effect on the $B\to K^*\ell\,\bar\ell$ decay rate. 
Consequently, a determination of $|V_{ub}|$ from experimental data on $D\to
K^*\bar\ell\,\nu_\ell$, $D\to\rho\,\bar\ell\,\nu_\ell$, $B\to
K^*\ell\,\bar\ell$ and $B\to\rho\,\ell\,\bar\nu_\ell$ with an uncertainty from
theory of about 10\% is feasible.

\section{Chiral perturbation theory for the form factor ratio}

\begin{figure}[tb]  
\centerline{\epsfysize=2.5truecm \epsfbox{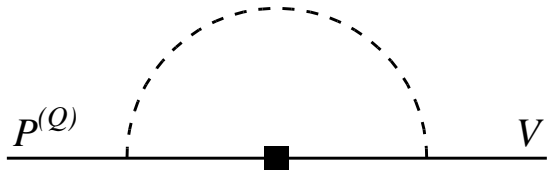}}
\caption[Feynman diagram that gives the leading contribution to $R(1)-1$, where
$R$ is defined in Eq.~(\ref{Gtdr}). The dashed line is a $\pi$ or an $\eta$.
The black square indicates insertion of the weak current.]{Feynman diagram that gives the leading contribution to $R(1)-1$.
The dashed line is a $\pi$ or an $\eta$.
The black square indicates insertion of the weak current.} \label{fig_R}
\end{figure}

The leading deviation of $R$ from unity can be calculated using a combination of
heavy hadron chiral perturbation theory for the mesons containing a heavy quark
(section~\ref{HQS}) and for the lowest lying vector mesons (section~\ref{HV}).  We
adopt the notations and conventions of Refs.~\cite{BGetal,jmw}.  The weak current
transforms as $(\bar3_L,1_R)$, and at the zero recoil kinematic point there are two
operators that are relevant for $P^{(Q)} \to V$ transition matrix elements (where
$P^{(b)}=B$, $P^{(c)}=D$, and $V$ is one of the lowest lying vector mesons $\rho,
\omega, K^*, \phi$).  Demanding that the Zweig suppressed
$D_s\to\omega\,\bar\ell\,\nu_\ell$ process vanishes relates the two operators,
yielding \cite{Hooman}
\begin{equation}\label{current}
\bar q_a\, \gamma_\mu (1-\gamma_5)\, Q = 
  \beta\, {\rm Tr} [ {N\!\!\!\!\slash}_{cb}^{\,\dag}
  \gamma_\mu (1-\gamma_5) H_c^{(Q)} \xi_{ba}^\dag ] \,,
\end{equation}
where $N_{cb}$ is given in Eq.~(\ref{HVN}).
Here repeated $SU(3)$ indices are summed and the trace is over Lorentz indices.
$H^{(Q)}$ contains the ground state heavy meson doublet, $N$ is the nonet
vector meson matrix \cite{jmw}, and $\beta$ is a constant.  The leading
contribution to $R(1)-1$ arises from the Feynman diagrams in Fig.~\ref{fig_R}.  Diagrams
with a virtual kaon cancel in the double ratio $R$.  Neglecting the vector
meson widths,\footnote{The only significant width is that of the $\rho$ meson. 
Since it occurs in the loop graph involving an $\eta$, neglecting the $\rho$
width amounts to treating $\Gamma_\rho/2m_\eta \ll 1$, which is a reasonable
approximation.} these diagrams yield
\begin{equation}\label{R1}
R(1) - 1 = -\frac{g\,g_2}{12\,\pi^2\,f^2}\, \Big[ 
  G(m_\pi,\Delta^{(b)}) - G(m_\eta,\Delta^{(b)}) - 
  G(m_\pi,\Delta^{(c)}) + G(m_\eta,\Delta^{(c)}) \Big] \,,
\end{equation}
where $\Delta^{(b)}=m_{B^*}-m_B$, $\Delta^{(c)}=m_{D^*}-m_D$, and for 
$m\ge\Delta$,
\begin{equation}\label{GmD}
G(m,\Delta) = \frac{\pi\,m^3}{2\,\Delta} - {(m^2-\Delta^2)^{3/2}\over\Delta}\, 
  \arctan \left( {\sqrt{m^2-\Delta^2}\over \Delta}\right) - \Delta^2 \ln m\,.
\end{equation}
Here $g_2$ is the $\rho\,\omega\,\pi$ coupling, $g$ is the $DD^*\pi$ coupling,
and $f\simeq131\,$MeV is the pion decay constant.  In the nonrelativistic
constituent quark model $g=g_2=1$ \cite{BGetal}, while in the chiral quark
model \cite{GM} $g=g_2=0.75$.  Experimental data on
$\tau\to\omega\,\pi\,\nu_\tau$ in the region of low $\omega\,\pi$ invariant
mass gives $g_2\simeq0.6$ \cite{dw}.  In chapter 3 we saw that the
measured branching ratios for $D^*$ decays give $g=0.27$ or $g=0.76$.

For small $\Delta$, Eq.~(\ref{R1}) for $R(1)-1$ has a non-analytic $\sqrt{m_q}$
dependence on the light quark masses.  This cannot arise from corrections to the
current in Eq.~(\ref{current}) or to the chiral Lagrangian, and must come from
1-loop diagrams involving the pseudo-Goldstone bosons $\pi,\,K,\,\eta$.  Using
the measured values of the pion and eta masses, Eqs.~(\ref{R1}) and (\ref{GmD})
imply $R(1)=1-0.035\,g\,g_2$.  There may be significant corrections from analytic
terms of order $m_s/m_c\sim 1/10$ or from higher orders in chiral perturbation
theory.  However, the smallness of our result lends support to the expectation that
$R(1)-1$ is very close to zero.  There is no reason to expect any different
conclusion over the kinematic range $1<y<1.5$.

\section{Long distance effects and extracting $S^{(B\to K^*)}$}


The decay rate for $B \to K^*\,\nu_\ell\,\bar\nu_\ell$ could determine
$S^{(B\to K^*)}$ free of theoretical uncertainties.  However, experimental
study of this decay is very challenging.  A more practical approach to
extracting this quantity is to use $B \to K^*\ell\,\bar\ell$.  The differential
decay rate is
\begin{eqnarray}\label{Rrate2}
{{\rm d}\Gamma(B\to K^*\ell\,\bar\ell)\over{\rm d}y} &=& 
  {G_F^2\, |V_{ts}^*V_{tb}|^2 \over 24\,\pi^3} 
  \left({\alpha\over4\pi}\right)^2 m_B\,m_{K^*}^2\, 
  \Big[|\widetilde C_9(y)|^2 + |C_{10}|^2\Big]\, [1+\Delta(y)] \nonumber\\*
&&\phantom{} \times S^{(B\to K^*)}(y)\, [1+d(y)] \,.
\end{eqnarray}
This and Eq.~(\ref{magic}) allow us to rewrite Eq.~(\ref{SLrate}) as 
\begin{eqnarray}\label{rewrite}
{{\rm d}\Gamma(B\to\rho\,\ell\,\bar\nu_\ell) \over {\rm d}y }
&=& {|V_{ub}|^2\over|V_{ts}^*V_{tb}|^2}\, {8\,\pi^2\over\alpha^2}\, 
  {1\over |\widetilde C_9(y)|^2 + |C_{10}|^2}\, 
  \frac1{1+\Delta(y)}\, \frac1{1+d(y)} \,{ m_\rho^2 \over m_{K^*}^2} \nonumber\\*
&&\phantom{} \times \bigg({m_B-m_\rho\over m_B-m_{K^*}}\bigg)^2
  \left|{f^{(D\to\rho)}(y)\over f^{(D\to K^*)}(y)}\right|^2 
  {{\rm d}\Gamma(B\to K^*\ell\,\bar\ell) \over {\rm d}y } \,.
\end{eqnarray}
which can be directly used to extract $|V_{ub}|$.  Unitarity of the CKM matrix
implies that $|V_{ts}^* V_{tb}| \simeq |V_{cs}^* V_{cb}|$ with less than a 3\%
uncertainty.  The fine structure constant, $\alpha=1/129$, is evaluated at the
$W$-boson mass.  $d(y)$ parameterizes long distance effects, and will be
discussed below.  $\Delta(y)$ takes into account the contribution of the
magnetic moment operator, $O_7=(e/16\pi^2)\,m_b\,(\bar
s_L\,\sigma_{\mu\nu}\,b_R)\, F^{\mu\nu}$ (a factor of $-4 G_FV_{ts}^*
V_{tb}/\sqrt{2}$ has been extracted out in the definition of operator
coefficients).  Ref.~\cite{lw} (see also Ref.~\cite{SaYa}) found using heavy
quark symmetry that $\Delta(y) \simeq -0.14-0.08(y-1)$ in the region $1<y<1.5$.
Corrections to this are expected to be small since there are no $1/m_c$
corrections to $\Delta(1)$.  $C_{10}$ is the Wilson coefficient of the operator
$O_{10} = (e^2/16\pi^2)\, (\bar s_L\,\gamma_\mu\,b_L)
(\bar\ell\,\gamma^\mu\gamma_5\,\ell)$.  $\widetilde C_9(y)$ takes into account
the contribution of the four-quark operators, $O_1-O_6$, and the operator $O_9
= (e^2/16\pi^2)\, (\bar s_L\,\gamma_\mu\,b_L)\, (\bar\ell\,\gamma^\mu\,\ell)$. 
In perturbation theory using the next-to-leading logarithmic approximation
\cite{BuMu,Misiak}
\begin{eqnarray}\label{C9eff}
\widetilde C_9(y) &=& C_9 + h(z,y)\, (3C_1+C_2+3C_3+C_4+3C_5+C_6)
  -\frac12\,h(0,y)\, (C_3+3C_4)  \nonumber\\*
&&\phantom{} -\frac12\,h(1,y)\, (4C_3+4C_4+3C_5+C_6) 
  + \frac29\, (3C_3+C_4+3C_5+C_6) \,,
\end{eqnarray}
where $z=m_c/m_b$.  Here 
\begin{equation}\label{qloops}
h(u,y) = -\frac89\ln u + \frac8{27} + \frac{4}9\,x
  - \frac29\, (2+x) \sqrt{|1-x|} \cases{ 
  \ln{\displaystyle 1+\sqrt{1-x}\over\displaystyle 1-\sqrt{1-x}} - i\pi \,;  
    &  $x<1$ \cr
  2\arctan(1/\sqrt{x-1}) \,; &  $x>1$\,, \cr} \nonumber
\end{equation}
where $x \equiv 4u^2m_b^2/(m_B^2+m_{K^*}^2-2m_B\,m_{K^*}\,y)$.  Using
$m_t=175\,$GeV, $m_b=4.8\,$GeV, $m_c=1.4\,$GeV, $\alpha_s(m_W)=0.12$, and
$\alpha_s(m_b)=0.22$, the numerical values of the Wilson coefficients are
$C_1=-0.26$, $C_2=1.11$, $C_3=0.01$, $C_4=-0.03$, $C_5=0.008$, $C_6=-0.03$,
$C_7=-0.32$, $C_9=4.26$, and $C_{10}=-4.62$.  Of these, $C_9$ and $C_{10}$ are
sensitive to $m_t$ (quadratically for $m_t\gg m_W$).

In Eq.~(\ref{C9eff}) the second term on the right-hand-side, proportional to
$h(z,y)$ comes from charm quark loops.  Since the kinematic region we are
interested in is close to $q^2=4m_c^2$, a perturbative calculation of the
$c\,\bar c$ loop cannot be trusted.  Threshold effects which spoil local
duality are important.  It is these long distance effects that give rise to the
major theoretical uncertainty in the extraction of $|V_{ub}|$ from the $B\to
K^*\ell\,\bar\ell$ differential decay rate using
Eq.~(\ref{rewrite}).\footnote{The four-quark operators involving light $u$,
$d$, and $s$ quarks also have uncertainty from long distance physics.  However,
this is expected to have a very small effect on the $B\to K^*\ell\,\bar\ell$
rate.}  The influence of this long distance physics on the differential decay
rate is parameterized by $d(y)$ in Eq.~(\ref{Rrate2}), where setting $d(y)=0$
gives the perturbative result.

\begin{figure}[tb]  
\centerline{\epsfysize=8truecm \epsfbox{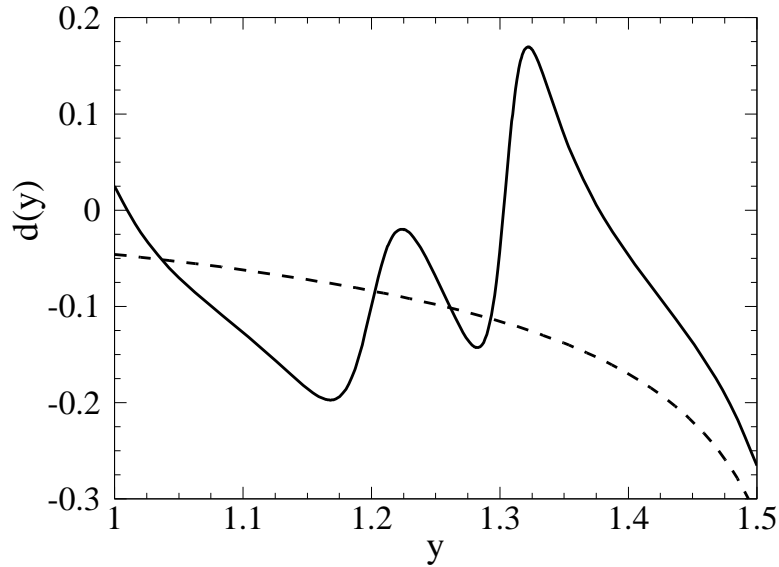}}
\caption[An estimate of long distance corrections to the $B\to K^* \ell \bar\ell$ rate
using $d(y)$ defined in Eq.~(\ref{dydef}).  The solid curve takes into
account all six $1^{--}$ $c\,\bar c$ resonances according to 
Eq.~(\ref{model}), whereas the dashed curve is obtained including only the 
three lightest ones.]{$d(y)$ defined in Eq.~(\ref{dydef}).  The solid curve takes into
account all six $1^{--}$ $c\,\bar c$ resonances according to 
Eq.~(\ref{model}), whereas the dashed curve is obtained including only the 
three lightest ones.}  \label{fig_d}
\end{figure}

For the part of the $c\,\bar c$ loop where the charm quarks are not far
off-shell, a model for $h(z,y)$ which sums over $1^{--}$ $c\,\bar c$ resonances
is more appropriate than the perturbative calculation.  Consequently, we model
the part of $h(z,y)$ with explicit $q^2$-dependence in Eq.~(\ref{qloops}) with
a sum over resonances \cite{Lim,Desh1,Desh2,ODonnell} calculated using 
factorization
\begin{equation}\label{model}
h(z,y) \to -\frac89 \ln{z} + \frac8{27} - \frac{3\pi\kappa}{\alpha^2}\,
  \sum_n { \Gamma_{\psi^{(n)}}\, {\cal B}(\psi^{(n)}\to\ell\,\bar\ell) \over 
  (q^2-M_{\psi^{(n)}}^2)/M_{\psi^{(n)}} + i\Gamma_{\psi^{(n)}}} \,.
\end{equation}
The resonances $\psi^{(n)}$ have masses $3.097\,$GeV, $3.686\,$GeV,
$3.770\,$GeV, $4.040\,$GeV, \\ $4.160\,$GeV, and $4.415\,$GeV, respectively, and
their widths $\Gamma_{\psi^{(n)}}$ and leptonic branching ratios ${\cal
B}(\psi^{(n)}\to\ell\,\bar\ell)$ are known \cite{PDG96}.  The factor $\kappa=2.3$
takes into account the deviation of the factorization model \cite{NeSt}
parameter $a_2$ from its perturbative value.  Denoting the value of $\widetilde
C_9(y)$ in this model by $\widetilde C_9'(y)$, its influence on the
differential decay rate is given by $d(y)$ defined as
\begin{equation}\label{dydef}
|\widetilde C_9'(y)|^2+|C_{10}|^2 =
  (|\widetilde C_9(y)|^2+|C_{10}|^2)\, [1+d(y)] \,.
\end{equation}
$d(y)$ is plotted in Fig.~\ref{fig_d} (solid curve).  The physical interpretation of the
$1^{--}$ resonances above $4\,$GeV is not completely clear.  It might be more
appropriate to treat them as $D\bar D$ resonances than as $c\,\bar c$ states. 
It is possible that for these resonances factorization as modeled by
Eq.~(\ref{model}) with $\kappa=2.3$ is not a good approximation.  Including
only the first three $1^{--}$ resonances in Eq.~(\ref{model}), yields $d(y)$
plotted with the dashed curve in Fig.~\ref{fig_d}.  

This estimate of $d(y)$ based on factorization and resonance saturation differs
from that in Ref.~\cite{lw} in two respects.  Firstly, the phase of $\kappa$ is
viewed as fixed because recent data has determined the sign of the ratio of
factorization model parameters, $a_2/a_1$, and the phase of $a_1$ is expected
to be near its perturbative value \cite{BHP}.  Secondly, since the resonance
saturation model only represents the $c\,\bar c$ loop for charm quarks that are
not far off-shell, we have only used it for the part of $h(z,y)$ in
Eq.~(\ref{qloops}) with explicit $q^2$ dependence, retaining the perturbative
expression for the first two terms, $-(8/9)\ln z+8/27$.  The $\ln z$ term has
dependence on $m_b$, which is clearly short distance in origin.  This reduces
somewhat the magnitude of $d(y)$ and makes it more symmetric about zero
(compare Fig.~\ref{fig_d} with Fig.~6 of Ref.~\cite{lw}).  It would be interesting to
have a more physical separation between the long and short distance parts of
the amplitude.

\begin{figure}[tb]  
\centerline{\epsfysize=8truecm \epsfbox{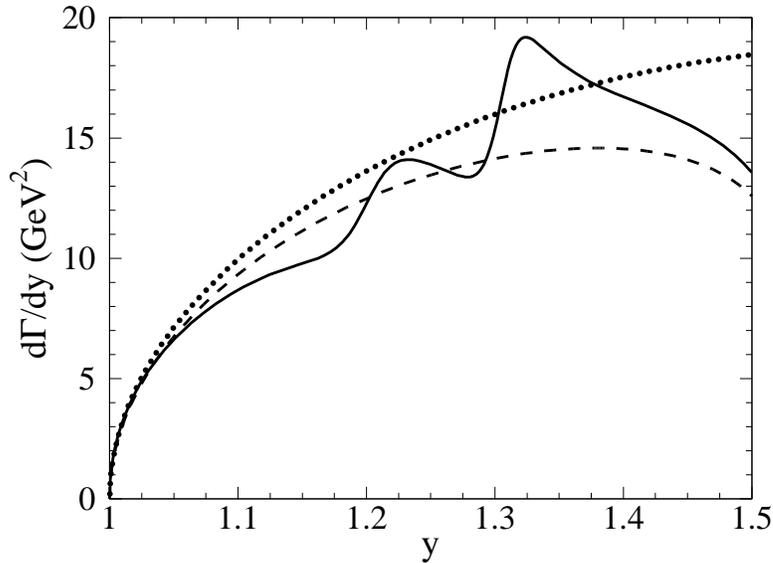}}
\caption[${\rm d}\Gamma(B\to K^*\ell\,\bar\ell)/{\rm d}y$ as given in 
Eq.~(\ref{Rrate2}).  The solid curve
takes into account all six $1^{--}$ $c\,\bar c$ resonances, the dashed curve
includes only the three lightest ones, and the dotted curve is the perturbative
result.]{${\rm d}\Gamma(B\to K^*\ell\,\bar\ell)/{\rm d}y$ in units of $[|\widetilde
C_9(1)|^2+|C_{10}|^2]\, m_B\,m_{K^*}^2\, G_F^2\, \alpha^2$ $\times
|V_{ts}^*V_{tb}|^2/(384\pi^5)$ as given in Eq.~(\ref{Rrate2}).  The solid curve
takes into account all six $1^{--}$ $c\,\bar c$ resonances, the dashed curve
includes only the three lightest ones, and the dotted curve is the perturbative
result (i.e., $d(y)=0$).} \label{fig_Kll}
\end{figure}

Whether it is reasonable to use factorization for the resonances above $4\,$GeV
can be tested experimentally, since these states cause a very distinctive
pattern in ${\rm d}\Gamma/{\rm d}y$.  In Fig.~\ref{fig_Kll} the shape of ${\rm
d}\Gamma/{\rm d}y$ is plotted in the region $1<y<1.5$ using the resonance
saturation model for $d(y)$ (solid curve).  Experimental support for this shape
would provide evidence that this model correctly describes the long distance
physics parameterized by $d(y)$.  Although $d(y)$ gets as large as $\pm0.2$,
since it oscillates, its influence on the $B\to K^*\ell\,\bar\ell$ decay rate
in the region $1<y<1.5$ is about $-8$\% compared to the perturbative result
(which is plotted with the dotted curve in Fig.~\ref{fig_Kll}).  Even if our estimates of
this long distance physics based on factorization and resonance saturation has
a 100\% uncertainty (a prospect that we do not consider particularly unlikely),
it will only cause about a 4\% uncertainty in this determination of $|V_{ub}|$.
Including only the first three $1^{--}$ resonances in the sum in
Eq.~(\ref{model}) yields the dashed curve in Fig.~\ref{fig_Kll}.  In this case $d(y)$
causes a $-13$\% change in the $B\to K^*\ell\,\bar\ell$ decay rate in the
region $1<y<1.5$.

\section{Nearer term prospects}


Without information on the $y$ spectrum for the $B$ decay rates in
Eq.~(\ref{rewrite}), it is still possible to determine $|V_{ub}|$ by comparing
the branching ratios for $B\to \rho\,\ell\,\bar\nu_\ell$ and $B\to
K^*\ell\,\bar\ell$ in the region $1<y<1.5$.  Integrating Eq.~(\ref{rewrite})
over $1<y<1.5$ we can write
\begin{eqnarray}\label{crts}
\Gamma(\bar B^0\to \rho^+\,\ell\,\bar\nu_\ell) \Bigl|_{y<1.5} &=&
  {|V_{ub}|^2\over|V_{ts}^*V_{tb}|^2}\, {8\pi^2\over\alpha^2}\,
  {1\over \overline C{}_9^2+C_{10}^2}\,
  \frac1{(1+\overline\Delta)}\, \frac1{(1+\overline d)} \\*
&&\phantom{}\!\! \times \frac{m_\rho^2}{m_{K^*}^2}\, 
  \bigg({m_B-m_\rho\over m_B-m_{K^*}}\bigg)^2\,
  \left|{f^{(D\to\rho)}(1)\over f^{(D\to K^*)}(1)}\right|^2
  \Gamma(B\to K^*\ell\,\bar\ell)\Big|_{y<1.5} \nonumber\,. 
\end{eqnarray}
Here the barred quantities, $\overline C{}_9^2$, $\overline\Delta$, and
$\overline d$ denote the averages of $|\widetilde C_9(y)|^2$, $\Delta(y)$, and
$d(y)$ weighted with $S^{(B\to K^*)}(y)$.  Using the shape for $S^{(B\to K^*)}$
predicted from heavy quark symmetry, we find $\overline C_9=4.58$,
$\overline\Delta=-0.16$, and $\overline d=-0.08$.  Note that the $y$-dependence
of $\widetilde C_9$ is small and $\overline C_9$ is close to $C_9$.  In
Eq.~(\ref{crts}) the $y$-dependence of the ratio $f^{(D\to\rho)}(y)/f^{(D\to
K^*)}(y)$ has been neglected.  If the shape of these form factors can be
approximated with a pole form, then the pole masses of $2.56\,{\rm GeV}$ for
$f^{(D\to K^*)}$ and $2.45\,{\rm GeV}$ for $f^{(D\to\rho)}$ (corresponding to
$D^{**}_s$ and to $D^{**}$) imply that $|f^{(D\to\rho)}(y)/f^{(D\to
K^*)}(y)|^2$ varies by less than 1\% over the range $1<y<1.5$.  $SU(3)$
symmetry and the measured $D\to K^*$ form factors imply that
$\delta^{(D\to\rho)}$ contributes only about 23\% of the
$D\to\rho\,\bar\ell\,\nu_\ell$ decay rate.  Using this prediction for
$\delta^{(D\to\rho)}$, and assuming that $f^{(D\to\rho)}$ and $f^{(D\to K^*)}$
have the same $y$-dependence, yields ${\cal
B}(D\to\rho^0\,\bar\ell\,\nu_\ell)/{\cal B}(D\to\bar
K^{*0}\,\bar\ell\,\nu_\ell)=0.044\,|f^{(D\to\rho)}(1)/f^{(D\to K^*)}(1)|^2$.

In the region $q^2=(p_\ell+p_{\bar\ell})^2<m_{J/\psi}^2$ (corresponding roughly
to $y>2$), one cannot use the double ratio and Eq.~(\ref{rewrite}).  Moreover,
the $O_7$ contribution to the $B\to K^*\ell\,\bar\ell$ rate is large and
proportional to $1/q^2$, so the (leading order) heavy quark symmetry relations
between the tensor and (axial-)vector form factors\footnote{It was argued in
Ref.~\cite{largew} that heavy quark symmetry can be used even at small $q^2$.}
introduce a significant uncertainty.  For $q^2<m_{J/\psi}^2$, one can do better
using $SU(3)$ flavor symmetry alone to predict ${\rm
d}\Gamma(B\to\pi\,\ell\,\bar\nu_\ell)/{\rm d}q^2$ from a measurement of ${\rm
d}\Gamma(B\to K\ell\,\bar\ell)/{\rm d}q^2$.  Since this region is far from
$q^2_{\rm max}$, the $B^*$ pole contribution 
\cite{Bpi1,ChptW,Burdman,Wolf,yan,BLNY} is unlikely to upset
the $SU(3)$ relations.  The $O_7$ contribution to ${\rm d}\Gamma(B\to
K\ell\,\bar\ell)/{\rm d}q^2$ is at the $10-15$\% level, fairly independent of
$q^2$.  In the region $(1-2){\rm GeV}^2<q^2<m_{J/\psi}^2$, neglecting
$m_{K,\pi}^2/m_B^2$,
\begin{equation}\label{Kpi}
{ {\rm d}\Gamma(\bar B^0 \to \pi^+\,\ell\,\bar\nu_\ell) \over {\rm d}q^2 } =
  {|V_{ub}|^2\over|V_{ts}^*V_{tb}|^2}\, {8\pi^2\over\alpha^2}\,
  {1\over |\widetilde C_9(q^2)+2C_7|^2+|C_{10}|^2}\,
{ {\rm d}\Gamma(B\to K\,\ell\,\bar\ell) \over {\rm d}q^2 } \,. 
\end{equation}
A similar relation also holds for integrated rates.  

A measurement of the $B\to K^*\ell\,\bar\ell$ decay rate is unlikely before the
Tevatron run II.  Without this measurement, one has to rely on predicting the
$B\to\rho$ form factors from $D\to\rho$ using heavy quark symmetry, or from $D\to
K^*$ using both chiral and heavy quark symmetries.  As discussed following
Eq.~(\ref{magic}), recent experimental data \cite{E791a} suggests that the $SU(3)$
relation between $f^{(D\to K^*)}$ and $f^{(D\to\rho)}$ is not violated by more than
15\%.  Furthermore, a prediction for $B\to K^*\gamma$ can be made using heavy
quark symmetry and extrapolating the $D\to K^*\bar\ell \nu$ form factors \cite{lw3}.
The surprising agreement of this prediction with data may indicate that heavy
quark symmetry violation in the form factor relations is smaller than anticipated.
Heavy quark symmetry and the measured $D\to K^*$ form factors in
Eq.~(\ref{ffexp}) imply that the $\bar B^0\to\rho^+\ell\,\bar\nu_\ell$ branching ratio
in the region $1<y<1.5$ is $5.9\,|V_{ub}|^2$.  The measured decay rate ${\cal B}(\bar
B^0\to\rho^+\ell\,\bar\nu_\ell)=(2.5\pm0.4^{+0.5}_{-0.7}\pm0.5)\times10^{-4}$
\cite{excl} together with $|V_{ub}|\sim0.003$ imply that about 20\% of $\bar
B^0\to\rho^+\ell\,\bar\nu_\ell$ decays are in the range $1<y<1.5$.

Despite the presence of long distance effects associated with the $c\,\bar c$
resonance region, the $B\to K^*\ell\,\bar\ell$ rate can be used in
Eq.~(\ref{rewrite}) to determine $|V_{ub}|$ with a theoretical uncertainty that
is about 10\%.  Experimental verification of the distinctive $y$-dependence of
the differential rate associated with the $1^{--}$ resonances above $4\,$GeV
(see Fig.~\ref{fig_Kll}) would reduce the theoretical uncertainty from long distance
effects.  A precise value of $|V_{ub}|$ may be available from other processes,
e.g., the hadronic invariant mass spectrum in inclusive $\bar B\to
X_u\ell\,\bar\nu_\ell$ decay \cite{FLW,du} or from lattice QCD results on
exclusive form factors \cite{Flynn} before the $B\to K^*\ell\,\bar\ell$ decay
rate is measured.  In that case, Eq.~(\ref{rewrite}) gives an accurate standard
model prediction for the $B\to K^*\ell\,\bar\ell$ decay rate in the region
$1<y<1.5$.  Comparison with data may signal new physics or provide stringent
constraints on extensions of the standard model.



\chapter{$NN$ Scattering}

\section{Introduction}

Effective field theory is a useful tool for studying nuclear interactions.  To
describe low energy processes involving nucleons and pions in a model
independent way, all possible operators consistent with the symmetries of QCD
are included in an effective Lagrangian.  A further advantage of effective field
theory over potential models is that theoretical errors can be estimated in a
systematic way.  Contributions to an observable are organized by a power
counting in $Q/\Lambda$, where $Q$ is a momentum scale which characterizes
the process under consideration, and $\Lambda$ is the range of validity of the
effective theory.  A disadvantage of the effective field theory method is that the
expansion parameter may not be very small, so that the description is not
precise at low orders.  

In an effective field theory, ultraviolet divergences must be regulated and a
renormalization scheme defined.  The ultraviolet divergences give a constraint on
the power counting, because when a divergent loop graph occurs a contact
operator that can absorb the divergence must be included at the same or lower
order in $Q$.  This is familiar from pion chiral perturbation theory as discussed in
section~\ref{ChS}. The choice of regulator cannot affect physical results, but may
make implementing a renormalization scheme easier.  The renormalization scheme
and power counting are also tied together.  In a natural scheme, the renormalized
coefficients of the operators in the Lagrangian are normal in size based on
dimensional analysis with $\Lambda$.  Once a power counting is established one
can translate between different renormalization schemes at a given order in $Q$
without changing the physical predictions.  Recall from section~\ref{NN} that
counting powers of $Q/\Lambda$ in the nuclear effective theory is a subtle issue
because of the large S-wave scattering length, $a$.  In Refs.~\cite{ksw1,ksw2}
Kaplan, Savage, and Wise (KSW) devised a power counting to take this into
account.

Two different calculational techniques for the effective theory of nucleons are used
in the literature.  In one approach, the power counting is applied to regulated
N-nucleon potentials and the Schroedinger equation is solved
\cite{W1,W2,Ordonez1,Ordonez2,Ordonez3,vanKolck1,Lepage,park}.  Solving the
Schroedinger equation is equivalent to including all ladder graphs with the
potential as the two-particle irreducible kernel (see, for e.g., \cite{ksw0}).  The
second approach, advocated by KSW, is like ordinary chiral perturbation theory in
that the power counting is applied directly to the Feynman graphs which
contribute to the amplitude.   As discussed in section~\ref{NNintro}, a
non-relativistic propagator is used which includes the kinetic energy term to
regulate the infrared divergence at zero kinetic energy.  In the Feynman diagram
approach, dimensional regularization is the most convenient regulator, and analytic
results are readily obtained.  In the potential method, the Schroedinger equation is
usually solved numerically.  In practice, divergences are regulated and
renormalized couplings are defined using a finite cutoff scheme.  In Ref.~\cite{sf}, it
has been explicitly shown that without pions the potential method can deal with
large scattering lengths, and gives an expansion in $Q/\Lambda$.

An important aspect of the KSW analysis is the use of a novel renormalization
scheme, power divergence subtraction (PDS).  In PDS, loop integrals in Feynman
graphs are regulated using dimensional regularization, and poles in both $d=3$
and $d=4$ are subtracted.   The subtraction of $d= 3$ poles gives a power law
dependence on the renormalization point, $\mu_R$, to the coefficients of
four-nucleon operators.   Choosing $\mu_R\sim Q$, graphs with an arbitrary
number of $C_0^{(s)}(\mu_R)$ ($s=^1S_0$, or $^3S_1$) vertices scale as $1/Q$
and must be summed to all orders.  This is precisely the set of graphs that sums
corrections that scale like $(Qa)^n$.  Higher order contributions form a series in
$Q/\Lambda$.  In Ref.~\cite{Bira}, it is emphasized that it is possible to phrase the
power counting in a scheme independent manner.  The choice of scheme is
simply to give natural sized coefficients which make the power counting
manifest.  PDS is one example of such a scheme.  In Ref.~\cite{Cohen1}, it is
shown how the KSW power counting can be implemented by solving the
Schroedinger equation in a finite cutoff scheme.

Pions can be added to the effective field theory by identifying them as the
pseudo-Goldstone bosons of the spontaneously broken chiral symmetry of QCD. 
All operators with the correct transformation properties are added to the effective
Lagrangian.  This includes operators with insertions of the light quark mass matrix
and derivatives, whose coefficients are needed to cancel ultraviolet divergences
from loop graphs.  In dimensional regularization, these ultraviolet divergences are
of the form $p^{2n} m_\pi^{2m}/\epsilon$.  For instance, for nucleons in the
${}^3\!S_1$ channel, the two loop graph with three pions and a two loop graph
with two pions and one $C_0$ have ultraviolet divergences of the form
$p^2/\epsilon$.  This pole must be cancelled by a counterterm involving a
four-nucleon operator with 2 derivatives.   Because divergences of the form
$p^{2n}/\epsilon$ must be cancelled by local counterterms, pion exchange can
only be calculated in a model independent way if higher derivative contact
interactions are included at the same order that these divergence
occur\cite{lm,Savage}.  In Weinberg's \cite{W1} power counting, pion exchange is
included in the leading order potential.  Therefore, graphs with arbitrary numbers of
pions are leading order, while the counterterms necessary to cancel the ultraviolet
divergences in these graphs are subleading.  However, the potential method can
still be used.  As higher order derivative operators are added to the potential the
accuracy is systematically improved, because the onset of the model dependence
of the pion summation appears at higher order in $Q/\Lambda$.  For example, the
cutoff dependence of the two pion graph with one $C_0$ will be cancelled by
cutoff dependence in $C_2$.  At a given order, the left over cutoff dependence in
this method is a measure of the size of higher order corrections.

Different estimates of the range, $\Lambda_\pi$, of an effective theory of
nucleons with perturbative pions exist in the literature.  Some authors
\cite{Gegelia2,sf2,Cohen2} argue that $\Lambda_\pi$ is as small as $m_\pi$, so
that including perturbative pions is superfluous.    One estimate of the range is
given by KSW who conclude that $\Lambda_\pi \sim 300\,{\rm MeV}$.  They point
out that in PDS the renormalization group equation for the coefficient
$C_0(\mu_R)$ is modified by the inclusion of pions in such a way that for $\mu_R
\gtrsim 300\,{\rm MeV}$, $C_0(\mu_R)$ scales like $\mu_R^0$ instead of
$\mu_R^{-1}$.  Since the power counting is no longer manifest above this scale,
KSW conclude that the effective theory breaks down at this point.  In
Ref.~\cite{sf2} different renormalized couplings are obtained.  Here a breakdown
of the power counting for $C_2(\mu_R)$ at $\mu_R\sim m_\pi$ is observed.  A
crucial question is whether a breakdown in the running of the coupling
constants is a physical effect or simply an artifact of the renormalization scheme.
It is dangerous to draw conclusions based on the large momentum behavior of
the coupling constants because the beta functions of the couplings are scheme
dependent\footnote{This is in contrast with dimensionless coupling constants
like $g$ in QCD.  In that case the first two coefficients of the beta function are
scheme independent, so conclusions based on the behavior of the running
coupling constant at small coupling (e.g., asymptotic freedom) are physical.}.  In
section~\ref{pcrs}, a momentum subtraction scheme is introduced where the power
law dependence of the coupling constants persists even in the presence of
pions, and for all values of $\mu_R > 1/a$.  This scheme is called the OS scheme,
since in a relativistic theory it might be called an off-shell momentum
subtraction scheme.  In Ref.~\cite{Gegelia1}, a similar scheme is applied to the
spin singlet channel in the theory without pions, where it is shown to give
results identical to the PDS scheme.  The OS scheme is a natural scheme that
works with arbitrary partial waves and with pions.  Thus, the range of validity
of the effective theory is not limited by the large $\mu_R$ behavior of the
couplings.  PDS is still a useful scheme in which to calculate observables.  If one
splits $C_0(\mu_R)$ into a non-perturbative and perturbative part,
$C_0(\mu_R)=C_0^p(\mu_R) + C_0^{np}(\mu_R)$, then $C_0^{np}(\mu_R)\sim
1/\mu_R$ for all $\mu_R > 1/a$.  We will see that this split is also necessary if we
wish to avoid having the location of the pole in the amplitude shifted by chiral 
$m_\pi/\Lambda$ corrections.  Once this split has been performed, it is
straightforward to establish relations between the OS and PDS schemes order
by order in perturbation theory, and any prediction for an observable will be
identical in the two schemes up to the order in $Q/\Lambda_\pi$ to which it is
calculated.  Since in both schemes there is no scale where the power counting
breaks down, it is possible that $\Lambda_\pi > 300\,{\rm MeV}$.  The importance
of looking at results in several schemes is that it allows us to disentangle which
results are physical and which are scheme dependent.

Physically, one expects the effective theory to be valid up to a threshold where
new degrees of freedom can be created on-shell. For elastic nucleon scattering,
the relevant physical threshold is production of $\Delta$ resonances which
occurs at $p = \sqrt{M_N(M_{\Delta} - M_N)} = 525~{\rm MeV}$ (the S-wave
channels couple only to the $\Delta\,\Delta$ intermediate state so $p = \sqrt{2
M_N(M_{\Delta} - M_N)} = 740~{\rm MeV}$ \cite{Savage2}).  Above this scale, the
$\Delta$ must be included as an explicit degree of freedom.  Below this scale, the
$\Delta$ can be integrated out leaving an effective theory of pions and nucleons. 
Rho exchange becomes relevant at a scale, $p\sim m_\rho = 770\,{\rm MeV}$.
There is also a $N^*(1440) N$ intermediate state with a threshold of $p=685\,{\rm
MeV}$.  One might expect $\Lambda_\pi$ to be of order these thresholds. 
However, there is an intermediate scale of $300\,{\rm MeV}$ associated with short
distance contributions from potential pion exchange\footnote{ The phrase
``potential pion exchange'' will be used for a perturbative pion with energy
independent propagator.  This is sometimes called static pion exchange.}.  Using
dimensional analysis, a graph with the exchange of $n+1$ potential pions is
suppressed by $p/300\,{\rm MeV}$ relative to a graph with $n$ potential pions. 
Comparison of the size of individual graphs is scheme dependent (for example
the size of graphs differ in MS and in $\overline{\rm MS}$).  The $300\,{\rm MeV}$
scale applies only to a subset of graphs, and may change once all graphs at a
given order in $Q$ are included in the estimate.  Therefore, $300\,{\rm MeV}$ can
be taken as an order of magnitude estimate for the range of the theory, but the
actual range may be enhanced or suppressed by an additional numerical factor.

This then motivates the important question: How does one determine the range of
the effective field theory? This is obviously a question of great practical
importance.  Theoretical arguments can only give an approximate estimate for the
range.  A good example comes from SU(3) chiral perturbation theory.  In this
strong coupling theory, it is natural to expect that the range of the theory is the
chiral symmetry breaking scale $\Lambda_\chi \sim 2\sqrt{2} f_\pi = 1200\,{\rm
MeV}$ \cite{GM,Georgi1,georgi2}.  However, the convergence of the momentum
expansion will depend on the particular process under consideration.  For
instance, in $\pi-\pi$ scattering the range of the expansion is set by the threshold
for $\rho$ production, $m_\rho = 770\,{\rm MeV}$.  In this chapter, the range of the
two nucleon effective theory will be estimated using nucleon-nucleon scattering
data.  Our results are consistent with $\Lambda_\pi \sim 500\, {\rm MeV}$.  As we
will explain in section~\ref{range}, the error analysis is applied to $\delta$ rather
than to $p\cot\delta$ as in Ref.~\cite{sf2}.  This range does not depend on the
value of the renormalization point chosen, and is found in both the OS and PDS
schemes.  However, only next-to-leading order calculations have been used so it
is hard to estimate the error in this value.  When higher order corrections are
computed, it should be possible to obtain a reasonably accurate estimate of the
range of the two nucleon effective field theory with perturbative pions.  This $500\,
{\rm MeV}$ estimate is based solely on the phase shift data.  The accuracy of 
predictions for deuteron observables \cite{KaplanEFT} indicates 
$Q/\Lambda\sim 1/3$, which for $Q\sim m_\pi$ is $\Lambda\sim 400\,{\rm MeV}$.

In section~\ref{pcrs}, we review the power counting method of KSW
\cite{ksw1,ksw2}, and the PDS scheme.  The importance of being able to count
factors of the large nucleon mass in a non-relativistic effective field theory is
discussed.  We review the OS scheme,  which is compatible with the KSW power
counting.  We describe the procedure for defining the renormalized couplings using
local counterterms for each of these schemes.
  
In section~\ref{nopion}, we discuss the theory with only nucleons, where
$\Lambda\sim m_\pi$.   Local counterterms for both the PDS and OS schemes are
computed.  These counterterms are used to obtain the beta functions for the
four-nucleon operators, and we explain why the beta functions for the most
relevant operators in this theory are one-loop exact.  In section~\ref{pole} we
explain how treating part of $C_0$ perturbatively allows us to reproduce the
effective range expansion with an amplitude that has its pole in the physical
location at every order in perturbation theory.  In section~\ref{cutoff} we show how
calculations with a cutoff regulator reproduce the dimensional regularization
results.

The theory with nucleons and pions is analyzed in section~\ref{withpion}.  In the
${}^3\!S_1$ channel, there are corrections to the PDS beta functions at all orders
in $Q$.    As examples, we compute the PDS beta functions for
$C_0^{({}^3\!S_1)}(\mu_R)$ to order $Q$, and for $C_2^{({}^3\!S_1)}(\mu_R)$ to
order $Q^0$.  In this channel, even in the limit $m_\pi\to 0$, there are logarithmic
divergences (poles of the form $p^2/\epsilon$ in dimensional regularization).   In
the OS scheme, the ${}^3\!S_1$ beta functions can be calculated exactly.   We
compute the exact beta functions for $C_0(\mu_R)$, $C_2(\mu_R)$, and
$C_4(\mu_R)$ in the OS scheme in the ${}^1\!S_0$ and ${}^3\!S_1$ channels.  In
section~\ref{D2coupling}, the counterterms for the coupling constant
$D_2(\mu_R)$ are derived in the OS and PDS schemes.

In section~\ref{samp},  we discuss why it is important to have $\mu_R$ independent
amplitudes order by order in the expansion. In the OS scheme amplitudes are
$\mu_R$ independent, while in PDS $\mu_R$ independent amplitudes can be
obtained by treating part of $C_0(\mu_R)$ perturbatively.  If this is not done then
the sensitivity to $\mu_R$ is larger than one might expect \cite{Gegelia2}, for
reasons we explain.  Fits to the data are presented for different values of
$\mu_R$ and the coupling constants in both OS and PDS are shown to evolve
according to the renormalization group equations.

In section~\ref{range},  an error analysis similar to a method due to Lepage
\cite{Lepage} is used to investigate the range of the effective field theory with
perturbative pions at next-to-leading order.  Weighted fits are performed for the
scattering data in both the ${}^1\!S_0$ and ${}^3\!S_1$ channels.  Our results rule
out $\Lambda_\pi \sim m_\pi$, and are consistent  with $\Lambda_\pi \sim 500\, {\rm
MeV}$.


\section{Power counting and renormalization schemes}  \label{pcrs}

In this section, the KSW power counting and compatible renormalization schemes
are discussed.  The theory containing only nucleon fields is considered first.  The
renormalized couplings are then defined in terms of local counterterms, and the
KSW power counting for coefficients of four-nucleon operators is reviewed.  Next,
we consider the theory including pions.  We review the power counting for
potential pions, and explain the origin of the $300\,{\rm MeV}$ scale associated
with potential pion exchange. The PDS renormalization scheme is then discussed
and we introduce the OS momentum subtraction scheme, which is also compatible
with the power counting.  

Recall from section~\ref{NNintro} that the Lagrangian 
in the two nucleon sector is given by:
\begin{equation} \label{LN}
  {\cal L}_{NN} =  N^\dagger \Big[ i\partial_t + \overrightarrow\nabla^2/(2M) 
   + \ldots \Big] N - \sum_s\,\sum_{m=0}^{\infty} C^{(s)}_{2m}\,
   {\cal O}^{(s)}_{2m}   \,.
\end{equation} 
The $C_{2 m}$ appearing in Eq.~(\ref{LN}) are bare parameters. To
renormalize the theory, the bare coupling is separated into a renormalized
coupling and counterterms as follows:
\begin{eqnarray}
  C_{2m}^{\rm bare} &=&C_{2m}^{\rm finite} - \delta^{\rm uv}C_{2m} \,,
 \qquad\quad C_{2m}^{\rm finite} =C_{2m}(\mu_R) -
   \sum_{n=0}^\infty \delta^n C_{2m}(\mu_R) \,.  \label{ctexpn}
\end{eqnarray}
Note that we divide the counterterms into two classes. The first, which have
the superscript uv, contain all genuine ultraviolet divergences. These include
$1/\epsilon$ poles, if dimensional regularization is used, or powers and
logarithms of the cutoff if a hard cutoff is used. We will also include some finite
constants (e.g., the $-\gamma + {\rm ln}(4 \pi)$ that is subtracted in
$\overline{MS}$) if this proves to be convenient for keeping expressions
compact.  By construction, these counterterms are $\mu_R$ independent, but
will depend on $C_{2m}^{\rm finite}$.  The renormalized coupling is denoted
$C_{2m}(\mu_R)$. The remaining counterterms, $\delta^nC_{2m}(\mu_R)$,
contain no ultraviolet divergences and will be referred to as the finite
counterterms.  The choice of the finite counterterms differentiates between the
schemes studied here. An infinite number of finite counterterms are needed
because an infinite number of loop graphs are included at leading order. The 
renormalization is carried out order by order in the loop expansion. The
superscript $n$ indicates that $\delta^nC_{2m}$ is included at tree level for a
graph with $n$ loops.  When higher loop graphs are considered, the $\delta^n
C_{2m}$ counterterm takes the place of $n$ loops \cite{Collins}.  For example,
at three loops we have three loop diagrams with renormalized couplings at the
vertices, two loop diagrams with a $\delta^1 C$ counterterm, one loop 
diagrams with either one $\delta^2 C$ or two $\delta^1 C$'s, and a tree level
diagram with $\delta^3 C$.  Examples are given in section~\ref{nopion}.

For the nucleon theory, the kinematic part of the power counting  is very
simple \cite{W1,ksw3}.  $Q$ is identified with a typical external momentum
characterizing the process under consideration.  For instance, in elastic
nucleon-nucleon scattering $Q\sim p$, where $p$ is the center of mass
momentum\footnote{ For the scattering $N(\vec q+\vec p\,) + N(\vec
q-\vec p\,) \rightarrow N(\vec q+\vec p\,') + N(\vec q-\vec p\,')$ it is useful to
define $p = \sqrt{ ME_{tot}-\vec q^2 +i\epsilon }$, where $E_{tot}$ is the total
incoming energy, and $M$ is the nucleon mass. To simplify the notation we will
work in the center of mass frame, $\vec q=0$, where $p^2 = \vec p\,^2 = \vec
p\,'\,^2 = ME$, and $E$ is the center of mass energy.  For external particles,
one can always translate between $E$ and $p$ using the equations of
motion.}.  Each nucleon propagator gives a $Q^{-2}$, each spatial derivative a
$Q$, each time derivative a $Q^2$, and each loop integration a $Q^5$.  

In the theory with only nucleons, the only graphs relevant to $2 \to 2$ scattering
are bubble chains. Consider a graph $\cal{G}$ with $L$ loops in the
non-relativistic limit. In dimensional regularization, each loop will give a factor $M
p/4\pi$, and there are $L+1$ vertices, each giving a factor $-i C_{2 m}^{\rm finite}
p^{2 m}$. If the operator $O_{2m}$ appears $n_m$ times in the graph ($L+1 =
\sum_m n_m$) the result is:
\begin{eqnarray}
 {\cal{G}} &=& {4 \pi \over M} \prod_{m=0}^\infty \bigg( {-i M C_{2 m}^{\rm finite}
  \over  4 \pi}  \bigg)^{n_m}  p^{\,j},   \qquad\quad 
 \mbox{where} \quad  j=\sum_{m=0}^\infty 2\,m\,n_m+ L \,. \label{pscale}
\end{eqnarray}
If one matches onto the effective range expansion in $\overline{\rm MS}$ one
finds \cite{ksw0}
\begin{eqnarray}
  C_{2m}^{\rm finite} \sim 4\pi\,a^{m+1}/(M\,\Lambda^m) \, .
\end{eqnarray}
Note that all graphs ${\cal G}$ are proportional to $1/M$ in agreement with the 
discussion in section~\ref{NNintro}.  The  large S-wave scattering lengths 
enhance the importance of some
graphs compared to the $p$ power counting.  This affects the power counting for
S-wave couplings, and through the mixing, couplings with $L$ and/or $L' =2$
and $S=1$.  For other channels we have the usual chiral power counting of
$p$'s.  The power counting for insertions of four-nucleon operators is \cite{ksw3}
\begin{eqnarray} \label{pc} 
  C_{2m}^{(s)}(\mu_R)\,&{\cal O}_{2m}^{(s)}& \sim  C_{2m}^{(L-L')}(\mu_R)\,
  p^{2m} 
 \sim Q^{\,q(s,m)}\ ,\qquad\quad\mbox{where} \nn \\[10pt]
  q(s,m) &=& \left\{  \begin{array}{ccl}  {m-1} & & \mbox{for}\ L=L'=0 \\
  {m} & & \mbox{for}\ S=1 \ \mbox{and}\ (L,L'=0,2) \,, \mbox{ or }\ (L,L'=2,0) \\
  {m+1} & & \mbox{for}\ S=1 \ \mbox{and}\ L,L'=2,2  \\
  {2m}& &\mbox{for all other } S, L, \mbox{ and } L' \end{array}  \right.  \,.
\end{eqnarray}
With the coefficients $C_{2m}$ scaling as in Eq.~(\ref{pc}), the graph 
${\cal G}$ scales as
\begin{eqnarray}
 {\cal G}\sim Q^{\,i} \qquad\qquad \mbox{where}\quad 
   i= \sum_m  n_m\, q(s,m) \:+\:L \,.  \label{Qscale}
\end{eqnarray}
Note that the power of $Q$ is less than or equal to the power of $p$, $i \le j$. 
A useful mnemonic for this power counting is $1/a \sim Q$, however, the 
power counting is still valid for $Qa \gg 1$.

This $Q$ power counting  will be manifest in any renormalization
scheme\footnote{Although the power counting can be implemented in different
schemes, the PDS scheme introduced in Ref.~\cite{ksw1} was very useful for
initially working out the power counting.} in which the $C_{2m}(\mu_R)$
scale with $\mu_R\sim Q$ in such a way that Eq.~(\ref{pc}) is true.  At leading
order the counterterms $\delta^nC_{2m}(\mu_R)$ will have the same $Q$ scaling
as the coefficient $C_{2m}(\mu_R)$.  These schemes may differ by contributions in
$C_{2m}(\mu_R)$ that scale with a larger power of $\mu_R/\Lambda$, since this
will not change the power counting  at low momentum.  

Let us now discuss the theory with pions. To add pions, we identify them as
the three pseudo-Goldstone bosons which arise from the breaking of chiral
symmetry, $SU(2)_L\times SU(2)_R \to SU(2)_V$.  With the pions included in
this way, we are doing an expansion in $m_\pi/\Lambda_\pi$ and
$p/\Lambda_\pi$. Note that in this theory, no matter how small $p$ is made the
expansion parameter will never be smaller than $m_\pi/\Lambda_\pi$.  This
theory still includes the four-nucleon operators in Eq.~(\ref{L2s}), but the short
distance physics parameterized by the coefficients $C_{2m}$ is different
because the pion is no longer integrated out.  In the pion theory, the short
distance $C_{2m}$ coefficients should be independent of the scale $m_\pi$. 
All the $m_\pi$ dependence  is now contained explicitly in powers of the light
quark mass matrix in the Lagrangian.

Pions will be encoded in the representation, $\Sigma=\xi^2=\exp{(2i\Pi/f)}$,
where
\begin{eqnarray}{
  \Pi = \left(  \begin{array}{cc} \pi^0/\sqrt{2} & \pi^+ \\ 
    \pi^- & -\pi^0/\sqrt{2} \end{array} \right) \,, }
\end{eqnarray}
and $f=130\, {\rm MeV}$ is the pion decay constant.  Under $SU(2)_L\times 
SU(2)_R$ the fields transform as $\Sigma \to L \Sigma R^\dagger$, $\xi \to 
L\xi U^\dagger = U\xi R^\dagger$, and $N \to U N$. The chiral covariant 
derivative is $D^\mu = \partial^\mu + \frac12(\xi\partial^\mu \xi^\dagger +
\xi^\dagger\partial^\mu \xi )$.
With pions we have the following Lagrangian with terms involving 0, 1 and 2
nucleons:
\begin{eqnarray}
 {\cal L}_\pi &=& \frac{f^2}{8} {\rm Tr}\,( \partial^\mu\Sigma\: \partial_\mu
  \Sigma^\dagger )+ \frac{f^2 w}{4}\, {\rm Tr} (m_q \Sigma+m_q
  \Sigma^\dagger) \nn \\*
 &+&
   \frac{i g_A}2\, N^\dagger \sigma_i (\xi\partial_i\xi^\dagger - 
   \xi^\dagger\partial_i\xi) N  + N^\dagger \bigg( i D_0+\frac{\vec D^2}{2M} 
   \bigg) N \label{Lpi} \\*
 &-& \sum_{s,m} C^{(s)}_{2m}\,  {\cal O}^{(s)}_{2m}   
  -{D_2^{(s)}} \omega {\rm Tr}(m^\xi ) ( N^T P^{(s)}_i N)^\dagger ( N^T 
  P^{(s)}_i N) \nn + \ldots \,.
\end{eqnarray}
Here $m^\xi=\frac12(\xi m_q \xi + \xi^\dagger m_q \xi^\dagger)$, $m_q={\rm
diag}(m_u,m_d)$ is the quark mass matrix,  $m_\pi^2 = w(m_u+m_d)$ where
$w$ is a constant,  and $g_A=1.25$ is the nucleon axial-vector coupling.   The
ellipsis in Eq.~(\ref{Lpi}) denote terms with more derivatives and more powers
of $m^\xi$.

With pions there are additional complications to the power counting
\cite{lm,ls,Savage} which are similar to those encountered in Non-Relativistic QED
and QCD \cite{Caswell,Bodwin,labelle}.  The complications arise because there are
two relevant energy scales for the pions, $E_\pi \sim Q^2/M$ for potential pions,
and $E_\pi \sim Q$ for radiation pions.  When the energy integral in loops is
performed via contour integration, the graphs with potential pions come from terms
in which one keeps the residue of a nucleon propagator pole.  In these loops, the
energy of the loop momentum is $\sim Q^2/M$ and the energy dependent pieces of
the pion propagator are suppressed by an additional $Q^2/M^2$.  Nucleon
propagators give a $Q^{-2}$ and the loop integrals give $Q^5$.  There are also
radiative pion graphs, in which the residue of the pion pole is kept. The power
counting for these graphs is discussed in Chapter 6.  In general graphs with
radiation pions are higher order than those with potential pions. The combined
propagator and vertices for a single pion exchange give $Q^0$, so the pions can
be treated perturbatively \cite{ksw2}.  

In general, pion exchange gives both long and short distance contributions. 
The short distance contributions from potential pions are important since they
may limit the range of the effective field theory.  A single potential pion
exchange gives
\begin{eqnarray}
 i \frac{g_A^2}{2f^2}\: {\vec q \cdot \vec \sigma_{\alpha\beta}\ 
  \vec q \cdot \vec \sigma_{\gamma\delta} \over  \vec q\,^2 +m_\pi^2 }\ \vec 
  \tau_1\,\cdot\,\vec \tau_2  = 
 i \frac{g_A^2}{2f^2}\: \left[ {\vec q \cdot  \vec \sigma_{\alpha
  \beta}\ \vec q \cdot \vec \sigma_{\gamma\delta} \over 
  \vec q\,^2 }\ - { m_\pi^2\ \vec q \cdot \vec \sigma_{\alpha
  \beta}\ \vec q \cdot \vec \sigma_{\gamma\delta} \over 
  \vec q\,^2\ (\vec q\,^2+ m_\pi^2 )} \right] \vec \tau_1\,\cdot\,\vec \tau_2 
  \ \,, \nn\\  \label{ppip}
\end{eqnarray}
where the spin indices connect to nucleon fields $N^\dagger_\alpha N_\beta
N^\dagger_\gamma N_\delta$ which belong to external lines or propagators.  The
first term dominates for $\vec q^2 \gg m_\pi^2$, and can be isolated by taking the
limit $m_\pi \to 0$.  To study the dominant short distance contribution of pion
exchange the $m_\pi$ dependence can be neglected.  Graphs with radiation pions
are suppressed by powers of $(m_\pi/M)^{1/2}$.   In the non-relativistic limit, with
only potential pions, the only loop diagrams are ladders.  Consider an arbitrary
graph ${\cal G}$ with $n_m$ four point vertices, $C_{2m}^{\rm finite}$, and $k$
potential pions.  For $L$ loops, this graph has a total of $L+1= k +\sum_m n_m$
vertices, and with $m_\pi=0$
\begin{eqnarray} \label{p2scale} 
  {\cal G} &\propto& \bigg(\frac{M}{4\pi}\bigg)^{L}\:  \bigg( \frac{-i g_A^2}{2\,f^2}
   \bigg)^k\: p^{\,j}\: \prod_{m=0}^\infty (-i C_{2m}^{\rm finite})^{n_m} \ \  \\
  &=&\ \ 
  \frac{4\pi}{M} \:  \bigg( \frac{-i M g_A^2}{8\,\pi f^2}   \bigg)^k\: p^{\,j}\: 
  \prod_{m=0}^\infty  \bigg( \frac{-i MC_{2m}^{\rm finite}}{4\pi}\bigg)^{n_m} \nn \,, 
  \quad \mbox{where} \quad  j=\sum_{m=0}^\infty 2\,m\,n_m+L \nn\,.
\end{eqnarray} 
In the ${}^1\!S_0$ channel, the relation in Eq.~(\ref{p2scale}) becomes an
equality.  The graph ${\cal G}\sim Q^{\,i}$ where $i$ is given in Eq.~(\ref{Qscale}).
The power counting of the $\delta^{\rm uv} C_{2m}$ counterterms is determined
by the need to cancel ultraviolet divergences, and will not spoil the scaling for
the renormalized coefficients, since $i \le j$.  For graphs with only potential pions
($n_m=0$), it appears that our expansion is in $p/(300\,{\rm MeV})$ since 
\begin{eqnarray}
 {1\over \Lambda_{NN}}\equiv \frac{M g_A^2}{8\pi f^2} \sim (300\,{\rm MeV})^{-1}   \,.
\end{eqnarray}
Comparing the size of potential pion graphs therefore predicts a range of $300\,{\rm
MeV}$, but the size of these graphs may change depending on the renormalization
scheme (i.e., the finite subtractions). As mentioned in section~\ref{ChS} the value
of the loop graphs and contact interactions separately do not have unique values. 
It is not known {\em a priori} how the contact interactions will affect the range of
the effective theory.  The scale $300\,{\rm MeV}$ is therefore an approximate
estimate for the range of the effective field theory with perturbative pions.  A
further discussion of this issue will be taken up in section IV.

Next, consider the power counting for coefficients that multiply operators with
powers of $m_q$.  If we are interested in momenta of order $m_\pi$, then one
counts $m_q \sim m_\pi^2 \sim Q^2$.  Therefore, any interaction term that has
an operator with a total of $2m$ powers of $p$ and $m_\pi$ will scale as
$Q^{\,q(s,m)}$ where $q(s,m)$ is given in Eq.~(\ref{pc}).  For example,
$D_2^{({}^1\!S_0)}m_\pi^2 \sim Q^0$.  It is important to understand that in the
KSW power counting $D_2$ should be treated perturbatively even though the
structure of the operator it multiplies is similar to that of the leading four nucleon
operator with no derivatives.  Graphs with radiation pions will also give
contributions with powers of $m_\pi^2$. 
  
\subsection{The PDS scheme}

PDS is one scheme in which the KSW power counting is manifest.  In PDS, we first
let $d=4$ and take the $\delta^{\rm uv}C_{2m}$ counterterms to subtract
$1/\epsilon$ poles as in $\overline{\rm MS}$.  We use the notation $\mu_R$ for the
renormalization point, and $\mu$ for the dimensional regularization parameter.  In
PDS, like in the $\overline{\rm MS}$ scheme, one takes $\mu=\mu_R$.  In a
momentum subtraction scheme this is not necessary.  The next step in PDS is to
take $d=3$ and define the finite counterterms, $\delta^nC_{2m}(\mu_R)$, to
subtract the $1/(d-3)$ poles in the amplitude.   Graphs which contribute to
$\delta^nC_{2m}(\mu_R)$ are those whose {\em{vertices}} have a total of $2m$
derivatives.  When calculating the $\delta^nC_{2m}(\mu_R)$ we can take $m_\pi=0$
since, for instance, counterterms proportional to $m_\pi^2$ renormalize coefficients
like $D_2(\mu_R)$.   After making these subtractions everything is continued back
to four dimensions.  It is this second set of finite subtractions that gives the right
power law dependence on $\mu_R$.  To define the coefficients that multiply
operators with powers of $m_q$, a similar procedure is followed except we count
the powers of $m_\pi^2$ at the vertices.  In PDS with just nucleons, all the graphs
that affect the running of $C_{2m}(\mu_R)$ are order $Q^{\,q(s,m)}$, except for
those with intermediate states of different orbital angular momentum. For example,
the beta function for $C_4^{({}^3\!S_1)}$ has contributions $\sim Q$
($q({}^3\!S_1,4) = 1$), as well as contributions $\sim Q^3$ from graphs with two
$C_2^{({}^3\!S_1- {}^3\!D_1)}$ vertices.  When pions are included there are
additional graphs that are sub-leading in the power counting and affect the
running of the couplings.  In fact, in section~\ref{withpion} we will show that there
will be corrections to the PDS beta function for $C_0^{({}^3\!S_1)}(\mu_R)$ at all
orders in $Q$.  

\subsection{The OS scheme}

Another renormalization scheme that can be used to reproduce the power
counting in Eq.~(\ref{pc}) is a momentum subtraction scheme.  A simple
physical definition for the renormalized couplings can be made by relating the
couplings to the amplitude evaluated at the unphysical momentum
$p=i\mu_R$.  This scheme will be called the OS scheme, since in a
relativistic field theory this would be referred to as an off-shell momentum
subtraction scheme.  We start by dividing up the full amplitude as
\begin{eqnarray}
  i\,{\cal A}^s =  i\, \sum_{m=0}^\infty  {\cal A}^s_{2m} +\ldots \,. \label{Asplit}
\end{eqnarray}
Here ${\cal A}^s_{2m}$ contains the Feynman diagrams that will be used to
define the coupling $C_{2m}^{(s)}(\mu_R)$ (or equivalently the counterterms
$\delta^nC_{2m}$).  The ellipsis in Eq.~(\ref{Asplit}) denotes pieces that vanish
as $m_\pi\to 0$ which are not needed to define $C_{2m}(\mu_R)$. $A_{2m}^s$ 
is defined to contain the remaining graphs that scale as $Q^{\,q(s,m)}$, where
$q(s,m)$ is defined in Eq.~(\ref{pc}).  The definition for the renormalized coupling 
is then 
\begin{eqnarray}
 i  \left.{\cal A}^s_{2m} \right|_{\mbox{\scriptsize$\begin{array}{c} \ p=i\mu_R
 \\[-4pt] \!\!\!\! m_\pi=0 \end{array}$}} = -i  C_{2m}^{(s)}(\mu_R) \:
  (i\mu_R)^{2m}\,.  \label{rc2m}
\end{eqnarray} 
As we will see, this ensures that $C_{2m}(\mu_R)$ scales in the desired way.  In
general, there may be divergent graphs scaling as $Q^{\,i}$ and $p^{2m}$ ($i \le
2m$) whose $1/\epsilon$ poles need to be absorbed by a $\delta^{\rm
uv}C_{2m}$ counterterm.  For example, consider the graph with two pions and
one $C_0$ shown in row four of Fig.~\ref{fig_2ppi}.  This graph has a
$p^2/\epsilon$ pole which is cancelled by a counterterm $\delta^{\rm uv}C_2$.
The finite part of this graph is used in Eq.~(\ref{rc2m}) to define $C_4(\mu_R)$
because the graph is order $Q$.  The key point is that since $q(s,m)\le 2m$, an
ultraviolet divergence that appears in a graph of a given order can always be
absorbed into a coefficient that appeared at the same or lower order in the power
counting.  Therefore, we will define $\delta^{\rm uv}C_{2m}$ in $\overline{\rm
MS}$ to subtract all four dimensional $1/\epsilon$ poles so that these
subtractions are independent of the renormalization point.  The finite
counterterms are then fixed by the renormalization condition in Eq.~(\ref{rc2m}).

 \begin{figure}[t!]  
  \centerline{\epsfysize=8.2truecm \epsfxsize=16.0truecm 
	\epsfbox{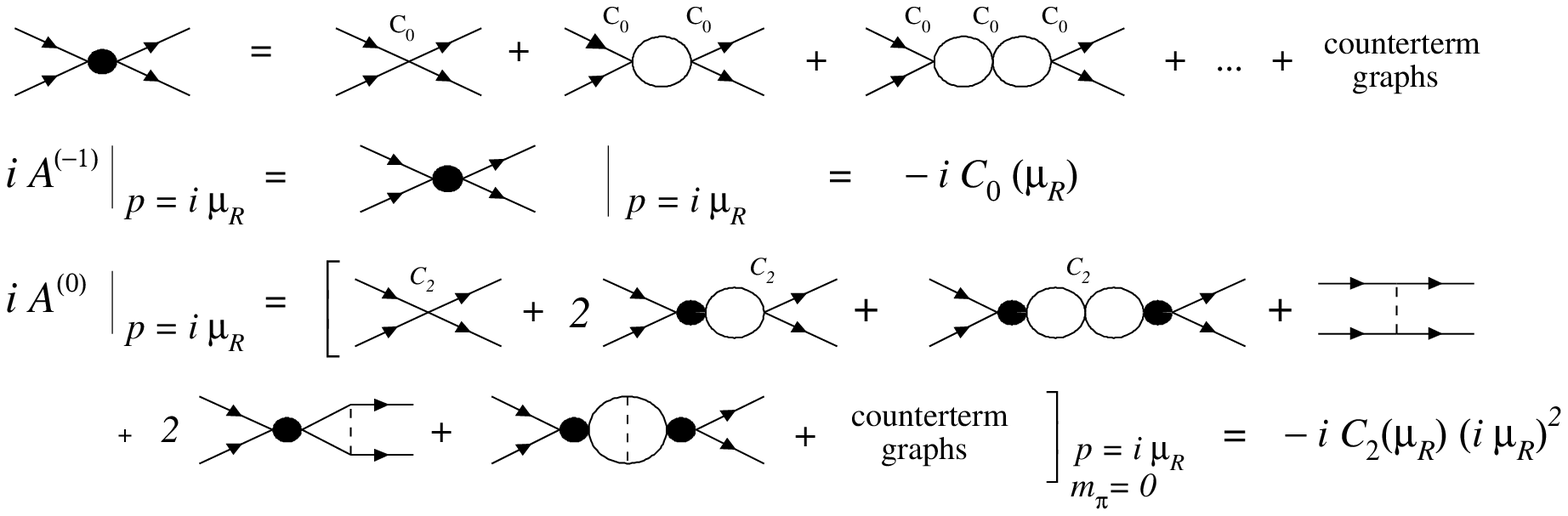}} 
{ \caption{Renormalization conditions for $C_0(\mu_R)$ and $C_2(\mu_R)$ in 
the OS scheme.  $i\,A^{(-1)}$ is the four point function with $C_0(\mu_R)$
 and $\delta^n C_0(\mu_R)$ vertices, evaluated between incoming and 
 outgoing  ${}^1S_0\mbox{ or }{}^3S_1$ states.  The amplitude $A^{(0)}$ 
 contains graphs with one $C_2$ or one potential pion dressed with $C_0$ 
 bubbles.}  \label{fig_C0} }  
\end{figure}  
In the OS scheme, the couplings $C_0(\mu_R)$ and $C_2(\mu_R)$ are defined by
the renormalization condition in Fig.~\ref{fig_C0}.  This
condition is to be imposed order by order in the loop expansion so that graphs
with $n$ loops determine $\delta^nC_0(\mu_R)$.  The $m_\pi=0$ part of pion
graphs contribute to $C_{2m}(\mu_R)$ for $m\ge1$ in which case the
condition $m_\pi=0$ in Eq.~(\ref{rc2m}) is important.  In the theory with pions, we
also need to define couplings multiplying powers of $m_q$, like $D_2$ in
Eq.~(\ref{Lpi}).  To define these couplings we will not include all the terms in the
amplitude proportional to $m_\pi^2$.  In particular, pion exchange graphs give long
distance non-analytic contributions which will not be used to define the running of
the short distance coupling $D_2(\mu_R)$.  The idea that long distance physics
must be excluded from the short distance coefficients is discussed in
Ref.~\cite{sf2}.  A detailed discussion of how we define $D_2(\mu_R)$ in the OS
scheme will be left to section V.

Note that in the OS scheme there is another approach for calculating an
amplitude in terms of renormalized couplings. One can calculate all loop
graphs in ${\cal A}^s_{2m}$ in terms of the finite (or $\overline {\rm MS}$)
parameters and then demand that the renormalization condition in
Eq.~(\ref{rc2m}) is satisfied. This gives expressions for the renormalized
couplings in terms of the constants $C_{2m}^{\rm finite}$.  The amplitude can
then be written in terms of renormalized couplings by inverting these
equations.  This simplifies higher order calculations.

In the OS scheme, when an amplitude is written in terms of renormalized couplings
it will be explicitly $\mu_R$ independent at each order in $Q$.  The $\mu_R$
dependence in PDS with pions is cancelled by higher order terms.  It is possible to
obtain $\mu_R$ independent amplitudes in PDS if part of $C_0(\mu_R)$ is treated
perturbatively\cite{ms0}.  Consequences of this $\mu_R$ dependence will be
discussed in section~\ref{samp}.  In section~\ref{nopion} we will see that for the
theory with just nucleons the OS scheme gives very similar definitions for the
renormalized couplings to those in PDS.  In section~\ref{withpion}, we investigate
the running couplings in both schemes in the theory with pions.


\section{Theory with pions integrated out} \label{nopion}

In this section, we compute the renormalized couplings in the non-relativistic
nucleon effective theory without pions.  We expect $\Lambda\sim m_\pi$.  This
theory will be examined in both PDS and the OS scheme.  The renormalization
program is implemented by explicitly calculating the local counterterms.  In
Ref.~\cite{Gegelia1}, it is shown that the PDS and OS schemes give the same
renormalized coupling constants in the $^1S_0$ channel.  Here we also consider
the spin-triplet channel and higher derivative operators.  Divergences in loop
integrals are regulated using dimensional regularization.  For the OS scheme, the
same renormalization program can be carried out using a momentum cutoff
regulator as shown in section~\ref{cutoff}.  Following Ref.~\cite{ksw2}, we will
multiply each loop integral by $(\mu/2)^{(4-d)}$, and define $d=4-2\epsilon$. 
Since there are no logarithmic divergences in the nucleon theory, $\delta^{\rm
uv}C_{2m}=0$ in dimensional regularization.

\begin{figure}[t!]  
 \epsfysize=7.0truecm \epsfbox{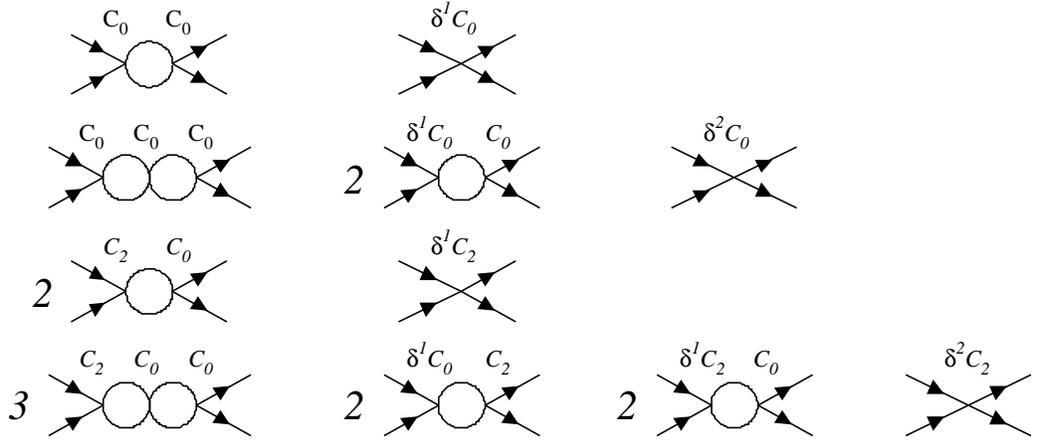}  
{  \caption{One and two loop counterterms for $C_0$ and $C_2$. 
The solid lines are nucleon propagators, and symmetry factors are shown 
explicitly. The generalization to higher loops is straightforward.}   \label{fig_ct0} }
\end{figure}

In both PDS and OS scheme, it is straightforward to derive the finite
counterterms, $\delta^n C_{2m}(\mu_R)$.  The tree level graphs with
$C_0(\mu_R)$ and $C_2(\mu_R)$ satisfy the renormalization condition in
Eq.~(\ref{rc2m}).  Therefore, in both PDS and OS, $\delta^0 C_0 = \delta^0 C_2
= 0$.  At one and two loops we have the graphs in Fig.~\ref{fig_ct0}.
In $d$ dimensions, the two graphs in the first row give
\begin{eqnarray}  \label{2c0}
 && (-i C_0)^2 \Big(\mbox{\small$\frac{-iM}{4\pi}$}\Big)
  \Gamma(\mbox{\small$\frac{3-d}2$}) \Big(\frac{\mu}2\Big)^{4-d}
  \bigg( \frac{-p^2-i\varepsilon}{4\pi} \bigg)^{\frac{d-3}2}  + i\:\delta^1 C_0 \,,
\end{eqnarray}
determining $\delta^1C_0$.  In PDS, we define the counterterm to cancel the $d=3$
pole in Eq.~(\ref{2c0}) and then continue back to four dimensions.  In the OS
scheme, we take $d=4$ and demand that the contribution to the amplitude in
Eq.~(\ref{2c0}) satisfies the condition in Fig.~\ref{fig_C0}.  The counterterms
calculated in each scheme are the same (with $\mu=\mu_R$ in PDS).  In both
schemes the counterterms determined from the graphs in Fig.~\ref{fig_ct0} are
\begin{eqnarray}  
 \delta^1 C_0(\mu_R) \!\!&=&\!\! \bigg(\frac{M\mu_R}{4\pi}\bigg)\,C_0(\mu_R)^2 \ , 
   \qquad\quad\quad
 \delta^2 C_0(\mu_R) =-\bigg(\frac{M\mu_R}{4\pi}\bigg)^2\,C_0(\mu_R)^3 \ ,\\
 \delta^1 C_2(\mu_R) \!\!&=&\!\! 2 \bigg(\frac{M\mu_R}{4\pi}\bigg)\,C_2(\mu_R)
   C_0(\mu_R)\ ,\quad
 \delta^2 C_2(\mu_R) = -3 \bigg(\frac{M\mu_R}{4\pi}\bigg)^2\, C_2(\mu_R)
   C_0(\mu_R)^2\ . \nn
\end{eqnarray}
Note that it is essential that loop graphs also have vertices with insertions
of the counterterms.  For instance, the contribution to the amplitude from all
the graphs in the second row of Fig.~\ref{fig_ct0} is
\begin{eqnarray}
   -i C_0(\mu_R)^3  \bigg( \frac{M(ip+\mu_R)}{4\pi} \bigg)^2 \,.
\end{eqnarray}
If the one-loop graph with a $\delta^1 C_0$ counterterm had been left out then
the answer would have been proportional to $(p^2 + \mu_R^2)$ which is not
correct.  Since the loops in the nucleon theory factorize, the renormalized
$n$-loop graph gives $(ip+\mu_R)^n$.  Loop graphs will not always factorize
once pions are included.  

It is straightforward to extend this calculation to $n$ loops and to include 
higher derivatives.  In both the OS and PDS schemes, this gives the following 
counterterms ($s= {}^1\!S_0, {}^3\!S_1$, $n\ge 1$):
\begin{eqnarray}\label{ct024}
\lefteqn{{}^1S_0:\ \ \ \  } \nn \\
 &&\qquad\qquad \delta^nC_0^{({}^1\!S_0)}(\mu_R) = (-1)^{n+1}\,\left(
 {M\mu_R\over 4\pi}  \right)^n \  C_0^{({}^1\!S_0)}(\mu_R)^{\,n+1} \,, \nn \\
 &&\qquad\qquad  \delta^nC_2^{({}^1\!S_0)}(\mu_R) = (-1)^{n+1}\,\left( 
  {M\mu_R\over 4\pi} \right)^n\:  (n+1)\  C_0^{({}^1\!S_0)}(\mu_R)^{\,n} \: 
  C_2^{({}^1\!S_0)}(\mu_R) \,, \nn \\ 
 &&\qquad\qquad  \delta^nC_4^{({}^1\!S_0)}(\mu_R) = (-1)^{n+1}\,\left( 
   {M\mu_R\over 4\pi} \right)^n (n+1) C_0^{({}^1\!S_0)}(\mu_R)^{n-1} \nn \\
 &&\qquad\qquad  \qquad\qquad \times \Big[ C_4^{({}^1\!S_0)}(\mu_R)\: 
  C_0^{({}^1\!S_0)}(\mu_R) + \frac{n}2 C_2^{({}^1\!S_0)}(\mu_R)^2  \Big] \,, \nn \\  
\lefteqn{{}^3\!S_1,{}^3\!D_1:\ \ \ \  }  \\
 &&\qquad\qquad  \delta^nC_0^{({}^3\!S_1)}(\mu_R) = (-1)^{n+1}\,\left( 
  {M\mu_R\over 4\pi} \right)^n \  C_0^{({}^3\!S_1)}(\mu_R)^{\,n+1} \,, \nn \\
 &&\qquad\qquad  \delta^nC_2^{({}^3\!S_1)}(\mu_R) = (-1)^{n+1}\,\left( 
  {M\mu_R\over 4\pi} \right)^n\:  (n+1)\  C_0^{({}^3\!S_1)}(\mu_R)^{\,n} \: 
  C_2^{({}^3\!S_1)}(\mu_R) \,, \nn \\ 
 &&\qquad\qquad  \delta^nC_2^{({}^3\!S_1- {}^3\!D_1)}(\mu_R) = (-1)^{n+1}\,
   \left( {M\mu_R\over 4\pi} \right)^n \: C_0^{(^3S_1)}(\mu_R)^{\,n} \:
   C_2^{({}^3\!S_1- {}^3\!D_1)}(\mu_R)  \,, \nn \\ 
 &&\qquad\qquad  \delta^nC_4^{({}^3\!D_1)}(\mu_R) = (-1)^{n+1}\,\left( 
  {M\mu_R\over 4\pi} \right)^n C_0^{(^3S_1)}(\mu_R)^{\,n-1} \, \Big[ 
  C_2^{({}^3\!S_1- {}^3\!D_1)} (\mu_R)\Big]^2 \,. \nn 
\end{eqnarray}
Note that with $\mu_R \sim Q$, the counterterms have the same $Q$ scaling as
their corresponding coupling constant.  In the PDS scheme, there are also 
subleading terms that come from the mixing of angular momentum 
states.  In PDS
\begin{eqnarray}
 \delta^nC_4^{({}^3\!S_1)}(\mu_R) &=& (-1)^{n+1}\,\left( {M\mu_R\over 4\pi} 
   \right)^n C_0^{({}^3\!S_1)}(\mu_R)^{n-1}\Big[ (n+1)C_4^{({}^3\!S_1)}(\mu_R)\: 
  C_0^{({}^3\!S_1)}(\mu_R) \nn\\
   &&\qquad\qquad+ \frac{n(n+1)}2 C_2^{({}^3\!S_1)}(\mu_R)^2 
  +  n C_2^{({}^3\!S_1- {}^3\!D_1)}(\mu_R)^2 \Big]    \,, 
\end{eqnarray}
where the last term is suppressed by $Q^2$.  In the OS scheme
\begin{eqnarray}
 \delta^nC_4^{({}^3\!S_1)}(\mu_R) &=& (-1)^{n+1}\,\left( {M\mu_R\over 4\pi} 
   \right)^n  C_0^{({}^3\!S_1)}(\mu_R)^{n-1} \nn\\
  &\times&\!\!\!\!   \Big[ (n+1) C_4^{({}^3\!S_1)}(\mu_R)\: 
  C_0^{({}^3\!S_1)}(\mu_R) + \frac{n(n+1)}2 C_2^{({}^3\!S_1)}(\mu_R)^2  \Big] \,,  
\end{eqnarray}
which is the same as the ${}^1\!S_0$ channel.  In the OS scheme, graphs with
two $C_2^{({}^3\!S_1- {}^3\!D_1)}$ couplings and any number of
$C_0^{({}^3\!S_1)}$'s contribute to the beta function for $C_8^{({}^3\!S_1)}$
since they are order $Q^3$.  One might also ask about channels where the
large scattering length does not effect the power counting.  In this case
$C_{2m}^{(s)}(\mu_R) \sim Q^0$, and we recover the usual chiral power 
counting.   In our OS scheme, the counterterms $\delta^nC_{2m}^{(s)}(\mu_R)$ 
in these channels are either zero or a constant independent of $\mu_R$.

From Eq.~(\ref{ctexpn}) one can derive the beta functions using
\begin{eqnarray}
  \beta_{2m} \equiv \mu_R { \partial \over \partial {\mu_R} } C_{2m}(\mu_R)
  = \sum_{n=0}^\infty \mu_R { \partial \over \partial {\mu_R} } \delta^n C_{2m}
  (\mu_R) \,.  \label{beta2m}
\end{eqnarray}
The first few beta functions are 
\begin{eqnarray} \label{bet02}
\lefteqn{{}^1\!S_0:\ \ \ \  } \nn \\
  &&\quad\qquad \beta_0^{({}^1\!S_0)} = \bigg( {M\mu_R\over 4\pi} \bigg) 
  C_0^{({}^1\!S_0)}(\mu_R)^2 \,,  \nn\\
  &&\quad\qquad \beta_2^{({}^1\!S_0)} =2 \bigg( {M\mu_R\over 4\pi} \bigg) 
  C_0^{({}^1\!S_0)}(\mu_R)\: C_2^{({}^1\!S_0)}(\mu_R)\,, \nn \\
  && \quad\qquad \beta_4^{({}^1\!S_0)} =  \bigg( {M\mu_R\over 4\pi} \bigg) 
 \bigg( 2 C_4^{(^1S_0)}(\mu_R)  \: C_0^{(^1S_0)}(\mu_R)+  
 C_2^{(^1S_0)}(\mu_R)^2 \bigg)  \,, \nn \\
\lefteqn{{}^3\!S_1,{}^3\!D_1:\ \ \ \  }  \\
  &&\quad\qquad \beta_0^{({}^3\!S_1)} = \bigg( {M\mu_R\over 4\pi} \bigg) 
  C_0^{({}^3\!S_1)}(\mu_R)^2 \,, \nn\\
  &&\quad\qquad \beta_2^{({}^3\!S_1)} =2 \bigg( {M\mu_R\over 4\pi} \bigg) 
  C_0^{({}^3\!S_1)}(\mu_R)\: C_2^{({}^3\!S_1)}(\mu_R)\,, \nn \\
  && \quad\qquad \beta_2^{({}^3\!S_1- {}^3\!D_1)} =  \bigg( {M\mu_R\over 4\pi} 
 \bigg) C_0^{({}^3\!S_1)}(\mu_R)\:C_2^{({}^3\!S_1- {}^3\!D_1)}(\mu_R)\,, \nn
\end{eqnarray}
in agreement with Refs.~\cite{ksw1,ksw2}.  For $S=0$  states the beta
functions are one loop exact in the sense that the contribution in
Eq.~(\ref{bet02}) comes from the one-loop graphs, with the higher order
graphs giving contributions which cancel.  The reason for this cancellation is
that the only loop corrections are in the bubble chain, and they form a
geometric series.  The sum of bubble graphs is just the chain of irreducible one
loop bubbles for the full (point-like) propagator.  An analogy would be QED, if
the only possible graphs were the two point photon graphs with 
electron loops.  In this case the beta function would also be one-loop exact
because the graphs that are not 1PI do not contribute. 
In general, the beta functions of higher order couplings may have contributions
beyond one-loop in cases where angular momentum mixing is present.

 Expressions for the running coupling constants can 
be derived by summing the counterterms in Eq.~(\ref{ctexpn}) or by solving 
renormalization group equations.  For $s={}^1\!S_0$ or ${}^3\!S_1$ this gives
\begin{eqnarray} \label{rcouplings}
  C_0^{(s)}(\mu_R) &=& {1 \over \frac{1}{C_0^{\rm finite}} - 
	\frac{M\mu_R}{(4\pi)} } \,, \quad
  C_2^{(s)}(\mu_R) = {C_2^{\rm finite} \over (C_0^{\rm finite})^2 } {1
      \over \bigg[ \frac1{C_0^{\rm finite}} - \frac{M\mu_R}{4\pi}  \bigg]^2 }\,, 
\end{eqnarray}
where $C_0^{\rm finite}$ and $C_2^{\rm finite}$ are constants which can be
determined by specifying boundary conditions.  Since the theory should be
good for arbitrarily small momenta, one possibility is to demand that the
amplitude reproduces the effective range expansion, $p \cot{(\delta)} = -1/a + 
\frac12 r_0 p^2 + {\cal O}(p^4)$.  In Refs.~\cite{ksw1,ksw2} this matching was 
done at $\mu_R=0$ giving $C_0^{\rm finite} = \frac{4\pi a}{M}$, $C_2^{\rm finite} =
\frac{4\pi a}{M}\: \frac{a\, r_0}{2}$, etc.  We could equally well have chosen a
different matching point (such as $\mu_R=1/a$), and obtained the same results.
For $\mu_R\sim Q$, the running couplings in Eq.~(\ref{rcouplings}) have the 
scaling in Eq.~(\ref{pc}).  Written in terms of renormalized couplings the 
amplitude in the ${}^1\!S_0$ or ${}^3\!S_1$ channels is \cite{ksw2}
\begin{eqnarray}
  {\cal A}= -\frac{4\pi}{M} \left[ {1\over \frac{4\pi}{MC_0(\mu_R)}+\mu_R+i p} + 
  \frac{4\pi}{M} \frac{C_2(\mu_R)}{C_0(\mu_R)^2} \frac{p^2}{(\frac{4\pi}
  {MC_0(\mu_R)}+\mu_R+i p)^2} + {\cal O}(Q) \right],   \label{rAmp}
\end{eqnarray}
and satisfies Eq.~(\ref{rc2m}).  The amplitude ${\cal A}$ is $\mu_R$
independent. It is interesting to note that we 
can choose a  renormalization point where all loop corrections vanish giving
\begin{eqnarray}
  {\cal A}^s &=& \sum_{m=0}^\infty {\cal A}^s_{2m} = -\sum_{m=1}^\infty \, 
     C_{2m}^{(s)}(\mu_R=-ip)\,p^{2m} \nn\\
  &=& - \frac{4\pi}{M} \frac1{1/a +ip} - \frac{4\pi}{M}
    \left(\frac1{1/a +ip}\right)^2\, \frac{r_0}2 \: p^2 + \ldots \,.  \label{Atrick}
\end{eqnarray}
The amplitude exactly reproduces the effective range expansion by
construction.   From Eq.~(\ref{Atrick}) the range of the effective field theory can
be estimated as $\Lambda \sim 2/r_0 \sim m_\pi$ as expected.

\section{Reproducing the pole in the amplitude}  \label{pole}

It is possible to choose the boundary condition for $C_0(\mu_R)$ to change the
location of the pole that appears at each order in the expansion. For instance, 
consider the following expansion of the amplitude in the theory without pions:
\begin{eqnarray}  \label{nwexpn}
 A &=& \frac{4\pi}{M} \left[ \frac{1}{-1/a + \frac{r_0}{2} p^2 +... -ip} \right] =  
   \frac{4\pi}{M} \left[ \frac{1}{-1/a -\Delta + \Delta + \frac{r_0}{2} p^2 +... -ip}
   \right]  \nn \\
 & =& \frac{-4\pi}{M} \left[\frac{1}{1/a+\Delta+ip} +\frac{ \frac{r_0}{2} p^2 +
    \Delta}{(1/a+\Delta+ip)^2} + ... \right] \, , 
\label{newexp}
\end{eqnarray}
where $\Delta \lesssim 1/a$.  The series with $\Delta = 0$ and with $\Delta \neq 0$
will both reproduce effective range theory, but differ in the location of the pole that
appears at each order in the perturbative expansion.   In the $^3S_1$ channel, the
pole of the physical amplitude is at 
\begin{eqnarray}
 -ip=\gamma=\sqrt{M E_d}= 45.7\, {\rm MeV} \,,
\end{eqnarray}
where $E_d$ is the binding energy of the deuteron.  For comparison, $1/a =
36.3\,{\rm MeV}$ in this channel. For $\Delta =0$, the pole that appears at each
order in the perturbative expansion will be off by 30\%.   For some calculations,
such as processes involving the deuteron \cite{ksw3,chen,SS,chen2,Kaplan1}, a
better behaved perturbation series is obtained by choosing $1/a +\Delta =\gamma$.
The pole in the amplitude occurs at $p\cot{\delta(p)}= i p$ so 
\begin{eqnarray}  \label{Dar0}
  \Delta = \gamma-1/a = \frac12 r_0 \gamma^2 \,.
\end{eqnarray}
Therefore, although the second term in Eq.~(\ref{nwexpn}) has a double pole, the 
residue of this unphysical double pole is zero.

If we want to reproduce the expansion in Eq.~(\ref{newexp}) in the theory without
pions then part of $C_0(\mu_R)$ must be treated perturbatively,
$C_0(\mu_R)=C_0^{np}(\mu_R) + C_0^{p}(\mu_R)$, where $C_0^{np}(\mu_R)\sim
1/Q$ and $C_0^{p}(\mu_R)\sim Q^0$.   Choosing the pole to be at $p=i\gamma$
gives $C_0^{\rm finite}=4\pi/(M\gamma)$. In this case the amplitude becomes
\begin{eqnarray}
  {\cal A}^s =  - \frac{4\pi}{M} \bigg[ \frac1{\gamma +ip} + \frac{4\pi}{M}
   \frac{C_0^{p}(\mu_R)}{(C_0^{np}(\mu_R))^2} \frac1{(\gamma +ip)^2} +
    \frac{4\pi}{M} \frac{C_2(\mu_R)}{(C_0^{np}(\mu_R))^2} \frac{p^2}{(\gamma+ip)^2}
    \bigg]  \,, \label{Ad}
\end{eqnarray}
where the first term is order $1/Q$, and the second and third terms are order
$Q^0$.  This is simply a reorganization of the perturbative series.  The RGE's are
\begin{eqnarray}
  \mu_R \frac{\partial}{\partial\mu_R} C_0^{np}(\mu_R) &=& \frac{M\mu_R}{4\pi} 
     C_0^{np}(\mu_R)^2  \,,\\
  \mu_R \frac{\partial}{\partial\mu_R} C_0^{p}(\mu_R) &=& 2\, \frac{M\mu_R}{4\pi} 
     C_0^{np}(\mu_R) C_0^p(\mu_R) + {\cal O}(Q) \nn \,.
\end{eqnarray}
These can be derived by substituting $C_0(\mu_R)=C_0^{np}(\mu_R) + 
C_0^{p}(\mu_R)$ into the renormalization group equation for $C_0(\mu_R)$.  They
can also be derived using the counterterm method described above. If we demand 
that the observed scattering length and effective range are reproduced at this 
order then we find 
\begin{eqnarray} \label{C0pbc}
 \frac{4\pi}{M} \frac{C_0^{p}(\mu_R)}{(C_0^{np}(\mu_R))^2} = 
    \gamma-\frac1a \,, \qquad\qquad 
 \frac{4\pi}{M} \frac{C_2(\mu_R)}{(C_0^{np}(\mu_R))^2} = \frac{r_0}2  \,.
\end{eqnarray}
In order for the power counting of $C_0^p(\mu_R)$ to be consistent we must treat 
$\gamma-1/a\sim Q^2$.  From Eq.~(\ref{Dar0}) we see that the first relation in 
Eq.~(\ref{C0pbc}) could have been derived by demanding that the residue of the
double pole in the amplitude vanishes
\begin{eqnarray}  \label{polec}
    \left.  \frac{A^{(0)}}{[A^{(-1)}]^2} \ \right|_{-ip = \gamma} = 0\,.
\end{eqnarray}
At higher orders there will be order $Q^n$ parts to $C_0$ whose values are fixed
by conditions analogous to Eq.~(\ref{polec}).  These conditions ensure that the
position of the pole is not shifted by perturbative corrections.  This is analogous 
to what we do in computing perturbative corrections to the electron mass in QED.

\section{Loop integrals with a momentum cutoff regulator} \label{cutoff}

Although the analysis in section~\ref{nopion} used dimensional regularization to
regulate divergent loop integrals, the results for the coefficients $C_{2m}(\mu_R)$
in our momentum subtraction scheme are independent of this choice.  As an
exercise we will derive the counterterms for $C_0(\mu_R)$ and $C_2(\mu_R)$
using a momentum cutoff regulator, $\Lambda$.  This will give us the chance to
see what type of complications can arise using a different regulator.  Note that this
is not the same as using a {\em{finite}} cutoff scheme.  There the momentum cutoff
plays a double role as both a regulator and as part of the subtraction scheme.

The graph in the first row first column of Fig.~\ref{fig_ct0} gives
\begin{eqnarray}
 &&  i C_0^2  M  \int_0^\Lambda \frac{d^3 q}{(2\pi)^3} 
     \frac1{\vec q\,^2-p^2}     \nn \\ 
&&  = \frac{iM}{2\pi^2} C_0^2 \bigg[ \Lambda + \frac{i\pi p}2 
     - p\,{\rm tanh}^{-1}{\left(\frac{p}{\Lambda}\right)} \bigg]   \\
&&  = \frac{iM}{2\pi^2} C_0^2 \bigg[ \Lambda + \frac{i\pi p}2
     - \frac{p^2}{\Lambda} -\frac{p^4}{3\Lambda^2} -\ldots \bigg]   \nn  \,.
\end{eqnarray}
An ultraviolet counterterm cancels the linear divergence,
\begin{eqnarray}
 \delta^{1,{\rm uv}}C_0 = \frac{M}{4\pi} {C_0^{\rm finite}}^2 \left( - 
   \frac{2\Lambda}{\pi}  \right) \,, \label{dc0L}
\end{eqnarray}
and the same finite counterterm, $\delta^1C_0(\mu_R)$ in 
Eq.~(\ref{ct024}) 
is used to satisfy the condition in Fig.~\ref{fig_C0}.  The renormalized graph 
is then the same as calculated in dimensional regularization in section III.
Note that contributions of order $p^2$ have been neglected in defining 
$C_0(\mu_R)$ as required by our renormalization condition.  An
added complication with a cutoff is that graphs with only $C_0$'s give a 
contribution to the amplitude proportional to $p^2$.  However, as 
$\Lambda \to \infty$, $p\:{\rm tanh}^{-1}{(p/\Lambda)}\to 0$, so these terms
can be completely neglected.  This will remain true even for higher loops 
since the counterterms will always cancel dangerous powers of $\Lambda$
that appear in the numerator.  At $n$ loops we find an ultraviolet 
counterterm of the form
\begin{eqnarray}
  \delta^{n,{\rm uv}}C_0 = -\bigg(\frac{-M}{4\pi}\bigg)^n C_0(\mu_R)^{n+1} 
    \left(  - \frac{2\Lambda}{\pi} \right)^n \,,  \label{dc0nL}
\end{eqnarray}
while the finite counterterms are given by Eq.~(\ref{ct024}).

The graph in the third row first column of Fig.~\ref{fig_ct0} gives
\begin{eqnarray}
  && 2 i C_0 C_2 \frac{M}2 \int_0^\Lambda \frac{d^3 q}
    {(2\pi)^3} \frac{\vec q\,^2+p^2}{\vec q^2-p^2}   \nn \\ 
&&  = 2\,\frac{iM}{2\pi^2} C_0 C_2 \left\{ \frac{\Lambda^3}{6} 
  + p^2\left[ \Lambda +  \frac{i\pi p}2 - p {\rm tanh}^{-1}
  {\left(\frac{p}{\Lambda}\right)} \right]  \right\} \,. \label{c0c2L}
\end{eqnarray}
Note that there are different contributions from this graph when the
vertices are in the order $C_0\,C_0\,C_2$ or $C_0\,C_2\,C_0$.
At order $p^2$, this graph gives a correction to the counterterm
$\delta^{1,{\rm uv}}C_0$, i.e., $\delta^{1,{\rm uv}}C_0 \to \delta^{1,{\rm uv}}
C_0 + \delta^{1*,{\rm uv}}C_0$, where
\begin{eqnarray}
\delta^{1*,{\rm uv}}C_0 &=& - \frac{M}{4\pi} 2\, C_0^{\rm finite}\, C_2^{\rm finite}
    \frac{\Lambda^3}{3\pi} \,. 
\end{eqnarray}
Unlike the contribution to $\delta^{1,{\rm uv}}C_0$ in Eq.~(\ref{dc0nL}), 
$\delta^{1*,{\rm uv}}C_0$ is to be treated perturbatively, so that at it only 
appears once in any graph.  The justification of this fact is 
that this contribution to the counterterm appeared at order $Q^0$ (a purely 
formal trick to recover this counting is to take $\Lambda\sim \mu_R \sim Q$).  
The counterterm $\delta^{1,{\rm uv}}C_2$ is fixed by considering the order 
$p^2$ terms in Fig.~\ref{fig_ct0}, row 3.  From Eq.~(\ref{c0c2L}) 
(the ${\rm tanh}^{-1}$ piece can again be thrown away) we have
\begin{eqnarray}
 \delta^{1,{\rm uv}}C_2 &=& \frac{M}{4\pi} 2\, C_0^{\rm finite}\, C_2^{\rm finite} 
   \left( - \frac{2\Lambda}{\pi}  \right)   \,.  \label{dC2L}
\end{eqnarray}
The calculation for higher loops is similar and there are again corrections 
$\delta^{n*,{\rm uv}} C_0$ to $\delta^{n,{\rm uv}} C_0$
\begin{eqnarray}
 \delta^{n*,{\rm uv}} C_0 &=&  \bigg(\frac{-M}{4\pi}\bigg)^n n(n+1)\, 
  ({C_0^{\rm finite}})^n\, C_2^{\rm finite} \left( - \frac{2\Lambda}{\pi} 
   \right)^{n-1} \frac{\Lambda^3}{3\pi} \,, \nn \\
 \delta^{n,{\rm uv}}C_2 &=& -\bigg(\frac{-M}{4\pi}\bigg)^n (n+1)\, 
  ({C_0^{\rm finite}})^n\, C_2^{\rm finite} \left( - \frac{2\Lambda}{\pi}  
  \right)^n   \,.  
\end{eqnarray}
The finite counterterms are the same as in Eq.~(\ref{ct024}).
Thus the running couplings and amplitudes with a cutoff are the same as
found using dimensional regularization.


\section{Theory with nucleons and pions}  \label{withpion}

\begin{figure}[t!]  
  \epsfysize=9.0truecm \epsfbox{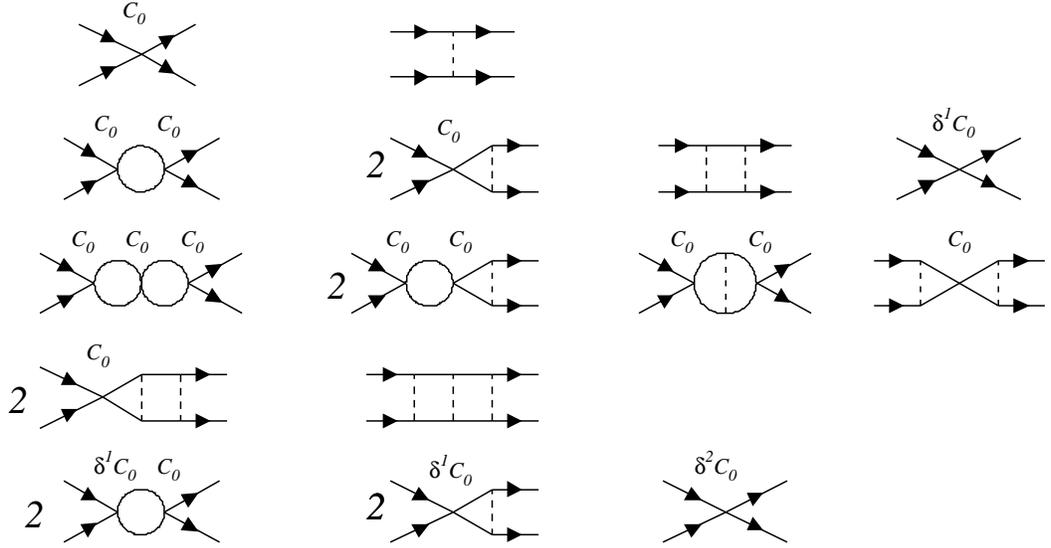}  
{  \caption{Zero, one, and two-loop graphs with $C_0$ and 
$\delta^nC_0$ vertices and potential pion exchange.  The dashed lines 
denote potential pion propagators.} 
\label{fig_ct0pi} }
\end{figure}  
In this section, we study the renormalization of contact interactions in the effective
field theory with pions.  In the ${}^3\!S_1$ channel, graphs with two or more
consecutive potential pions do not factorize and give poles of the form $p^{2m}
/\epsilon$ where $d=4-2\epsilon$.  We explicitly compute these poles for two loop
pion graphs.  There are also $m_\pi^2/\epsilon$ poles in both the ${}^1\!S_0$ and
${}^3\!S_1$ channels at order $Q^0$ \cite{ksw1,ksw2}.  Because of these
$1/\epsilon$ poles, pions cannot be summed to all orders in a model independent
way.  The finite counterterms in PDS and OS are different in this theory.  
Throughout this section we will take $m_\pi=0$, since we are only interested in the
couplings $C_{2m}(\mu_R)$.  The $D_2(\mu_R)$ counterterms will be considered in
section V.  We compute the PDS counterterms and beta functions for
$C_0(\mu_R)$ and $C_2(\mu_R)$ to order $Q$.  In PDS, $C_0(\mu_R)$ no longer
obeys the $Q$ scaling for $\mu_R \gtrsim 300\,{\rm MeV}$\cite{ksw2}.  This can be
fixed by treating part of the coupling $C_0(\mu_R)$ perturbatively as discussed in
Section VI. The exact expressions for $C_0(\mu_R)$, $C_2(\mu_R)$, and
$C_4(\mu_R)$ are given in the OS scheme and exhibit the correct $Q$ scaling for
all $\mu_R > 1/a $.  Therefore, it is no longer apparent that the power counting
breaks down at $300\,{\rm MeV}$.  The $300\,{\rm MeV}$ scale does appear in the
short distance contribution to the amplitude from pion exchange, however, it can
only be taken as an estimate for the range of the effective field theory once pion
and contact interactions are both included.  In section VI, we will discuss how
experimental data suggests that $\Lambda_\pi\gtrsim 300\,{\rm MeV}$.

To determine how the pions contribute to the beta functions for
$C_{2m}(\mu_R)$, we use the rules in section II.   Some of the pion graphs that
will be needed are shown in Fig.~\ref{fig_ct0pi}.  

\begin{figure}[t!]  
  \centerline{\epsfxsize=15.0truecm \epsfbox{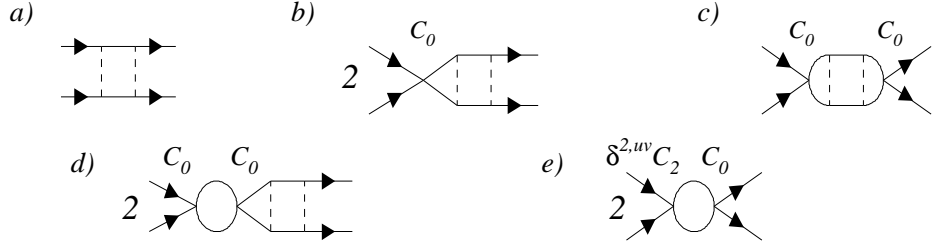}}
{
 \caption{The basic order $Q$ graphs in the ${}^3S_1$ channel whose loop
  integrals do not factorize even for $m_\pi=0$. }  \label{fig_2ppi} }
\end{figure} 
In both PDS and OS, the first step is to subtract $1/\epsilon$ poles with 
$d=4-2\epsilon$.  For two
nucleons in the ${}^1\!S_0$ channel the spinor indices in Eq.~(\ref{ppip}) are
dotted into $\delta_{\alpha\delta}\, \delta_{\beta\gamma}$.  Therefore the
$m_\pi=0$ piece of pion exchange reduces to a contact interaction and gives
no $1/\epsilon$ poles.  In the ${}^3\!S_1$ channel, graphs with two or more
consecutive pions do not factorize and may have $1/\epsilon$ poles.
Order $Q$ graphs with two consecutive potential pions are shown in the 
first row of Fig.~\ref{fig_2ppi}, and labeled $a)$, $b)$, and $c)$.  We find
\begin{eqnarray} \label{2ppians}
  a)\  &=&\ -i\: \frac32\: \bigg(\frac{g_A^2}{2f^2}\bigg)^2 \bigg(
	\frac{-ip\,M}{4\pi} \bigg) \,, \\ 
  b)\ &=&\ -3 i\: C_0^{\rm finite}\: \bigg(\frac{g_A^2}{2f^2}\bigg)^2 \:\bigg(
	\frac{-ip\,M}{4\pi}\bigg)^2 \left[ \frac1{\epsilon}-2\gamma+\frac{14}{3}
	-4\ln{(2)} + 2\,\ln{\bigg(\frac{\pi\mu^2}{-p^2-i\varepsilon}\bigg)}\right]
	 \,, \nn \\
  c)\ &=& -3 i\:(C_0^{\rm finite})^2\: \bigg(\frac{g_A^2}{2f^2}\bigg)^2 \: 
	\bigg( \frac{-ip\,M}{4\pi}\bigg)^3 \left[ \frac1{\epsilon}-3\gamma-
	6\ln{2} +\frac{37}{6} + 3\,\ln{\bigg(\frac{\pi\mu^2}{-p^2-
	i\varepsilon}\bigg)} \right] \,.\nn
\end{eqnarray}
Graphs $b)$ and $c)$ have been written with $C_0^{\rm finite}$ vertices to 
emphasize that the uv counterterm which cancels 
their divergent part is independent of $\mu_R$.   The divergence in $b)$ is 
cancelled by a tree level graph with the counterterm
\begin{eqnarray} \label{C2infct}
  \delta^{2,\,\rm uv}C_2 = -\, 6\, C_0^{\rm finite} \bigg( \frac{M g_A^2}
  {8\pi f^2} \bigg)^2 \left[ \frac1{2\epsilon} -\gamma + \ln{(\pi)} +2-2\ln{(2)} 
  \right] \,,
\end{eqnarray}
where the superscript $2$ indicates that the counterterm comes in at two
loops. The extra factor $2-2\ln{(2)}$ is included because this leads to
simpler analytic expressions.  Expanding the $C_0$ bubble graph
(second row, first column of Fig.~\ref{fig_ct0pi}) in $\epsilon$ gives
\begin{eqnarray}  \label{c0expn}
   - \frac{p M}{4\pi} (C_0)^2 \left\{ 1 + \epsilon\bigg[ 2-\gamma-2\ln{(2)}+
     \ln{\bigg( \frac{\pi\mu^2}{-p^2-i\epsilon} \bigg) \bigg]}  \right\}  \,.
\end{eqnarray} 
When graphs with $1/\epsilon$ poles are dressed with $C_0$ bubbles, the
factors of $[2-\gamma-2\ln{2} + \ln{(\pi)}]$ that appear are cancelled by similar
factors from the counterterms.  In fact, $\delta^{2,\,\rm uv}C_2$ is the only uv 
counterterm we need for two potential pion exchange
with $m_\pi=0$.  The $1/{\epsilon}$ pole in $c)$ is nonanalytic since it is
proportional to $p^3$.  When graph $c)$ is added to graphs $d)$ and $e)$ the
poles cancel.  These cancellations continue to occur when more $C_0$
bubbles are added to $b)$ and $c)$.  After including graphs with 
$\delta^{2,\rm uv}C_2$ we find
\begin{eqnarray}  \label{newbc}
  b)+i\delta^{2,\,\rm uv}C_2 \:p^2 &=&-3 i\: C_0^{\rm finite}\: 
    \bigg(\frac{g_A^2}{2f^2}\bigg)^2 \:\bigg(
    \frac{-ip\,M}{4\pi}\bigg)^2 \left[ \frac23+ 
    2\ln{\bigg(\frac{\mu^2}{-p^2-i\varepsilon}\bigg)} \right] , \\[5pt]
  c)+\frac12 e) &=& -3 i\:(C_0^{\rm finite})^2\: \bigg(\frac{g_A^2}{2f^2}
     \bigg)^2 \: \bigg( \frac{-ip\,M}{4\pi}\bigg)^3  \left[\frac16 + 2 
     \ln{\bigg(\frac{\mu^2}{-p^2-i\varepsilon}\bigg)}\right] . \nn
\end{eqnarray}
Note that for $\mu \sim p$ there are no large numerical factors from these
graphs.  

In the $^3S_1$ channel, potential pion graphs without contact interactions 
also have $p^2/\epsilon$ poles.  The two loop graph with three 
potential pions (fourth row, second column in Fig.~\ref{fig_ct0pi}) is equal to
\begin{eqnarray}
   \frac{4\pi i }{M} \bigg( \frac{M g_A^2}{8\pi f^2} \bigg)^3 \: p^2\ \bigg[ 
 	\frac3{\epsilon} + \ldots \bigg] \,.
\end{eqnarray}
In the $Q$ power counting, this graph is order $Q^2$ and will not be
considered here.  Because of these $1/\epsilon$ poles it is not possible to sum
pion ladder graphs to all orders.  Now that the ultraviolet divergences have been 
removed from graphs $b)$ and $c)$, the finite subtractions can be performed.  

\subsection{PDS}

For PDS in the $^1S_0$ channel, we can compute the effect of potential pions
on the $C_{2m}(\mu_R)$ counterterms to all orders in $Q$ (neglecting
relativistic corrections).  For $C_0(\mu_R)$, the relevant zero, one, and two 
loop graphs are  shown in Fig.~\ref{fig_ct0pi}.  The $C_0(\mu_R)$  and 
$C_2(\mu_R)$ counterterms are
\begin{eqnarray}
 && \delta^nC_0^{({}^1\!S_0)}(\mu_R) =(-1)^{n+1} \bigg( \frac{M\mu_R}{4\pi}
  \bigg)^n  \bigg[ C_0(\mu_R)+\frac{g_A^2}{2f^2} \bigg]^{n+1} \,, \nn \\
 && \delta^nC_2^{({}^1\!S_0)}(\mu_R) = (-1)^{n+1} (n+1) \bigg( \frac{M\mu_R}
  {4\pi} \bigg)^n  \bigg[ C_0(\mu_R)+\frac{g_A^2}{2f^2} \bigg]^{n} 
   C_2(\mu_R) \,. \label{1s0ct} 
\end{eqnarray}

The PDS counterterms in the ${}^3\!S_1$ channel will only be computed to
order $Q$ since the loop graphs with consecutive pions do not factorize.  For
this case it is essential to use the counterterms to carry out the PDS
renormalization program.  To define $C_0(\mu_R)$ at order $Q$, we set up the
finite subtractions as in Fig.~\ref{fig_ct0pi}, but leave out all graphs with more
than two potential pions since they are ${\cal O}(Q^2)$ (we also neglect
relativistic corrections that are order $Q$ but come with an additional
$1/M^2$).  Note that in $d=3$ only the overall divergence ($\propto 1/(d-3)^n$
for $n$ loops) is needed since loops with counterterms will cancel the
sub-divergences.  Evaluating  the graphs in Fig.~\ref{fig_2ppi} with $d=3$ and
then continuing back to $d=4$ gives
\begin{eqnarray}
  a)&=& -9\, i\, \bigg(\frac{g_A^2}{2f^2}\bigg)^2 \bigg(\frac{\mu_R\,M}{4\pi} \bigg),
  \qquad
  b) = -12\, i\, C_0(\mu_R) \bigg(\frac{g_A^2}{2f^2}\bigg)^2 
	\bigg(\frac{\mu_R\,M}{4\pi} \bigg)^2  \,, \nn\\
  c) &=& -5\, i\, C_0(\mu_R)^2 \bigg(\frac{g_A^2}{2f^2}\bigg)^2 
	\bigg(\frac{\mu_R\,M}{4\pi} \bigg)^3 \,.
\end{eqnarray}
Using these values we find
\begin{eqnarray}   \label{3s1ct0} 
 \delta^1C_0^{(^3S_1)} &=&  \bigg( \frac{M\mu_R}{4\pi}\bigg) \bigg[
   C_0(\mu_R)^{2}+2 C_0(\mu_R) \frac{g_A^2}{2f^2}  + 9\, 
    \Big(\frac{g_A^2}{2f^2}\Big)^2   \bigg] \,, \\
 \delta^nC_0^{({}^3\!S_1)} &=& (-1)^{n+1} \bigg( \frac{M\mu_R}{4\pi}\bigg)^n
   \bigg[ C_0(\mu_R)^{n+1}+(n+1) C_0(\mu_R)^n \frac{g_A^2}{2f^2} 
  \nn\\
& &\qquad\qquad+\frac12 (n+1)
   (n+4) C_0(\mu_R)^{n-1} \Big(\frac{g_A^2}{2f^2}\Big)^2
   \bigg] \,. \qquad\mbox{for $n\ge2$}  \nn
\end{eqnarray}
Note that for graphs with two consecutive potential pions, the $\mu_R$ 
dependence does not come in the linear combination $\mu_R+ip$.  For 
instance, adding the PDS counterterm to graph a) in Fig.~\ref{fig_2ppi} gives
the linear combination $3 ip/2 + 9\mu_R$.

In PDS, like in $\overline{\rm MS}$, the renormalized coupling $C_2(\mu_R)$ will
depend on $\ln(\mu_R^2/\mu_0^2)$ in such a way that the $\ln(\mu_R^2)$
dependence in the amplitudes in Eq.~(\ref{2ppians}) is cancelled.  Here $\mu_0$ is
an arbitrary scale expected to be of order $\Lambda_\pi$.  At order $Q$ we find 
\begin{eqnarray} \label{3s1ct2}
  \delta^1C_2^{(^3S_1)}(\mu_R) &=&  2\bigg( \frac{M\mu_R}{4\pi} \bigg)
      \bigg[ C_0(\mu_R)+ \frac{g_A^2}{2f^2} \bigg] C_2(\mu_R)\,, \\
  \delta^nC_2^{(^3S_1)}(\mu_R) &=& (-1)^{n+1} \left\{ (n+1) \bigg( 
     \frac{M\mu_R}{4\pi} \bigg)^n  \bigg[ C_0(\mu_R)^n+ n \frac{g_A^2}{2f^2} 
    C_0(\mu_R)^{n-1} \bigg] C_2(\mu_R) \right. \nn  \\
 && \quad \left. + 6 \bigg( \frac{M\mu_R}{4\pi}
     \bigg)^{n-2} C_0(\mu_R)^{n-1} \bigg( \frac{M g_A^2}{8\pi f^2} \bigg)^2  
     \ln{\Big(\frac{\mu_R^2}{\mu_0^2}\Big)} \right\}
\qquad\mbox{for $n\ge2$} \nn \,.
\end{eqnarray}
Note that the part of $\delta^nC_2(\mu_R)$ proportional to 
$\ln{(\mu_R^2/\mu_0^2)}$ has a coefficient that sums up to $C_0^{\rm finite}$ at
this order.  From Eqs.~(\ref{1s0ct}), (\ref{3s1ct0}), and (\ref{3s1ct2}) we find
\begin{eqnarray}  \label{PDSbeta}
  \beta_0^{({}^1\!S_0)} &=&  \frac{M\mu_R}{4\pi}\: \bigg[C_0(\mu_R)+
   \frac{g_A^2}{2f^2}  \bigg]^2  \,,  \\[5pt]
  \beta_2^{({}^1\!S_0)} &=&  2\: \frac{M\mu_R}{4\pi}\:  \bigg[C_0(\mu_R)+
   \frac{g_A^2}{2f^2}  \bigg] C_2(\mu_R) \,, \nn\\[5pt]
  \beta_0^{({}^3\!S_1)} &=&  \frac{M\mu_R}{4\pi} \left\{ C_0^2+
   2 \frac{g_A^2}{2f^2} C_0 + \bigg[9+4\bigg(\frac{\mu_R M C_0}{4\pi}\bigg) 
   +2\bigg(\frac{\mu_R M C_0}{4\pi}\bigg)^2 \bigg]
   \Big(\frac{g_A^2}{2f^2}\Big)^2 \right\} \nn\\
 && + {\cal O}(Q^2) \,, \nn \\[5pt]
  \beta_2^{({}^3\!S_1)} &=& 2\: \frac{M\mu_R}{4\pi}\: \bigg[C_0(\mu_R)+
   \frac{g_A^2}{2f^2}  \bigg] C_2(\mu_R) -12 \bigg(\frac{M g_A^2}{8\pi f^2}
   \bigg)^2 C_0(\mu_R) \bigg[ 1+ \mu_R \frac{M}{4\pi} C_0(\mu_R) \bigg] \nn\\
  &&  +{\cal O}(Q^0) \nn \,.
\end{eqnarray}
Note that in the ${}^1\!S_0$ channel all contributions to the beta functions
beyond one-loop cancel, leaving them one-loop exact.  In  Ref.~\cite{ksw2},
the last two terms in $\beta_0^{(^3S_1)}$ are absent, but should be included in
the complete order $Q$ calculation.  Dimensional analysis implies that the
${}^3\!S_1$ beta functions can have corrections at all higher orders in Q, since
there is nothing to prevent the dimensionless factor $(\mu_R\,g_A^2 \,M)/(8\pi
f^2) \sim Q$ from appearing.  In Ref.~\cite{sf2}, expressions for the beta
functions are derived by demanding that $\partial{\cal A}/\partial\mu_R =0$, but
these are not the PDS beta functions.  Since in all renormalization schemes
$\partial{\cal A}/ \partial\mu_R=0$, this condition is not sufficient to fix the
renormalization scheme uniquely.  

When the beta function is not exactly known, the large $\mu_R$ behavior is
ambiguous.  For example, the PDS beta function for $C_0(\mu_R)$ is \cite{ksw2}
\begin{eqnarray}   \label{b0} 
\beta_0 &=&  \mu_R \frac{\partial C_0(\mu_R)}{\partial \mu_R} = 
     \frac{M\mu_R}{4\pi}\:  \left[ C_0^2(\mu_R)+ 2\frac{g_A^2}{2f^2} C_0(\mu_R) 
     \right] + {\cal O}(Q)  \,.  
\end{eqnarray} 
Two solutions which satisfy this equation to order $Q^0$ are 
\begin{eqnarray}   \label{gsln}
  C_0(\mu_R) &=& -\,\frac{4\pi}{M} \frac{\Big[ 1-2 a \mu_R-2\mu_g - \sqrt{1+4
    \mu_g^2 -4 \mu_g (1-a\mu_R)} \ \Big]}{2\mu_R (1-a\mu_R -\mu_g)} \,,  \nn \\
  C_0(\mu_R) &=& -\,\frac{g_A^2}{f^2} \frac{1}{1- \Big[1+2/(a \Lambda_{NN} )
  \Big] \exp{(-2\mu_g)} } \,, 
\end{eqnarray}
where $\mu_g = \mu_R/\Lambda_{NN}$ and we have chosen $C_0(0)=4\pi a/M$. 
The first solution is obtained by computing the counterterms $\delta^n
C_0(\mu_R)$ to order $Q^0$ and summing them.  This solution falls as $1/\mu_R$
for all $\mu_R > 1/a$, and is numerically close to the $g_A\to 0$ solution.  The
second solution is obtained by truncating and solving Eq.~(\ref{b0}).  This solution
approaches a constant as $\mu_R\to \infty$.  The two solutions both solve the
beta function to order $Q^0$ but have very different large $\mu_R$ behavior. 
Since the beta function for $C_0^{(^3S_1)}(\mu_R)$ can have corrections at any
order in $Q$ its large $\mu_R$ behavior is unknown and not meaningful. 
This issue can be dealt with by expanding the $C_{2m}$ coupling constants in a  
series in $Q$ and solving the beta function for each order separately.  

\subsection{OS}

In the OS scheme, there is no such ambiguity since at a given order in $Q$ the
running of all the coupling constants that enter at that order are known
exactly.  The coupling $C_0^{(s)}(\mu_R)$ has contributions only from the
nucleon graphs discussed in section II and therefore has the same beta
function.  For $C_2^{(s)}(\mu_R)$, the order $Q^0$ graphs in ${\cal A}_2$
include the nucleon graphs from section II, as well as the graphs with one
potential pion and any number of $C_0$ vertices.   At tree level we add a finite
counterterm to cancel the $m_\pi=0$ part of the tree level pion interaction at
$p=i\mu_R$
\begin{eqnarray}
  \delta^0 C_2^{(s)}(\mu_R) = -\ \frac{g_A^2}{2f^2}\ \frac1{\mu_R^2} \,.
\end{eqnarray}
This counterterm is order $Q^{-2}$ like $C_2(\mu_R)$ itself.  Since all the 
graphs in ${\cal A}_2$ factorize the higher loop counterterms are the same as 
in the theory without pions, so $\delta^n C_2$ for $n\ge 1$ are given in 
Eq.~(\ref{ct024}).  The exact beta function is then
\begin{eqnarray}
  \beta_2^{(s)} = 2\: \frac{M\mu_R}{4\pi}\: C_0(\mu_R) C_2(\mu)+
    2\:\frac{g_A^2}{2f^2} \bigg( 1+\frac{M\mu_R}{4\pi}C_0(\mu_R)
  \bigg)^2\: \frac1{{\mu_R}^2}  \,.
\end{eqnarray}
Note that the finite $\ln(\mu^2/(-p^2-i\epsilon))$ terms in Eq.~(\ref{2ppians}) 
are order $Q$ and in the OS scheme do not affect the running of 
$C_2(\mu_R)$, but rather $C_4(\mu_R)$.  In terms of the finite 
constants $C_0^{\rm finite}$ and $C_2^{\rm finite}$ we have solutions 
\begin{eqnarray} \label{OSc0c2}
  C_0^{(s)}(\mu_R) &=& {1 \over \frac{1}{C_0^{\rm finite}} - 
	\frac{M\mu_R}{(4\pi)} } \,, \qquad\quad
  C_2^{(s)}(\mu_R) = {C_2^{\rm finite} - \frac{g_A^2}{2f^2\mu_R^2} 
      \over \left[ 1 - \mu_R \frac{C_0^{\rm finite} M}{4\pi}  \right]^2 }\,. 
\end{eqnarray}
Although it may seem that the piece of $C_2^{(s)}(\mu_R)$ that goes as
$1/\mu_R^4$ will spoil the power counting for low momentum, in fact, the
$1/\mu_R^2$ part dominates entirely until $\mu_R\sim 1/a$, since $C_0^{\rm
finite} \sim a$, $C_2^{\rm finite}\sim a^2$.    Written in terms of
renormalized couplings the $m_\pi=0$ part of the next-to-leading order
OS amplitude is
\begin{eqnarray}
 { - C_2(\mu_R) \:p^2 \over \bigg[1 + (\mu_R+ip)\frac{M C_0(\mu_R)}
   {4\pi} \bigg]^2 } \ -\  \frac{g_A^2}{2f^2}\: {\mu_R^2 + p^2 \over \mu_R^2} \:
    { \bigg[ 1+ \mu_R \frac{C_0(\mu_R) M}{4\pi}  \bigg]^2 \over \bigg[ 1+
   (\mu_R+ip) \frac{C_0(\mu_R) M}{4\pi} \bigg]^2 } \,,
\end{eqnarray}
which is order $Q^0$ as desired.  

One might still ask if the problem with the $300\,{\rm MeV}$ scale will reappear
in higher order coefficients.  To check that this is not the case we compute
the running of the coupling $C_4(\mu_R)$ in the OS scheme.  
The  easiest way to compute this running coupling constant is to
compute the order $Q$ amplitude in terms of the finite couplings,
$C_{2m}^{\rm finite}$, and then demand that the amplitude satisfies the
renormalization condition in Eq.~(\ref{rc2m}).  The graphs we need to compute 
include those with
\begin{eqnarray} \begin{array}{rl}
    i)\ &\mbox{ one $C_4$ and any number of $C_0$'s}  \,, \\
    ii)\  &\mbox{ two $C_2$'s and any number of $C_0$'s}  \,, \\
    iii)\  &\mbox{ one $C_2$, one potential pion and any number of $C_0$'s}\,, \\
    iv)\  &\mbox{ two potential pions and any number of $C_0$'s}  \,.
\end{array}
\end{eqnarray}
Computing these graphs in terms of the finite couplings and then demanding
that the amplitudes satisfy the renormalization condition gives the OS 
couplings
\begin{eqnarray}  \label{C4sln}
  C_4^{({}^1\!S_0)}(\mu_R) &=& {C_4^{\rm finite}\over\Big[1-\mu_R\frac{M}{4\pi}
     C_0^{\rm finite}\Big]^2 } + \frac{\mu_R M}{4\pi} { \Big[C_2^{\rm finite}  -
     g_A^2/(2f^2)\, \frac1{{\mu_R}^2} \Big]^2 \over \Big[1-\mu_R\frac{M}{4\pi} 
     C_0^{\rm finite}\Big]^3 } \,, \\
  C_4^{({}^3\!S_1)}(\mu_R) &=&{C_4^{\rm finite}\over\Big[1-\mu_R\frac{M}{4\pi}
     C_0^{\rm finite}\Big]^2 } + \frac{\mu_R M}{4\pi} { \Big[C_2^{\rm finite}  -
     g_A^2/(2f^2)\, \frac1{{\mu_R}^2} \Big]^2 \over \Big[1-\mu_R\frac{M}{4\pi} 
     C_0^{\rm finite}\Big]^3 }  \nn\\
  + \frac12 \bigg( \frac{g_A^2}{2f^2} \bigg)^2 \!\!\!&&\!\!\!\!
     \frac{M}{4\pi}\, \frac1{{\mu_R}^3}\, {\bigg[ 1-2\mu_R \frac{M}{4\pi} 
      C_0^{\rm finite} \bigg] \over \bigg[1-\mu_R  \frac{M}{4\pi} C_0^{\rm finite} 
      \bigg]^2 } -  
  6\,\bigg( \frac{M g_A^2}{8\pi f^2}\bigg)^2 \frac{ C_0^{\rm finite} 
  \ln{( \mu_R^2 /\mu^2 )}} {\mu_R^2 \bigg[1-\mu_R \frac{M}{4\pi} 
   C_0^{\rm finite}\bigg]  }  \,, \nn
\end{eqnarray}
where here $\mu$ is an unknown scale expected to be of order $\Lambda_\pi$. 
Again the pion contributions do not spoil the $\mu_R$ scaling behavior,
since they are suppressed by factors of the large scattering length.  Note that
at order $Q$ the PDS coupling $C_4(\mu_R)$ \cite{ksw2} is the $g_A\to 0$ limit of 
Eq.~(\ref{C4sln}). 

In this section, expressions for the renormalized couplings $C_0(\mu_R)$,
$C_2(\mu_R)$, and $C_4(\mu_R)$ were derived in the PDS and OS schemes
working to order $Q$.   For the $^3S_1$ channel, we have shown that
$C_0(\mu_R)$ has corrections at all orders in $Q$ in PDS.  Unlike PDS, the OS
couplings $C_{2m}(\mu_R)$ can be computed exactly because they only have
contributions at one order in $Q$.  The OS couplings exhibit the correct
$\mu_R$ scaling for all $\mu_R>1/a$.


\section{The coupling $D_2(\mu_R)$}  \label{D2coupling}

In this section, the OS and PDS counterterms for $D_2(\mu_R)$ are computed.
To define $D_2(\mu_R)$ in the OS scheme, we take
\begin{eqnarray}
   &&  i\,A^s(D_2) \Big|_{p=i\mu_R,} =  -i\, D_2^{(s)}(\mu_R)\, m_\pi^2 \,,
\end{eqnarray}
where ${\cal A}^s(D_2)$ contains terms in the amplitude that are analytic in
$m_\pi^2$ and proportional to $m_\pi^2$.  Only terms that are analytic in
$m_\pi^2$ are kept because it is unnatural to put  long-distance nonanalytic
contributions that come from pion exchange into the definition of the short
distance coupling \cite{sf2}.  For example, one potential pion exchange gives a
$m_\pi^2/p^2 \: \ln(1+4p^2/m_\pi^2)$ term. Including this in $A^s(D_2)$ would give
$D_2(\mu_R)$ both a branch cut at $\mu_R=m_\pi/2$ as well as explicit
dependence on the scale $m_\pi$.  In the OS scheme, $D_2(\mu_R)$ will be
calculated as follows.  First $m_\pi^2/\epsilon$ poles are subtracted.  The finite
counterterms are then determined by including graphs with a single $D_2(\mu_R)$
or potential pion and any number of $C_0(\mu_R)$ vertices in ${\cal A}^s(D_2)$. 
Contributions from these graphs that are non-analytic in $m_\pi^2$ are dropped.

There is a $m_\pi^2/\epsilon$ pole in the ${\cal O}(Q^0)$ graph in the third row
and third column of Fig.~\ref{fig_ct0pi} \cite{ksw1,ksw2}, so we have a
counterterm
\begin{eqnarray}
  \delta^{2,{\rm uv}}D_2 = - i \bigg(\frac{MC_0^{\rm finite}}{4\pi}\bigg)^2 \:
   \frac{g_A^2}{4f^2} \:  \left[ \frac1{2\epsilon} -\gamma + \log{(\pi})  \right] \,.
\end{eqnarray}
Note that when this counterterm is dressed with $C_0$ bubbles the  
extra factors of $2-\ln 2$ from Eq.~(\ref{c0expn}) will cancel without the need for
an additional finite term in $\delta^{2,{\rm uv}}D_2$.  After subtracting this 
counterterm the value of the two-loop graph is 
\begin{eqnarray} \label{tlp}
    i \bigg(\frac{MC_0^{\rm finite}}{4\pi}\bigg)^2 \:
   \frac{g_A^2}{2f^2} \:  \left[ -(ip)^2 + \frac{m_\pi^2}{2} +\frac{m_\pi^2}{2}
    \ln{\bigg(\frac{\mu^2}{m_\pi^2}\bigg)}- m_\pi^2 
    \ln{\bigg(1-\frac{2ip}{m_\pi}\bigg)} \right] \,.
\end{eqnarray}
For PDS we set $\mu=\mu_R$ and then find finite counterterms 
\begin{eqnarray} \label{3s1ctd2}
  \delta^1D_2(\mu_R) &=&  2\bigg( \frac{M\mu_R}{4\pi} \bigg)
    C_0(\mu_R) D_2(\mu_R) \,, \\
  \delta^nD_2(\mu_R) &=& (-1)^{n+1} \bigg[ (n+1) \bigg( \frac{M\mu_R}{4\pi}
     \bigg)^n  C_0(\mu_R)^n D_2(\mu_R) \nn \\
  && \quad - \frac{(n-1)}2 \bigg( \frac{M\mu_R}{4\pi}
     \bigg)^{n-2} C_0(\mu_R)^{n} \bigg( \frac{M}{4\pi}\bigg)^2 
     \frac{g_A^2}{2 f^2}   \ln{\Big(\frac{\mu_R^2}{\mu_0^2}\Big)} \bigg]
\qquad\mbox{for $n\ge2$} \nn \,.
\end{eqnarray}
Here $\mu_0$ is an unknown scale expected to be of order $\Lambda_\pi$.

In the OS scheme, the $\delta^1D_2(\mu_R)$ counterterm is the same as in PDS. 
In dimensional regularization logarithms of the form $\ln(\mu^2/m_\pi^2)$ will
appear in loop graphs.  To make the $\mu^2$ dependent part analytic in 
$m_\pi^2$ we write
\begin{eqnarray}
  \ln\left(\frac{\mu^2}{m_\pi^2}\right) = \ln\left(\frac{\mu^2}{\mu_R^2}\right) + 
	\ln\left(\frac{\mu_R^2}{m_\pi^2}\right) \,,
\end{eqnarray}
and then subtract the $\ln(\mu^2/\mu_R^2)$ term with the counterterms.  This
will give $D_2(\mu_R)$ a $\mu_R$ dependence which cancels the 
$\ln(\mu_R^2/m_\pi^2)$ in the amplitude. In the OS scheme, the $m_\pi^2/2$ in
Eq.~(\ref{tlp}) gets subtracted along with the logarithm.  We  find
\begin{eqnarray} \label{3s1ctd2os}
  \delta^nD_2(\mu_R) &=& (-1)^{n+1} \Bigg\{ (n+1) \bigg( \frac{M\mu_R}{4\pi}
     \bigg)^n  C_0(\mu_R)^n D_2(\mu_R) \\
  && \quad - \frac{(n-1)}2 \bigg( \frac{M\mu_R}{4\pi}
     \bigg)^{n-2} C_0(\mu_R)^{n} \bigg( \frac{M}{4\pi}\bigg)^2 
     \frac{g_A^2}{2 f^2} \bigg[-1+  \ln{\Big(\frac{\mu_R^2}{\mu^2}\Big)} 
     \bigg] \ \Bigg\}  \nn 
\end{eqnarray}
for $n\ge2$.
Summing the counterterms the solutions for $D_2(\mu_R)$ are then
\begin{eqnarray} \label{D2sln}
 {D_2^{(s)}(\mu_R)\over C_0^{(s)}(\mu_R)^2} &=& { D_2^{\rm finite} \over 
   (C_0^{\rm finite})^2}  +\frac{M}{8\pi} \bigg(\frac{M g_A^2}{8\pi f^2} \bigg) 
   \ln\bigg({\mu_R^2 \over \mu_0^2}\bigg)\qquad\qquad\quad\mbox{in PDS} \,,\\
 {D_2^{(s)}(\mu_R)\over C_0^{(s)}(\mu_R)^2} &=& { D_2^{\rm finite} \over 
   (C_0^{\rm finite})^2}  +\frac{M}{8\pi} \bigg(\frac{M g_A^2}{8\pi f^2} \bigg) 
   \bigg[-1+  \ln\bigg({\mu_R^2 \over \mu^2}\bigg)  \bigg]
   \qquad \mbox{in OS} \,, \nn
\end{eqnarray}
which can be written as
\begin{eqnarray}
{D_2^{(s)}(\mu_R)\over C_0^{(s)}(\mu_R)^2}  &=& \frac{M}{8\pi} 
  \bigg(\frac{M g_A^2} {8\pi f^2} \bigg)  \ln\bigg({\mu_R^2  \over \tilde\mu^2}
  \bigg) \,, \\[5pt] \mbox{  where    } 
  && \tilde\mu^2 = \mu_0^2 \, \exp{\bigg(\frac{-64\pi^2 f^2  D_2^{\rm finite}}
  {M^2  g_A^2(C_0^{\rm finite})^2} \bigg)}   \qquad\mbox{in PDS} \nn \,, \\
  && \tilde\mu^2 = \mu^2 \, \exp{\bigg(1-\frac{64\pi^2 f^2  D_2^{\rm finite}}
  {M^2  g_A^2(C_0^{\rm finite})^2} \bigg)}   \quad\mbox{in OS} \nn \,.  
   \label{D2ln} 
\end{eqnarray} 
The parameter $\tilde \mu$ must be determined by fitting to data.  
With $m_\pi\sim Q\sim \mu_R$, $D_2(\mu_R)m_\pi^2 \sim Q^0$ in both OS and
PDS, implying that $D_2(\mu_R)$ should be treated perturbatively.


\section{Schemes and amplitudes} \label{samp}

In this section, the amplitudes in the $^1S_0$ and $^3S_1$ channels are
presented to order $Q^0$, both in PDS \cite{ksw1,ksw2} and  OS. Fits to the
$^1S_0$ and $^3S_1$ phase shift data are done in both schemes for different
values of $\mu_R$.  As pointed out in section~\ref{pole}, one  has the freedom to
split $C_0(\mu_R)$ into perturbative and nonperturbative pieces:
$C_0(\mu_R)=C_0^{np}(\mu_R) + C_0^p(\mu_R)$, where $C_0^{np}(\mu_R) \sim
Q^{-1}$ and $C_0^p(\mu_R) \sim Q^0$.  This division is necessary in PDS in
order to obtain $\mu_R$ independent amplitudes at each order.  Furthermore,
$C_0^{np}(\mu_R)\sim 1/\mu_R$ so the coefficients scale in a manner consistent
with the power counting for all $\mu_R>1/a$.  For convenience we will drop the
superscript $np$ in what follows.  Some issues that arise in matching the pion
theory onto the effective range expansion are also discussed. 

First, we give the nucleon-nucleon scattering amplitudes in the PDS and OS
schemes.  In PDS, the amplitudes were calculated to order $Q^0$ in
Refs.~\cite{ksw1,ksw2}. At this order, amplitudes in the  ${{}^1\!S_0}$ and
${{}^3\!S_1}$ channels have the same functional form,
\begin{eqnarray}
  A &=& A^{(-1)} + A^{(0,a)} + A^{(0,b)} + {\cal O}(Q^1) \,.
\end{eqnarray} 
In both OS and PDS we have
\begin{eqnarray}  \label{PDSA}
 {A^{(-1)}} &=& -\frac{4\pi}{M}\: \frac{1} {\frac{4\pi}{M C_0(\mu_R)} +
	\mu_R + ip } \,,  \\[10pt]  
 \frac{A^{(0,a)}}{\Big[{A^{(-1)}}\Big]^2} &=&  \frac{ g_A^2 m_\pi^2 }{2 f^2} \bigg( \frac{M}{4\pi} \bigg)^2 \left\{ 
  \frac12 \ln{\bigg({ \mu_R^2\over m_\pi^2} \bigg)} - \bigg(\frac{4\pi}
  {MC_0(\mu_R)} + \mu_R\bigg) \frac{1}{p} \tan^{-1}\bigg(\frac{2p}{m_\pi}\bigg)
  \right. \nn \\[5pt]
&&\left.\  + \bigg[ \bigg(\frac{4\pi}{MC_0(\mu_R)} 
  +\mu_R\bigg)^2-p^2 \bigg] \frac1{4p^2} \ln{\bigg(1+ \frac{4p^2}{m_\pi^2} 
  \bigg)} \right\}  - \frac{D_2(\mu_R)\: m_\pi^2 }{C_0(\mu_R)^2} \nn \,.
\end{eqnarray}
The remaining part of the order $Q^0$ PDS amplitude is
\begin{eqnarray}  
 \frac{A^{(0,b)}}{\Big[{A^{(-1)}}\Big]^2} &=& -\frac{ C_2(\mu_R)\: p^2}
  {C_0(\mu_R)^2} + \frac12 \: \frac{ g_A^2 m_\pi^2 }{2 f^2} \bigg(\frac{M}{4\pi} 
  \bigg)^2 -\: \frac1{C_0(\mu_R)^2} \bigg[\frac{g_A^2}{2 f^2} + C_0^p(\mu_R) 
  \bigg] \,. \nn \\
\end{eqnarray}
Note that since we have made a different finite subtraction than KSW the second
term has a prefactor of $1/2$, rather than a $1$ as in Ref.~\cite{ksw2}.  In the OS 
scheme 
\begin{eqnarray}  \label{OSA}
\frac{A^{(0,b)}}{\Big[{A^{(-1)}}\Big]^2} &=& 
  -\frac{ C_2(\mu_R)\: p^2} {C_0(\mu_R)^2}  -\frac{g_A^2}
  {2 f^2} \bigg(1+\frac{p^2}{\mu_R^2}\bigg) \bigg( \frac1{C_0(\mu_R)} + 
  \frac{M\mu_R}{4\pi} \bigg)^2 -\: \frac{C_0^p(\mu_R)}{C_0(\mu_R)^2}
   \label{Amp4} \,. \nn \\
\end{eqnarray}
The OS and PDS couplings that appear at this order are related by a change 
of variables.  Couplings on the left are in PDS while those on the right are in the 
OS scheme.	
\begin{eqnarray}
   C_0(\mu_R) &=& C_0(\mu_R) \,, \nn\\
   \frac{C_2(\mu_R)}{C_0(\mu_R)^2} &=& \frac{C_2(\mu_R)}{C_0(\mu_R)^2}  +
	\frac{g_A^2}{2f^2} \frac1{\mu_R^2} \bigg[ \frac1{C_0(\mu_R)} + 	
	\frac{M\mu_R}{4\pi} \bigg]^2 \,, \nn \\
  \frac{D_2(\mu_R)}{C_0(\mu_R)^2}-\frac{M}{8\pi} \bigg( \frac{Mg_A^2}{8\pi f^2} 
      \bigg)  &=& \frac{D_2(\mu_R)}{C_0(\mu_R)^2}  \,,  \\
   \frac{C_0^p(\mu_R)}{C_0(\mu_R)^2} + \frac{g_A^2}{2f^2} \frac1{C_0(\mu_R)^2}
	&=& \frac{C_0^p(\mu_R)}{C_0(\mu_R)^2} + \frac{g_A^2}{2f^2} 
	\bigg[ \frac1{C_0(\mu_R)} + \frac{M\mu_R}{4\pi} \bigg]^2 \nn \,.
\end{eqnarray}

In the PDS scheme, there are order $Q^0$ contributions to $\beta_0$ (c.f.,
Eq.~(\ref{PDSbeta})).  If the order $Q^0$ contributions are separated from the 
order $1/Q$ pieces, the beta function for $C_0^p(\mu_R)$ is
\begin{eqnarray}  \label{C0pb}
  \mu_R \frac{\partial C_0^p(\mu_R)}{\partial \mu_R}  =  2\,\frac{M\mu_R}{4\pi} 
    C_0(\mu_R) \bigg[ C_0^p(\mu_R) + \frac{g_A^2}{2f^2} \bigg] +{\cal O}(Q)\,.
\end{eqnarray}
This equation has the solution
\begin{eqnarray} \label{C0ps}
    \frac{C_0^p(\mu_R)}{C_0(\mu_R)^2} = \frac{M}{4\pi} K 
	-\frac{g_A^2}{2f^2} \frac1{C_0(\mu_R)^2} \,,
\end{eqnarray}
where $K$ is a constant which must be order $Q^2$ for $C_0^p(\mu_R)\sim 
Q^0$.  (Recall, from Eq.~(\ref{C0pbc}) that $K=\gamma-1/a \lesssim 1/a$ in the 
pure nucleon theory.) Including $C_0^p(\mu_R)$ makes the PDS amplitudes
explicitly $\mu_R$ independent.  In performing fits to the data the constant $K$
and coupling $D_2(\mu_R)$ cannot be determined independently.  In what
follows we will drop $K$ when fitting and simply remember that the values of 
$D_2(\mu_R)$ extracted from the fits may differ from the renormalized coupling in 
the Lagrangian.  In PDS, if $C_0^p(\mu_R)$ is omitted from our expressions then 
$D_2(\mu_R)$ does not follow the renormalization group equation, as we will see 
below.

In the OS scheme, the constant $g_A^2/(2f^2)$ in Eqs.(\ref{C0pb}) and (\ref{C0ps})
is not present, so $C_0^p(\mu_R)/C_0(\mu_R)^2$ is $\mu_R$ independent.   The
OS scheme amplitudes $A^{(-1)}$ and $A^{(0)}$ are therefore $\mu_R$
independent without $C_0^p(\mu_R)$ as can be seen by examining
Eqs.~(\ref{OSc0c2}), and (\ref{D2sln}).  In OS the constant $K$ will also be
absorbed into $D_2(\mu_R)$ for fitting.
 
\begin{figure}[!t]  
  \centerline{\epsfxsize=8.0truecm \epsfbox{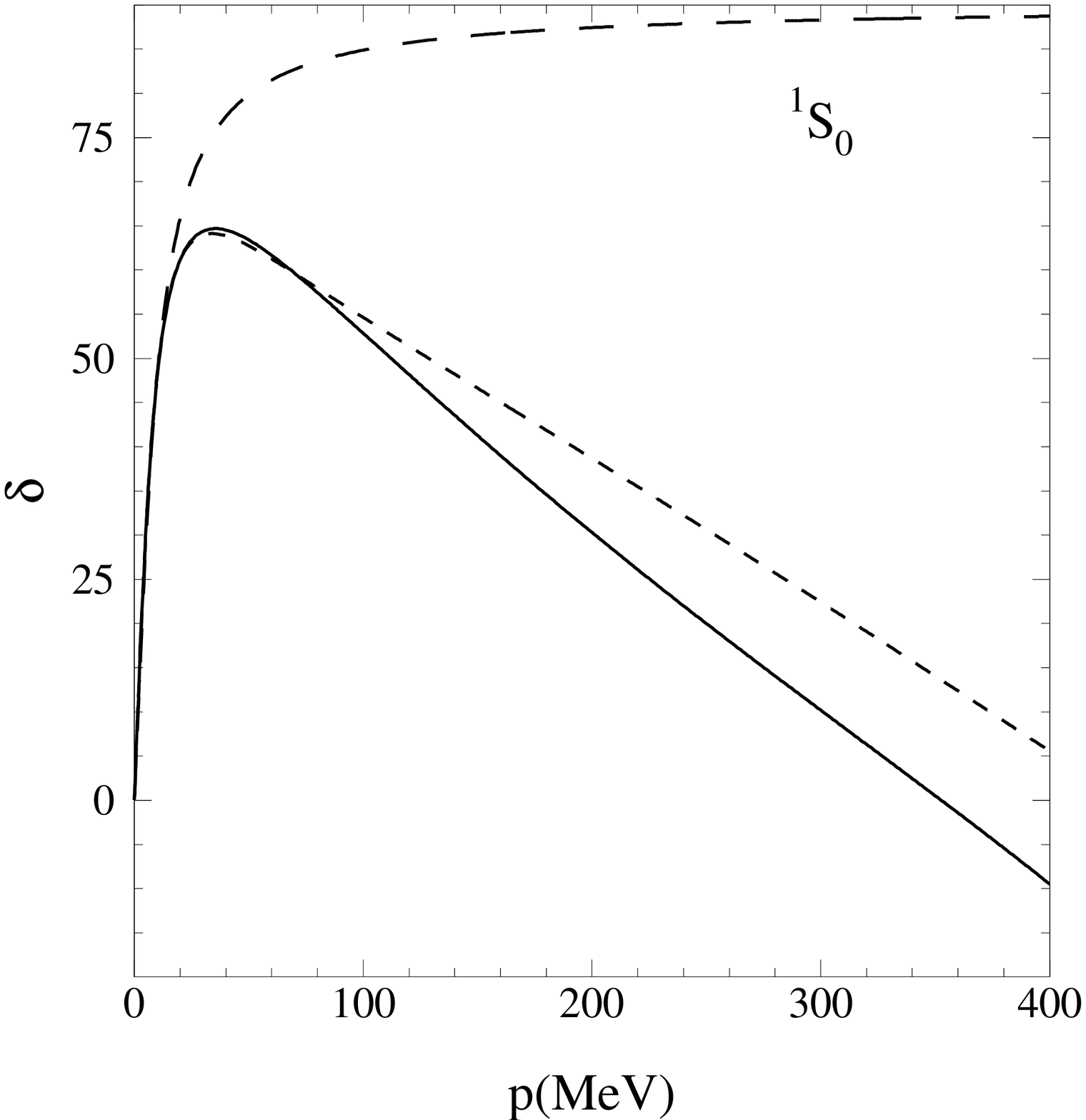}
        \epsfxsize=8truecm \epsfbox{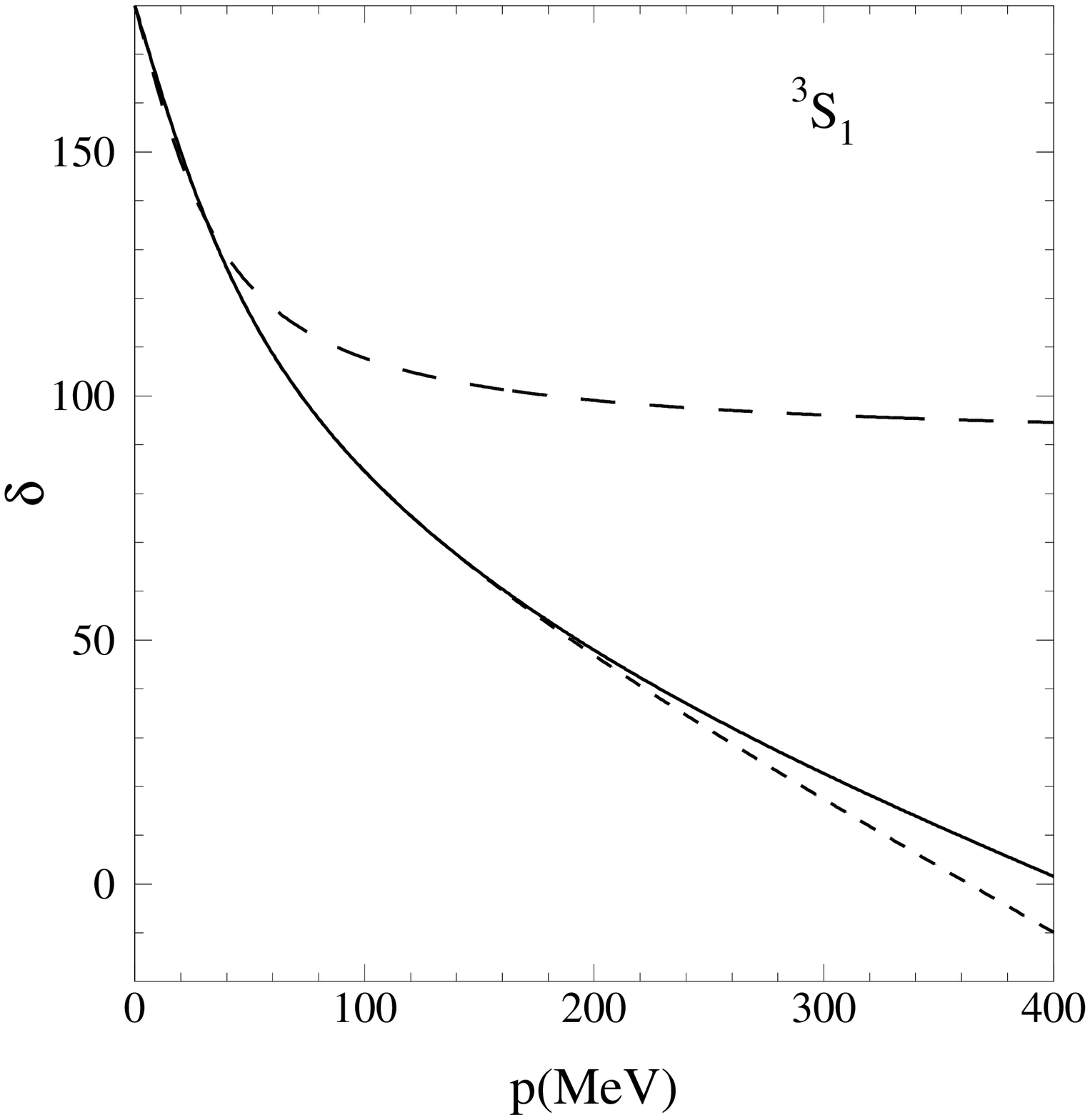} }
 {  
\caption{Fit to the phase shift data emphasizing the low momentum 
region.  The solid line is the Nijmegen fit to the data \cite{Nij}, the long dashed 
line is the order $1/Q$ result, and the short dashed line is the order $Q^0$ 
result. } \label{fig_fits} }
\end{figure}

Using the Nijmegen phase shifts \cite{Nij} between $7$ and $100\,{\rm MeV}$, we fit
the coefficients $C_0(\mu_R)$, $C_2(\mu_R)$ and $D_2(\mu_R)$.  We took
$m_\pi=137\,{\rm MeV}$. Clearly we would like to bias the fit towards the low
momentum points since that is where the theoretical error is smallest.  This can be
accomplished by assigning a percent error, $\simeq p/(300\,{\rm MeV})$, to the data
and then minimizing the $\chi^2$ function.  In Tables~\ref{tble_sPDS} and
\ref{tble_sOS} we show the values\footnote{ The coefficients extracted from our
fits differ from those in Ref.\protect\cite{ksw2} because we have emphasized the
low energy data as opposed to doing a global fit.  It is interesting to note that
using our PDS value $C_2(\mu_R =137\,{\rm MeV})=11.5\,{\rm fm^4}$, the prediction
for the RMS charge radius of the deuteron \protect\cite{ksw3} becomes $1.966\,{\rm
fm}$ which is within 1\% of the experimental result.} of $C_0(\mu_R)$,
$C_2(\mu_R)$ and $D_2(\mu_R)$ extracted from the fits for $\mu_R =
70,100,137,160,280\,{\rm MeV}$.  These values exhibit the $\mu_R$ dependence
predicted by the RGE's to $\sim1\%$ in the $^1S_0$ channel and $\sim 4\%$ in the
$^3S_1$ channel.   Since the amplitudes are explicitly $\mu_R$ independent this
deviation is a measure of the accuracy of the fitting routine. In Fig.~\ref{fig_fits} the
results of the fits are shown. The results of the fits shown in the figure are identical
in both schemes.  Higher order corrections will give contributions to $\delta$ of the
form $p^2/\Lambda_\pi^2$.  The error in $\delta$ at $p=300\,{\rm MeV}$ is
consistent with $\Lambda_\pi \gtrsim 500\,{\rm MeV}$.

For processes involving the deuteron it is convenient to fix $C_0(\mu_R)$ using
the deuteron binding energy, $C_0(m_\pi)=-5.708\, {\rm fm}^2$.  Fitting with this
constraint we find $C_2(m_\pi)=10.80\,{\rm fm^4}$ and $D_2(m_\pi)=1.075\,{\rm
fm^4}$ in the PDS scheme. The fit to the phase shift data with these values is as
good as that in Fig.~\ref{fig_fits}.  

{ \begin{table}[!t]
\begin{center} \begin{tabular}{ccccccccccc} 
& & \multicolumn{3}{c}{Fit to ${}^1\!S_0$} &&  
	\multicolumn{3}{c}{Fit to ${}^3\!S_1$} &\\ 
$\mu_R({\rm MeV}) $  && $C_0(\mu_R)$ & $C_2(\mu_R)$ & $D_2(\mu_R)$ & 
    & $C_0(\mu_R)$ & $C_2(\mu_R)$ &  $D_2(\mu_R)$ &  \\ \hline
$70$  &&  $-6.48$ & $10.11$ & $-0.532$    &&  
	 $-22.73$ & $171.$  & $-70.41$  &   \\ 
$100$  &&  $-4.71$ & $5.36$ & $1.763$    && 
	$-9.93$ & $32.7$ & $-4.157$ &  \\ 
$137$ &&  $-3.53$ & $3.01$ & $2.000$   && 
	$-5.88$ & $11.5$ & $1.500$ &  \\ 
$160$  && $-3.05$ & $2.25$ & $1.869$   && 
	$-4.69$ & $7.32$ & $1.897$ &  \\ 
$280$  &&  $-1.79$ & $0.772$ & $1.105$   && 
	$-2.19$ & $1.57$ & $1.004$ &  
\end{tabular} \end{center} 
{ \caption{$^1S_0$ and $^3S_1$ couplings in the PDS
scheme.  $C_0(\mu_R)$ (in ${\rm fm}^2$), $C_2(\mu_R)$ (in ${\rm fm}^4$), and 
$D_2(\mu_R)$ (in ${\rm fm}^4$) are fit to the
Nijmegen data at different values of $\mu_R$.  } \label{tble_sPDS} } 
\end{table} } 

{
\begin{table}[!t]  
\begin{center} 
\begin{tabular}{ccccccccccc} & &
\multicolumn{3}{c}{Fit to ${}^1\!S_0$} &&  \multicolumn{3}{c}{Fit to ${}^3\!S_1$} &\\
  $\mu_R({\rm MeV}) $  && $C_0(\mu_R)$ & $C_2(\mu_R)$ & $D_2(\mu_R)$ &
  &  $C_0(\mu_R)$ & $C_2(\mu_R)$ &  $D_2(\mu_R)$ & \\   \hline 
$70$  && $-6.50$ & $9.75$ & $-6.047$   &&  
	 $-24.1$ & $121.$ & $-170.1$ &   \\ 
$100$  && $-4.73$ & $5.33$ & $-1.143$    && 
	$-10.0$ & $27.3$ & $-20.18$ &  \\ 
$137$ &&   $-3.54$ & $3.00$ & $0.378$  && 
	$-5.92$ & $10.5$ & $-4.124$ &  \\ 
$160$  && $-3.06$ & $2.25$ & $0.658$  && 
	$-4.74$ & $6.89$ & $-1.671$ &  \\ 
$280$  &&  $-1.80$ & $0.779$ & $0.692$  && 
	$-2.23$ & $1.61$ & $0.2985$ &  
\end{tabular}  
\end{center} 
{ \caption{$^1S_0$ and $^3S_1$ couplings in the OS scheme. 
$C_0(\mu_R)$ (in ${\rm fm}^2$), $C_2(\mu_R)$ (in ${\rm fm}^4$), and 
$D_2(\mu_R)$ (in ${\rm fm}^4$) are fit to the Nijmegen data at
different values of $\mu_R$.  }
\label{tble_sOS} } 
\end{table} }

In PDS, it is necessary to break $C_0(\mu_R)$ into perturbative and
non-perturbative parts to obtain amplitudes that are $\mu_R$ independent
order-by-order.  If $C_0^p(\mu_R)$ is omitted then the values of $D_2(\mu_R)$
determined from the fit will not follow the RGE. To see this we define the $\mu_R$
independent quantity
\begin{eqnarray} 
  R &=& c \bigg[ \frac{-D_2(\mu_R)}
  {C_0(\mu_R)^2} + \frac{M}{8\pi} 
   \bigg(\frac{M g_A^2}{8\pi f^2}\bigg) \ln{\bigg({\mu_R^2 \over  \mu_0^2}
   \bigg)} \bigg] \,, 
\end{eqnarray} 
and choose the constant $c$ so that $R=1$ for $\mu_R=137\,{\rm MeV}$.  For
other values of $\mu_R$ the deviation of $R$ from $1$ gives the discrepancy
between the values predicted by the RGE and those extracted from the fit.  For
$\mu_R=70,280\,{\rm MeV}$ we find $R=-0.53,\,7.25$ in the $^1S_0$ channel and
$R=-0.52,\,11.4$ in the $^3S_1$ channel.  These large deviations disappear if
$C_0^p(\mu_R)$ is included.  Without $C_0^p(\mu_R)$, the PDS amplitude is 
$\mu_R$ independent to the order that one is working.  However, this residual 
$\mu_R$ dependence gives larger corrections than expected 
\cite{Gegelia2}.  The reason for this (as explained below) is that the residual
$\mu_R$ dependence makes the tuning that was setup to give the large scattering
length $\mu_R$ dependent.  

Integrating out the pion gives low-energy theorems for the coefficients $v_i$ in 
the effective range expansion\cite{Cohen2},
\begin{eqnarray} \label{erexpn}
  p\cot{(\delta)} =  -\frac{1}{a} + \frac{r_0}{2}\, p^2 + 
     v_2\, p^4 + v_3\, p^6 + v_4\, p^8 + \ldots \,.
\end{eqnarray}
Performing a matching calculation between the two theories gives expressions
for $a$, $r_0$ and the $v_i$ in terms of the parameters in the pion theory. 
Since the theory with pions is an expansion in $Q$ these predictions take the 
form of Taylor series in $Q/\Lambda_\pi$
\begin{eqnarray}  \label{erce}
  \frac1a = \gamma +  \sum_{i=2}^\infty B_0^{(i)} \,, \qquad\quad 
   \frac{r_0}{2} = \sum_{i=0}^\infty  B_1^{(i)} \,, \qquad\quad
    v_n = \sum_{i=2-2n}^\infty  B_n^{(i)} \,,
\end{eqnarray}
where $B_n^{(i)} \sim Q^{\,i}$.  At this time only the first coefficient in each series
is known since $p\cot\delta$ has only been calculated\footnote{A calculation of
the low energy theorems at order $Q^3$ in the $^1S_0$ channel was presented at 
the INT Workshop on Nuclear Physics with Effective Field Theory, Seattle, WA, 
25-26 Feb 1999 (S. Fleming, T. Mehen, and I.W. Stewart) \cite{msconf}.  This result 
will not be discussed here.} to order $Q^2$.  The notation
\begin{eqnarray}  \label{C0g}
 \gamma = \frac{4\pi}{M C_0(\mu_R)} + \mu_R \,
\end{eqnarray}
will be used to denote the location of the perturbative pole in the amplitudes. In 
PDS
\begin{eqnarray} \label{B01PDS}
  B_0^{(2)} &=& \bigg( \frac{-4\pi }{M} \bigg) {m_\pi^2 D_2(\mu_R)   \over 
    C_0(\mu_R)^2} +\frac{m_\pi^2\,M g_A^2}{8\pi f^2}
    \Bigg[ \frac12 + \frac12 \ln{\bigg({\mu_R^2 \over
    m_\pi^2}\bigg)} -\frac{2\gamma}{m_\pi}  
  +\frac{\gamma^2}{m_\pi^2} \ \Bigg]   - K \nn \,, \\
 B_1^{(0)} &=& \bigg( \frac{4\pi}{M} \bigg) 
  \frac{C_2(\mu_R)}{\Big[C_0(\mu_R) \Big]^2} + \frac{Mg_A^2}{8\pi f^2} 
  \bigg[  1 - \frac{8\gamma}{3\,m_\pi} +   \frac{2\gamma^2}{ m_\pi^2 } \bigg]  \,.
\end{eqnarray}
Note that if $C_0^p(\mu_R)$ had been neglected then $B_0^{(2)}$ would not be
$\mu_R$ independent.  With  $\mu_R=m_\pi$ Eq.~(\ref{B01PDS}) agrees with 
Ref.~\cite{ksw2} if their definition of $D_2(\mu_R)$ is adopted. In the OS scheme 
we have 
\begin{eqnarray}  \label{B01OS}
  B_0^{(2)} &=& \bigg( \frac{-4\pi}{M}\bigg)  { m_\pi^2 D_2(\mu_R)\over 
  C_0(\mu_R)^2} +\frac{ m_\pi^2 \,Mg_A^2}{8\pi f^2}
    \bigg[ \frac12 \ln{\bigg({\mu_R^2 \over
    m_\pi^2}\bigg)} -\frac{2\gamma}{m_\pi}  \bigg]  -K \,, \nn \\
 B_1^{(0)} &=&  \bigg( \frac{4\pi}{M} \bigg) 
  \frac{C_2(\mu_R)}{C_0(\mu_R)^2} + \frac{Mg_A^2}{8\pi f^2} 
  \bigg[ \frac{\gamma^2}{\mu_R^2 } +1- \frac{8\gamma}{3\,m_\pi} + \frac{2\gamma^2}
  {m_\pi^2} \bigg]  \,.
\end{eqnarray}
The value of the remaining $B_n^{(i)}$ determined at this order are the same in
both schemes 
\begin{eqnarray}  \label{Bvn}
  B_n^{(2-2n)} &=& - \frac{M g_A^2}{8\pi f^2} \left(\frac{-4}{m_\pi^2}\right)^n 
  \bigg[\frac{1}{4n} - \frac{2\gamma}{(2n+1)m_\pi} + \frac{\gamma^2}{(n+1)m_\pi^2}
    \bigg]\, m_\pi^2 \,.
\end{eqnarray}
For $n=2,3,4,$ Eq.~(\ref{Bvn}) gives the low-energy theorems derived in
Ref.~\cite{Cohen2}.

\begin{figure}[!t]  
  \centerline{\epsfxsize=8.0truecm \epsfbox{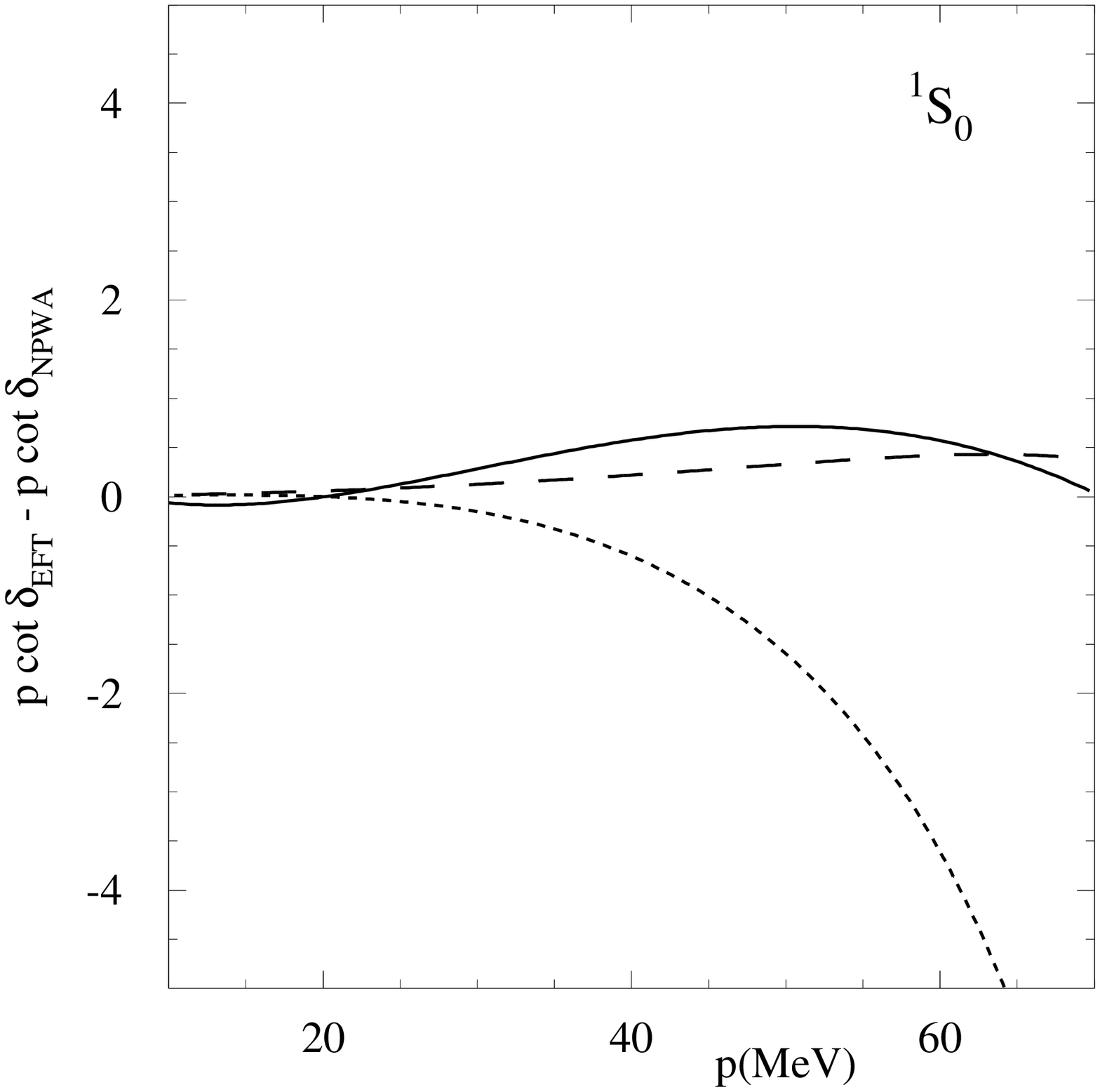}
        \epsfxsize=8truecm \epsfbox{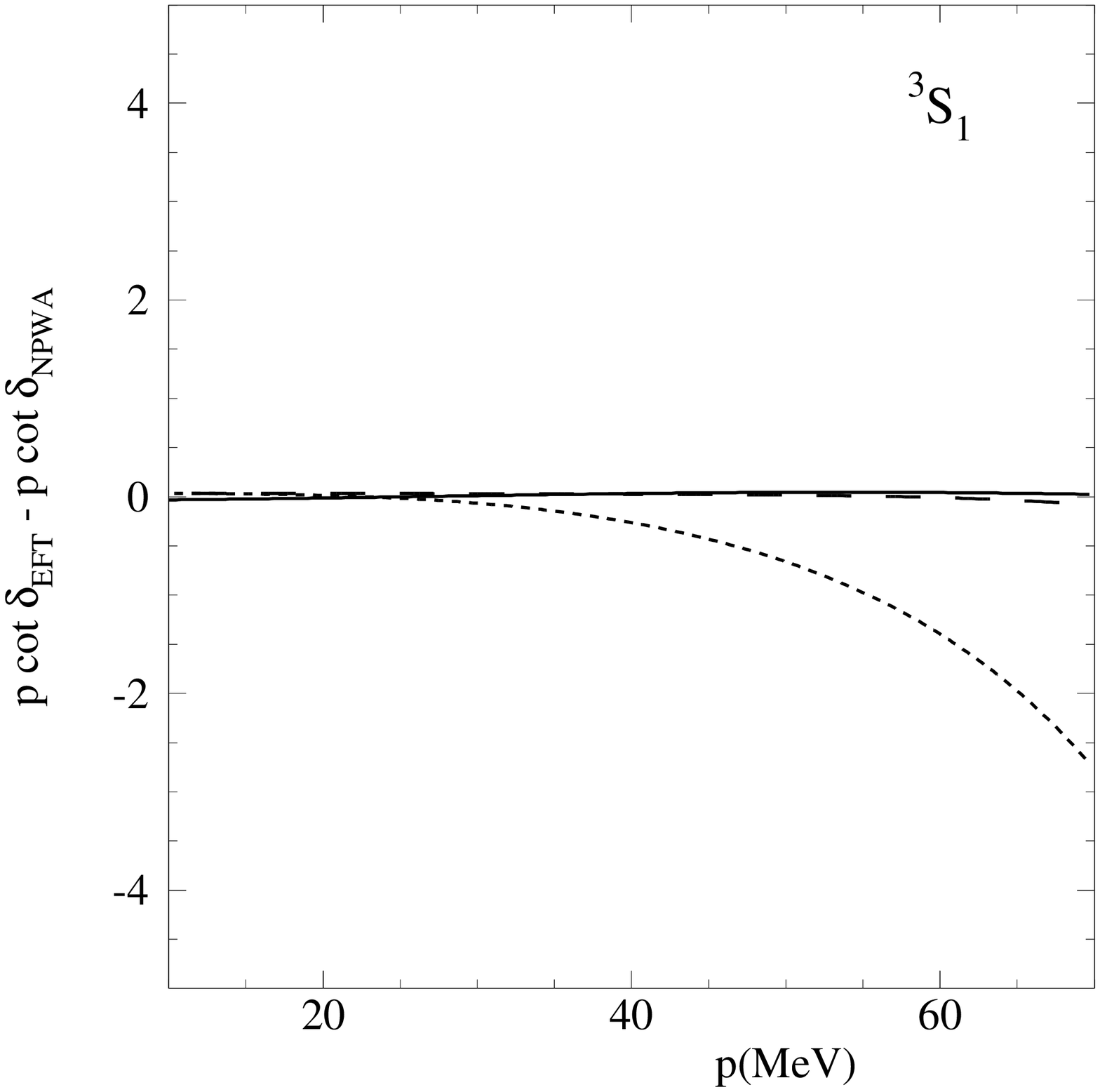} }
 {  \caption{The effective field theory  and Nijmegen Partial Wave analysis 
    \cite{Nij}
values of $p\cot\delta$ are compared.  The solid lines use $p\cot(\delta^{(0)}+
\delta^{(1)})$, the dashed lines use Eq.~(\ref{erexpn}) with the 
$v_i$ from Ref.~\cite{Cohen2}, and the dotted lines use the values of $v_i$ 
from the low-energy theorems.  } \label{fig_ch} }
\end{figure}

The $v_i$ in Eq.~(\ref{erexpn}) can be predicted in terms of one parameter,
$C_0^{np}(\mu_R)$, which is fixed in Ref.~\cite{Cohen2} by the condition $4\pi/[M
C_0^{np}(\mu_R)]+\mu_R = 1/a$.  Corrections to these predictions are expected to
be $30-50\%$ due to higher order $Q/\Lambda$ terms.  The $v_i$ extracted from
the phase shift data \cite{Cohen2,Nij2} disagree with the low-energy theorems by
factors of order $5$.  In Fig.~\ref{fig_ch}, we see that the agreement of
$p\cot(\delta^{(0)} + \delta^{(1)})$ (solid lines)\footnote{ Note that when expanded
in $Q$, $p\cot\delta= ip +4\pi/[MA^{(-1)}] -4\pi A^{(0)}/[M(A^{(-1)})^2]+{\cal
O}(Q^3)$, which differs from $p\cot(\delta^{(0)}+\delta^{(1)})$ by terms of order
$Q^3$. The latter expression is used since the parameters in
Table~(\ref{tble_sPDS}) were fit using Eq.~(\ref{delta01}).} with the Nijmegen partial
wave analysis is comparable to that of the effective range expansion with the $v_i$
from the fits in Refs.~\cite{Cohen2,Nij2} (dashed lines).  Note that our fit is more
accurate at low momentum than the global fit in Ref.~\cite{ksw2}.  However,
keeping only the first five terms from the low-energy theorems (dotted lines) gives
larger disagreement at $70\,{\rm MeV}$.  This is not surprising since the pion
introduces a cut at $p= i \,m_\pi/2$, so the radius of convergence of the series
expansion of $p \cot{(\delta)}$ in Eq~(\ref{erexpn}) is $\simeq 70\, {\rm MeV}$.  At
$p=70\,{\rm MeV}$, one expects large corrections from the next term in the series. 
However, the fit values of $v_i$ give good agreement with the data even at
$70\,{\rm MeV}$.  It is possible that uncertainty from higher order terms in the
Taylor series has been absorbed into $v_2$, $v_3$, and $v_4$ in the process of
performing the fits.  For this reason, the uncertainty in the values of $v_i$ that were
found from fitting to the data may be considerable.  

To get an idea of the error in $v_2$, we will specialize to the $^3S_1$ channel. 
The Nijmegen phase shift analysis \cite{NijPW} lists two data points for $p <
70\,{\rm MeV}$: $ p= 21.67 \,{\rm MeV}$ where $\delta^{(^3S_1)}= 147.747 \pm
0.010\,^\circ$, and $p= 48.45 \,{\rm MeV}$, where $\delta^{(^3S_1)}= 118.178 \pm
0.021\,^\circ$.  Using $a=5.420 \pm 0.001 \,{\rm fm}$ and $r_0 = 1.753 \pm
0.002\,{\rm fm}$ \cite{Nij2} in the effective range expansion and fitting to the
lowest momentum data point, we find $v_2 = -0.50\, \pm\, 0.52 \,{\rm fm^3}$,
where the error in $a$, $r_0$, and $p\cot\delta $ have been added in quadrature. 
This differs by one sigma from both the value predicted by the low-energy
theorem, $v_2^{\rm thm}=-0.95 \,{\rm fm^3}$, and the value from the fit,
$v_2^{\rm fit} = 0.04 \,{\rm fm^3}$.  Since the range of the pure nucleon theory is
$70\,{\rm MeV}$, there will also be a $\simeq 0.1 \,{\rm fm^3}$ error in this
extraction from $v_3$ and higher coefficients.  This error was estimated by
comparing the theoretical expression for $p\cot\delta$ with the first three terms
in its series expansion.  If we instead use the higher momentum point we find
$v_2 = 0.03\, \pm\, 0.04 \,{\rm fm^3}$ with $\simeq 0.5\,{\rm fm^3}$ theoretical
uncertainty.  The uncertainty in these values of $v_2$ is too large to make a
definitive test of the low-energy theorems.  

Recall that the unnaturally large scattering length $a$ is a fine tuning that was
accounted for by demanding that in Eq.~(\ref{C0g}), $C_0(\mu_R)$ is close to its
ultraviolet fixed point, and $\gamma\approx 1/a$.  Examining the expression for
$1/a$ in Eq.~(\ref{erce}) it may seem that this could be destroyed by chiral
corrections.  If $D_2(\mu_R)\sim C_0(\mu_R)^2$ then the first term gives
$B_0^{(2)}\sim 205\,{\rm MeV}$.  In fact from Table~\ref{tble_ar0}, we see that the fit
gives $B_0^{(2)}\lesssim 1/a$.  The reason for this small value is that since
$A^{(0)} \propto (A^{(-1)})^2$ the amplitude has a double pole.  Since this pole is
spurious (occurring from the perturbative expansion) the residue of the double
pole must be small in order to fit the data.  This leads to a good fit
condition\cite{ms0} which will be approximately satisfied 
\begin{eqnarray}  \label{condition}
     \left.  \frac{A^{(0)}}{[A^{(-1)}]^2} \ \right|_{-ip = \gamma} = 0\,.
\end{eqnarray}
The condition in Eq.~(\ref{condition}) implies $B_0^{(2)}\simeq 4\pi\gamma^2/M$. 
In fact this gives the right order of magnitude for the values of $B_0^{(2)}$
determined from the fits in Table~\ref{tble_ar0}.  Similar good fit conditions occur at
higher order keeping the coefficients $B_0^{(i)}$ small.  The division of
$C_0(\mu_R)$ into nonperturbative and perturbative pieces is arbitrary, allowing us
to set up the theory so that the $Q$ expansion for $1/a$ is well behaved.  By
imposing conditions like Eq.~(\ref{condition}), chiral corrections to the location of
the pole are absorbed into perturbative pieces of $C_0(\mu_R)$ order by order. 
Thus, we choose to spoil the short distance nature of $C_0$ by giving it $m_\pi$
dependence \cite{dkap} to keep the pole in the right place.

{
\begin{table}[!t] 
\begin{center} 
\begin{tabular}{ccccccccccc} & &
\multicolumn{4}{c}{${}^1\!S_0$ Fit}  && \multicolumn{4}{c}{${}^3\!S_1$ Fit}  
  \vspace{.05in} \\ 
  $\mu_R({\rm MeV}) $  && $\gamma$ & $B_0^{(2)}$ & $1/a$ &
  $r_0$ &&  $\gamma$ & $B_0^{(2)}$ & $1/a$ &  $r_0$  \\   \hline 
 $70$  && $-10.18$ & $2.05$ & $-8.124$ & $0.01468$ && 
	$48.39$ & $-15.82$ & $32.57$ & $0.01101$ \\ 
$137$ && $-10.16$ & $2.04$ & $-8.121$ & $0.01480$ &&
	$48.96$ & $-16.76$ & $32.19$ & $0.01098$ \\ 
$280$  && $-10.23$ & $2.12$ & $-8.105$ & $0.01484$ &&
	$46.39$ & $-12.64$ & $33.76$ & $0.01111$
\end{tabular}  
\end{center} 
{ \caption{Values of $\gamma$, $B_0^{(2)}$, $1/a$, and $r_0$ (in MeV) 
obtained from our fits. Three values of $\mu_R$ are shown to emphasize that the 
value of the extracted parameters depends weakly on $\mu_R$.}
\label{tble_ar0} } 
\end{table} }

In Table~\ref{tble_ar0} we see that when $B_0^{(2)}$ is added to $\gamma$, values
of $1/a$ are obtained which are close to the physical values, $1/a(^1S_0)=
-8.32\,{\rm MeV}$ and $1/a(^3S_1)=36.4\,{\rm MeV}$.  It is encouraging that the
value of $\gamma$ found from fits in the $^3S_1$ channel are close to the physical
pole in the amplitude which corresponds to the deuteron, $\gamma=45.7\,{\rm
MeV}$.  Values for $r_0$ can also be predicted from the fits using
Eq.~(\ref{B01OS}).  Experimentally, $r_0(^1S_0)=0.0139\,{\rm MeV}^{-1}$ and
$r_0(^3S_1)=0.00888\,{\rm MeV}^{-1}$, so the values in Table~\ref{tble_ar0} agree
to the expected accuracy.

\section{Determining the range $\Lambda_\pi$ }  \label{range}

Here we will examine the phase shift data to see what it tells us about the range
of the effective field theory with perturbative pions.  In Ref.~\cite{sf2}, a Lepage
analysis is performed on the observable $p\cot{\delta(p)}$ in the $^1S_0$
channel.  Near $350\,{\rm MeV}$ the experimental $^1S_0$ phase shift passes
through zero.  Therefore, the error $|p\cot{\delta}^{\rm NPWA}-p\cot{\delta}^{\rm
EFT}|$ is greatly exaggerated since $p\cot\delta(p)\to \infty$. To avoid this
problem we will use the $^1S_0$ and $^3S_1$ phase shifts as our observables,
since $\Delta \delta= |\delta^{\rm NPWA}-\delta^{\rm EFT}|$ remains finite for all
$p$.  The next-to-leading order amplitudes given in section V will be used. The
phase shifts have an expansion of the form $\delta = \delta^{(0)} + \delta^{(1)} +
{\cal O}(Q^2/^2)$, where \cite{ksw2} 
\begin{eqnarray}
  \delta^{(0)} &=& -\frac{i}{2} \ln{\bigg[ 1 + i\frac{pM}{2\pi} {\cal A}^{(-1)}
	\bigg]} \,,\qquad
   \delta^{(1)} = \frac{pM}{4\pi} \frac{ {\cal A}^{(0)} }{ 1 + i\frac{pM}{2\pi}
  {\cal A}^{(-1)} }\,.  \label{delta01}
\end{eqnarray}
Recall that a momentum expansion of $\delta$ would result in terms with only odd
powers of $p$.  However, the expansion for $\delta$ in Eq.~(\ref{delta01}) is not
simply a momentum expansion, so the next-to-leading order calculation can have
errors which scale as $p^2/\Lambda_\pi^2$.  For example, once pions are included we
can have a term $p^2 \tan^{-1}(2 p/m_\pi)$ which is odd in $p$, order $Q^2$, and
scales as $p^2$ for large momenta. 

\begin{figure}[!t]  
  \centerline{\epsfxsize=8.0truecm \epsfbox{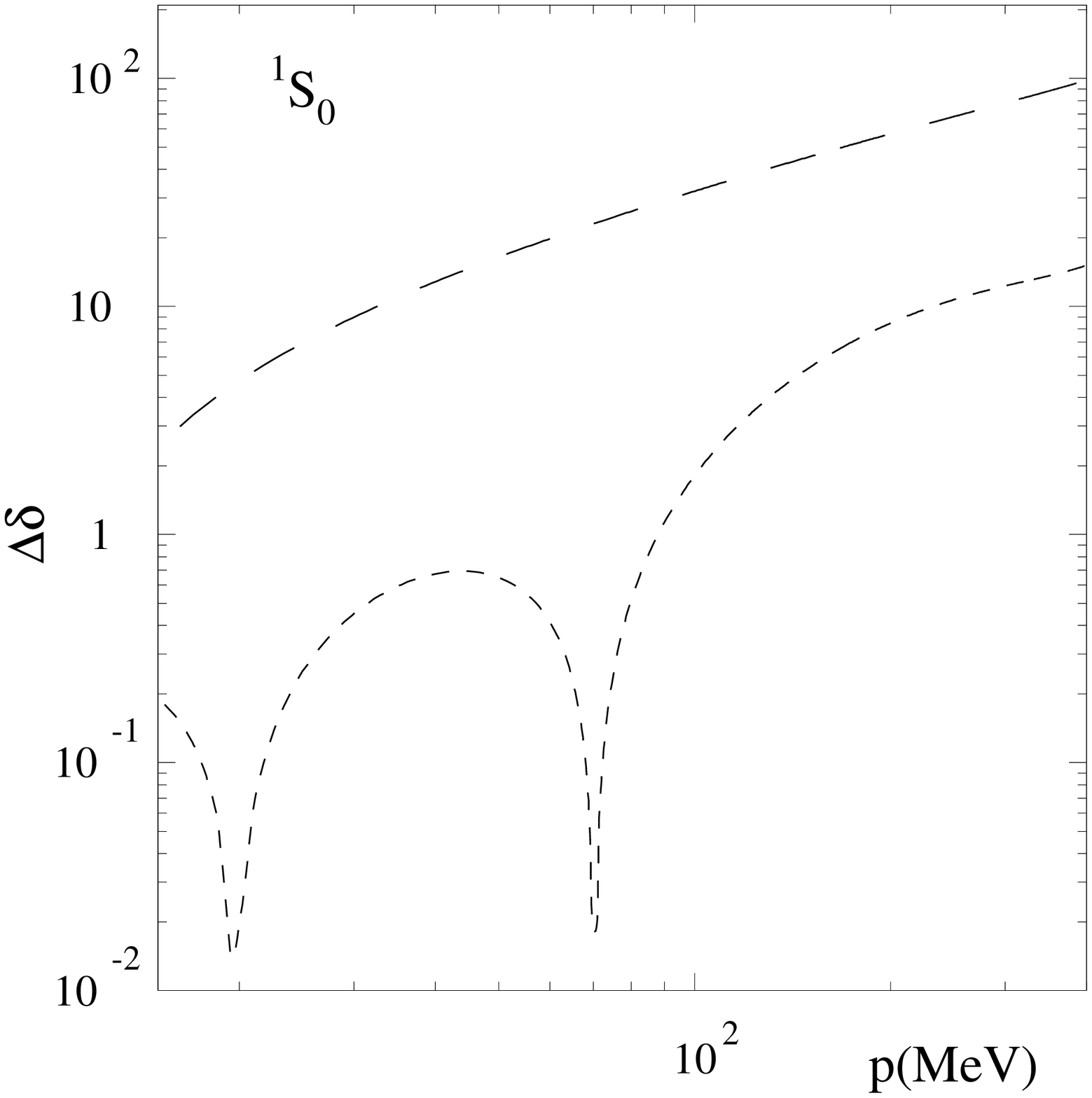}
        \epsfxsize=8truecm \epsfbox{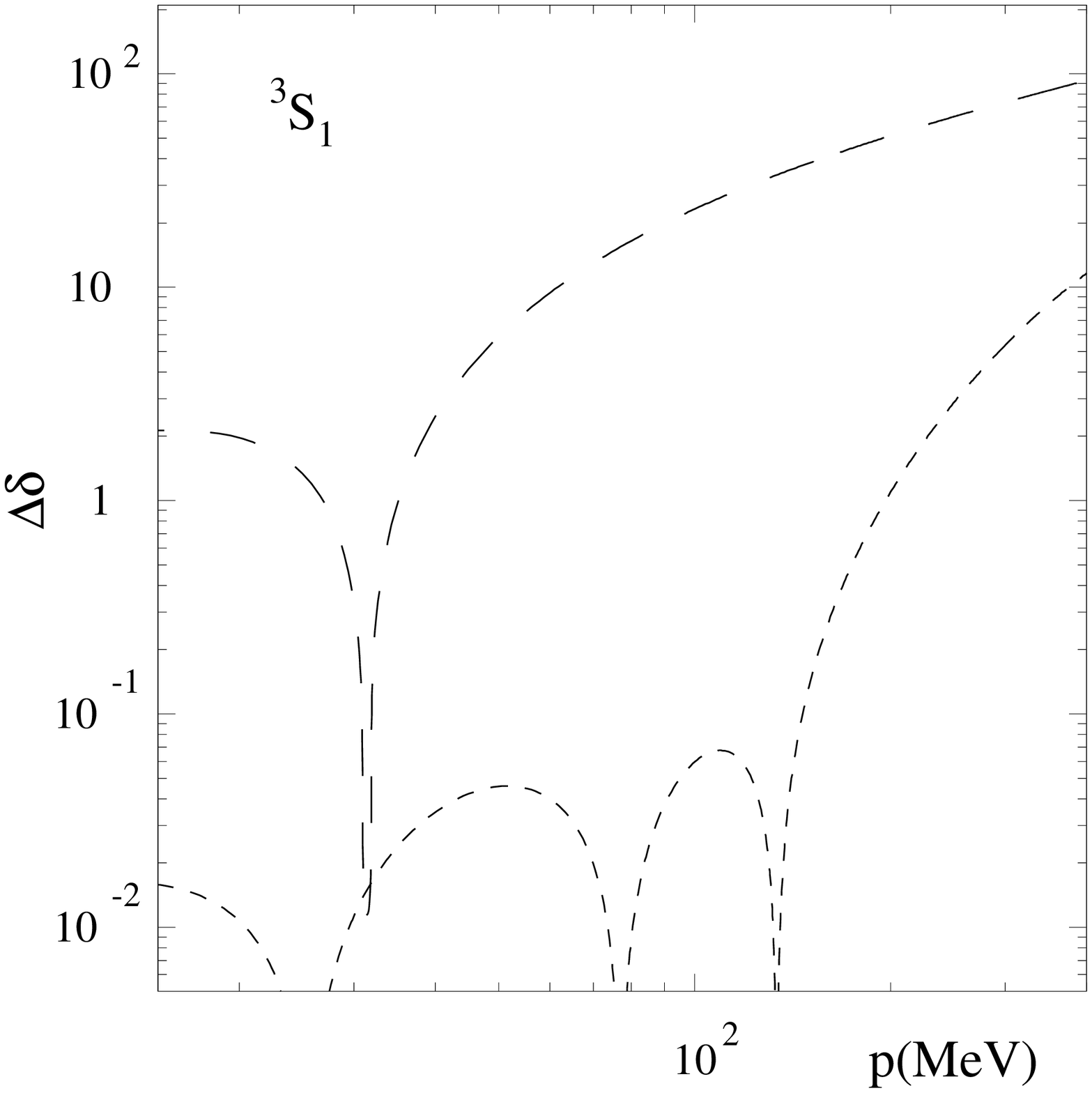} }
 {  \caption{Error analysis for the phase shifts in the ${}^1\!S_0$ and
${}^3\!S_1$ channels.  $\Delta \delta$ is the difference between the
the effective field theory prediction and the Nijmegen partial wave 
analysis\cite{Nij}. 
The long and short dashed lines use the ${\cal O}(Q^0)$ and ${\cal O}(Q)$ 
theoretical phase shifts respectively.} \label{fig_lep} }
\end{figure}

In Fig.~\ref{fig_lep}, we plot $\Delta \delta$ versus $p$ using log-log axes.  Note
that the sharp dips in Fig.~\ref{fig_lep} are just locations where the theory happens
to agree with the data exactly.  The Nijmegen data\cite{Nij} is available up to
$p=405\,{\rm MeV}$.  
In a theory with just a momentum expansion the errors will appear as straight lines
on the log-log plot as pointed out by Lepage \cite{Lepage}.  In the pion theory the
expansion is in both $m_\pi/\Lambda_\pi$ and $p/\Lambda_\pi$, so this is no longer
true.   For $p>m_\pi$ we expect the errors to be of the form\footnote{ At
momenta $1/a \ll p \ll m_\pi$ we could have $\Delta \delta^{(0)} \sim
B_0^{(2)}/p\sim m_\pi^2/\Lambda_\pi p$.  However, as explained in section VI,
$B_0^{(2)}\lesssim 1/a$ so this term is very small.}
\begin{eqnarray} 
 \Delta \delta^{(0)} &\sim& \bigg(1+\frac{m_\pi}{\Lambda_\pi}+\ldots\bigg) {p\over
    \Lambda_\pi} +\ldots\,, \\ \Delta \delta^{(1)} &\sim&
   \bigg(\frac{m_\pi}{\Lambda_\pi}+\frac{m_\pi^2}{\Lambda_\pi^2}+ \ldots \bigg)
  {p\over \Lambda_\pi} + \bigg(1+\frac{m_\pi}{\Lambda_\pi}+\ldots \bigg) 
  {p^2\over \Lambda_\pi^2} +\ldots \nn \,.
\end{eqnarray}
The fact that there is always a $p/\Lambda_\pi$ error arises from the fact that, as
seen in Eq.~(\ref{erce}), $r_0$ is reproduced in the effective field theory as an
expansion in $m_\pi/\Lambda_\pi$.  For $p/\Lambda_\pi \gg m_\pi/\Lambda_\pi$ the
slope of the lines on the plot should indicate the lowest power of $p$ that has not
been included.  At low momentum the error in $\Delta \delta^{(n)}$ is dominated by
the $p\, m_\pi^n/\Lambda_\pi^{n+1}$ term and the lines should be parallel.  From
Fig.~\ref{fig_lep} we see that the error is smallest at low momentum and increases
as the momentum increases, which is how the theoretical error is expected to
behave.  

It is clear that even for $p \sim 400\,{\rm MeV}$ the next-to-leading order
calculations are reducing the error in the phase shift.  Because two new
parameters are added at next-to-leading order it is always possible to force exact
agreement at some value of p.  However, if one were to force the data to agree too
well at high momentum then this would destroy the agreement at low momentum. 
Since the improvement of the fit in Fig.~\ref{fig_lep} at high momentum does not
come at the expense of the fit at low momentum this is evidence that the error is
being reduced in a systematic way.  At high momentum one expects that the error
is $\sim p^2/\Lambda_\pi^2$.  From Fig.~\ref{fig_lep}, $\Delta \delta\sim 0.26\,{\,\rm
radians}$ for $p=400\,{\rm MeV}$, implying $\Lambda_\pi\sim 800\,{\rm MeV}$.  This
is only a rough estimate for the range because we cannot yet exclude the
possibility that the next-to-next-to-leading order phase shift has an anomalously
small coefficient.  Even though the lines in Fig.~\ref{fig_lep} are not straight they
should still cross at approximately the range of the theory since at this point
higher order corrections do not improve the agreement with the data.   This error
analysis is consistent with the possibility that the range is $\gtrsim 500\,{\rm
MeV}$.  

Further information on the range of the effective field theory can be obtained by
examining electromagnetic processes involving the deuteron 
\cite{ksw3,chen,SS,chen2,Kaplan1},
such as the deuteron charge radius, electromagnetic form factors, deuteron
polarizability, and deuteron Compton scattering.  For these observables errors are
typically $\sim 30-40\%$ at leading order and $\sim 10\%$ at next-to-leading
order.  This is what one would expect if the expansion parameter
$m_\pi/\Lambda_\pi\sim 1/3$, implying $\Lambda_\pi\sim 410\,{\rm MeV}$.  This is
consistent with our previous estimate for $\Lambda_\pi$.  If the range is this large
one should expect that the error in deuteron properties will be at the few percent
level once  next-to-next-to-leading order calculations are performed\footnote{
Recently calculations of the deuteron quadrapole moment \cite{binger} and the
$^1S_0$ phase shift \cite{rupak,msconf} have been carried out at this order.}.

\section{Summary}

In this chapter the structure of the effective field theories of nucleons with and
without pions is studied.  We discuss a momentum subtraction scheme, the OS
scheme, which obeys the KSW power counting.  The method of local counterterms
is used to obtain the renormalization group equations for the coupling constants in
these theories.  Using local counterterms defines the OS and PDS renormalization
schemes unambiguously.  Two-loop graphs with potential pions in the $^3S_1$
channel are computed and shown to have $p^2/\epsilon$ poles.  The presence of
$1/\epsilon$ poles implies that the only model independent piece of pion exchange
is the part that can be treated perturbatively.  We obtain the renormalized
couplings $C_0(\mu_R)$, $C_2(\mu_R)$ and $C_4(\mu_R)$ at order Q in the OS
and PDS schemes.  

We have emphasized why it is important to have $\mu_R$ independent amplitudes
order by order in $Q$.  Such amplitudes are obtained automatically in the OS
scheme.  In PDS $\mu_R$ independent amplitudes may be obtained by treating
part of $C_0(\mu_R)$ perturbatively.  It is also necessary to treat part of
$C_0(\mu_R)$ perturbatively if we wish to keep the pole in the amplitude in a fixed
location order by order in the chiral expansion. Another result concerns the large
$\mu_R$ behavior of the couplings in this theory. In the OS scheme the coupling
constants obey the KSW power counting for all $\mu_R>1/a$.  In PDS the
breakdown in the power counting for $C_0(\mu_R)$ is avoided if $C_0(\mu_R)$ is
split into non-perturbative and perturbative parts.  Therefore, the breakdown of the
scaling in the coupling constants is artificial.

Next-to-leading order calculations of nucleon-nucleon phase shift data
\cite{ksw2} provide fits to data at large momenta which are far more accurate than
one would expect if the theory broke down completely at $300 {\rm\,MeV}$.  Of
course, this does not mean that nucleon effective theory can be applied at
arbitrarily high energies.  The scale, $M g_A^2/(8 \pi f^2) \sim 300\,{\rm MeV}$, is
associated with short distance contributions from pion exchange and provides an
order of magnitude estimate for the range.  In the S-wave channel, $\Delta$
production and $\rho$ exchange become relevant at $\sim 700\,{\rm MeV}$, which
sets an upper limit on the range of the expansion.  To get a better understanding of
the range of the nucleon effective theory with perturbative pions one must examine
experimental data. An error analysis of the S wave phase shifts with
next-to-leading-order calculations seems to be consistent with a range of
$500\,{\rm MeV}$.  Though next-to-next-leading order corrections need to be
compared with data and other processes investigated, we remain cautiously
optimistic that the range could be as large as $500\,{\rm MeV}$.


\chapter{Radiation and Soft Pions}

In the KSW power counting for nucleon-nucleon interactions, pions are included
perturbatively.  In evaluating diagrams with pions, three types of contributions can
be identified: potential, radiation, and soft.  These pion effects differ in size and
therefore each have a different power counting.  In this chapter the power counting
for radiation and soft pions are discussed. The distinction between pion
contributions arises because there are several scales associated with two nucleon
systems.  In this respect the theory is similar to NRQCD and NRQED \cite{Caswell,
Bodwin}.

In NRQCD there are three mass scales associated with non-relativistic systems
containing two heavy quarks: the heavy quark mass $M$, momenta $\sim M v$, and
kinetic energy $\sim M v^2$, where $v$ is the relative velocity.  QCD effects at the
scale $M$ are integrated out and appear as local operators in NRQCD.  The
remaining low energy contributions can be divided into potential, radiation
(sometimes referred to as ultra-soft), and soft pieces
\cite{lm,gr,ls,labelle,beneke,gries1,gries2}.  Potential gluons have energy of order
$M v^2$ and momentum of order $M v$, radiation gluons have energy and
momentum of order $M v^2$, and soft gluons have energy and momentum of order
$M v$.  The power counting for radiation gluons requires the use of a multipole
expansion at a quark-gluon vertex \cite{gr,labelle}.  The $v$ power counting of
potential and radiation gluons can be implemented in the effective Lagrangian by
introducing separate gluon fields and rescaling the coordinates and fields by
powers of $v$ \cite{lm,ls}.  In Ref.~\cite{beneke} the separation of scales was
achieved on a diagram by diagram basis using a threshold expansion.  The
potential, radiation, and soft regimes were shown to correctly reproduce the low
energy behavior of relativistic diagrams in a scalar field theory.  In
Ref.~\cite{gries1,gries2} it was pointed out that these effects may be reproduced by
an effective Lagrangian in which separate fields are also introduced for the soft
regime.  Note that soft contributions come from a larger energy scale than
potential and radiation effects.  The heavy quark system does not have enough
energy to radiate a soft gluon, so they only appear in loops.  In Ref.~\cite{beneke}
it was shown that soft contributions to scattering do not appear until graphs with
two or more gluons are considered.  

In the nucleon theory there is another scale because the pions are massive.  For
the purpose of power counting it is still useful to classify pion contributions as
potential, radiation, or soft.  For a pion with energy $q_0$ and momentum $q$, a
potential pion has $q_0\sim q^2/M$ where $M$ is the nucleon mass, while a
radiation or soft pion has $q_0\sim q \ge m_\pi$.  In a non-relativistic theory,
integrals over loop energy are performed via contour integration.  Potential pions
come from contributions from the residue of a nucleon pole and give the
dominant contribution to pion exchange between two nucleons.  For these pions,
the energy dependent part of the pion propagator is treated as a perturbation
because the loop energy, $q_0\sim q^2/M \ll q$.  The residues of pion propagator
poles give radiation or soft pion contributions.  The power counting for soft and
radiation pions differs.  For instance, the coupling of radiation pions to nucleons
involves a spatial multipole expansion, while the coupling to soft pions does not.

\section{Radiation pions}

In chiral perturbation theory the expansion is in powers of momenta and the pion
mass $m_\pi$.  For power counting potential pions it is convenient to take the
nucleon momentum $p=M v\sim m_\pi$ \cite{ksw2,lm}, so $v=m_\pi/M\sim 0.15$. 
The situation is different for radiation pions.  There is a new scale associated with
the threshold for pion production, which occurs at energy $E=m_\pi$ in the center
of mass frame. This corresponds to a nucleon momentum $p=Q_r$, where
$Q_r\equiv \sqrt{M m_\pi}=360\,{\rm MeV}$.  Because the radiation pion fields
cannot appear as on-shell degrees of freedom below the threshold $E= m_\pi$,
one expects that the radiation pion can be integrated out for $p \ll Q_r$. (Potential
pions should be included for $p\gtrsim m_\pi/2$.)  Another way to see that
radiation pions require $p\sim Q_r$ is to note that in order to simultaneously
satisfy $k_0^2= k^2 + m_\pi^2$ and $k_0\sim k\sim M v^2$ requires $v\sim
\sqrt{m_\pi/M} \sim 0.38$.  

The full theory with pions has operators in the Lagrangian with powers of $m_q$
which give all the $m_\pi$ dependence.  If the radiation pions are integrated out,
then the chiral expansion is no longer manifest because there will be $m_\pi$
dependence hidden in the coefficients of operators in the Lagrangian.  One is still
justified in considering the same Lagrangian, but predictive power is lost since it is
no longer clear that chiral symmetry relates operators with a different number of
pion fields.  Also, the $m_\pi$ dependence induced by the radiation pions may
affect the power counting of operators. For example, as shown in
Ref.~\cite{bkm0,bkm1,bkm2}, integrating out the pion in the one-nucleon sector
induces a nucleon electric polarizability $\alpha_E \propto 1/m_\pi$.  Alternatively,
one can keep chiral symmetry manifest by working with coefficients in the full
theory and including radiation pion graphs.  This is the approach we will adopt.

\begin{figure}[!t]
  \centerline{\epsfysize=7.5truecm \epsfbox{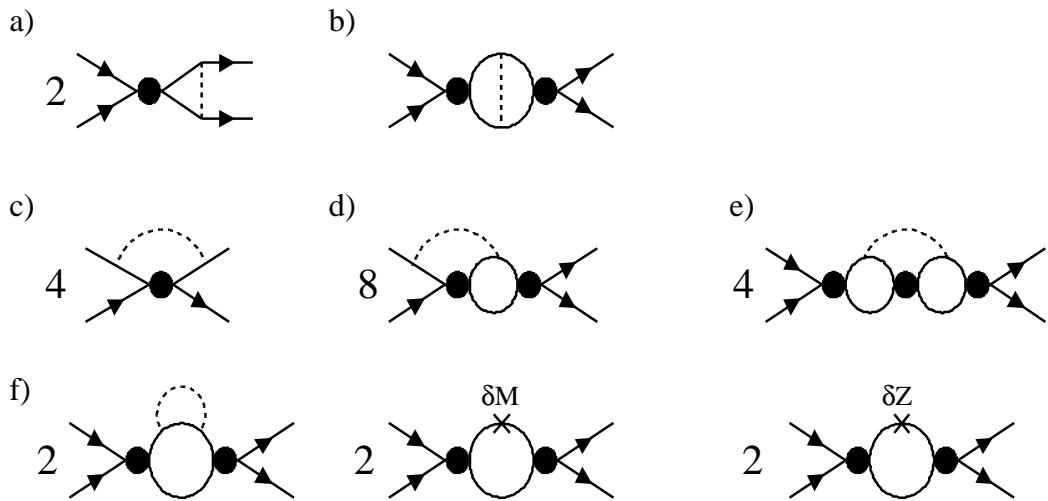}  }
 {\caption{Leading order radiation pion graphs for $NN$ scattering.  The solid
lines are nucleons, the dashed lines are pions and $\delta M$, $\delta Z$ are the
mass and field renormalization counterterms. The filled dot denotes the
$C_0(\mu_R)$ bubble chain.  There is a further field renormalization contribution
that is calculated in text, but not shown.} \label{Qr3} }
\end{figure}

The presence of the scale $Q_r$ modifies the power counting of the theory with
radiation pions.  In the KSW power counting, one begins by taking external
momenta $p \sim m_\pi \sim Q$. The theory is organized as an expansion in
powers of $Q$.  To estimate the size of a graph, loop 3-momenta are taken to be
of order $Q$.  However, potential loops within graphs with radiation pions can
actually be dominated by three momenta of order $Q_r$.  To see how this comes
about, consider as an example the graph shown in Fig.~\ref{Qr3}c.  Let $q$ be the
momentum running through the pion propagator, and let $k$ be the loop
momentum running through a nucleon bubble inside the radiation pion loop. 
The poles from the pion propagator are
\begin{eqnarray}
 {i\over q_0^2-{\vec q}\,^2 -m_\pi^2+i\epsilon} = {i\over
 (q_0-\sqrt{{\vec q}\,^2+m_\pi^2}+i\epsilon)(q_0+\sqrt{{\vec
 q}\,^2+m_\pi^2}-i\epsilon)} \,,
\end{eqnarray}
so the radiation pion has $|q_0| \ge m_\pi$. This energy also goes into the
nucleon bubbles. The $k$ integrand is largest when the nucleons are close to
their mass shell. But since the energy going into the loop is $\sim m_\pi$, this
occurs when $k^2/M \sim m_\pi$, i.e., $k \sim Q_r$.  We will begin by considering
the contribution of radiation pions to elastic nucleon scattering at the threshold,
$E = m_\pi$.  At this energy, external and potential loop momenta are of the
same size and power counting is easiest. Because $p \sim Q_r$ it is obvious that
we want to count powers of $Q_r$ rather than $Q$

The power counting rules at the scale $Q_r$ are as follows.  A scheme with
manifest power counting will be used, so that $C_0(\mu_R) \sim 1/(M \mu_R),
C_2(\mu_R) \sim 1/(M\Lambda \mu_R^2)$, etc., where $\Lambda$ is the range of
the theory. We will take $p \sim \mu_R \sim Q_r$.  A radiation loop has $q_0\sim q
\sim m_\pi$ so $d\,^4q\sim Q_r^8/M^4$, where $q$ is the momentum running
through the pion propagator.  A radiation pion propagator gives a $M^2/Q_r^4$,
while the derivative associated with a pion-nucleon vertex gives $Q_r^2/M$. A
nucleon propagator gives a $M/Q_r^2$.  External energies and momenta are kept in
the nucleon propagator since $E\sim p^2/M \sim Q_r^2/M$.  Furthermore, it is
appropriate to use a multipole expansion for radiation pion-nucleon vertices which
is similar to the treatment of radiation gluons in NRQCD \cite{gr}.  Therefore,
radiation pions will not transfer three-momenta to a nucleon. This is usually
equivalent to expanding in powers of a loop momentum divided by $M$ before
doing the loop integral.  The multipole expansion is an expandion in
$v=\sqrt{m_\pi/M}$.  A potential loop will typically have running through it either an
external or radiation loop energy $\sim Q_r^2/M$.  Therefore, in these loops the
loop energy $k_0\sim Q_r^2/M$, while the loop three momentum $k\sim Q_r$, so
$d^4k\sim Q_r^5/M$.  It is not inconsistent for $k\sim Q_r$ while $q\sim Q_r^2/M$,
since three momenta are not conserved at the nucleon-radiation pion vertices.   At
the scale $Q_r$ potential pion propagators may still be treated in the same way,
$i/(k_0^2-k^2-m_\pi^2)= -i/(k^2+m_\pi^2)+ {\cal O}(k_0^2/k^4)$, which has an
expansion in $Q_r^2/M^2$.  We will see through explicit examples that this power
counting correctly estimates the size of radiation pion graphs.

Note that only the potential loop measure gives an odd power of $Q_r$, so
without potential loops the power counting reduces to power counting in powers
of $m_\pi$. The power counting here therefore correctly reproduces the usual
chiral power counting used in the one nucleon sector\cite{jm,ulf}. 

Graphs with one radiation pion and additional higher order contact interactions or
potential pions are suppressed by factors of $Q_r/\Lambda$ relative to graphs with
a single radiation pion and $C_0$ vertices. The $Q_r$ expansion is a chiral
expansion about $m_\pi=0$, so there is a limit of QCD where it is justified.  The
scale $\Lambda$ is unknown. One possible estimate is $\Lambda_{\rm NN}=8\pi
f^2/(M g_A^2)= 300\,{\rm MeV}$ since a graph with $m+1$ potential pions is
suppressed by $p/300\,{\rm MeV}$ relative to a graph with $m$ potential pions. 
However, this order of magnitude estimate only takes into consideration a partial
subset of the graphs of the theory. As argued in chapter 5, it is possible that the
range is of order the scale of short range interactions that are integrated out,
implying $\Lambda \sim 500\,{\rm MeV}$. In fact, the accuracy of NLO computations
of nucleon-nucleon phase shifts is in agreement with this physically motivated
estimate of the range.  We will assume in this section that an expansion in
$Q_r/\Lambda$ is valid.  This hypothesis will be tested further by seeing how well
the effective theory makes predictions at $p\sim 300\,{\rm MeV}$.  For example,
processes with external pions could be considered.   If the $Q_r/ \Lambda$
expansion is not convergent, then application of the theory is restricted to $p <
Q_r$. 

The radiation pion graphs that give the leading order contribution to
nucleon-nucleon scattering are shown in Fig.~\ref{Qr3}. The filled dot denotes
the leading order interaction between nucleons, a $C_0(\mu_R)$ bubble sum. 
We illustrate the power counting with an example, the graph in Fig.~\ref{Qr3}d. For 
the moment, replace the $C_0$ bubble sums with single $C_0$ vertices.  Each 
$C_0$ gives a factor of $1/M Q_r$ and each nucleon line gives a factor $M/Q_r^2$. 
The derivatives from the pion couplings combine with the radiation pion propagator
to give a factor of unity. The radiation loop gives $Q_r^8/M^4$, while the nucleon
bubble loop gives $Q_r^5/M$. There is also a factor of $1/f^2$ from pion
exchange, and two factors of $1/4 \pi$ from the radiation loop giving a
$1/\Lambda_\chi^2$.  ($\Lambda_\chi \sim 1 \,{\rm GeV}$ is the chiral symmetry
breaking scale.) Combining all factors, we find that this graph scales like
$Q_r^3/(M^3 \Lambda_\chi^2)$. This graph is suppressed relative to the leading
order amplitude, $A^{(-1)}$, by a factor of $Q_r^4/(M^2 \Lambda_\chi^2) =
m_\pi^2/\Lambda_\chi^2$.  Note that $C_0$ bubbles are summed on external
nucleon lines as well as in the interior of the radiation loop, and each graph in
the sum has the same size.  It is straightforward to verify that all graphs in 
Fig.~\ref{Qr3} scale the same way. For external bubble sums we can simply use the
vertex $i A^{(-1)}$ where $A^{(-1)}$ is the leading order S-wave amplitude,
\begin{eqnarray}\label{LO}
  A^{(-1)} = -{4\pi \over M} {1 \over \gamma+ip}  \,,
\end{eqnarray}
and the pole $\gamma = 4\pi/MC_0(\mu_R) + \mu_R \sim 1/a$. 
Graphs with two radiation pions are suppressed by at least 
$Q_r^8/(M^4 \Lambda_\chi^4) = m_\pi^4/\Lambda_\chi^4$ and will not be
considered.

The first graphs we consider are those in Fig.~\ref{Qr3}a,b. These graphs
have contributions from potential and radiation pions, and it may not
be obvious that a clean separation occurs. Here the energy integrals will be
evaluated without any approximations, after which the graphs split into
radiation and potential parts.  The graph in Fig.~\ref{Qr3}a gives:
\begin{eqnarray}\label{oneloop} 
iA^{(-1)} {g_A^2 \over 2 f^2} \int {d^d q \over
  (2 \pi)^d} {i \over \frac{E}2+q_0 -{({\vec q}-{\vec p})^2 \over 2M} +i
  \epsilon}\: {i \over \frac{E}2-q_0 -{({\vec q}-{\vec p})^2 \over 2M} +i
  \epsilon}\: {i\: {\vec q\,}^2 \over q_0^2 - {\vec q}\,^2 - m_\pi^2 + i \epsilon} \,. \nn\\ 
\end{eqnarray} 
(Throughout this chapter we will include a factor of $(\mu/2)^{4-d}$ in the loop
measures.) Performing the $q_0$ integral gives a term from the residue of the
nucleon pole and a term from the pion pole, 
\begin{eqnarray}
{ -iA^{(-1)} {g_A^2 \over 2 f^2} \int {d^n q \over (2 \pi)^n} 
{M \over {({\vec q}-{\vec p})^2 }-M E}\: {{\vec q}\,^2 \over {\vec q}\,^2 + 
m_\pi^2 -[\frac{E}2-{({\vec q}-{\vec p})^2 \over 2M}]^2 }\: }
\phantom{ccccccccccccccccccccc}\label{bme}
\\  -iA^{(-1)} {g_A^2 \over 4 f^2} \int {d^n q \over (2 \pi)^n} {{\vec q}\,^2
\over \sqrt{{\vec q}\,^2 + m_\pi^2}}\: {1 \over {E\over 2}+\sqrt{{\vec q}\,^2 
+ m_\pi^2}-{({\vec q}-{\vec p})^2 \over 2M} }\: 
{1 \over {E\over 2}-\sqrt{{\vec q}\,^2 + m_\pi^2}-{({\vec q}-{\vec p})^2 
\over 2M}} \,, \nn\\[5pt]  \phantom{c} \label{bme2} 
\end{eqnarray} 
where $n=d-1$.  Eq.~(\ref{bme}) is the potential pion contribution. Expanding in
$[\frac{E}2-{({\vec q}-{\vec p})^2 \over 2M}]^2=[{ 2{\vec q}.{\vec p}-{\vec q}\,^2
\over 2M}]^2$ gives the result in Ref.  \cite{ksw1,ksw2}.  The subleading terms in
this expansion are suppressed by\footnote{  Note that $m_\pi^2/q^2\sim
m_\pi/M$, but we have kept the $m_\pi^2$ term in the potential pion propagator in
Eq.~(\ref{bme}).  We could consider expanding in $m_\pi/q$ using the asymptotic
expansion techniques discussed in section B, but for $p\sim m_\pi$ these terms
would have to be resummed.  Unlike the soft and radiation contributions, there is
no issue of double counting for potential pions, so for simplicity we will simply
keep the $m_\pi^2$ in the propagator.} $m_\pi^2/M^2$.  Eq.~(\ref{bme2}) is the
radiation pion contribution.  With $|\vec q|<M$, we may take $({\vec q}-{\vec
p})^2/M\to p^2/M$ in the last two propagators, which is the same approximation
that is made by performing the multipole expansion.  Finally, we use the equations
of motion to set $E - p^2/M = 0$. It is important to note that we have not neglected
$E$ relative to $|{\vec q}|$.  For $n=3-2\epsilon$, Eq.~(\ref{bme2}) becomes 
\begin{eqnarray}\label{gr1} 
a) =\, i A^{(-1)} {g_A^2 \over 4 f^2} \int {d^n q \over (2 \pi)^n}{{\vec q\,}^2 \over 
  ({\vec q}\,^2 +  m_\pi^2)^{3/2}} = 
  -3 i A^{(-1)} { g_A^2 m_\pi^2\over (4 \pi f)^2} \bigg[ {1 \over \epsilon} + 
  {1\over 3} - {\rm ln}\Big({m_\pi^2 \over {\overline{\mu}}^2}\Big) \bigg] \,, 
\end{eqnarray} 
where $\overline{\mu}^2=\mu^2 \pi e^{-\gamma_E}$.  Note that this integral is 
finite in three dimensions ($n=2$).

The next graph we consider is shown in Fig.~\ref{Qr3}b.  We have chosen to
route loop momenta so that $q$ runs through the pion and $\pm k$ and
$\pm(k+q)$ run through the nucleon lines. The momentum $k$ is potential, while
$q$ can be potential or radiation. Doing the $k_0$ contour integral and
combining the two terms gives:
\begin{eqnarray}
&& -2\, [A^{(-1)}]^2\, {g_A^2 \over 2 f^2} \int {d^{n} k \over (2
 \pi)^{n}} \int {d^d q \over (2 \pi)^d} {{\vec q}\,^2\ \over q_0^2 -
 {\vec q}\,^2 - m_\pi^2 + i \epsilon}\: {1 \over E - {{\vec k}^2
 \over M}+ i\epsilon}\: {1 \over E - {({\vec k} + {\vec q})^2 \over
 M}+i \epsilon} \nn \\ &&\qquad\qquad\qquad\qquad\qquad \times {
 E-{({\vec k}+{\vec q})^2 \over 2M}-{{\vec k}\,^2\over 2M} \over
 [E-{({\vec k}+{\vec q})^2 \over 2M}-{{\vec k}\,^2\over 2M}-q_0+ i
 \epsilon] [E-{({\vec k}+{\vec q})^2 \over 2M}-{{\vec k}\,^2\over
 2M}+q_0+ i \epsilon]} \nn \,. \\
\end{eqnarray}
Doing the $q_0$ integral gives two terms, but the radiation and
potential contributions are still mixed.  Combining these gives
\begin{eqnarray}
&& i\, [A^{(-1)}]^2\, {g_A^2 \over 2 f^2} \int {d^{n} k \: d^nq
\over (2 \pi)^{2n}} {{\vec q}\,^2\ \over \sqrt{{\vec q}\,^2
 + m_\pi^2} }\: {1 \over E - {{\vec k}^2 \over M}}\: {1 \over E -
 {({\vec k} + {\vec q})^2 \over M}}\: \nn\\ 
  &&\qquad\qquad\qquad\qquad\quad\times { 1 \over E-{({\vec
k}+{\vec q})^2 \over 2M}-{{\vec k}\,^2\over 2M}-\sqrt{{\vec q}\,^2
 + m_\pi^2} }   \,, 
\end{eqnarray}
which can be split into potential and radiation parts
\begin{eqnarray}\label{twoloop}
i [A^{(-1)}]^2\, {g_A^2 \over 2f^2} \int {d^n k \over (2 \pi)^n}
\int {d^n q \over (2 \pi)^n} {{\vec q}\,^2 \over \sqrt{ {\vec
q}\,^2 + m_\pi^2 }}\: {1 \over E - {{\vec k}^2 \over M}}\: {1
\over \sqrt{ {\vec q}\,^2 + m_\pi^2}- {(2{\vec k}\cdot{\vec q} +
{\vec q}\,^2)\over 2 M}  }
\\ \times\Bigg[ {-1 \over E - {({\vec k}+{\vec q})^2 \over M}}
+{1 \over E - {{\vec k}^2 \over M} - {(2{\vec k}\cdot{\vec q} +
 {\vec q}\,^2)\over 2 M}-\sqrt{{\vec q}\,^2 + m_\pi^2}} \Bigg] \nn \,.
\end{eqnarray}
The first term in Eq.~(\ref{twoloop}) is the two-loop potential pion graph
evaluated in Ref.\cite{ksw2}. The factors of $(2{\vec k}\cdot{\vec q} + {\vec
q}\,^2)/(2M)$ appearing in the denominators can be dropped because the loop
integral is dominated by $k,q \ll M$ and therefore $(2{\vec k} \cdot {\vec q} + {\vec
q}\,^2)/(2M) \ll \sqrt{{\vec q}\,^2 + m_\pi^2}$.  For the second term, which is the
radiation pion contribution, this is equivalent to the multipole expansion. 
Momenta $k\sim \sqrt{M m_\pi}$ and $q\sim m_\pi$  dominate the integrals in the
second term.  In Ref.~\cite{beneke}, the potential and radiation parts of the graph
in Fig.~\ref{Qr3}b were evaluated in the limit $m_\pi=0$, and shown to correctly
make up the corresponding part of the fully relativistic calculation.  The
calculation in Ref.~\cite{beneke} agrees with Eq.~(\ref{twoloop}) for $m_\pi=0$. 
Note that the radiation part would not agree if we assumed $k\sim m_\pi$ and
used static nucleon propagators in the radiation loop\footnote{
Furthermore, if static nucleon propagators are used one obtains a linear 
divergence requiring a non-analytic counterterm $\propto m_\pi$ \cite{ngpc}.}.  
For $n=3-2\epsilon$ the radiation part of Eq.~(\ref{twoloop}) is
\begin{eqnarray}\label{gr2}
b) &=&\, i [A^{(-1)}]^2\, {g_A^2 \over 2f^2} \int {d^n k \over (2
\pi)^n} \int {d^n q \over (2 \pi)^n} {{\vec q}\,^2 \over {\vec
q}\,^2 + m_\pi^2 }\: {1 \over E - {{\vec k}\,^2 \over M}}\: {1
\over E - {{\vec k}\,^2 \over M} -\sqrt{{\vec q}\,^2 + m_\pi^2}} \nn
\\ &=&[A^{(-1)}]^2\, {g_A^2 M m_\pi^2 \over (4 \pi f)^2}\, \Bigg\{
{3\,p \over 4 \pi}\, \bigg[ {1 \over \epsilon} + {7\over 3} -
2\,{\rm ln}\, 2 - {\rm ln}\Big({m_\pi^2 \over {\overline{\mu}}^2}
\Big) - {\rm ln}\Big({-p^2 \over {\overline{\mu}}^2}\Big) \bigg] \nn \\
&& +
{i\sqrt{M m_\pi} \over 4\sqrt{\pi}}\,I_1 \Big({E\over m_\pi}\Big)
\Bigg\} \,, 
\end{eqnarray}
where
\begin{eqnarray}\label{I1}
I_1(x) &=& {3 \over 2} {\Gamma(-{5\over 4}) \over \Gamma({5 \over
  4})} \, {}_3 F_2 \bigg( \{-{5\over 4},-{1\over 4},{1\over 4}\}, \{
  {1\over 2},{5\over 4}\},x^2 \bigg)  \nn \\ && + {x \Gamma({1 \over 4})\over
  \Gamma({7 \over 4})}  \, {}_3 F_2 \bigg( \{-{3\over 4},{1\over
  4},{3\over 4}\}, \{ {3\over 2},{7\over 4}\} ,x^2 \bigg)\,. 
\end{eqnarray}
For $n=2$ the loop integral in Eq.~(\ref{gr2}) is finite, except for the $I_1$ term
which has a $p^2/(n-2)$ pole.  In PDS this pole would effect the running of
$C_2(\mu_R)$, but as we will see, contributions proportional to $I_1$ will cancel
between graphs.  Since $A^{(-1)}\sim 1/(M Q_r)$ the results in Eqs.~(\ref{gr1},
\ref{gr2}) are order $Q_r^3/(M^3 \Lambda_\chi^2)$ as expected. At one-loop the
$1/\epsilon$ pole in Eq.~(\ref{gr1}) is cancelled by a counterterm
\begin{eqnarray}\label{cntr}
   \delta^{{\rm uv},1a}D_2=-3\, C_0^{\rm finite}\,{g_A^2 \over (4
   \pi f)^2} \bigg[ {1 \over \epsilon}-\gamma_E +\ln{(\pi)} \bigg]  \,.
\end{eqnarray}
For higher loops the $1/\epsilon$ poles in Eq.~(\ref{gr1}), Eq.~(\ref{gr2}), and
$\delta^{{\rm uv},1a}D_2$ dressed with $C_0$ bubbles cancel. Note that the
$O(\epsilon)$ piece of the bubbles give a finite contribution,
\begin{eqnarray}
  -C_0(\mu_R) \left( {M p \over 4 \pi} \right) \left\{ 1 +\epsilon
  \left[ 2-2\,{\rm ln}\,2 - {\rm ln} \left( {-p^2-i\epsilon \over
  {\overline{\mu}}^2}\right) \right] \right\} \,.
\end{eqnarray}
The result of combining Figs.~\ref{Qr3}a,b and $\delta^{{\rm uv},1a}D_2$ dressed
by $C_0$ bubbles is:
\begin{eqnarray}\label{pt1}
a)+b)&=&\,  3i\,[A^{(-1)}]^2\,{ g_A^2 m_\pi^2 \over (4 \pi f)^2}
  \,{M\gamma\over 4\pi}\,\bigg[ {1\over 3} - {\rm ln}\Big({m_\pi^2
   \over \mu^2}\Big)\bigg]  \nn\\
 && + i\,[A^{(-1)}]^2\,{ g_A^2 M m_\pi^2
   \over (4 \pi f)^2} {\sqrt{M m_\pi} \over 4\sqrt{\pi}}\, I_1
   \Big({E\over m_\pi}\Big) \,.
\end{eqnarray}

Next we consider the graphs in Fig.~\ref{Qr3}c,d,e. The loop integrals in these
graphs vanish if the pion pole is not taken so there is no potential pion
contribution. As pointed out in Ref.~\cite{ksw2}, emission of the radiation pion in
these graphs changes the spin/isospin of the nucleon pair. Therefore, if the
external nucleons are in a spin-triplet(singlet) state, then the coefficients
appearing in the internal bubble sum are $C_0^{(^1S_0)}(\mu_R)\:
(C_0^{(^3S_1)}(\mu_R))$.  The notation $C_0$ ($C_0^{\,\prime}$) will be used for
vertices outside (inside) the radiation pion loop. We begin with Fig.~\ref{Qr3}c. 
The contribution from the graph with $m$ nucleon bubbles in the internal bubble
sum is
\begin{eqnarray}\label{gr3a}
 && 4\,{g_A^2 \over 2 f^2} \int {d^d q \over (2
  \pi)^d} {i \over q_0 - i \epsilon}\: {i \over q_0 - i \epsilon}\:
  {-i \,{\vec q}\,^2 \over q_0^2 -{\vec q}\,^2 - m_\pi^2 +
  i\epsilon} [- i C_0^{\,\prime}(\mu_R)]^{m+1} \nonumber \\ 
 &&\qquad \times
  \left[ \int {d^d k \over (2 \pi)^d} {i \over -k_0 +q_0
  +\frac{E}2-{({\vec k}-{\vec q})^2 \over 2M} +i \epsilon}\: {i
  \over k_0 +\frac{E}2-{{\vec k}^2 \over 2M} +i \epsilon} \right]^m ,
\end{eqnarray}
where we used the multipole expansion and then the equations of motion to
eliminate $E$ and $p$ from the first two propagators.  All nucleon propagators
have a $q_0$ pole above the real axis, while the pion propagator has one pole
above and one below.  Therefore, the $q_0$ contour is closed below. The $d^d
k$ integrals are also easily performed giving
\begin{eqnarray} \label{gphcm}
&& {-i}{ g_A^2 C_0^{\,\prime}(\mu_R) \over f^2}\left[ {-C_0^{\,\prime}(\mu_R) M
\Gamma({1-n/2}) \over (4 \pi)^{n/2}}\right]^m\! \int\! {d^n q \over (2
\pi)^n} {{\vec q}\,^2 \over ({\vec q}\,^2 +
m_\pi^2)^{3/2} } \\
 && \qquad\qquad\qquad\qquad\qquad\qquad \times 
  \left[(-p^2 +M \sqrt{{\vec q}\,^2 + m_\pi^2}\, )^{n/2-1}-\mu_R\right]^m \,. \nn 
\end{eqnarray}
Note  that the size of the loop momenta $k$ in the nucleon bubbles is $\sim \sqrt{M
m_\pi}$ even for $p < \sqrt{M m_\pi}$. The $\mu_R$ inside the brackets comes from
inclusion of the PDS or OS $\delta^n C_0(\mu_R)$ counterterm graphs for the
internal bubble sum. The integral will be dominated by ${\vec q} \sim m_\pi$ so the
graph will scale as
\begin{eqnarray}
{1\over \Lambda_\chi^2}\:{m_\pi^2 \over M
\mu_R}\bigg({\sqrt{M m_\pi} \over \mu_R }\bigg)^m \,.
\end{eqnarray}
Since $\mu_R\sim \sqrt{M m_\pi}$, all graphs in the sum are of order
$Q_r^3/(M^3 \Lambda_\chi^2)$.

For Figs.~\ref{Qr3}c,d,e the sum over bubbles should be done before the radiation
loop integral. The reason is that an arbitrary term in the bubble sum has a much
different dependence on the energy flowing through it than the sum itself. This can
be seen in the $\vec q$ dependence in Eq.~(\ref{gphcm}). If we integrate over $\vec
q$ then terms in the sum may diverge whereas the integral of the complete sum is
finite.  In fact, for $n=3$, Eq.~(\ref{gphcm}) has divergences of the form
$\Gamma(-1-m/4) {\rm F}(E^2/m_\pi^2)$ and $\Gamma(-1/2-m/4) E\, {\rm
F}(E^2/m_\pi^2)$ where ${\rm F}$ is a hypergeometric function. These divergences
are misleading because for momenta $>1/a$ we know that the correct form of the
leading order four point function falls off as $1/p$.  Recall from 
section~\ref{NNbeta} that for $p\sim \mu_R \gg 1/a$ the summation is infrared
physics that we have built into our theory.  For this reason the summation
is performed before introducing counterterms to subtract divergences. (This
approach is also taken in the analysis of three body interactions in
Ref.~\cite{Bedaque3,Bedaque4}). Summing over $m$, Eq.~(\ref{gphcm}) becomes:
\begin{eqnarray}\label{gr3}
c) &=&\, {- i}\, {g_A^2 \over f^2} {4 \pi \over M} \int {d^{n} q \over (2
   \pi)^{n}}{{\vec q}\,^2 \over ({\vec q}\,^2 + m_\pi^2)^{3/2}}\: { 1
   \over \gamma' - \left[-p^2 +M \sqrt{{\vec q}\,^2 +
   m_\pi^2}\right]^{n/2-1}}  \nn\\
&=&\,  {ig_A^2 \over \sqrt{\pi} f^2}  \Big({m_\pi \over M}\Big)^{3/2} 
    I_2 \Big( {E \over m_\pi} \Big)\,,
\end{eqnarray}
where $\gamma'=4\pi/MC_0^{\,\prime}(\mu_R)+\mu_R \sim 1/a$. As expected the
graph scales as $Q_r^{3}$. In the limit $n\to3$, $I_2$ is finite and given by
\begin{eqnarray}\label{I2}
I_2(x) \!\!\!\!&=&\!\!\!\!  {\Gamma(-{3\over 4}) \over \Gamma({3 \over 4})} \,
{}_3 F_2 \left( \{-{3\over 4},{1\over 4},{3\over 4}\}, \{ {1\over
2},{7\over 4}\},x^2 \right) - {3 x \over 2} {\Gamma({3 \over
4})\over \Gamma({9 \over 4})} \, {}_3 F_2 \left( \{-{1\over
4},{3\over 4},{5\over 4}\}, \{ {3\over 2},{9\over 4}\} ,x^2 \right
)  \nn \\ &&\quad + {\cal O}(\gamma'/\sqrt{Mm_\pi}) \,.
\end{eqnarray}
$I_2$ is manifestly $\mu_R$ independent and is also finite as
$n\to 2$. 

Next we consider the graph in Fig.~\ref{Qr3}d. Integrals are done
in the same manner as that of Fig.~\ref{Qr3}c. For
$n=3-2\epsilon$, Fig.~\ref{Qr3}d is
\begin{eqnarray}\label{gr4}
d) &=& {-4i\, A^{(-1)} } {g_A^2 \over 2 f^2}{\Gamma(n/2-1)\over  (4\pi)^{n/2-1}} 
 \int{d^{n} q \over (2 \pi)^{n}}{{\vec q}\,^2 \over ({\vec q}\,^2 + m_\pi^2)^{3/2}} \nn \\
 && \qquad\qquad\qquad\qquad\qquad \times {
  \left(-p^2 +M \sqrt{{\vec q}^2 + m_\pi^2}\,\right)^{n/2-1}-(-p^2)^{n/2-1} \over
  {\gamma'} - \left(-p^2 +M \sqrt{{\vec q}\,^2 + m_\pi^2}\,\right)^{n/2-1}}\nn \\
&=&-12 i\, A^{(-1)} { g_A^2 m_\pi^2 \over (4 \pi f)^2} \bigg[{1\over
  \epsilon} + {1\over 3} - {\rm ln}\Big({m_\pi^2 \over {\overline \mu}^2}\Big) \bigg]
  \nn \\
  && -4 {(p-i\gamma')\over \sqrt{\pi}} {M A^{(-1)} \over 4 \pi} 
  {g_A^2 \over 2 f^2}\Big({m_\pi \over M} \Big)^{3/2} I_2 \Big({E\over m_\pi} \Big)\,.
\end{eqnarray}
Fig.~\ref{Qr3}d is finite for $n=2$. The $1/\epsilon$ pole in Eq.~(\ref{gr4}) is
cancelled by a new tree level counterterm $i \delta^{\rm uv,1d}D_2 m_\pi^2$
where $\delta^{\rm uv,1d}D_2$ has the same form as Eq.~(\ref{cntr}) except with
a $-12$ instead of a $-3$.

Evaluation of Fig.~\ref{Qr3}e is also similar to Fig.~\ref{Qr3}c.
For $n=3-2\epsilon$ we find:
\begin{eqnarray} \label{gr5}
e)&=&{-2\,i }\,{(p-i\gamma')^2\over\sqrt{\pi}}  \bigg[{ M A^{(-1)} \over 4
  \pi}\bigg]^2 {g_A^2 \over 2 f^2} \bigg({m_\pi \over M}\bigg)^{3/2}
  I_2\Big({E \over m_\pi} \Big) \nn\\
 && + i[A^{(-1)}]^2\, {g_A^2M m_\pi^2
  \over (4 \pi f)^2}\, {\sqrt{M m_\pi} \over 2\sqrt{\pi}}\,
  I_1\left({E \over m_\pi} \right) \nn \\
&&+12\,[A^{(-1)}]^2\, {M p \over 4 \pi}\,{g_A^2 m_\pi^2
  \over (4\pi f)^2}\bigg[{1\over\epsilon}+ {7 \over 3}-2\ln{2} -
  {\rm ln}\Big({m_\pi^2 \over {\overline{\mu}}^2}\Big) - {\rm
  ln}\Big({-p^2 \over {\overline{\mu}}^2}\Big)\bigg] \nn \\
&& -6\, i\,[A^{(-1)}]^2\, {M \gamma' \over 4 \pi}\,{g_A^2 m_\pi^2
  \over (4\pi f)^2}\bigg[{1\over\epsilon}+ {1 \over 3}-
  {\rm ln}\Big({m_\pi^2 \over {\overline{\mu}}^2}\Big) \bigg] \,. 
\end{eqnarray}
This graph is finite for $n=2$ except for the $I_1$ term.  A $D_2$ counterterm 
cancels the divergence in the last line,
\begin{eqnarray}  \label{ctgr5}
  \delta^{{\rm uv},1e}D_2=6 \, (C_0^{\rm finite})^2\,{ M\gamma'\over 4\pi} 
  \,{g_A^2 \over (4 \pi f)^2} \bigg[ {1 \over \epsilon}-\gamma_E +\ln{(\pi)} \bigg] \,.
\end{eqnarray}
For two and higher loops the remaining
$1/\epsilon$ poles cancel between Eqs.~(\ref{gr4},\ref{gr5},\ref{ctgr5}) and
$\delta^{{\rm uv},1d}D_2$ dressed with $C_0$ bubbles, so no
new counterterms need to be introduced. The $O(\epsilon)$ piece of the bubbles
again give a finite contribution.  Combining Figs.~\ref{Qr3}c,d,e, and 
$\delta^{{\rm uv},1d}D_2$ and $\delta^{{\rm uv},1e}D_2$ dressed with $C_0$ 
bubbles gives
\begin{eqnarray}\label{ans3}
&& c)+d)+e) =2i\, [ A^{(-1)}]^2 {g_A^2 \over (4 \pi f)^2} \Bigg\{ 6 m_\pi^2\,
{M(\gamma - \gamma'/2)\over 4\pi\,} \bigg[ {1 \over 3} - {\rm ln}\Big({m_\pi^2
\over {\mu}^2}\Big) \bigg] \nn\\
 && \qquad + { M^{3/2}m_\pi^{5/2} \over 4 \sqrt{\pi}} I_1\Big( {E \over m_\pi} \Big) 
+{(\gamma - \gamma')^2\over 2\sqrt{\pi}}\,
{(Mm_\pi)^{3/2} \over M}\, I_2\Big({E \over  m_\pi} \Big) \Bigg\}  \,.
\end{eqnarray}

Fig.~\ref{Qr3}f shows a two loop graph with a nucleon self energy on an internal
line. It is important to also include graphs with the one-loop wavefunction and
mass renormalization counterterms, $\delta Z,\delta M$ inserted on the internal
nucleon line. We will use an on-shell renormalization scheme for defining these
counterterms, which ensures that the mass, $M$, appearing in all expressions is
the physical nucleon mass.  The counterterms are:
\begin{eqnarray}
\delta M &=& {3 g_A^2 m_\pi^3 \over 16 \pi f^2}\ , \qquad\qquad \delta Z =
{9\over 2}{g_A^2 m_\pi^2 \over (4 \pi f)^2} \left({1 \over
\epsilon} +{1\over 3} - {\rm ln}\left({m_\pi^2 \over
{\overline{\mu}}^2}\right) \right) \,.
\end{eqnarray}
The result from the graphs in Fig.\ref{Qr3}f is then
\begin{eqnarray}\label{gph6}
 f) = -3i [A^{(-1)}]^2 { g_A^2 \over (4 \pi f)^2} {M^{3/2}m_\pi^{5/2}
\over 4\sqrt{\pi}} I_1 \left({E \over m_\pi} \right) \,.
\end{eqnarray}
When Eq.~(\ref{gph6}) is added to Eqs.~(\ref{pt1},\ref{ans3}) the terms
proportional to $I_1$ cancel.  

To implement PDS we must consider the value of
the graphs in Fig.~\ref{Qr3}f using Minimal Subtraction with $n=2$. For
$n=2+\epsilon$ we have $\delta M= 3g_A^2 m_\pi^2\mu/(16\pi f^2\,\epsilon)$ and
$\delta Z=0$, which makes the sum of graphs in Fig.~\ref{Qr3}f finite except for the 
$I_1$ term.  Finally, renormalization of the bare nucleon fields in the Lagrangian, 
$N^{\rm bare}=\sqrt{Z} N$, $Z=1+\delta Z$, induces a four-nucleon term
\begin{eqnarray} \label{dL}
  \delta {\cal L} = - {C_0^{(s),{\rm finite}}}\,(2\delta Z)\, ( N^T P^{(s)}_i
N)^\dagger ( N^T P^{(s)}_i N)\,.
\end{eqnarray}
Since $\delta Z\sim Q_r^4\sim Q^2$ this term is treated
perturbatively.  A tree level counterterm
\begin{eqnarray}
 \delta^{\rm uv,0}D_2 = 9\, C_0^{\rm finite} {g_A^2 \over (4\pi
 f)^2} \Big[ \frac1\epsilon -\gamma_E +\ln(\pi) \Big]
\end{eqnarray}
is introduced to cancel the $1/\epsilon$ pole.  Dressing the
operator in Eq.~(\ref{dL}) with $C_0$ bubbles gives
\begin{eqnarray}  \label{wfnren}
  -9i \:[A^{(-1)}]^2\,{M\gamma \over 4\pi }\,{g_A^2 m_\pi^2 \over (4\pi
 f)^2}\, \bigg[\frac13 + \ln\Big({\mu^2 \over m_\pi^2}\Big)
 \bigg]\,.
\end{eqnarray}
Again, for $n=2$ we have $\delta Z=0$ so no new PDS counterterms were added. 
Note that if we had instead used bare nucleon fields then there would be no
correction of the form in Eq.~(\ref{dL}).  However, Eq.~(\ref{gph6}) would be
modified because the last graph in Fig.~\ref{Qr3}f is no longer present.  When
this is combined with the  contribution from the LSZ formula the sum of
Eq.~(\ref{gph6}) and Eq.~(\ref{wfnren}) is reproduced.

For PDS, the sum of graphs in Figs.\ref{Qr3}a-f are finite for $n=2$ so no new finite
subtractions were introduced. For $n=3$, counterterms are introduced to
renormalize the terms with $\ln(\mu^2)$ in
Eqs.~(\ref{pt1},\ref{ans3},\ref{wfnren}) (in PDS $\mu=\mu_R$). In OS only terms
analytic in $m_\pi^2$ are subtracted \cite{ms1} (including $m_\pi^2
\ln(\mu^2)$).  We find $D_2(\mu_R)\to D_2(\mu_R)+\Delta D_2(\mu_R)$, with
\begin{eqnarray} \label{DD2}
  \Delta D_2(\mu_R)=  6\, C_0(\mu_R)^2\, {g_A^2 \over (4\pi f)^2}\,
    {M (\gamma-\gamma')\over 4\pi}\, \bigg[-\frac13 +\kappa+ \ln\Big({\mu_R^2 
    \over  \mu_0^2} \Big) \bigg]\,.
\end{eqnarray}
Here $\kappa=1/3$ in PDS and $\kappa=0$ in OS, and $\mu_0$ is an unknown
scale.  Note that the logarithm in Eq.~(\ref{DD2}) gives a contribution to the
beta function for $D_2(\mu_R)$ of the form
\begin{eqnarray}
  \beta_{D_2}^{(rad)} =
 { 3 g_A^2 \over  4\pi^2 f^2} \,  {M (\gamma-\gamma')\over 4\pi} \, C_0(\mu_R)^2\,.
\end{eqnarray}
This disagrees with the beta function of Ref.~\cite{ksw2}, because in that paper
the beta function was calculated including only the one-loop graphs.

Adding the contributions in Eqs.~(\ref{pt1},\ref{ans3},\ref{gph6},\ref{wfnren})
gives the total radiation pion contribution to the amplitude at order $Q_r^3$:
\begin{eqnarray} \label{fans}
i A^{rad} &=&6i\, [ A^{(-1)}]^2 {g_A^2 m_\pi^2 \over (4 \pi f)^2}\, 
  {M (\gamma-\gamma') \over 4\pi}\, \bigg[   \kappa + 
  {\rm ln}\Big({ {\mu_R}^2\over m_\pi^2 }\Big) \bigg]  
 -i \, [ A^{(-1)}]^2\: {\Delta D_2(\mu_R)\, m_\pi^2 \over C_0(\mu_R)^2} \nn \\ 
&&+i\, [ A^{(-1)}]^2 \bigg[{M(\gamma -\gamma')\over 4\pi}\bigg]^2 
  {g_A^2 \over \sqrt{\pi} f^2} \Big({m_\pi \over M}\Big)^{3/2} I_2 
  \Big( {E \over m_\pi} \Big)\,.
\end{eqnarray}
The first term here has the same dependence on the external momentum as an
insertion of the $D_2$ operator dressed by $C_0$ bubbles. Its $\mu_R$
dependence is cancelled by $\mu_R$ dependence in $\Delta D_2(\mu_R)$. Note
that due to cancellations between graphs, this term is actually suppressed by a
factor of $\gamma/Q_r$ relative to what one expects from the power counting. 
The second term in Eq.~(\ref{fans}) has a nontrivial dependence on $E$ and is
suppressed by an even smaller factor of $\gamma^2/Q_r^2$.  These cancellations
were not anticipated by the power counting.  In the next chapter it will be shown
that the dependence of Eq.~(\ref{fans}) on $(\gamma-\gamma')$ follows from
the fact that the nucleon theory obeys Wigner's SU(4) symmetry in limit that
$\gamma,\gamma' \to 0$. Therefore terms at order $Q_r^4$ will likely give
the leading contribution of radiation pions to NN scattering.

If we now consider momenta $p\sim m_\pi \sim Q\ll Q_r$, we should fix $\mu_R$
at the threshold, $\mu_R=\sqrt{M m_\pi}$, and expand in $E/m_\pi$ giving
$I_2(E/m_\pi)=-3.94 +{\cal O}(E/m_\pi)$.  Therefore, the dominant effect of the
graphs that occur at order $Q_r^3$ is indistinguishable from a shift in
$D_2(\mu_R)$. Integrating out the radiation pions amounts to absorbing their
effects into the effective $D_2$ in the low energy theory.  The
result in Eq.~(\ref{fans}) is suppressed relative to the NLO contributions in
Ref.~\cite{ksw2} by a factor of roughly $2(\gamma-\gamma')/M\sim 1/10$.  Since
this is smaller than the expansion parameter, $Q/\Lambda\sim 1/3$, it can be
neglected at NNLO.  

It is useful to consider more closely how the size of a calculation at $p\sim Q_r$
changes when we take $p\sim m_\pi$.   At $p\sim Q_r$ the power counting
gave contributions of size
\begin{eqnarray}
   {Q_r^3 \over M^3 (4\pi f)^2 } \Big( {p \over Q_r} \Big)^k  =  {1 \over (4\pi f)^2} 
      \Big({m_\pi \over M}\Big)^{3/2} \, \Big( {p \over Q_r} \Big)^k \,,
\end{eqnarray}
where $k$ is an integer.  Taking $p\sim m_\pi$ gives $p/Qr=\sqrt{M/m_\pi}$, so
terms with $k \le -1$ will grow in size.  It is the external bubble sums which are
responsible for the factors of momentum in the denominator since
$A^{(-1)}\sim1/p$.  The graphs in Fig.~\ref{Qr3} have either zero, one, or two
external bubble sums.  Thus, there can be at most two factors of p in the
denominator, and the biggest possible enhancement is by $M/m_\pi$.  Therefore,
the $Q_r^3$ graphs are expected to have terms of order $m_\pi^{1/2}$, $m_\pi$, 
and $m_\pi^{3/2}$ which agrees with what was found above.  However, since the
terms in Eqs.~(\ref{pt1}) ,(\ref{ans3}), and (\ref{gph6}) that are proportional to 
$I_1$ cancel, there is no term in the answer proportional to $m_\pi^{1/2}$.  There is
no obvious reason why this cancellation had to occur.  It is interesting to note that
the $I_1$ terms scale as $1/\sqrt{M}$, which is a higher power of $M$ than the
leading order amplitude.  

Recall, that in evaluating the graphs in Fig.~\ref{Qr3} a mulitpole expansion was
used, which in this case is an expansion in $v=\sqrt{m_\pi/M}$.  If the first
correction in the multipole series were to multiple a nonzero term of order
$m_\pi^{1/2}$ then this would give an order $m_\pi$ contribution.  However, we
have checked that for all the graphs in Fig.~\ref{Qr3} the first correction in the
multipole  expansion gives a vanishing contribution.  This occurs because these
terms involve loop integrals with numerators that are odd in the loop momentum.

\begin{figure}[!t]
  \centerline{\epsfysize=3.0truecm \epsfbox{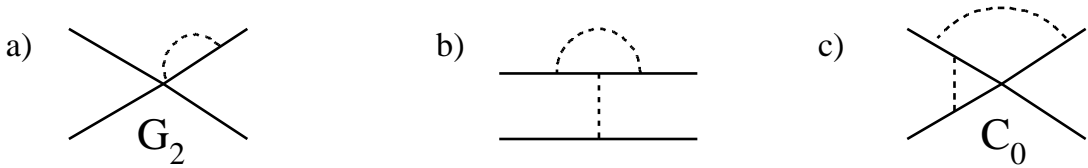}  }
 {\caption{Examples of order $Q_r^4$ radiation pion graphs for NN scattering.  
} \label{ho} }
\end{figure}
At order $Q_r^4$ graphs such as those in Fig.~\ref{ho} will contribute to NN 
scattering.  The graph in Fig.~\ref{ho}a includes an insertion of the operator
\begin{eqnarray}
{\cal L} = i\, G_2 \, [ N^T P^{(s)}_i
N ]^\dagger\, [ N^T P^{(s)}_i  \sigma_j (\xi\partial_j\xi^\dagger -
\xi^\dagger\partial_j\xi) N] +h.c. \,.
\end{eqnarray}
(Note that because of the hermitian conjugate this operator is the same for
$s=^1\!\!\!S_0$ and $s=^3\!\!\!S_1$.) This graph will be dressed with $C_0$ 
bubbles inside and outside the radiation pion loop.  The renormalization group
equation for $G_2$ gives $G_2(\mu_R) \sim 1/(M\mu_R^2) \sim 1/(M Q_r^2)$.  When
this is combined with the remaining factors of $Q_r$,  Fig.~\ref{ho}a is of order
$Q_r^4/(M^4 \Lambda_\chi^2)$ and is therefore suppressed by $Q_r/M$ relative to
a graph in Fig.~\ref{Qr3}.  Power counting the graphs in Fig.~\ref{ho}b,c gives
$Q_r^4/(M^3\Lambda_{\rm NN} \Lambda_\chi^2)$, yielding a factor of
$Q_r/\Lambda_{\rm NN}$ relative to a graph in Fig.~\ref{Qr3}.  This provides an
example of how graphs with potential pions seem to restrict the range of the
effective field theory to $ \Lambda_{\rm NN}\sim 300\,{\rm MeV}$.  The $300\,{\rm
MeV}$ scale applies only to a subset of graphs and may change once all graphs at
this order are included.

A complete NNLO calculation of the NN phase shifts at $p\sim m_\pi$ requires a
calculation of terms of order $m_\pi$.  For graphs with one radiation pion at order
$Q_r^4$, an enhancement by $1/m_\pi$ could give an order $m_\pi$ contribution. 
Therefore, it would be interesting to learn if the same cancellation that occurred at
order $Q_r^3$ continues at higher orders.  Although terms proportional to
$[A^{(-1)}]^3$ will appear at this order, they are factorizable. Therefore, the same
cancellation that occured for the order $Q_r^3$ graphs will occur in these terms.

\section{Soft pions}

In this section soft pion contributions will be discussed.  We will see that there are
graphs with non-vanishing soft contributions that should be included for $p
\gtrsim m_\pi$.  In soft loops, two scales appear: $m_\pi$ and $p=M v$.  It will be
shown below that it is necessary to take $p\sim Q_r$ when power counting graphs
with soft loops in order to avoid double counting.  In other words $v$ should have
the same value as in the radiation pion calculation. A soft loop has energy and
momentum $q_0\sim q\sim Q_r$, so $d^4 q\sim Q_r^4$.  The mass of the soft pion
is smaller than its momentum, and is treated perturbatively.  Nucleon propagators
in a soft loop are static (like in heavy quark effective theory, see Eq.~(\ref{LH}))
since the loop energy is greater than the nucleon's kinetic energy
\cite{gries1,gries2}.  Therefore, these propagators count as $1/Q_r$.  This power
counting is identical to that proposed in Ref.~\cite{ksw2} except powers of $Q_r$
are counted rather than $Q$.
 
Unlike potential pions, both soft and radiation pion pieces come from taking the
pole in a pion propagator.  Therefore, care must be taken not to double count when
adding these contributions.  This is accomplished by taking $p\sim Q_r$ when
evaluating both soft and radiation pion graphs.  This ensures that the soft and
radiation modes have different momenta ($\sim Q_r$ and $\sim m_\pi$
respectively).  Integrals involving the scales $Q_r$ and $m_\pi$ can be separated
using the method of asymptotic expansions and dimensional regularization
\cite{beneke,Gorishnii,Pivovarov,Smirnov,Czarnecki,Smirnov2}.  Consider splitting a loop integral into two regimes by
introducing a momentum factorization scale $L$ such that $m_\pi < L < Q_r$.  After
the pion pole is taken in an energy integral over $q_0$, the remaining integral is of
the form 
\begin{eqnarray} \label{split}
  \int d^n q \ = \ \int_0^L d^n q \mbox{ (radiation)} + 
	\int_L^{\infty} d^nq \mbox{ (soft)} \,,
\end{eqnarray}
which is obviously independent of $L$.  In Eq.~(\ref{split}) the power counting
dictates that expansions in $m_\pi^2/Q_r^2$ should be made so that each integral
becomes a sum of integrals involving only one scale ($m_\pi$ for radiation and
$Q_r$ for soft).  In dimensional regularization power divergences vanish, while
logarithmic divergences show up as $1/\epsilon$ poles. Therefore, after expanding
we can take $L\to \infty$ in the radiation integral and $L\to 0$ in the soft integral. 
Taking the $L\to \infty$ and $L\to 0$ limits may introduce ultraviolet divergences
for the radiation integral and infrared divergences for the soft integral.   When the
radiation and soft contributions are added any superfluous $1/\epsilon$ poles will
cancel.  This will be illustrated with an explicit example below.  The asymptotic
expansion procedure has been rigorously proven for Feynman graphs with large
external Euclidean momenta and large masses
\cite{chetyrkin,chetyrkin2,Chetyrkin3,smirnov3,smirnov4}.  It has also been shown
to work for the non-relativistic threshold expansion of one and two-loop graphs
\cite{beneke}.

Notice that it is crucial to expand the soft pion propagator in powers of
$m_\pi^2/Q_r^2$, because otherwise the radiation pion contribution may be double
counted.  As an example, consider the graph in Fig.~\ref{Qr3}a.  Taking $p\sim
Q_r$ implies $M v^2\sim m_\pi$.  For the radiation pion contribution $q_0
\sim q \sim M v^2 \sim m_\pi$, so we keep the $m_\pi^2$ in the denominator of
Eq.~(\ref{gr1}).  When computing the soft contribution, we assume $q_0\sim q\sim
Q_r \gg m_\pi$, and must expand the denominator in powers of $m_\pi^2/Q_r^2$. 
The $\vec q$ integration is now scaleless so the soft contribution to Fig.~\ref{Qr3}a
vanishes in dimensional regularization.  If we did not expand in $m_\pi/Q_r$ when
evaluating the soft contribution, we would have double counted the radiation
contribution.  The same argument can be applied to all the diagrams in
Fig.~\ref{Qr3}.  In each case the soft contribution vanishes. 

\begin{figure}[!t]
  \centerline{\epsfysize=3.0truecm \epsfbox{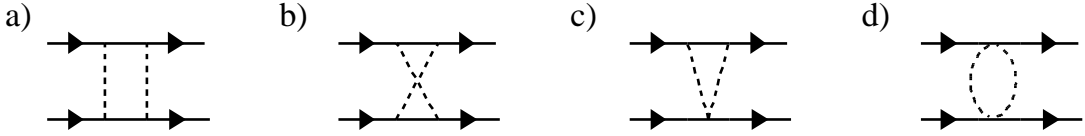}  }
 {\caption{Examples of one-loop graphs which have soft pion contributions. 
Graphs a)-d) also have a radiation pion contribution, while in addition
graph a) has a potential pion contribution. } \label{fig_soft} }
\end{figure}
Examples of graphs which have a non-vanishing soft contribution are shown in
Fig.~\ref{fig_soft}.  These diagrams were calculated in Ref.~\cite{kaiser} (although
the S-wave channels were not analyzed there).  In the KSW power counting they
must be dressed on the outside with $C_0$ bubbles.  (If Fig.~\ref{fig_soft}a or
\ref{fig_soft}b are dressed on the inside with $C_0$ bubbles then the soft
contribution vanishes.)  To see how these graphs obtain contributions from the
soft and radiation regimes consider Fig.~\ref{fig_soft}a.  Unlike the massless case
\cite{beneke}, this graph has a radiation like contribution.  In the $^1S_0$ channel 
the loop integral for Fig.~\ref{fig_soft}a is
\begin{eqnarray}  \label{s1}
&& \bigg({-i g_A^2 \over 2 f^2}\bigg)^2 \int {d^d q \over
  (2 \pi)^d} {i \over \frac{E}2+q_0 -{({\vec q}+{\vec p})^2 \over 2M} +i
  \epsilon}\ {i \over \frac{E}2-q_0 -{({\vec q}+{\vec p})^2 \over 2M} +i
  \epsilon}\ { {\vec q\:}^2 \over q_0^2 - {\vec q}\:^2 - m_\pi^2 + i \epsilon}\nn\\ 
&& \qquad\qquad\qquad \times { ({\vec q-\vec t\,})^2 \over q_0^2 - 
   ({\vec q-\vec t\,})^2 - m_\pi^2 +  i \epsilon} \,,
\end{eqnarray}
where $\vec t = {\vec p\,}' -\vec p$ and $\pm \vec p$ and $\pm \vec p\,'$ are the
incoming and outgoing nucleon momenta.  Unlike the graphs in Fig.~\ref{Qr3}, we
are forced to route an external momentum, $\vec t$, through a pion propagator. 
Taking a nucleon pole in Eq.~(\ref{s1}) gives the potential pion contribution
proportional to $M p$.  Taking the contribution from the pion poles gives soft and
radiation contributions.  Our power counting tells us that the leading order soft
contribution will be $\sim Q_r^2$, while the leading order radiation contribution will
be $\sim Q_r^4/M^2$.  In Eq.~(\ref{s1}) the factors of $E/2-(\vec q+\vec p)^2/(2M)$ 
can be dropped.  In the soft regime the factors of $E/2-(\vec q+\vec p)^2/(2M)$ are
order $Q_r^2/M$, and are dropped relative to $q_0 \sim Q_r$ leaving static
nucleon propagators.  In the radiation regime $E/2-(\vec q+\vec p)^2/(2M)\to 0$ 
after using the multipole expansion and equations of motion.  This leaves
\begin{eqnarray}  \label{s2}
 && \frac{i}2 \bigg({g_A^2 \over 2 f^2}\bigg)^2 \int {d^n q \over
  (2 \pi)^n}\: {{\vec q}\:^2 (\vec q-\vec t\,)^2 \over 
 {\vec q}\,^2 - (\vec q -\vec t\,)^2}\: \Bigg\{ {1 \over 
 [{\vec q}\:^2 + m_\pi^2]^{3/2} } -{1\over [ ({\vec q -\vec t\,})^2 + m_\pi^2]^{3/2} }
  \Bigg\} \nn\\[5pt]
 &&=  \frac{i}2 \bigg({g_A^2 \over 2 f^2}\bigg)^2 \int {d^n q \over
  (2 \pi)^n}\: {{\vec q}\:^2 (\vec q-\vec t\,)^2 \over [{\vec q}\:^2 + m_\pi^2]^{3/2} }
  \bigg\{ {1\over {\vec q}\:^2 - (\vec q -\vec t\,)^2 + i\epsilon } + 
  {1\over {\vec q}\:^2 - (\vec q -\vec t\,)^2 - i\epsilon } \bigg\} \,, \nn\\
\end{eqnarray}
where $n=d-1$.  The singularity at ${\vec q}\:^2 = (\vec q -\vec t\,)^2$ is cancelled
only in the sum of terms in the first line of Eq.~(\ref{s2}).  These terms can be
calculated separately by introducing an $i\epsilon$ in this denominator
\cite{beneke}, giving an average over $\pm i\epsilon$ as indicated\footnote{
The second line of Eq.~(\ref{s2}) is more easily split into soft and radiation
contributions.  If we had used the integrand on the first line we would also have to
consider $(\vec q-\vec t)\,^2\sim m_\pi^2$.}.  The factor of $(\vec q-\vec t\,)^2$ in
the numerator can be removed by partial fractioning.  For the soft contribution the
scale of the loop momentum is set by the external momentum, $q_0 \sim |\vec q|
\sim |\vec t| \sim Q_r$. Expanding Eq.~(\ref{s2}) in $m_\pi^2/{\vec q}\:^2$ gives
\begin{eqnarray}  \label{boxs}
  && i \bigg({g_A^2 \over 2 f^2}\bigg)^2 \int {d^n q \over (2 \pi)^n}\: { |\vec q| \over
  (2 \vec q \cdot \vec t - {\vec t\,}^2 \pm i\epsilon) } \sum_{m=0}^\infty  
  {\Gamma(-1/2)  \over \Gamma(-1/2-m) \Gamma(m+1)} \bigg( \frac{m_\pi^2}
  {{\vec q\:}^2} \bigg)^m  \nn\\[5pt] 
  && = {-i \over 192 \pi^2} \bigg({g_A^2 \over 2 f^2}\bigg)^2 \Bigg\{ \bigg[ 
  \frac{1}{\epsilon} + \ln\bigg(\frac{\bar \mu^2}{{t}^2}\bigg) \bigg] \Big(t^2 
  -18\, m_\pi^2\Big) \nn\\ &&- \bigg[ 
  \frac{1}{\epsilon} + \ln\bigg(\frac{\bar \mu^2}{{t}^2}\bigg) \bigg] \bigg( 90\, 
  {m_\pi^4 \over t^2} - 140\, {m_\pi^6 \over t^4} +\ldots \bigg) 
   + \frac83\, t^2 -36\, m_\pi^2 +280\, 
      \frac{m_\pi^6}{t^4} + \ldots \Bigg\} \,, \nn\\
\end{eqnarray}
where we have kept the first few terms in the expansion.  The soft
contribution starts at order $Q_r^2$ as expected.  The first $1/\epsilon$ pole in
Eq.~(\ref{boxs}) is an ultraviolet divergence, while the second is an infrared
divergence.  For the radiation contribution $q_0 \sim |\vec q| \sim m_\pi \ll |\vec t|$. 
Expanding in $(2\vec t \cdot \vec q) / {\vec t}\:^2$ gives
\begin{eqnarray}  \label{boxr}
  && -i \bigg({g_A^2 \over 2 f^2}\bigg)^2 \int {d^n q \over (2 \pi)^n}\: { 1 \over
  [{\vec q}\:^2 + m_\pi^2]^{3/2}  } \Bigg[\ {\vec q\:}^2 + { {\vec q\:}^4 \over 
  {\vec t\:}^2 } \sum_{m=0}^\infty \bigg( \frac{2 \vec q \cdot \vec t } 
  {{\vec t\:}^2\pm i\epsilon} \bigg)^m \ \Bigg] \nn\\[5pt] 
  && = {-i \over 192 \pi^2} \bigg({g_A^2 \over 2 f^2}\bigg)^2 \Bigg\{ -72\, m_\pi^2
  \bigg[ \frac{1}{\epsilon} + \ln\bigg(\frac{\bar \mu^2}{m_\pi^2}\bigg) \bigg] \nn\\
  &&+ \bigg[  \frac{1}{\epsilon} + \ln\bigg(\frac{\bar \mu^2}{m_\pi^2}\bigg) \bigg] 
  \bigg( 90\, {m_\pi^4 \over t^2} - 140\, {m_\pi^6 \over t^4} +\ldots \bigg)
   -24\, m_\pi^2 +39\, \frac{m_\pi^4}{t^2} -
  \frac{482}{3} \, \frac{m_\pi^6}{t^4} + \ldots \Bigg\} \,. \nn\\
\end{eqnarray}
The radiation contribution starts out as order $Q_r^4/M^2$ as expected.  Note that
only powers of $m_\pi=Q_r^2/M$ appear.  The $1/\epsilon$ poles in Eq.~(\ref{boxr})
are ultraviolet divergences.  When the soft and radiation contributions are added
the infrared poles in Eq.~(\ref{boxs}) cancel a subset of the ultraviolet poles in
Eq.~(\ref{boxr}).  Adding Eq.~(\ref{boxs}) and Eq.~(\ref{boxr}) we find
\begin{eqnarray}  \label{boxsr}
  &&  {-i \over 192 \pi^2} \bigg({g_A^2 \over 2 f^2}\bigg)^2 \Bigg\{ t^2 \bigg[ 
  \frac{1}{\epsilon} + \ln\bigg(\frac{\bar \mu^2}{{t}^2}\bigg) \bigg]+\frac83\,t^2 
  -m_\pi^2  \bigg[ \frac{90}{\epsilon} +18 \ln\bigg(\frac{\bar \mu^2}{{t}^2}\bigg)
  +72 \ln\bigg(\frac{\bar \mu^2}{{m_\pi}^2}\bigg) \bigg]  \nn\\
 &&  -60\, m_\pi^2+\ln\Big(\frac{t^2}{m_\pi^2}\Big) \bigg( 90\, 
  {m_\pi^4 \over t^2} - 140\, {m_\pi^6 \over t^4} +\ldots \bigg) +39\, 
      \frac{m_\pi^4}{t^2} +\frac{358}{3} \,  \frac{m_\pi^6}{t^4} + \ldots \Bigg\} \,,\nn\\
\end{eqnarray}
where the remaining ultraviolet $1/ \epsilon$ poles are cancelled by counterterms
for $C_2$ and $D_2$.

If we are interested in making predictions for $p\sim m_\pi$, then powers of 
$m_\pi^2/t^2$ must be summed up.  Summing the series in Eq.~(\ref{boxsr}) gives
\begin{eqnarray}  \label{boxsum}
  &3a)=&  {-i \over 192 \pi^2} \bigg({g_A^2 \over 2 f^2}\bigg)^2 \Bigg\{ \Big(t^2
  -90\,m_\pi^2\Big) \bigg[ \frac{1}{\epsilon} + \ln\bigg(\frac{\bar \mu^2}{{m_\pi}^2}
  \bigg) \bigg]+\frac83\,t^2 -58 m_\pi^2 \nn\\  
&& \qquad\qquad - {(128 m_\pi^4 +16 m_\pi^2 t^2 -t^4)\over
   t \sqrt{t^2 + 4m_\pi^2} } \ln\bigg({\sqrt{t^2 + 4m_\pi^2}-t \over 
   \sqrt{t^2 + 4m_\pi^2} +t }\bigg) \Bigg\} \,. 
\end{eqnarray}
Since the coefficients in the series in Eqs.~(\ref{boxs},\ref{boxr}) diverge for $\vec
p\:' = \vec p$, Eq.~(\ref{boxsum}) should be used when integrating over $dt$ to
obtain the $^1S_0$ partial wave amplitude (even for $p\sim Q_r$). 
Eq.~(\ref{boxsum}) agrees with the result of evaluating Eq.~(\ref{s2}) exactly. 
Although the asymptotic expansion is not necessary for evaluating
Fig.~\ref{fig_soft}a, it allows us to identify the radiation and soft contributions to
this graph and verify that the power counting for each regime works.  We also see
that adding soft and radiation pions reproduces the correct answer without double
counting.  Recall that power counting with $p\sim Q_r$ was necessary to avoid
double counting for graphs like those in Fig.~\ref{Qr3}.  

For $p\sim Q_r$ the diagrams in Fig.~\ref{fig_soft} are order $Q_r^2/(f^2
\Lambda_\chi^2)$, and are larger than the order $Q_r^3$ graphs with a single
radiation pion in Fig.~\ref{Qr3}.  Decreasing $p$ to $p\sim m_\pi$ the graphs in
Fig.~\ref{fig_soft} give contributions of the form
\begin{eqnarray}
  {  {\vec t\,}^2 \over f^2 \, (4\pi f)^2  } \
	F\Big({{\vec t\:}^2 / m_\pi^2}\Big) \,,
\end{eqnarray}
where $F$ is a function.  For $p\sim m_\pi$ the graphs in Fig.~\ref{fig_soft} are
order $m_\pi^2$ which is smaller than the graphs in Fig.~\ref{Qr3} which
include order $m_\pi^{1/2}$, $m_\pi$, and $m_\pi^{3/2}$ terms.  It is interesting
to note that the relative importance of these graphs changes with $p$.  The graphs
in Fig.~\ref{fig_soft} dressed by $C_0$ bubbles give a contribution that is the same
size as a four nucleon operator with 6 derivatives, $C_6(\mu_R) p^6$, and are
N$^3$LO in the KSW power counting.   Note that dressing these graphs with 
$C_0$ bubbles does not result in enhanced terms for $p\sim m_\pi$, unlike the
graphs in the previous section.

It would be nice to see the expansion used in evaluating the radiation contribution
to Fig.~\ref{fig_soft}a implemented at the level of the Lagrangian.  It is not clear to
us how to do this at the present time.  In the radiation regime, the pion whose pole
is taken can be thought of as a radiation pion.  However, the other propagator
gives factors of $1/t^2$, ${\vec q}\cdot {\vec t} /t^4$, etc., which look more like
insertions of non-local operators than the propagator of a field.  Also, since in
general $\vec p \ne \vec p\:'$, the couplings for this second propagator change the
nucleon momenta and therefore do not involve a multipole expansion.  Finally, the
result in Eq.~(\ref{boxsum}) does not have an expansion in $E/m_\pi$.  So unlike
the radiation pion contribution computed in section A, this contribution cannot be
integrated out for $p<\sqrt{M m_\pi}$.   For these reasons, the use of the term
radiation for this contribution differs somewhat from the usage in section A.

To summarize, a power counting in factors of $Q_r=\sqrt{M m_\pi}$ has been
introduced which is appropriate for graphs with radiation pions.  The order $Q_r^3$
radiation contributions to NN scattering were computed and found to be
suppressed by inverse powers of the scattering length.  Soft pion contributions
also have a power counting in $Q_r$.  For $p\sim Q_r$ they are $\sim Q_r^2$, but
are higher order than the radiation contributions for $p\sim m_\pi$.  Higher order
corrections are suppressed by factors of $Q_r/\Lambda$, and whether or not this
expansion is convergent is an open question.  If the range of the two-nucleon
effective field theory with perturbative pions is really $300\,{\rm MeV}$, then
contributions from radiation pions induce an incalculable error of order
$m_\pi^2/\Lambda_\chi^2$ to the NN scattering amplitude in this theory.  The
validity of the $Q_r/\Lambda$ expansion can be tested by examining processes at
$p\sim 300\,{\rm MeV}$ such as those with external pions.


\chapter{Wigner's SU(4) Symmetry from Effective Field Theory}

In this chapter it is shown that in the limit where the $NN$ $^1S_0$ and $^3S_1$
scattering lengths, $a^{(^1S_0)}$ and $a^{(^3S_1)}$, go to infinity, the leading
terms in the effective field theory for strong $NN$ interactions are invariant under
Wigner's SU(4) spin-isospin symmetry.  This explains why the leading effects of
radiation pions on the S-wave $NN$ scattering amplitudes, calculated in the
chapter 6, vanish as $a^{(^1S_0)}$ and $a^{(^3S_1)}$ go to infinity.  The
implications of Wigner symmetry for $NN \to NN\, \mbox{axion}$ and $\gamma\,
d\to n\, p$ are also considered.

Wigner's SU(4) spin-isospin transformations are
\cite{Wigner,Wigner2} 
\begin{eqnarray} 
  { \delta N = i  \alpha_{\mu\nu}\, \sigma^\mu \, \tau^\nu \, N \,, 
  \qquad N=\Bigg(\begin{array}{c} p \\ n \end{array} \Bigg)  \, .  } \label{trnfm} 
\end{eqnarray} 
In Eq.~(\ref{trnfm}), $\sigma^\mu = (1,\vec\sigma)$, $\tau^\nu=(1,\vec\tau)$, and
$\alpha_{\mu\nu}$ are infinitesimal group parameters (the notation here is 
that greek indices run over $\{0,1,2,3\}$, while roman indices run over $\{1,2,3\}$). 
The $\sigma$ matrices act on the spin degrees of freedom, and the $\tau$
matrices act on the isospin degrees of freedom.  (Actually the transformations in
Eq.~(\ref{trnfm}) correspond to the group SU(4)$\times$U(1). The additional U(1)
is baryon number and corresponds to the $\alpha_{00}$ term.)

Consider first the effective field theory for nucleon strong interactions with the
pion degrees of freedom integrated out.  The Lagrange density is composed of
nucleon fields and has the form ${\cal L}={\cal L}_1+{\cal L}_2 + \ldots$, where
${\cal L}_n$ denotes the n-body terms.  We have
\begin{eqnarray} \label{LN1}
  {\cal L}_1 &=&  N^\dagger \Big[ i\partial_t + {\overrightarrow\nabla^2/(2M) }
    \Big] N  + \ldots \,, \nn \\
  {\cal L}_2 &=& -\sum_{s}\,{C_0^{(s)} ( N^T P^{(s)}_i N)^\dagger 
  ( N^T P^{(s)}_i N)}   + \ldots \,, \nn \\
  {\cal L}_3 &=& -{D_0\over 6}\, (N^\dagger N)^3 + \ldots  \,, \nn\\
  {\cal L}_4 &=& E_0\, (N^\dagger N)^4 + \ldots \,,
\end{eqnarray}
where the ellipses denote higher derivative terms and $P^{(s)}_i$ was defined in
Eq.~(\ref{Sproj}) for $s=^1\!\!S_0,^3\!S_1$.  Without derivatives, higher body 
contact interactions vanish because of Fermi statistics.  Fermi statistics also
implies that there is only one four body term, which is invariant under Wigner
symmetry.  Furthermore, there is only one term in ${\cal L}_3$ which is also
invariant\footnote{P.F. Bedaque, H.W. Hammer, and U. van Kolck, private
communication and Ref.~\cite{bhkW}.}.  To count operators it is useful to classify 
them by their transformation properties under SU(4). The three nucleon and
anti-nucleon fields must be combined in an antisymmetric way, so the three $N$'s
($N^\dagger$'s) combine to a $\bar 4$ (4) of SU(4).  Combining the $4$ and $\bar
4$ gives $1 \oplus 15$. However, the $15$ does not contain a singlet under the
spin and isospin SU(2) subgroups so there is only one three body operator with no
derivatives.  A group theory argument can also be used to show that there are only
two two-body operators.  Combining two $N$'s in an antisymmetric manner gives
a $\bar 6$, while combining two $N^\dagger$'s gives a $6$.  Then $6 \times \bar 6 =
1 \oplus 15 \oplus 20$, and only the $1$ and $20$ contain singlets under the spin
and isospin subgroups.  The Lagrange density ${\cal L}_2$ can also be written in a
different operator basis:
\begin{eqnarray}  \label{LN2}
  {\cal L}_2 &=& -\frac12 \Big[ {C_0^{S}}  ( N^\dagger  N )^2 
    + {C_0^{T}}( N^\dagger \vec \sigma N )^2 \Big] + \ldots \,,
\end{eqnarray}
where $C_0^{(^1S_0)}=C_0^S-3 C_0^T$ and $C_0^{(^3S_1)}=C_0^S+C_0^T$.  In
this basis it is the $C_0^T$ term that breaks the SU(4) symmetry (as well as
some of the higher derivative terms).  The Lagrangian in Eq.~\ref{LN1} will be 
SU(4) symmetric at leading order if $C_0^T=0$ in the two-body sector and if 
higher derivative operators are suppressed in the three and four body sectors.

In the $\overline {\rm MS}$ scheme, 
neglecting two derivative operators the $NN$ scattering amplitude is
\begin{eqnarray}
 {\cal A}^{(s)} &=& { -\bar{C_0}^{(s)} \over 1 + i \frac{Mp}{4\pi} \,\bar{C_0}^{(s)}}\,,
\end{eqnarray}
where
\begin{eqnarray}
  \bar{C_0}^{(s)} &=& \frac{4\pi a^{(s)}}{M} \,,
\end{eqnarray}
as was discussed section~\ref{NNbeta}.  The scattering lengths are very large:
$a^{(^1S_0)}=-23.714 \pm 0.013 \,{\rm fm}$ and $a^{(^3S_1)}=5.425 \pm 0.001\, {\rm
fm}$ \cite{burcham}, and even have the opposite sign.  Therefore, the value of
$\bar{C_0}^{(^1S_0)}$ and $\bar{C_0}^{(^3S_1)}$ are very different.  In this scheme
the Lagrangian in Eq.~(\ref{LN2}) does not have Wigner symmetry.  Nonetheless,
for $p\gg 1/a^{(s)}$ the amplitudes become, ${\cal A}^{(s)}=4\pi i/(Mp)$.  The
equality of the $^1S_0$ and $^3S_1$ amplitudes is consistent with expectations
based on Wigner symmetry.  The $p$-dependence is consistent with expectations
based on scale invariance\footnote{The scale transformations appropriate for the
non-relativistic theory are $x\to \lambda x$, $t\to \lambda^2 t$, and $N \to
\lambda^{-3/2} N$.}, since the cross section $\sigma^{(s)}=4\pi/p^2$.

Recall from section~\ref{NNbeta} that in the $\overline {\rm MS}$ scheme each term
in the bubble sum in Fig.~\ref{C0bub} gives a factor of $a\,p$, so for $p >
1/a$ the series diverges. For these momenta minimal subtraction is not
natural from a power counting perspective.  In natural schemes like PDS and OS,
each term in the bubble sum is of the same order as the sum.  It is in these
``natural'' schemes that the fixed point structure of the theory and Wigner
spin-isospin symmetry are manifest in the Lagrangian.  In PDS or OS the
coefficients depend on the renormalization point $\mu_R$ and for 
$a^{(s)}\to \infty$ we have
\begin{eqnarray}
   C_0^{(s)}(\mu_R)=-\frac{4\pi}{M} \frac1{\mu_R -1/a^{(s)}} \ \to \ 
      -\frac{4\pi}{M\mu_R}\,,
\end{eqnarray}
which is the same in both channels.  In this limit
\begin{eqnarray}
C_0^T(\mu_R)=[C_0^{(^3S_1)}(\mu_R)-C_0^{(^1S_0)}(\mu_R)]/4=0 \,,
\end{eqnarray}
and
\begin{eqnarray}  \label{L3}
   {\cal L}_2 = -{2\pi \over M\mu_R} (N^\dagger N)^2 + \ldots \,.
\end{eqnarray}
The first term in Eq.~(\ref{L3}) is invariant under the Wigner spin-isospin
transformations in Eq.~(\ref{trnfm}).  The ellipses in Eq.~(\ref{L3}) denote terms
with derivatives which will not be invariant under Wigner symmetry even in
the limit $a^{(s)}\to \infty$.  However, these terms are corrections to the leading
order Lagrange density and their effects are suppressed by powers of
$p/\Lambda$ (where $\Lambda$ is a scale determined by the pion mass and
$\Lambda_{\rm QCD}$).  In the region $1/a^{(s)} \ll p \ll \Lambda$ Wigner
spin-isospin symmetry is a useful approximation and deviations from this
symmetry are suppressed by 
\begin{eqnarray}
  C_0^T(\mu_R) \propto (1/a^{(^1S_0)}-1/a^{(^3S_1)}) \,,
\end{eqnarray}
and by powers of $p/\Lambda$. The measured effective ranges are
$r_0^{(^1S_0)}\!=\! 2.73\pm 0.03\,{\rm fm}$ and $r_0^{(^3S_1)}\!=\! 1.749\pm
0.008\,{\rm fm}$ \cite{burcham}.  A rough estimate of the scale is $1/\Lambda \sim
[r_0^{(^1S_0)} - r_0^{(^3S_1)}]/2 = 0.49\,{\rm fm}$, or $\Lambda\sim 400\,{\rm
MeV}$. In PDS or OS, the limit $a^{(s)} \to \infty$ is clearly a fixed point of
$C_0^{(s)}(\mu_R)$ since $\mu_R\, {\partial}/{\partial\mu_R} \, [\mu_R\,
C_0^{(s)}(\mu_R)] =0$.  Also, scale invariance is manifest since $\mu_R\to
\mu_R/\lambda$ under scale transformations.  

In the KSW power counting potential pion exchange is order $Q^0$ and is treated 
perturbatively.  Thus, pion exchange is higher order than the iterated $C_0$ 
bubbles and the theory still has Wigner symmetry at leading order.

Wigner symmetry is useful in the two-body sector even though $a^{(^1S_0)}$ and
$a^{(^3S_1)}$ are very different.  This is because for $1/a^{(s)} \ll p \ll \Lambda$
corrections to the symmetry limit go as $(1/a^{(^1S_0)}-1/a^{(^3S_1)})$ rather
than $(a^{(^1S_0)}-a^{(^3S_1)})$.  This is similar to the heavy quark spin-flavor
symmetry of QCD \cite{HQS1,HQS2}, which occurs in the $m_Q\to \infty$ limit. 
Heavy quark symmetry is a useful approximation for charm and bottom quarks
even though $m_b/m_c\simeq 3$.

As an application of the symmetry consider $NN\to NN\,\mbox{axion}$, which is
relevant for astrophysical bounds on the axion coupling \cite{axionguy}.  The axion
is essentially massless.  If the axion has momentum $\vec k$, and the initial
nucleons have momenta $\vec p$ and $-\vec p$ then the final state nucleons have
momenta $\vec q-\vec k/2$ and $-\vec q-\vec k/2$.  Energy conservation implies
that $p^2/M=q^2/M+k^2/(4M)+k$ where $p=|\vec p|, q=|\vec q|$, and $k=|\vec k|$.  In
the kinematic region we consider $q,p \gg k$, and the axion momentum can be
neglected in comparison with the nucleon momenta. Again, this is just a multipole
expansion (see chapter 6). In this limit the terms in the Lagrange density
which couple the axion to nucleons take the form
\begin{eqnarray}
  {\cal L}_{int} &=& g_0  \Big(\nabla^j X^0\Big)\bigg|_{\vec x=0} \!\!\! N^\dagger 
  \sigma^j N +
   g_1  \Big(\nabla^j X^0 \Big)\bigg|_{\vec x=0}\!\!\!  N^\dagger \sigma^j\, \tau^3 N 
   \,,\nn\\
 & &
\end{eqnarray}

\begin{figure}[!t]
  \centerline{\epsfysize=5.8truecm \epsfbox{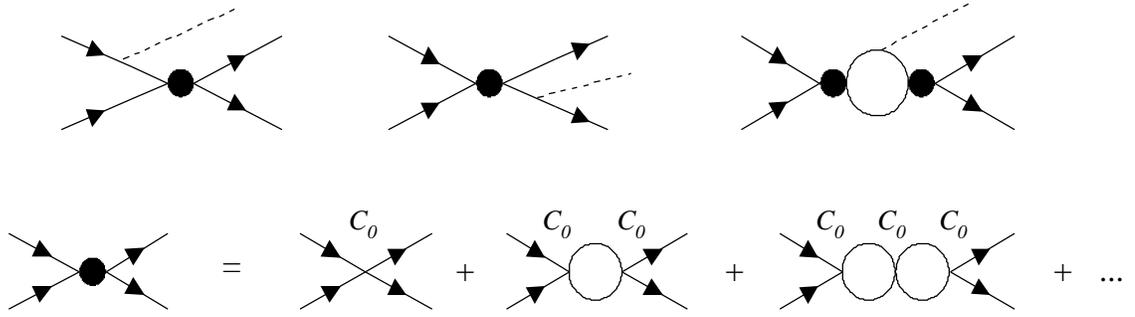}  }
 {
\caption{Graphs contributing to $NN \to NN\,\mbox{axion}$ at leading order.  
The solid lines denote nucleons and the dashed lines are axions.} \label{axion} }
\end{figure}  
\noindent where $X^0$ is the axion field and $g_0,g_1$ are the axion-nucleon
isosinglet and isovector coupling constants.  Associated with spin-isospin 
symmetry are the conserved charges
\begin{eqnarray}
   Q^{\mu\nu} = \int d^3x  N^\dagger \sigma^\mu \tau^\nu N \,,
\end{eqnarray}
and the axion terms in the action are proportional to these charges
\begin{eqnarray} \label{Saxion}
  S_{int} &=& g_0 \int\!\! dt \Big(\nabla^j X^0\Big)\bigg|_{\vec x=0}\!\! Q^{j0} +
    g_1 \int\!\! dt \Big(\nabla^j X^0\Big)\bigg|_{\vec x=0}\!\! Q^{j3} \,. \nn\\
 &&
\end{eqnarray}
The charge $Q^{j0}$ is the total spin of the nucleons which is conserved even
without taking the $a^{(s)}\to \infty$ limit, however $Q^{j3}$ is only conserved in
the $a^{(s)}\to \infty$ limit (and also in the limit $a^{(^1S_0)}\to a^{(^3S_1)}$). 
Since conserved charges are time independent, only a zero energy axion couples
in Eq.~(\ref{Saxion}), and these terms will not contribute to the scattering
amplitude.  We conclude that $NN(^1S_0)\to NN(^3S_1) X^0$ vanishes in the
limit $a^{(s)}\to \infty$ and that $NN(^3S_1)\to NN(^3S_1)X^0$ vanishes for all
scattering lengths\footnote{ $NN(^1S_0)\to NN(^1S_0)X^0$ vanishes due to
angular momentum conservation since the axion is emitted in a P-wave.}. 
Calculation of the Feynman diagrams in Fig.~\ref{axion} shows that the leading
order $^3S_1\to {^3S_1}$ scattering amplitude does indeed vanish, and the
$NN(^1S_0)\to NN(^3S_1) X^0$ amplitude is 
\begin{eqnarray}
  {\cal A} &=& g_1\, {4\pi \over M} \,  {\vec k \cdot \vec\epsilon\,^* \over k } 
  \bigg[ \frac1{a^{(^1S_0)}}-\frac1{a^{(^3S_1)}} \bigg] \bigg[{1 \over 
  1/{a^{(^1S_0)}}+i\,p}\bigg] \,\bigg[{1 \over 1/{a^{(^3S_1)}}+i\,q} \bigg] \,,
\end{eqnarray}
where $\vec \epsilon$ is the polarization of the final $^3S_1$ $NN$
state.  This is proportional to $(1/a^{(^1S_0)}-1/a^{(^3S_1)})$ and is consistent
with our expectations based on the Wigner symmetry.  The fact that the graphs
in Fig.~\ref{axion} vanish as $a^{(s)}\to\infty$ lends some support to statements
in the literature \cite{Turner} which claim that one-pion exchange is sufficiently 
accurate to describe the matrix element for $NN \to NN\,\mbox{axion}$ (at least 
for momenta $\gg 1/a$ such as in neutron stars).  This process has contributions 
from different partial waves, and in all but the S-wave a single perturbative pion
exchange is the leading order contribution in the KSW power counting.  For the
${}^1S_0 \to {}^3S_1$ transition the graphs in Fig.~\ref{axion} are small, so the first
sizeable S-wave contribution occurs at NLO (the same order as the other partial
waves).  It involves one-pion exchange and insertions of $C_2^{(s)}$ dressed by
$C_0$ bubbles.  Besides tree level pion exchange these NLO contributions to 
$NN\to NN\,\mbox{axion}$ have not been considered in the axion literature.  
Note that to properly treat the S-wave contribution in neutron stars Pauli 
blocking effects would have to be incorporated in the calculation of the loop 
graphs.

Coupling of photons to nucleons occurs by gauging the strong effective field
theory and by adding terms involving the field strengths $\vec E$ and $\vec B$. 
In the kinematic regime where the photon's momentum is small compared to the
nucleons' momentum the part of the action involving the field strengths is
\begin{eqnarray} \label{SEB}
  S_{int} = {e\over 2M} \int dt\, {B^j}\Big|_{\vec x=0} \bigg( \kappa_0 Q^{j0}+
     \kappa_1 Q^{j3}  \bigg) + \ldots \,,
\end{eqnarray}
where $\kappa_0$ and $\kappa_1$ are the isosinglet and isovector nucleon
magnetic moments in nuclear magnetons, and the ellipses denote subdominant
terms.  The term proportional to $\kappa_1$ in Eq.~(\ref{SEB}) gives the lowest
order contribution to the amplitude for $\gamma d \to n p(^1S_0)$.  The form of
the coupling above implies that like the axion case, this amplitude is proportional
to $(1/a^{(^1S_0)}-1/a^{(^3S_1)})$.

As our last example, we discuss the corrections to $NN$ scattering due to radiation
pions discussed in chapter 6.  As pointed out in chapter 6, one should perform a 
multipole expansion on the coupling of radiation pions to nucleons.  The first term 
in the multipole expansion is:
\begin{eqnarray}
  S_{int} = -{g_A \over \sqrt{2} f}  \int dt \,\Big(\nabla^i \pi^j\Big)\bigg|_{\vec x=0} 
    Q^{ij} \,,
\end{eqnarray}
where $g_A\simeq 1.25$ is the axial coupling and $f\simeq 131\,{\rm MeV}$ is the
pion decay constant.  Radiation pions also couple to a conserved charge of the
Wigner symmetry in the large scattering length limit.  (A multipole expansion is
not performed on the coupling to potential pions so they do not couple to a
conserved charge.) This implies that only a radiation pion with $k^0=0$ will
couple, which is incompatible with the condition $k^0 \sim \sqrt{k^2 + m_\pi^2}$,
so in the symmetry limit radiation pions do not contribute to the scattering
matrix element. In Eq.~(\ref{fans}) we saw by explicit computation that
graphs with one radiation pion and any number of $C_0^{(s)}$'s give a
contribution that is suppressed by at least one power of
$1/a^{(^3S_1)}-1/a^{(^1S_0)}$.  This suppression was the result of cancellations
between the Feynman diagrams shown in Fig.~\ref{Qr3}.  Wigner symmetry 
guarantees that the leading contribution of graphs with an arbitrary number of 
radiation pions are suppressed by inverse powers of the scattering lengths.

It has also been shown that Wigner symmetry is obtained in the large number of
colors limit of QCD \cite{LNC1,LNC2}.  The implications of Wigner symmetry in
nuclear physics were studied in Ref.~\cite{WN1,WN2,WN3,WN4,WN5,WN6,WN7}. So
far the applications in this chapter have been specific to the two-nucleon sector,
however Wigner symmetry is observed in some nuclei with many nucleons.  Recent
progress \cite{Bedaque1,Bedaque2,Bedaque3,Bedaque4} in the three body sector
suggests that the $(N^\dagger N)^3$ contact interaction is not subleading
compared with the effects of the first two body term in Eq.~(\ref{L3}). (This
conclusion is not uncontroversial, see Ref.~\cite{ggg}.)  If the higher body
operators with derivatives can be treated as perturbations, then the above
discussion shows that approximate Wigner symmetry in nuclear physics is a
consequence of the large $NN$ scattering lengths.


\chapter{Predictions for the $^3S_1-^3D_1$ Mixing Parameter, $\epsilon_1$}

This chapter briefly discusses results for the $NN\to NN$ $^3S_1-^3D_1$ mixing
parameter, $\epsilon_1$, at next-to-leading order (NLO) in the theory with pions. 
A number of observables have been computed at NLO in the KSW power counting.
These include, nucleon-nucleon phase shifts \cite{ksw1,ksw2,epel,rupak,Mnopi},
coulomb corrections to proton-proton scattering \cite{kong2,kong4},
proton-proton fusion \cite{kong1,kong3}, electromagnetic form factors for the
deuteron \cite{ksw3,Mnopi}, deuteron polarizabilities \cite{chen,Mnopi}, $np\to
d\gamma$ \cite{SSW,Mnopi,CRSsup}, Compton deuteron scattering
\cite{chen2,chen3}, and parity violating effects \cite{Kaplan1,SS}.  Typically errors
of order 30\%-40\% are found at LO and of order 10\% at NLO.  As mentioned in
chapter 5, this is consistent with an expansion parameter $Q/\Lambda\sim 1/3$. 
Since the expansion parameter is fairly large, calculations at
next-to-next-to-leading order (NNLO) are useful.  In the theory without pions,
calculations at this order can be carried out in a straightforward manner
\cite{Mnopi}.  With pions, the loop graphs become more difficult, but even two and
three loop graphs can be evaluated analytically\footnote{The basic reason that
loops in three dimensions are simpler, is that the integrals can be done in position
space where the Bessel functions reduce to exponentials.}.  To truly test the
convergence and range of the theory with pions, the above observables need to be
calculated to NNLO.  The phase shift in the $^1S_0$ channel has been calculated
at this order \cite{rupak} (and independently in Ref.~\cite{msconf}), while work in 
the $^3S_1$, $^3D_1-^3S_1$, and $^3D_1$ channels is near completion \cite{fms}.
 In this chapter a presentation of the prediction for $\epsilon_1$ at order $Q$ is
given\footnote{The material in this chapter appears in Ref.~\cite{fms1} with a more
detailed discussion}.

\begin{figure}[!t]
  \centerline{\epsfxsize=16.truecm \epsfysize=5.truecm\epsfbox{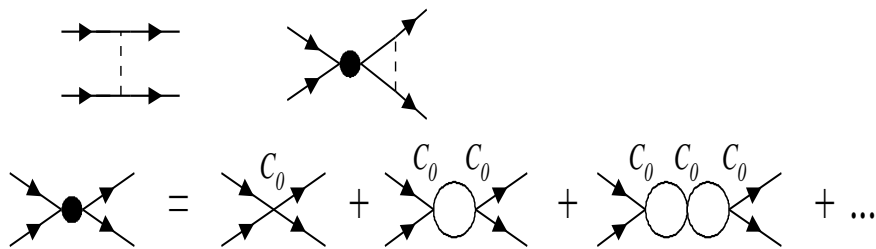}  }
\caption{The two order $Q^0$ diagrams that contribute to 
$\epsilon_1$ \cite{ksw2}.   The solid lines are nucleons and the dashed lines are 
potential pions.} \label{diag_lo}    
\end{figure}

\begin{figure}[!t]
  \centerline{\epsfxsize=18.truecm \epsfysize=7.truecm\epsfbox{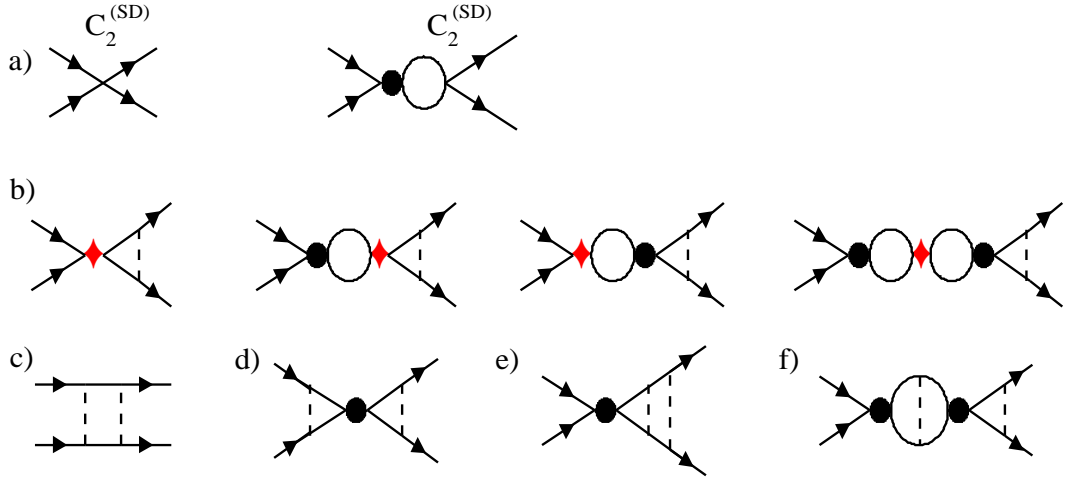}  }
\caption{Order $Q$ diagrams for $\epsilon_1$.  The filled circle is
defined in Fig.~\ref{diag_lo}, and the diamonds in c) denote insertions of the 
$^3S_1$ operators with $C_0^{(0)}$, $C_2$ or $D_2$ coefficients. }  \label{diag_Q}  
\end{figure}

At order $1/Q$ there is no contribution to $\epsilon_1$, which is consistent with
the fact that this angle is much smaller than the $^3S_1$ phase shift.  At order
$Q^0$ the graphs for $\epsilon_1$ include single potential pion exchange and pion
exchange dressed on one side by $C_0^{(^3S_1)}$ bubbles as shown in 
Fig.~\ref{diag_lo} \cite{ksw2}. 
This prediction does not involve any free parameters.  The order $Q$ calculation
includes a two derivative operator which causes transitions from $^3S_1$ to
$^3D_1$ with coefficient $C_2^{(SD)}$.  This operator is normalized so that 
on-shell in the center of mass frame the Feynman rule is
\begin{eqnarray}
  \begin{picture}(20,10)(1,1)
      \put(1,3){\line(1,1){10}} \put(1,3){\line(1,-1){10}} 
      \put(1,3){\line(-1,1){10}} \put(1,3){\line(-1,-1){10}}
      \put(-4,15){\mbox{\footnotesize $C_2^{(SD)}$}}
      \put(-24,0){\mbox{\scriptsize ${}^3S_1$}}  
      \put(15,0){\mbox{\scriptsize ${}^3D_1$}} 
  \end{picture} 
   \quad &=&  -i\,C_2^{(SD)}\, p^2 \,, 
\end{eqnarray}  
where $p$ is the center of mass momentum.  In this section the PDS 
renormalization scheme will be used.  It was shown in chapter 5, that in the PDS
scheme $C_2^{(SD)}(\mu_R)\sim 1/\mu_R$, so this operator enters at order
$Q$.  The Feynman diagrams that can contribute to $\epsilon_1$ at order $Q$
include:
\begin{eqnarray}
 \begin{array}{rl}
    i)\ &\mbox{ one $C_2^{(SD)}$ and any number of $C_0^{(^3S_1)}$'s}   \\
    ii)\  &\mbox{ one $C_2^{(^3S_1)}$, one potential pion and any number of 
	$C_0^{(^3S_1)}$'s}  \\
    iii)\  &\mbox{ one $D_2^{(^3S_1)}$, one potential pion and any number of 
	$C_0^{(^3S_1)}$'s}   \\
    iv)\  &\mbox{ two potential pions and any number of $C_0^{(^3S_1)}$'s}   \\
    v)\  &\mbox{ radiation pion corrections}  \,. \\
\end{array}
\end{eqnarray}
The graphs for i) through iv) are shown in Fig.~\ref{diag_Q}.
It can be shown that there are no radiation pion contributions to the mixing
parameter at this order.  

\begin{figure}[!t]
  \centerline{\epsfxsize=17.truecm \epsfysize=10.truecm\epsfbox{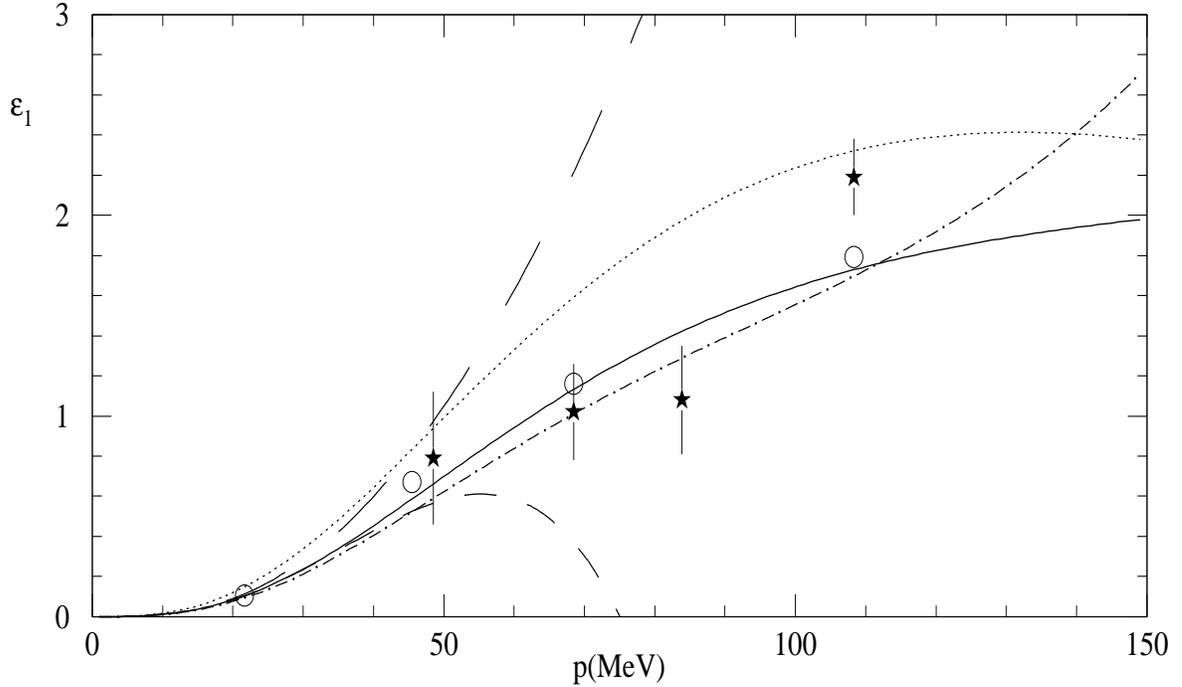}  }
{ \caption{Predictions for the $^3S_1-^3D_1$ mixing parameter $\epsilon_1$ in
degrees.  The solid line is the multi-energy Nijmegen partial wave analysis
\cite{NijPW}.  The long and short dashed lines are the LO and NLO predictions in
the theory without pions \cite{Mnopi}.  The dotted line is the LO prediction in the
theory with pions from Ref.~\cite{ksw2}.  The dash-dotted line is the NLO
prediction in the theory with pions.  The stars are data from Virginia Tech
\cite{vpiPWA} and the open circles are Nijmegen single energy data \cite{NijPW}
whose quoted errors are invisible on the scale shown.  }
\label{ep1a} }
\end{figure}

The value of $C_2^{(^3S_1)}$, $D_2^{(^3S_1)}$, $C_0^{(^3S_1)}$ and
$C_0^{p\,(^3S_1)}$ are fixed from the $^3S_1$ phase shift calculation at order
$Q^0$.   The only free parameter is $C_2^{(SD)}(\mu_R)$, which is varied to
give a reasonable fit.  The result of the $\epsilon_1$ calculation at order $Q$ in the
theory with pions is given by the dot-dashed line in Fig.~\ref{ep1a}.  The order
$Q^0$ result in the theory with pions \cite{ksw2} is shown by the dotted line.  The
stars in Fig.~\ref{ep1a} are data from Virginia Tech \cite{vpiPWA}.  The open circles
are the Nijmegen single energy fit to the data \cite{NijPW} whose quoted errors are
invisible on the scale shown.  The solid line is the Nijmegen multi-energy partial
wave analysis \cite{NijPW}.  The value of $C_2^{(SD)}(\mu_R)$ corresponding
to the fit in Fig.~\ref{ep1a} is not given since we have not specified what constants
are subtracted along with a $p^2/\epsilon$ pole that appears in one of the graphs
with two potential pions.

For comparison results have been shown for the theory without pions \cite{Mnopi}.
The long dashed line is the LO result and involves a one parameter fit.  The
smaller dashed line is the NLO result and involves fitting two free parameters. 
With fewer free parameters, the theory with pions does better than the theory
without pions for $p>60\,{\rm MeV}$.  In fact the theory without pions breaks down
around $m_\pi/2$, which is where we expect it to since this is where the pion cut
begins (see chapter 5).  It has been noted in the literature \cite{Cohen3} that many
observables do not test the power counting for perturbative pions.  As can be seen
from Fig.~\ref{ep1a}, the mixing parameter provides an example in which 
perturbative pions clearly give improved agreement with the data.

The dot-dashed line in Fig.~\ref{ep1a} improves over the order $Q^0$ result for
$p<140\,{\rm MeV}$.  For larger values of $p$, the order $Q$ prediction grows, while
the Nijmegen partial wave analysis has $\epsilon_1(p)\le 3^\circ$ for $p\le 300\,{\rm
MeV}$.  In Ref.~\cite{Ordonez3} the mixing angle was calculated using Weinberg's
power counting and solving with a potential.  In this approach graphs with
potential pions are summed up.  A direct comparison with this calculation is
difficult since graphs with $\Delta$'s were included, and more parameters were
varied in the fit.   When the potential method is used there is cutoff dependence in
the result which cannot be cancelled by cutoff dependence in the coefficients. 
This is because contributions that are formally higher order are included in the
answer.  Since the results are numerical these can not be thrown away.  For a
cutoff of order $m_\rho$, the prediction in Ref.~\cite{Ordonez3} also grows with
$p$.  In the potential approach the cutoff dependence gives a measure of the
uncertainty due to higher order corrections.  Varying the cutoff from $0.6\,m_\rho$
to $1.3\,m_\rho$ the prediction varies by $\sim 1.0^\circ$ at $p=150\,{\rm MeV}$. 
Therefore, at $150\,{\rm MeV}$ the theory with perturbative pions seems to be doing
no worse than a calculation where the pions are treated non-perturbatively.

\chapter{Conclusion}

In this thesis effective field theory techniques are used to describe the interaction
of heavy particles at low momentum in a model independent way.  The focus has
been on interactions with pions, described using chiral perturbation theory
techniques.  Chapter 2 gives an overview of the formalism for describing
interactions of one or two heavy particles.  

In chapter 3, the decays $D^* \to D\pi$ and $D^* \to D\gamma$ are studied using
heavy meson chiral perturbation theory.  With the recent measurement of ${\cal
B}(D^{*+} \to D^+ \gamma$), the $D^{*0}$, $D^{*+}$, and $D_s^*$ branching
fractions can be used to extract the $D^*D\pi$ and $D^*D\gamma$ couplings $g$
and $\beta$.  The $D^* \to D\gamma$ decays receive important corrections at order
$\sqrt{m_q}$ and, from the heavy quark magnetic moment, at order $1/m_c$.  Here
all the decay rates are computed to one-loop, to first order in $m_q$ and $1/m_c$,
including the effect of heavy meson mass splittings, and the counterterms at order
$m_q$.  A fit to the experimental data gives two possible solutions,
$g=0.27\,^{+.04}_{-.02} \,^{+.05}_{-.02}$, $\beta=0.85^{+.2}_{-.1} \,^{+.3}_{-.1}\,{\rm
GeV^{-1}}$ or $g=0.76\,^{+.03}_{-.03} \,^{+.2}_{-.1}$, \mbox{$\beta=4.90^{+.3}_{-.3}
\,^{+5.0}_{-.7}\,{\rm GeV^{-1}}$}. The first errors are experimental, while the second
are estimates of the uncertainty induced by the counterterms.  (The experimental
limit $\Gamma_{D^{*+}} < 0.13\,{\rm MeV}$ excludes the $g=0.76$ solution.)
Predictions for the $D^*$ and $B^*$ widths are given.  

In chapter 4, the prospects for determining $|V_{ub}|$ from exclusive $B$
semileptonic decay are discussed.  The double ratio of form factors
\begin{eqnarray}
   {f^{(B\to\rho)}/f^{(B\to K^*)} \over f^{(D\to\rho)}/f^{(D\to K^*)} }
\end{eqnarray}
is calculated using chiral perturbation theory.  Its deviation from unity due to
contributions that are non-analytic in the symmetry breaking parameters is very
small.  It is concluded that combining experimental data obtainable from
$B\to\rho\,\ell\,\bar\nu_\ell$, $B\to K^*\ell\,\bar\ell$ and $D\to\rho\,\bar\ell\,\nu_\ell$
can lead to a model independent determination of $|V_{ub}|$ with an uncertainty
from theory of about 10\%.

In chapter 5, an effective field theory description of nucleon nucleon interactions is
investigated.  A momentum subtraction scheme (OS) is introduced which obeys
the power counting of Kaplan, Savage, and Wise (KSW).  The KSW power counting
was developed for systems with large scattering lengths, $a$. Unlike the power
divergence subtraction scheme (PDS), coupling constants in this scheme obey the
KSW scaling for all $\mu_R > 1/a$.  This chapter explains in detail how the
renormalization in the OS and PDS schemes is implemented using local
counterterms. The main complication is the need to include an infinite number of
counterterms since the leading order result includes an infinite number of loop
graphs.  Fits to the NN scattering data are performed in the $^1S_0$ and $^3S_1$
channels.  An error analysis indicates that the range of the theory with perturbative
pions is consistent with $500\,{\rm MeV}$, so it can be concluded that there is no
obstruction to using perturbative pions for momenta $p>m_\pi$.  Some comments
are made on the low-energy theorems derived by Cohen and Hansen\cite{Cohen2}.

In chapter 6, radiative pion interactions are investigated.  For interactions involving
two or more nucleons it is useful to divide pions into three classes: potential,
radiation, and soft.  The momentum threshold for the production of radiation pions
is $Q_r = \sqrt{M_N m_\pi}$.  It is shown that radiation pions can be included
systematically with a power counting in $Q_r$. The leading order radiation pion
graphs which contribute to NN scattering are evaluated using the the PDS and OS
renormalization schemes and are found to give a small contribution which vanishes
as the singlet and triplet scattering lengths go to infinity.  The power counting for
soft pion contributions is also discussed.
  
Chapter 7 shows that in the limit where the $NN$ $^1S_0$ and $^3S_1$
scattering lengths, $a^{(^1S_0)}$ and $a^{(^3S_1)}$, go to infinity, the leading
terms in the effective field theory for strong $NN$ interactions are invariant under
Wigner's SU(4) spin-isospin symmetry.  This explains why the leading effects of
radiation pions on the S-wave $NN$ scattering amplitudes vanish as $a^{(^1S_0)}$
and $a^{(^3S_1)}$ go to infinity.  The implications of Wigner symmetry for $NN \to
NN\, \mbox{axion}$ and $\gamma\, d\to n\, p$ are also considered.

Finally, in chapter 8 results for the $^3S_1-^3D_1$ mixing parameter, $\epsilon_1$
are presented.  This observable provides an example where the theory with
perturbative pions gives better agreement with the data at $p\sim m_\pi$ with fewer
parameters than the theory without pions.

\bibliographystyle{h-physrev2}
\bibliography{pths}

\end{document}